\documentclass[a4paper, twoside, openright]{report}
\usepackage{a4wide}
\usepackage{amsmath}
\usepackage{amssymb}
\usepackage[hang, small, bf]{caption}
\usepackage{cite}
\usepackage{float}
\usepackage[dvips]{graphicx}
\usepackage{tocbibind}
\usepackage{fancyhdr}
\fancypagestyle{plain}{%
  \fancyhead{} 
}
\newcommand{\ud}{\mathrm{d}}
\newcommand{\dirac}{\partial\llap{$\diagup$\kern-2pt}}
\newcommand{\covariant}{D\llap{$\diagup$}}
\newcommand{\fett}[1]{\boldsymbol{#1}}
\newcommand{\fettu}[1]{\mathbf{#1}}
\newcommand{\quabla}{\square}
\newcommand{\diag}{\mathrm{diag}}
\newcommand{\sgn}{\mathrm{sgn}}
\newcommand{\Tr}{\mathrm{Tr}}
\newcommand{\D}{\mathcal{D}}
\newcommand{\e}{\mathrm{e}}
\newcommand{\Heaviside}{\theta}
\newcommand{\msun}{\mathrm{M}_\odot}
\newcommand{\Res}{\mathrm{Res}}
\newcommand{\be}{\begin{equation}}
\newcommand{\ee}{\end{equation}}
\newcommand{\bea}{\begin{eqnarray}}
\newcommand{\eea}{\end{eqnarray}}
\newcommand{\bsub}{\begin{subequations}}
\newcommand{\esub}{\end{subequations}}
\usepackage[american, german]{babel}
\begin{document}
\selectlanguage{german}
\begin{titlepage}
\begin{center}
\vspace*{0.8cm}
\begin{Huge}
\textbf{The Phase Diagram of Neutral Quark Matter} \\[2cm]
\end{Huge}
\begin{LARGE}
Dissertation \\
zur Erlangung des Doktorgrades \\
der Naturwissenschaften \\[2cm]
vorgelegt beim Fachbereich Physik \\
der Johann Wolfgang Goethe~-~Universit\"at \\
in Frankfurt am Main \\[2.5cm]
von \\
Stefan Bernhard R\"uster \\
aus Alzenau in Ufr. \\[2cm]
Frankfurt 2006 \\[4mm]
(D 30)
\end{LARGE}
\end{center}
\end{titlepage}
\vspace*{9cm}
\noindent
vom Fachbereich Physik der \\[4mm]
Johann Wolfgang Goethe~-~Universit\"at als Dissertation
angenommen. \\[10.1cm]
Dekan: Prof.\ Dr.\ A{\ss}mus \\[4mm]
Gutachter: Prof.\ Dr.\ Rischke und HD PD Dr.\ Schaffner-Bielich
\\[4mm]
Datum der Disputation: 14.\ Dezember 2006
\selectlanguage{american}
\chapter*{Abstract}
\addcontentsline{toc}{chapter}{Abstract}
In this thesis, I study the phase diagram of dense, locally
neutral three-flavor quark matter as a function of the strange
quark mass, the quark chemical potential, and the temperature,
employing a general nine-parameter ansatz for the gap matrix. At
zero temperature and small values of the strange quark mass, the
ground state of quark matter corresponds to the
color--flavor-locked (CFL) phase. At some critical value of the
strange quark mass, this is replaced by the recently proposed
gapless CFL (gCFL) phase. I also find several other phases, for
instance, a metallic CFL (mCFL) phase, a so-called uSC phase
where all colors of up quarks are paired, as well as the
standard two-flavor color-superconducting (2SC) phase and the
gapless 2SC (g2SC) phase.

I also study the phase diagram of dense, locally neutral
three-flavor quark matter within the framework of a
Nambu--Jona-Lasinio (NJL) model. In the analysis, dynamically
generated quark masses are taken into account self-consistently.
The phase diagram in the plane of temperature and quark chemical
potential is presented. The results for two qualitatively
different regimes, intermediate and strong diquark coupling
strength, are presented. It is shown that the role of gapless
phases diminishes with increasing diquark coupling strength.

In addition, I study the effect of neutrino trapping on the
phase diagram of dense, locally neutral three-flavor quark
matter within the same NJL model. The phase diagrams in the
plane of temperature and quark chemical potential, as well as in
the plane of temperature and lepton-number chemical potential
are presented. I show that neutrino trapping favors two-flavor
color superconductivity and disfavors the color--flavor-locked
phase at intermediate densities of matter. At the same time, the
location of the critical line separating the two-flavor
color-superconducting phase and the normal phase of quark matter
is little affected by the presence of neutrinos. The
implications of these results for the evolution of protoneutron
stars are briefly discussed.
\selectlanguage{american}
\chapter*{Acknowledgments}
\addcontentsline{toc}{chapter}{Acknowledgments}
I am very grateful to my advisor Prof.\ Dr.\ Dirk Rischke who
suggested the topic for my thesis. He introduced me to quantum
field theory and color superconductivity. I learnt a lot in his
lectures and in private communication. I thank him for his
suggestions and advices. I am very thankful to Prof.\ Dr.\ Igor
Shovkovy. I thank him for the excellent cooperation, the
discussions, suggestions, and advices. I am grateful to our
colleagues Verena Werth and PD Dr.\ Michael Buballa from the
Institut f\"ur Kernphysik at the Technische Universit\"at
Darmstadt for the teamwork. I thank Hossein Malekzadeh for the
cooperation concerning the spin-zero A-phase of
color-superconducting quark matter.

I am grateful to HD PD Dr.\ J\"urgen Schaffner-Bielich. I learnt
a lot in his lectures, seminars, and in our astro group
meetings. I also thank him and Matthias Hempel for the
cooperation and discussions concerning the outer crust of
nonaccreting cold neutron stars.

I am grateful to the computer trouble team for removing computer
problems. I am thankful for using the Center for Scientific
Computing (CSC) of the Johann Wolfgang Goethe~-~Universit\"at.

I am very grateful to my parents who supported me during the
whole time of my study.
\tableofcontents
\listoffigures
\listoftables
\selectlanguage{american}
\chapter{Introduction}
\label{Introduction}
The phase diagram of neutral quark matter was poorly understood
as I began with the research on this topic in 2003. The task of
my thesis was therefore to illuminate the phase structure of
neutral quark matter. A phase diagram is a two-dimensional
diagram with axes representing the temperature and the chemical
potential, the density, or other similar quantities. Therefore,
phase diagrams tell us in which state is a system for a given
temperature and a given chemical potential. Besides, they
contain the information at which temperatures and which chemical
potentials transitions to other phases occur. Such phase
transitions can be of first or second order, or simply
crossovers. This depends on the order parameter of the system.
If it changes discontinuously, then a first, otherwise a
second-order phase transition or a crossover appears.

In Sec.~\ref{The_phase_diagram_of_strongly_interacting_matter},
I show the status of knowledge of the phase diagram of strongly
interacting matter before I began with my research for this
thesis in 2003. As one can see, the phase diagram of neutral
quark matter was indeed poorly understood at that time.
In Sec.~\ref{Color_superconductivity}, I discuss the behavior
of sufficiently cold and dense quark matter, namely that quark
matter is color-superconducting and that color-superconducting
quark matter appears in several phases. The most important
color-superconducting phases are presented in
Sec.~\ref{Color_superconductivity}. In
Sec.~\ref{Stellar_evolution}, I give a short but comprehensive
introduction into stellar evolution because the cores of neutron
stars are the only places in nature where one expects neutral
color-superconducting quark matter. Thereby, it will become
clear what a star is, how it is formed, which processes happen
in a star, and what are the final stages of stellar evolution.
In Sec.~\ref{Neutron_stars}, I focus on neutron stars. It is
explained how neutron stars are formed, of which matter they
consist, and I present the structure of neutron stars. In
Sec.~\ref{Color_superconductivity_in_neutron stars}, it is
argued why neutral color-superconducting quark matter is
expected to occur in neutron star cores.

In Chapter~\ref{phase_diagram}, I present the calculations to
obtain the pressure for neutral color-superconducting quark
matter, and I show the phase diagram of neutral quark matter.
With this one can predict in which state neutral quark matter is
in the cores of neutron stars.

In Chapter~\ref{Conclusions}, I summarize the results and
conclude my thesis.

Important definitions and useful formulae can be found in the
Appendix.
\section{The phase diagram of strongly interacting matter}
\label{The_phase_diagram_of_strongly_interacting_matter}
The fundamental theory of the strong interaction is called
quantum chromodynamics (QCD). The participants of the strong
interaction are the elements of hadronic matter, the quarks and
gluons. Quarks interact via gluons because both particle species
carry so-called color charges (red, green, and blue) which are
responsible for the interaction. In our everyday life, we do not
see quarks or gluons because they are confined into hadrons.
This is because the quark interaction caused by gluons is so
strong that they cannot exist as free particles. QCD is an
asymptotically free theory~\cite{asymptotic_theory}. At high
temperatures or densities, the quarks are deconfined because
their mutual distances decrease and the exchanged momenta
increase so that the interaction becomes sufficiently
weak~\cite{Collins_Perry}. The state of deconfined quarks is
called the quark-gluon plasma (QGP). Such a phase certainly
existed in the early universe which was very hot, but close to
net-baryon free. Nowadays, the only place in nature where a QGP
may exist is in the interior of neutron stars. Here, the density
is extremely high and the temperature low. The third place where
an artificially created QGP could appear is in heavy-ion
collisions. The temperatures and densities which are reached by
the collisions depend on the bombarding energies.
\begin{figure}[H]
\begin{center}
\includegraphics[width=0.6\textwidth]{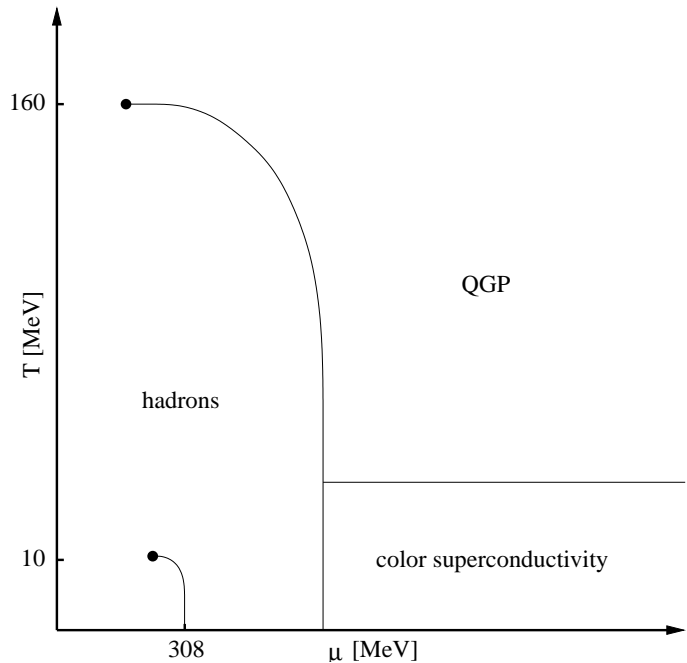}
\caption[The knowledge about the phase diagram of
strongly interacting matter in 2003.]{The status of knowledge of
the phase diagram of strongly interacting matter as I began with
the research on this topic for my thesis in
2003~\cite{CSCReviewRischke}. Note that this phase diagram is
drawn schematically.}
\label{phase_diagram_schematic}
\end{center}
\end{figure}
In Fig.~\ref{phase_diagram_schematic}, I show the status of
knowledge of the phase diagram of strongly interacting matter in
the plane of temperature $T$ and quark chemical potential $\mu$
before I began with the research for my thesis in
2003~\cite{CSCReviewRischke}. There is a phase transition at
the point $(T, \mu) \simeq (0, 308)$~MeV which separates the
gaseous nuclear phase at lower $\mu$ from the liquid nuclear
phase at higher $\mu$. The nuclear liquid-gas
transition~\cite{nuclear_liquid-gas_transition} which is a
first-order phase transition starts from this point and
disappears in a critical endpoint at $T \sim 10$~MeV and
slightly lower quark chemical potential. In this endpoint, the
transition is of second order. Above the endpoint, there is no
distinction between these two phases. The transition point at
zero temperature is easily found because nucleons of infinite
and isospin-symmetric nuclear matter in the ground state at
normal nuclear density $n_0 \simeq 0.15$~fm$^{-3}$ are bound
by 16~MeV (if one neglects the repulsive Coulomb forces). The
energy per baryon or the energy density per baryon density
respectively is given by $E / N_B = \epsilon / n_B = m_n -
16$~MeV where $m_n \simeq 940$~MeV is the rest mass of the
baryon. By using the thermodynamic relation $\epsilon = Ts - p +
\mu n$, one obtains for the ground state of nuclear matter where
the pressure $p = 0$ that the baryon chemical potential is
identical to the energy per baryon $\mu_B \equiv \epsilon / n_B
\simeq 924$~MeV. Since a baryon contains three quarks, the quark
chemical potential is one third of the baryon chemical
potential, $\mu = \mu_B / 3$. This leads to the result $\mu
\simeq 308$~MeV for the transition point at zero temperature.

Nuclear matter consists of droplets which is the most
energetically preferred form for nuclear matter in the
low-density and low-temperature regime of the phase diagram. In
the gaseous phase at low density and nonzero temperature,
nucleons will be evaporated from the surface of the droplets so
that there is a mixture of droplets and nucleons. As soon as the
chemical potential exceeds the one corresponding to the nuclear
liquid-gas phase transition, only droplets of nuclear matter but
no evaporated nucleons will appear.

At low quark chemical potentials, strongly interacting matter is
in the hadronic phase. At nonzero temperature, nuclear matter
not only consists of nucleons but also of thermally excited
hadrons. Therefore, at low quark chemical potentials, a
large amount of pions can be found. By increasing the
temperature, the system passes the quark-hadron transition line
and enters the regime of the QGP. The critical endpoint of the
transition line at $(T, \mu) = (162 \pm 2, 120 \pm 14)$~MeV
obtained by lattice QCD calculations~\cite{lattice} depends on
the value of the quark masses and is of second order. For
smaller quark chemical potentials, the transition becomes a
crossover, and there is no real distinction between hadronic
matter and the QGP. For larger quark chemical potentials, the
transition is a line of first-order phase transitions that
separates the hadronic phase from the QGP. But it is not known
whether the line of first-order phase transitions goes all
the way down to zero temperature. If so, also the precise value
of the quark chemical potential for the phase transition at zero
temperature is unknown. One should mention that these lattice
QCD calculations are not very reliable at nonzero quark chemical
potential. In addition, these calculations are done with
probably unrealistic large quark masses and on fairly small
lattice sizes. For smaller quark masses, the endpoint should
move towards the temperature axis. By increasing the quark
chemical potential, the nucleons will be packed denser and
denser until the quark-hadron phase transition is reached. As
soon as the quark chemical potential exceeds this transition
line, the system becomes a color superconductor at low
temperatures or a QGP which is in the normal conducting
phase (NQ) at high temperatures.

At large quark chemical potentials and low temperatures, quark
matter becomes a color superconductor. In this thesis, I shall
focus on this part of the phase diagram and its transitions to
the hadronic phase and to normal quark matter. The reader will
see that there exist various color-superconducting phases and
that this thesis will update the phase diagram of strongly
interacting matter in the quark regime.
\section{Color superconductivity}
\label{Color_superconductivity}
Quarks are spin-$\frac12$ fermions and therefore obey the Pauli
principle which requires that one quantum state is occupied by
only one fermion. At zero temperature, non-interacting quarks
occupy all available quantum states with lowest possible
energies. This behavior is expressed with the Fermi-Dirac
distribution function for zero temperature,
\be
\label{distribution_function}
f_F \left( \fettu{k} \right) = \Heaviside \left( \mu -
E_\fettu{k} \right) \; ,
\ee
where $E_\fettu{k} = \sqrt{ k^2 + m^2 }$ is the energy of a free
massive quark. All states with momenta $k \equiv |\fettu{k}|$
which are less than the Fermi momentum $k_F = \sqrt{ \mu^2 -
m^2 }$ are occupied. The states with momenta larger than the
Fermi momentum are empty. The pressure for massive
non-interacting quarks at zero temperature is given by
\be
p = \frac{g}{6 \pi^2} \int_0^\infty \ud k \, 
\frac{k^4}{E_\fettu{k}} f_F \left( \fettu{k} \right) - B
= \frac{g}{6 \pi^2} \int_0^{k_F} \ud k \, 
\frac{k^4}{E_\fettu{k}} - B \; ,
\ee
where $g = 2 N_c N_f$ is the degeneracy factor in which $N_c$ is
the number of colors, and $N_f$ the number of flavors. The
factor two comes because of spin degeneracy. The bag constant
$B$ assigns a nonzero contribution to the vacuum pressure and,
in this way, provides the simplest modelling of quark
confinement in QCD~\cite{Chodos}. A typical value for
the bag pressure is $B^{1/4} = 0.17$~GeV which is used
in Refs.~\cite{diploma_thesis_Ruester, RuesterCSCstars}. In the
limit of zero mass and zero temperature, the pressure of
non-interacting quarks reads
\be
\label{p00}
p = \frac{g}{24 \pi^2} \mu^4 - B \; .
\ee
This is the case of non-interacting quarks. What happens
when the quark interaction is switched on?

\begin{figure}[H]
\begin{center}
\includegraphics[width=0.55\textwidth]{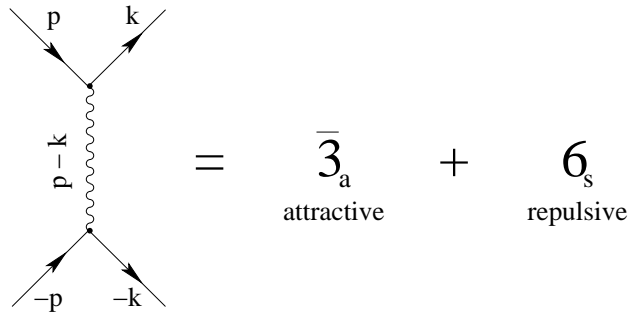}
\caption[The one-gluon exchange interaction between two quarks
in QCD.]{The diagrammatic representation of the one-gluon
exchange interaction between two quarks in
QCD~\cite{lecturesShovkovy}.}
\label{one-gluon_exchange}
\end{center}
\end{figure}

At asymptotically large quark chemical potentials, the strong
coupling constant becomes small so that the dominant interaction
between quarks is given by single-gluon exchange. The
quark-quark scattering amplitude in the one-gluon exchange
approximation is proportional to
\be
\sum_{A=1}^{N_c^2-1} T_{ki}^A T _{lj}^A = - \frac{N_c+1}{4N_c}
\left( \delta_{jk} \delta_{il} - \delta_{ik} \delta_{jl}
\right)
+ \frac{N_c-1}{4N_c} \left( \delta_{jk} \delta_{il} -
\delta_{ik} \delta_{jl} \right) \; ,
\ee
where $i$, $j$ are the colors of the incoming, and $k$, $l$
those of the outgoing channel. The first term in this equation
is antisymmetric and corresponds to the antitriplet channel
which is responsible for the dominant attractive interaction
while the second term is symmetric and corresponds to the sextet
channel which is responsible for the repulsive interaction, see
Fig.~\ref{one-gluon_exchange}.
Therefore, the antitriplet channel with its dominant attractive
interaction ensures that quarks with large momenta (quarks near
the Fermi surface) form bosonic quark Cooper
pairs~\cite{Barrois, Frautschi, Bailin_Love}. This state is
called a color superconductor in analogy to superconductivity of
electrons~\cite{BCS}. The arguments how color superconductivity
is created hold rigorously at asymptotically large densities.
The highest densities of nuclear matter that can be achieved
in colliders or that occur in nature in the cores of neutron
stars are of the order of ten times the nuclear matter ground
state density at which the quark chemical potential
approximately amounts to $\mu \simeq 500$~MeV. Nevertheless,
calculations in the framework of an NJL model~\cite{NJL} show
that color superconductivity also occurs at moderate densities
and is not limited to asymptotically large
densities~\cite{Alford_Rajagopal_Wilczek}.

As a consequence of (color) superconductivity, there exists at
least one gap in the quasiparticle spectra. Such a
color-superconducting gap (parameter) $\Delta$ is a diquark
condensate which is defined as an expectation value,
\be
\label{diquark_condensate}
\Delta \propto \langle \psi^T \mathcal{O} \psi \rangle \; ,
\ee
where the operator,
\be
\mathcal{O} = \mathcal{O}_\mathrm{color} \otimes
\mathcal{O}_\mathrm{flavor} \otimes \mathcal{O}_\mathrm{Dirac}
\; ,
\ee
acts on the quark spinor field $\psi$ in color, flavor, and
Dirac space. The color-superconducting gap parameters $\Delta_i$
are zero in normal quark matter and nonzero ($\Delta_i \sim
100$~MeV) in color-superconducting quark matter, and they are
equal to one half of the binding energy of a quark Cooper pair.
The values of the gap parameters can be obtained by solving the
gap equations
\be
\label{gap_eqn_intro}
\frac{\partial p_\mathrm{CSC}}{\partial \Delta_i} = 0 \; ,
\ee
where $p_\mathrm{CSC}$ is the pressure of color-superconducting
quark matter.

In ordinary superconductors, the $[U(1)_\mathrm{em}]$ gauge
symmetry is broken so that the photons become massive.
(Throughout this thesis, I indicate local, i.e., gauged,
symmetries by square brackets.) This leads to the so-called
Meissner effect, the expulsion of magnetic fields from the
superconducting region. In a color superconductor, the
$[SU(3)_c]$ color gauge symmetry is broken so that some of the
eight types of gluons become massive. In an ordinary
superconductor as well as in a color superconductor, thermal
motion will break up Cooper pairs and therefore destroy
the (color-)superconducting state. In a color superconductor,
this transition is of second order and happens at the critical
temperature
\be
T_c^\mathrm{BCS} = \frac{\e^\gamma}{\pi} \Delta_0 \simeq 0.567
\Delta_0
\ee
where $\Delta_0$ is the color-superconducting gap at zero
temperature, and $\gamma \simeq 0.577$ the Euler-Mascheroni
constant.

But there are also differences when one compares
superconductivity with color superconductivity: the electrons in
superconductors first have to overcome their repulsive Coulomb
forces in order to form Cooper pairs while in color
superconductors, the formation of quark Cooper pairs is much
simpler because there already exists the attractive interaction
in the antitriplet channel. Quarks, unlike electrons, come in
various flavors, see Table~\ref{quark_flavors}, and carry color
charges. Because of this latter quark property,
superconductivity of quarks is called color superconductivity.
\begin{table}[H]
\begin{center}
\begin{tabular}{|l||l|r|}
\hline
Flavor & Mass [MeV] & $Q$ [e] \\
\hline
\hline
\textbf{u}p      & $2 \ldots 8$       & $2/3$  \\
\textbf{d}own    & $5 \ldots 15$      & $-1/3$ \\
\textbf{s}trange & $100 \ldots 300$   & $-1/3$ \\
\textbf{c}harm   & $1300 \ldots 1700$ & $2/3$  \\
\textbf{b}ottom  & $4700 \ldots 5300$ & $-1/3$ \\
\textbf{t}op     & $174000 \pm 17000$ & $2/3$  \\
\hline
\end{tabular}
\end{center}
\caption[The masses and electric charges of the quark
flavors.]{The masses and electric charges of the quark
flavors~\cite{Taschenbuch_der_Physik}. The abbreviation for the
respective quark is written in the bold Roman font. There also
exists the corresponding antiquark for each flavor with charge
$-Q$. They are not shown here.}
\label{quark_flavors}
\end{table}
The quarks in color-superconducting quark matter are called
quasiparticles or quasiquarks, respectively. One only needs to
consider the lightest quarks (up, down, and strange) for
color-superconducting quark matter, because the heavy quarks
(charm, bottom, and top) are so massive that their occurrence in
quark matter is extremely unlikely for the densities and
temperatures under consideration.

The color and flavor structure of the condensate of quark Cooper
pairs which is also called the color-flavor gap matrix depends
on which quark flavors participate in pairing and which total
spin $J$ the Cooper pairs have. For $J = 0$, the spin part of
the quark Cooper pair wavefunction is antisymmetric, and
therefore the color-flavor part has to be symmetric in order to
fulfill the requirement of overall antisymmetry. Since quarks
pair in the antisymmetric color-antitriplet channel, the flavor
part of the wavefunction also has to be antisymmetric.
\subsection{The 2SC phase}
In order to fulfill the symmetry requirements, at least two
quarks of different flavor are needed for the $J = 0$ case.
Therefore, the simplest ansatz for the color-flavor gap matrix
has the form:
\be
\Delta_{ij}^{fg} = \epsilon_{ijk} \epsilon^{fg} \Delta_k \; ,
\ee
where the order parameter
\be
\Delta_k = \delta_{k3} \Delta
\ee
conventionally points in anti-blue color direction. Together
with the Dirac part, the gap matrix reads
\be
\Phi_{ij}^{fg} = \gamma_5 \Delta_{ij}^{fg} \; .
\ee
This is the so-called 2SC phase which is an abbreviation for
\textbf{2}-flavor color \textbf{s}uper\textbf{c}onductor. The
color indices $i$ and $j$ run from one to the number of colors
$N_c$ which is equal to three. The flavor indices $f$ and $g$
run from one to the number of flavors $N_f$ participating in
pairing which is equal to two in the 2SC phase. In this phase,
red up quarks pair with green down quarks, and red down quarks
pair with green up quarks, and form anti-blue quark Cooper
pairs.

The blue quarks remain unpaired and therefore cause gapless
quasiparticles. These quasiparticles give dominant contributions
to the specific heat and to the electrical and heat
conductivities. They are also responsible for a large neutrino
emissivity produced by the $\beta$ processes $\mathrm{d_b
\rightarrow u_b + e^- + \bar{\nu}_e}$ and $\mathrm{u_b + e^-
\rightarrow d_b + \nu_\mathrm{e}}$. The other four
quasiparticles and quasiantiparticles fulfill the dispersion
relation
\be
\label{epsilon_2SC_intro}
\epsilon_\fettu{k}^e = \sqrt{\left( E_\fettu{k} - e \mu
\right)^2 + \left| \Delta \right|^2} \; ,
\ee
where $\Delta$ is the gap, and $e = \pm$ stands for
quasiparticles and quasiantiparticles, respectively. At small
temperatures ($ T \ll \Delta$) , the contributions of these
quasiparticles to all transport and many thermodynamic
quantities are suppressed by the exponentially small factor
$\exp \left( - \Delta / T \right)$~\cite{lecturesShovkovy}. The
gluons are bosons and therefore, their number density is small
at low temperature. In the 2SC phase, the $[SU(3)_c]$ gauge
symmetry is broken to $[SU(2)_c]$. Consequently, there are
$8 - 3 = 5$ broken generators. They represent five gluons which
are gapped because of the color Meissner mass. Therefore, gluons
have only tiny influence on the properties of quark matter in
the 2SC phase. The unpaired blue quarks are responsible for the
absence of baryon superfluidity. Only the anti-blue
quasiparticles carry a nonzero baryon number. This can be seen
by the generator of baryon number conservation,
\be
\tilde B = B - \frac{2}{\sqrt{3}} T_8 = \diag_c
\left(0,0,1\right) \; ,
\ee
where $T_8$ is the eighth generator of the $[SU(3)_c]$ group.
The electromagnetic generator of the unbroken $[\tilde
U(1)_\mathrm{em}]$ gauge symmetry in the 2SC phase is
\be
\label{Qtildecharges}
\tilde Q = Q - \frac{1}{\sqrt{3}} T_8 \; , \quad
\tilde Q^r_u = \tilde Q^g_u = \frac12 \; , \quad
\tilde Q^b_u = 1 \; , \quad
\tilde Q^r_d = \tilde Q^g_d = -\frac12 \; , \quad
\tilde Q^b_d = 0 \; .
\ee
where $Q = \diag_f \left( \frac23,-\frac13 \right)$ is the
electromagnetic generator of the $[U(1)_\mathrm{em}]$ gauge
symmetry in vacuum. The gauge boson of $[\tilde
U(1)_\mathrm{em}]$ is the medium photon. Therefore, there exists
no electromagnetic Meissner effect in the 2SC phase and that is
the reason why a magnetic field would not be expelled from the
color-superconducting region. The 2SC phase is a so-called
$\tilde Q$-conductor because its electrical conductivity is
large. The $\tilde Q$ charge of the blue up quasiparticle is
responsible for this behavior, see Eq.~(\ref{Qtildecharges}).

At weak coupling, the difference of the color-superconducting
pressure to the pressure of normal-conducting quark matter for
massless non-interacting quarks at zero temperature~(\ref{p00})
amounts to
\be
\label{deltap}
\delta p = \frac{\mu^2 \Delta^2}{4 \pi^2}
\ee
per each gapped quasiparticle~\cite{deltap}. In the 2SC phase
without strange quarks, there are six quarks from which four of
them are gapped so that the pressure for color-superconducting
quark matter in the 2SC phase at zero temperature approximately
reads,
\be
p_\mathrm{2SC} \simeq \frac{\mu^4}{2 \pi^2} + \frac{\mu^2
\Delta^2}{\pi^2} - B \; .
\ee
\subsection{The CFL phase}
If the strange quark chemical potential exceeds the mass of
strange quarks, then there are also strange quarks in quark
matter at zero temperature, see
Eq.~(\ref{distribution_function}). Therefore, it is possible
that also the strange quarks participate in pairing. Since the
spin part of the quark Cooper pair wavefunction is antisymmetric
for $J = 0$ and quarks pair in the antisymmetric
color-antitriplet channel, also the flavor part has to be
antisymmetric in order to fulfill the requirement of overall
antisymmetry. This leads to the following color-flavor gap
matrix in the three-flavor case:
\be
\label{CFL_color-flavor_gap_matrix}
\Delta_{ij}^{fg} = \epsilon_{ijk} \epsilon^{fgh} \Delta_k^h \; ,
\ee
where the order parameter is given by
\be
\Delta_k^h = \delta_k^h \Delta \; .
\ee
Together with the Dirac part, the gap matrix reads
\be
\Phi_{ij}^{fg} = \gamma_5 \Delta_{ij}^{fg} \; .
\ee
There is one difference in comparison to the 2SC phase: the
antisymmetric tensor in Eq.~(\ref{CFL_color-flavor_gap_matrix})
now possesses three instead of two flavor indices because three
instead of two quarks participate in pairing, $N_f = 3$. The
condensate breaks $[SU(3)_c] \times SU(3)_{r + \ell}$ to the
vectorial subgroup $SU(3)_{c + r + \ell}$ and is still invariant
under vector transformations in color and flavor space. This
means that a transformation in color requires a simultaneous
transformation in flavor to preserve the invariance of the
condensate. Therefore, the discoverers~\cite{CFL_discoverers} of
this three-flavor color-superconducting quark state termed it
the CFL phase which is an abbreviation for
\textbf{c}olor--\textbf{f}lavor-\textbf{l}ocked phase. The CFL
phase is the true ground state of quark matter because all
quarks are paired which leads to the highest pressure of all
color-superconducting phases.

In contrast to the 2SC phase, the CFL phase also has gaps in the
repulsive sextet channel. The color-flavor gap
matrix~(\ref{CFL_color-flavor_gap_matrix}) can be extended by
the (small) symmetric sextet gaps,
\be
\label{CFL_33_66}
\Delta_{ij}^{fg} = \Delta_{\left(\bar 3,\bar 3\right)} \left(
\delta_i^f \delta_j^g - \delta_i^g \delta_j^f \right) +
\Delta_{\left(6,6\right)} \left( \delta_i^f \delta_j^g +
\delta_i^g \delta_j^f \right) \; ,
\ee
where $\Delta_{\left(\bar 3,\bar 3\right)}$ is the antitriplet
and $\Delta_{\left(6,6\right)}$ is the sextet gap. This can be
rewritten as
\be
\Delta_{ij}^{fg} = \Delta_1' \delta_i^f \delta_j^g + \Delta_2'
\delta_i^g \delta_j^f \; ,
\ee
where $\Delta_1' = \Delta_{\left(\bar 3,\bar 3\right)} +
\Delta_{\left( 6, 6 \right)}$ and $\Delta_2' = -
\Delta_{\left( \bar 3, \bar 3 \right)} + \Delta_{\left( 6, 6
\right)}$. By introducing the color-flavor
projectors~\cite{Shovkovy_Wijewardhana},
\be
\left[\mathcal{P}_1\right]_{ij}^{fg} = \frac13 \delta_i^f
\delta_j^g \; , \qquad
\left[\mathcal{P}_2\right]_{ij}^{fg} = \frac12 \delta_{ij}
\delta^{fg} - \frac12 \delta_i^g \delta_j^f \; , \qquad
\left[\mathcal{P}_3\right]_{ij}^{fg} = \frac12 \delta_{ij}
\delta^{fg} + \frac12 \delta_i^g \delta_j^f - \frac13 \delta_i^f
\delta_j^g \; ,
\ee
which fulfill the properties of completeness, $\sum_n
\mathcal{P}_n = 1$, and orthogonality, $\mathcal{P}_i
\mathcal{P}_j = \delta_{ij} \mathcal{P}_j$, the color-flavor gap
matrix assumes the form,
\be
\Delta_{ij}^{fg} = \sum_{n=1}^3 \Delta_n
\left[\mathcal{P}_n\right]_{ij}^{fg} = \frac13 \left( \Delta_1 +
\Delta_2 \right) \delta_i^f \delta_j^g - \Delta_2 \delta_i^g
\delta_j^f \; ,
\ee
where $\Delta_3 = -\Delta_2$. The singlet gap $\Delta_1$ and
the octet gap $\Delta_2$ appear in the spectra of quasiparticles
and quasiantiparticles,
\be
\label{epsilon_CFL_intro}
\epsilon_n^e \left( \fettu{k} \right) = \sqrt{\left( k -
e \mu \right)^2 + \left| \Delta_n \right|^2} \; ,
\ee
where $n = 1,2$, and all quarks are treated as massless for
simplicity. By neglecting the small repulsive sextet gap, one
finds that $\Delta_1 = 2 \Delta_2 \equiv 2 \Delta$.

In the CFL phase, there are no gapless quasiparticles. At small
temperatures ($T \ll \Delta$), the contributions of the quark
quasiparticles to all transport and many thermodynamic
quantities are suppressed by the small exponential factor $\exp
( -\Delta /T )$~\cite{lecturesShovkovy}. The influence of the
gluons on the CFL phase is negligible because all of them are
massive because of the color Meissner effect. In contrast to the
2SC phase, the CFL phase is superfluid because the $U(1)_B$
baryon number symmetry is broken, but it has an unbroken $[
U(1)_\mathrm{em} ]$ gauge symmetry and therefore it is, like the
2SC phase, not an electromagnetic superconductor. This is the
reason why the CFL phase does not expell a magnetic field from
its color-superconducting interior. The electromagnetic
generator of the CFL phase reads,
\be
\tilde Q = Q - T_3 - \frac{1}{\sqrt{3}} T_8 \; , \quad
\tilde Q^r_u = \tilde Q^g_d = \tilde Q^g_s = \tilde Q^b_d =
\tilde Q^b_s = 0 \; , \quad \tilde Q^g_u = \tilde Q^b_u = 1 \; ,
\quad \tilde Q^r_d = \tilde Q^r_s = -1 \; .
\ee
The CFL phase is a $\tilde Q$-insulator because all quarks are
gapped and there is no remaining electric charge as in the 2SC
phase. Therefore, the CFL phase is electrically charge neutral.
At zero temperature, there are no electrons
present~\cite{enforced_neutrality}. At small temperatures, the
electrical conductivity of the CFL phase is dominated by
thermally excited electrons and positrons~\cite{Shovkovy_Ellis1}
and becomes transparent to light~\cite{Shovkovy_Ellis1,
Vogt_Rapp_Ouyed}.

As in the case of chiral perturbation theory in vacuum
QCD~\cite{Gasser}, one could write down an effective low-energy
theory in the CFL phase. From the symmetry breaking pattern, it
is known that there are nine Nambu-Goldstone bosons and one
pseudo-Nambu-Goldstone boson in the low-energy spectrum of the
theory~\cite{Nambu-Goldstone}. Eight of the Nambu-Goldstone
bosons are similar to those in vacuum QCD: three pions ($\pi^0$
and $\pi^\pm$), four kaons ($K^0$, $\bar K^0$, and $K^\pm$) and
the eta-meson ($\eta$). The additional Nambu-Goldstone boson
($\phi$) comes from breaking the $U(1)_B$ baryon symmetry. In
absence of gapless quark quasiparticles, this Nambu-Goldstone
boson turns out to play an important role in many transport
properties of cold CFL matter~\cite{Shovkovy_Ellis1,
Shovkovy_Ellis2}. Finally, the pseudo-Nambu-Goldstone boson
($\eta'$) results from breaking of the approximate axial
$U(1)_A$ symmetry. A possible phase transition to the CFL phase
with a meson (e.g., kaon or eta) condensate could happen if $M_s
\gtrsim M_u^{1/3} \Delta^{2/3}$~\cite{Schaefer, Hong, Bedaque,
Kaplan_Reddy, Kryjevski_Kaplan_Schaefer}, where $M_u$ is the up
and $M_s$ the strange quark mass.

In the CFL phase, there are nine quarks, and all of them are
gapped: one quasiparticle with gap $\Delta_1$ and eight
quasiparticles with gap $\Delta_2$. Therefore, the pressure of
the CFL phase for massless quarks at zero temperature
approximately is given by
\be
p_\mathrm{CFL} \simeq \frac{3 \mu^4}{4 \pi^2} + \frac{\mu^2
\Delta_1^2}{4 \pi^2} + \frac{2 \mu^2 \Delta_2^2}{\pi^2} - B
\simeq \frac{3 \mu^4}{4 \pi^2} + \frac{3 \mu^2 \Delta^2}{\pi^2}
- B \; ,
\ee
where Eqs.~(\ref{p00}), (\ref{deltap}), and $\Delta_1 = 2
\Delta_2 \equiv 2 \Delta$ are used.
\subsection{Spin-one color superconductivity}
Since quarks pair in the antisymmetric color-antitriplet channel
and the spin part of the quark Cooper pair wavefunction is
antisymmetric for $J = 0$, condensation with only one flavor is
forbidden by the Pauli principle, but it is possible for the $J
= 1$ channel, where the spin part of the wavefunction is
symmetric. Thus, the Cooper pair wavefunction is, as required,
overall antisymmetric. Spin-one color
superconductivity~\cite{Bailin_Love, Bailin_Love2,
Pisarski_Rischke1, Pisarski_Rischke2, Iwasaki_Iwado, spin-1,
Schmitt_Wang_Rischke, Schmitt_PhD} is much weaker than spin-zero
color superconductivity. The gap in spin-one
color-superconducting systems is of the order of 100~keV. Such a
small gap will not have big influences on the transport and
many thermodynamic properties of the quark
matter~\cite{lecturesShovkovy}. Spin-one color superconductivity
is less favored than spin-zero color superconductivity since the
latter has a higher pressure because of the larger gap. This is
why one does not expect that spin-one color-superconducting
quark phases dominate in the phase diagram of neutral quark
matter. But they could be favored if it is not possible to form
a spin-zero color-superconducting state because of a too large
mismatch between the Fermi surfaces of different quark
flavors~\cite{Alford_Cowan}.

The general structure of the gap matrix for spin-one
color-superconducting systems reads~\cite{Schmitt_Wang_Rischke,
Schmitt_PhD},
\be
\Delta^{ab}=i \Delta_0 \sum_{c, i = 1}^3 \epsilon^{abc}
\mathcal{C}_{ci} \left[ \hat{k}^i \cos \theta + \gamma_\perp^i
\sin \theta \right] \; ,
\ee
where $\gamma_\perp^i \equiv \gamma^i - \hat{k}^i \left(
\fett{\gamma} \cdot \fettu{k} \right)$, $\hat{\fettu{k}} \equiv
\fettu{k}/k$, $\fettu{k}$ is the momentum vector, and $k$ its
absolute value. Spin-one color-superconducting phases are called
longitudinal if $\theta = 0$ and transverse if $\theta = \pi /
2$. Many different spin-one color-superconducting phases can be
constructed by choosing various specific $3 \times 3$-matrices
$\mathcal{C}$. The most important of them are the A-phase, the
\textbf{c}olor--\textbf{s}pin-\textbf{l}ocked (CSL) phase, the
polar phase, and the planar phase~\cite{Schmitt_Wang_Rischke,
Schmitt_PhD},
\be
\begin{aligned}
\mathcal{C}^\mathrm{\left( A-phase \right)} =
\frac{1}{\sqrt{2}}
\left(
\begin{array}{@{\extracolsep{1.5mm}}ccc}
0 & 0 & 0 \\
0 & 0 & 0 \\
1 & i & 0
\end{array}
\right) \; &, &
\mathcal{C}^\mathrm{\left( CSL \right)} &=
\frac{1}{\sqrt{3}}
\left(
\begin{array}{@{\extracolsep{1.5mm}}ccc}
1 & 0 & 0 \\
0 & 1 & 0 \\
0 & 0 & 1
\end{array}
\right) \; , \\
\mathcal{C}^\mathrm{\left( polar \right)} =
\left(
\begin{array}{@{\extracolsep{1.5mm}}ccc}
0 & 0 & 0 \\
0 & 0 & 0 \\
0 & 0 & 1
\end{array}
\right) \; &, &
\mathcal{C}^\mathrm{\left( planar \right)} &=
\frac{1}{\sqrt{2}}
\left(
\begin{array}{@{\extracolsep{1.5mm}}ccc}
1 & 0 & 0 \\
0 & 1 & 0 \\
0 & 0 & 0
\end{array}
\right) \; ,
\end{aligned}
\ee
which are characterized by different symmetries of their ground
state.

The original group $[SU(3)_c] \times SO(3)_J \times
[U(1)_\mathrm{em}]$ breaks down as
follows~\cite{Schmitt_Wang_Rischke, Schmitt_PhD}:
\begin{description}
\item[A-phase:] $SU(2)_c \times \widetilde{SO}(2)_J \times
\tilde{U}(1)_\mathrm{em} \; ,$
\item[CSL:] $\widetilde{SO}(3)_J \; ,$
\item[Polar:] $SU(2)_c \times SO(2)_J \times \tilde
U(1)_\mathrm{em} \; ,$
\item[Planar:] $\widetilde{SO}(2)_J \times \tilde
U(1)_\mathrm{em} \; .$
\end{description}
In spin-one color superconductors, there can exist an
electromagnetic Meissner effect in contrast to spin-zero color
superconductors. If so, magnetic fields will be expelled from
the color-superconducting region. This is, for example, the case
in the CSL phase. The most energetically preferred spin-one
color-superconducting quark phase is the transverse CSL phase
because it has the highest pressure~\cite{Schmitt}.
\section{Stellar evolution}
\label{Stellar_evolution}
Stars begin their life as objects which are formed out of
contracted interstellar gas clouds in galaxies. They are nuclear
burning factories: light nuclei, such as hydrogen, will be
burned to heavier nuclei by fusion reactions. After all fusion
reactions are completed, stars end their life as compact stars:
white dwarfs, neutron stars, or black holes.
\subsection{The formation of stars}
\label{star_formation}
\begin{figure}[H]
\begin{center}
\includegraphics[width=0.95\textwidth]{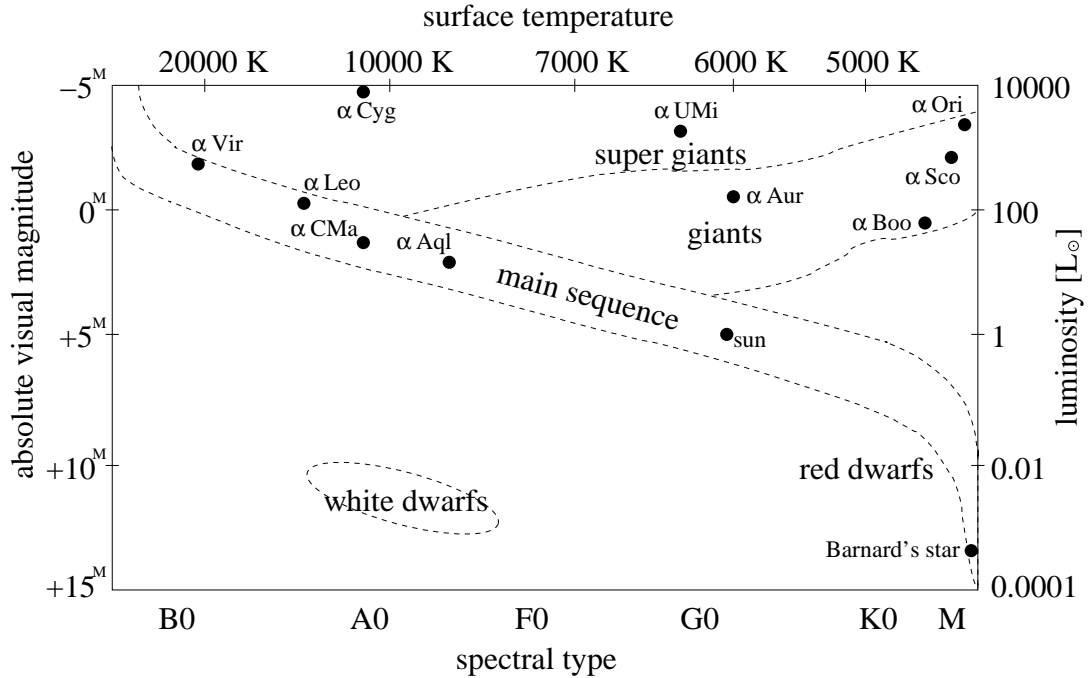}
\caption[The Hertzsprung-Russell diagram.]{The
Hertzsprung-Russell diagram (HRD)~\cite{dtv-Atlas}.}
\label{HRD}
\end{center}
\end{figure}

Interstellar clouds which mainly consist of hydrogen contract if
their gravitation exceeds the pressure from inside caused by
turbulence and temperature. In order to obtain such a large
gravitation, interstellar clouds have to possess big masses. The
so-called Jeans criterion has to be fulfilled so that an
interstellar cloud is able to contract~\cite{Voigt},
\be
M > M_\mathrm{cr} \left( T, \rho_c \right) \simeq 28 T^{\frac32}
{\rho_c}^{\!\!\!-\frac12} \; ,
\ee
where $T$ is the temperature in Kelvin and $\rho_c$ the central
density of the interstellar cloud in cm$^{-3}$. The critical
mass $M_\mathrm{cr}$ is obtained in units of the solar mass. In
Table~\ref{critical_masses}, some values for the critical mass
are shown.
\begin{table}[H]
\begin{center}
\begin{tabular}{|r||r|r|r|}
\hline
& 1 cm$^{-3}$ & 100 cm$^{-3}$ & 10$^4$ cm$^{-3}$ \\
\hline
\hline
 10 K &   880 $\msun$ &   88 $\msun$ & 8.8 $\msun$ \\
100 K & 28000 $\msun$ & 2800 $\msun$ & 280 $\msun$ \\
\hline
\end{tabular}
\end{center}
\caption[Values for the critical mass of interstellar
clouds.]{Values for the critical mass of interstellar
clouds~\cite{Voigt}.}
\label{critical_masses}
\end{table}
By the contraction of the interstellar cloud, first stars of
spectral type O are created in its center. Their ultraviolet
radiation ionizes the hydrogen gas around them so that it
becomes hot and is shining. This so-called H~II region has a
temperature $T \simeq 10000$~K and expands into the cool outer
regions of the interstellar cloud which consist of cold
hydrogen gas, so-called H~I regions which have a temperature $T
\simeq 100$~K. Wavy bays and globules are created by this
expansion. The globules have diameters of up to one parsec (pc)
and masses of up to 70~$\msun$. Within 500000 years, they
fragment and collapse to protostars which emit only infrared
radiation because there are dense dust clouds around them which
fall on them within several million years. In the meantime, the
contraction of the star continues until the pressure from inside
becomes as large as the gravitation. Then, the star is on the
main sequence in the Hertzsprung-Russell diagram (HRD), see
Fig.~\ref{HRD}.
The massive and bright stars are in the upper and the darker
stars with lower mass are on the lower part of the HRD. Hot blue
stars are in the left and cooler red stars are in the right part
of the HRD. During the contraction of gas balls to stars, their
rotation becomes faster and faster because of angular momentum
conservation. This could lead to a splitting of the gas balls
because of large centrifugal forces so that gravitationally
bound narrow double, multi-star, or solar systems will be
formed. Wider gravitationally bound double or multi-star systems
are formed if the stars of the interstellar cloud come close to
each other by their movement and by gravitational forces. In
this way, many stars and star systems are created inside
interstellar clouds so that finally there exists an open cluster
in their center from which an O association, a star cluster with
hundreds of stars of spectral type O until B2, expands into the
outer regions.
\subsection{Main sequence stars}
\begin{figure}[H]
\begin{center}
\includegraphics[width=0.85\textwidth]{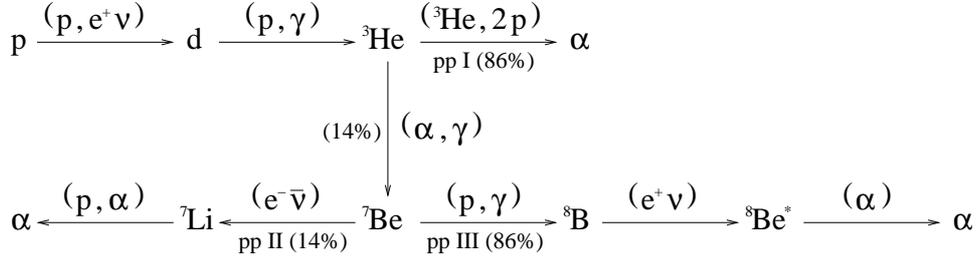}
\caption[The proton-proton cycles.]{The three proton-proton
cycles (pp cycles). Percent numbers specify how large the
probability of a reaction is. The star at ${}^8$Be means that it
is not stable and decays into two $\alpha$-particles.
Cycle~I was discovered in Ref.~\cite{Critchfield_Bethe}.
Positrons which are formed by fusion reactions will be
annihilated.}
\label{pp}
\end{center}
\end{figure}

\begin{figure}[H]
\begin{center}
\includegraphics[width=0.85\textwidth]{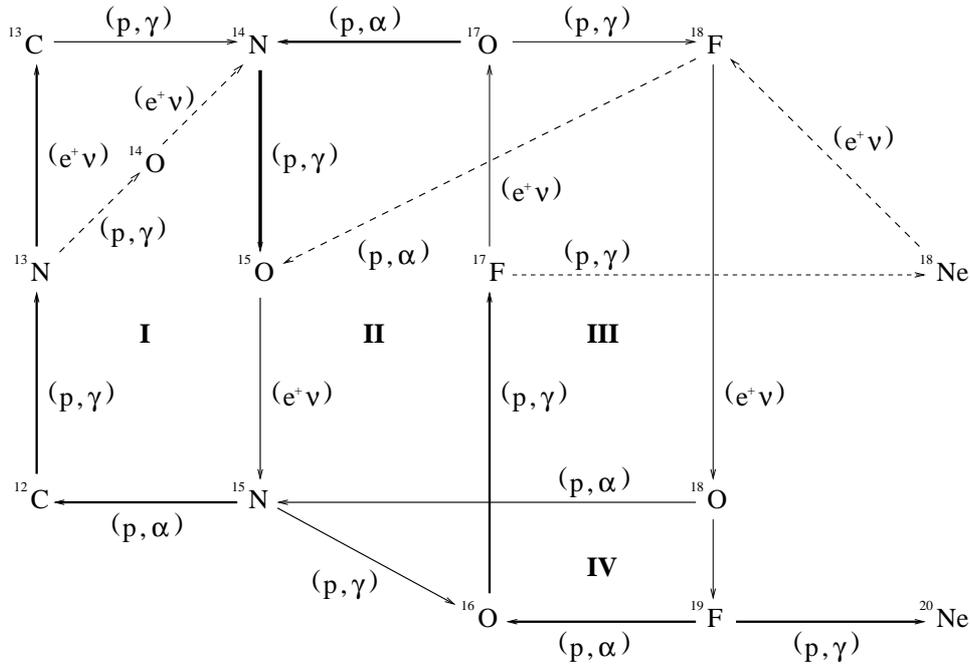}
\caption[The carbon-nitrogen-oxygen cycles.]{The
carbon-nitrogen-oxygen cycles (CNO cycles)~\cite{Adelberger}.
Cycle~I is named after their discoverors: Bethe-Weizs\"acker
cycle~\cite{Bethe_Weizsaecker}. Cycle~II is called CNO bi-cycle,
and cycle~III is termed CNO tri-cycle. The widths of the arrows
illustrate the significance of the reactions in determining the
nuclear fusion rates in the stellar CNO cycles. Certain
\textit{hot CNO processes} are indicated by dashed arrows.
Positrons which are formed by fusion reactions will be
annihilated.}
\label{cno}
\end{center}
\end{figure}

During the contraction, the central temperature $T_c$ of the
stars becomes higher and higher. At $T_c \simeq 5 \cdot 10^6$~K,
hydrogen burning is initiated in the cores of the stars. For
producing such large temperatures in the cores, the stars need
at least a mass $M \simeq 0.1$~$\msun$. Otherwise brown dwarfs
will be created in which hydrogen burning never occurs.
Hydrogen-burning stars are on the main sequence of the HRD. The
contraction is finished because the pressure is as large as the
gravitational attraction. There are two possibilities to burn
hydrogen into helium by fusion reactions~\cite{dtv-Atlas,
Wolschin}: these are the proton-proton cycles (pp cycles), see
Fig.~\ref{pp}, and the carbon-nitrogen-oxygen
cycles (CNO cycles), see Fig.~\ref{cno}.
The main difference between these two possibilities is that in
the pp cycles protons fuse directly with each other while in the
CNO cycles, they are always burnt into helium by using carbon,
nitrogen, oxygen, etc.\ as catalysts. In cooler yellow and red
stars, where $T_c \simeq 5 \cdot 10^6$~K, only the pp cycles are
present. The CNO cycles are initiated not until $T_c \simeq 9.5
\cdot 10^6$~K, but they are dominant in hot blue and white stars
where $T_c \gtrsim 1.5 \cdot 10^7$~K. The energy production of
the CNO cycles is then much larger as the energy production of
the pp cycles. The larger the mass of a (main sequence) star the
higher is its central temperature. Stars with big masses use the
CNO cycles to burn hydrogen into helium. But only twelve percent
of the total hydrogen is in the core of a main sequence star
which can be burnt into helium. Because of the much higher
central temperatures, blue and white stars with large masses
burn the hydrogen in their cores much faster into helium as cold
yellow and red stars with low masses, see Table~\ref{hydrogen}.
\begin{table}[H]
\begin{center}
\begin{tabular}{|c||c|c|}
\hline
Type & $M$ [$\msun$] & Million years \\
\hline
\hline
O5 & 39    &    0.5 \\
B0 & 20    &    4.6 \\
B5 &  6.7  &   46   \\
A0 &  3.5  &  319   \\
A5 &  2.2  & 1160   \\
F0 &  1.7  & 2700   \\
F3 &  1.26 & 3800   \\
F6 &  1.13 & 6000   \\
\hline
\end{tabular}
\end{center}
\caption[Burning period of hydrogen in the cores of main
sequence stars.]{Burning period of hydrogen in the cores of main
sequence stars~\cite{Voigt}.}
\label{hydrogen}
\end{table}
\subsection{Red giants and red super giants}
When the hydrogen in the core is used up, gravitation becomes
dominant so that the star contracts because there are no fusion
reactions any more which can maintain the pressure from inside.
This leads to higher central temperatures, $T_c \gtrsim
100$~million~Kelvin. Such high temperatures are needed to
initiate helium burning in the core of a star,
\be
\label{alpha}
\alpha \left( \alpha, \gamma \right) {}^8 \mathrm{Be}^\ast
\left( \alpha, \gamma \right) {}^{12} \mathrm{C} \left( \alpha,
\gamma \right) {}^{16} \mathrm{O} \; .
\ee
To produce such high temperatures in the core, a star needs at
least a mass $M \simeq 0.4$~$\msun$. Otherwise the star is only
able to burn hydrogen and then ends as a white dwarf. Helium
burning happens in a process of two stages, a so-called Salpeter
process~\cite{lecturesSchaffner}: two helium nuclei form an
intermediate stage, a ${}^8$Be nucleus which is unstable and is
therefore marked by a star in the fusion reaction~(\ref{alpha}).
It reacts with a third helium nucleus to produce ${}^{12}$C. The
Salpeter process of helium burning is called triple-$\alpha$
process, and it is much more likely as a direct reaction of
three helium nuclei to form carbon. A further reaction of
${}^{12}$C with a helium nucleus produces ${}^{16}$O.

Above the core of helium burning, there remains a shell of
hydrogen burning which expands until it reaches the star
surface. The star blows up because of the high pressure caused
by helium burning. Its diameter and luminosity are much larger
as in the period of a main sequence star but the surface
temperature is colder so that the star appears red. It has
become a red giant. When hydrogen burning ends and helium
burning is initiated, the stars go from the main sequence into
the giant region in the HRD. Red giants are not as stable as
main sequence stars because the pressure from inside is in
imbalance with the gravity of the star. This leads to
oscillations of red giants: the star contracts, the radius
becomes smaller and the star whitens a little. The interior
pressure increases, the star expands, the radius becomes larger
and the star becomes a little more red. This procedure repeats
again and again. When the helium in the core is used up, the red
giant contracts. If its mass exceeds eight solar masses, the
central temperature is hot enough to burn carbon in the core of
the star,
\be
{}^{12}\mathrm{C} + {}^{12}\mathrm{C} \longrightarrow
{}^{20}\mathrm{Ne} + \alpha \; , \qquad
{}^{12}\mathrm{C} + {}^{12}\mathrm{C} \longrightarrow
{}^{23}\mathrm{Na} + \mathrm{p} \; , \qquad
{}^{12}\mathrm{C} + {}^{12}\mathrm{C} \longrightarrow
{}^{23}\mathrm{Mg} + \mathrm{n} \; ,
\ee
otherwise the red giant ends up as a white dwarf. By carbon
burning, the red giant blows up to a red super giant because the
interior pressure becomes very high. If all carbon is used up,
the red super giant contracts. If the mass of the red super
giant exceeds ten solar masses, its central temperature becomes
high enough so that neon, oxygen, and silicon will be burnt in
the core,
\begin{align}
{}^{20}\mathrm{Ne} + \gamma &\longrightarrow {}^{16}\mathrm{O}
+ \alpha \; , &
{}^{20}\mathrm{Ne} + \alpha &\longrightarrow {}^{24}\mathrm{Mg}
+ \gamma \; , \qquad
{}^{24}\mathrm{Mg} + \alpha \longrightarrow {}^{28}\mathrm{Si}
+ \gamma \; , \\
{}^{16}\mathrm{O} + {}^{16}\mathrm{O} &\longrightarrow
{}^{31}\mathrm{P} + \mathrm{p} \; , &
{}^{16}\mathrm{O} + {}^{16}\mathrm{O} &\longrightarrow
{}^{28}\mathrm{Si} + \alpha \; , \\
\gamma + {}^{28}\mathrm{Si} &\longrightarrow \left(\mathrm{p},
\, \alpha, \, \mathrm{n} \right) \; , &
{}^{28}\mathrm{Si} & \left(\alpha, \gamma\right)
{}^{32}\mathrm{S} \left(\alpha, \gamma\right) {}^{36}\mathrm{Ar}
\, \ldots \, {}^{52}\mathrm{Fe} \left(\alpha, \gamma\right)
{}^{56}\mathrm{Ni} \; ,
\end{align}
otherwise the fusion reactions end with the fusion of carbon.
After all carbon is used up in the core of the red super giant
and its mass is lower than ten solar masses so that the red
super giant is not able to proceed with further fusion
reactions, it finally explodes by a supernova of type~II and
forms a neutron star. But if the mass of the red super giant
exceeds ten solar masses, fusion reactions can happen until
nuclei with mass number $A = 56$ are produced. These are iron
and nickel. For larger $A$, fusion reactions would need energy
to produce heavier nuclei. So, there is no other way for a
massive red super giant with an iron core as to collapse because
of the missing interior pressure. It will cause a supernova of
type~II and form a neutron star. If the mass is even larger than
approximately twenty solar masses, the star will collapse into a
black hole.
\subsection{Compact stars}
If the mass of a star is smaller than approximately
$0.1$~$\msun$, hydrogen burning is never initiated and the star
will end up as a brown dwarf. Otherwise, if the mass is larger,
the star will end up as a compact star: a white dwarf, a neutron
star, or a black hole. If the mass is larger than approximately
$0.1$~$\msun$ but does not exceed eight solar masses, the star
will end up as a white dwarf. The outer shells of red giants are
blown away and form a planetary nebula. There are no fusion
reactions any more so that the core of the red giant collapses
to a white dwarf in the center of the planetary nebula. Only the
pressure of degenerate electrons is able to stop the collapse
which is caused by gravity. White dwarfs have densities of about
one million grams per cubic centimeter, radii of a few thousand
kilometers, and cool approximately within ten billion years to
black dwarfs and become invisible.

If a star A is a star with a small mass in a narrow binary
system, matter from its massive companion star B is accreted by
star A if star B exceeds its Roche volume when it becomes a red
giant. This phenomenon has a big influence on the evolution of
stars in narrow binary or multi-star systems: star A becomes
much more massive than star B. If star A is a white dwarf and
star B is a red giant, the white dwarf accretes matter which can
be seen as nova bursts because the accreted matter causes
nuclear reactions. If the mass of the white dwarf exceeds the
Chandrasekhar mass which is approximately 1.44~$\msun$, the
white dwarf explodes in a supernova of type I because its
gravity becomes so large that the degenerate electron pressure
is not able to stabilize the star any more. Expanding gas shreds
are the only remnants of the white dwarf.

If the mass of the red super giant is larger than eight solar
masses and there are no further fusion processes in the star,
the pressure of degenerate electrons is not able to stop the
collapse any more. Such red super giants collapse in supernova
explosions of type~II to neutron stars. More details about
neutron stars will be presented in Sec.~\ref{Neutron_stars}.

A black hole will be formed if the progenitor star, a red
super giant, possesses a mass of about 20--25 solar masses. No
internal force is able to stop the collapse of the red super
giant because of the huge gravitation. A black hole is defined
as a region of space-time which cannot communicate with the
external universe which means that there is no chance for a
particle or even for light to escape from a black hole if it
has reached the region beyond its event horizon. The event
horizon is the boundary of a black hole at which gravity is so
strong that nothing has a chance to escape. The radius of the
event horizon for a spherical mass is called the Schwarzschild
radius, $r_S = 2 G M / c^2$.
\section{Neutron stars}
\label{Neutron_stars}
At the end of its life, a red super giant with a mass
larger than eight solar masses consists of many shells
of different nuclei which are created by fusion processes. 
They are arranged like onion shells: the surface shell of the
red super giant consists of the lightest nucleus, hydrogen.
Towards the center, the nuclei which were created by fusion
processes get heavier. In the center of red super giants with
masses larger than ten solar masses, there are iron nuclei. If
all fusion processes are finished, the red super giant will
collapse. Since gravity of such massive stars is so dominant,
not even the pressure of degenerate electrons is able to stop
the collapse. The iron nuclei in the center of the star break up
because of the high pressure and temperature. This process is
called photo dissociation,
\be
\gamma + {}^{56}\mathrm{Fe} \longrightarrow 13 \alpha + 4
\mathrm{n} \; ,
\ee
costs energy, and the thermal pressure from inside reduces so
that the star collapses in approximately $0.1$ seconds.
Electrons are captured by protons so that matter in the core
mostly consists of neutrons,
\be
\label{inversebetadecay}
\mathrm{p} + \mathrm{e}^- \longrightarrow \mathrm{n} +
\nu_\mathrm{e} \; .
\ee
The pressure of these degenerate neutrons stops the collapse.
The process~(\ref{inversebetadecay}) is called inverse $\beta$
decay in which a huge amount of neutrinos is produced. These
neutrinos are trapped for a while in the collapsed hot star core
which is called a protoneutron star. The outer shells fall down
on this protoneutron star, bounce back and thereby produce an
outgoing shock wave which is seen as a supernova explosion of
type~II in which nuclei with mass numbers $A > 56$ are formed.
\begin{figure}[H]
\begin{center}
\includegraphics[width=0.7\textwidth,angle=90]{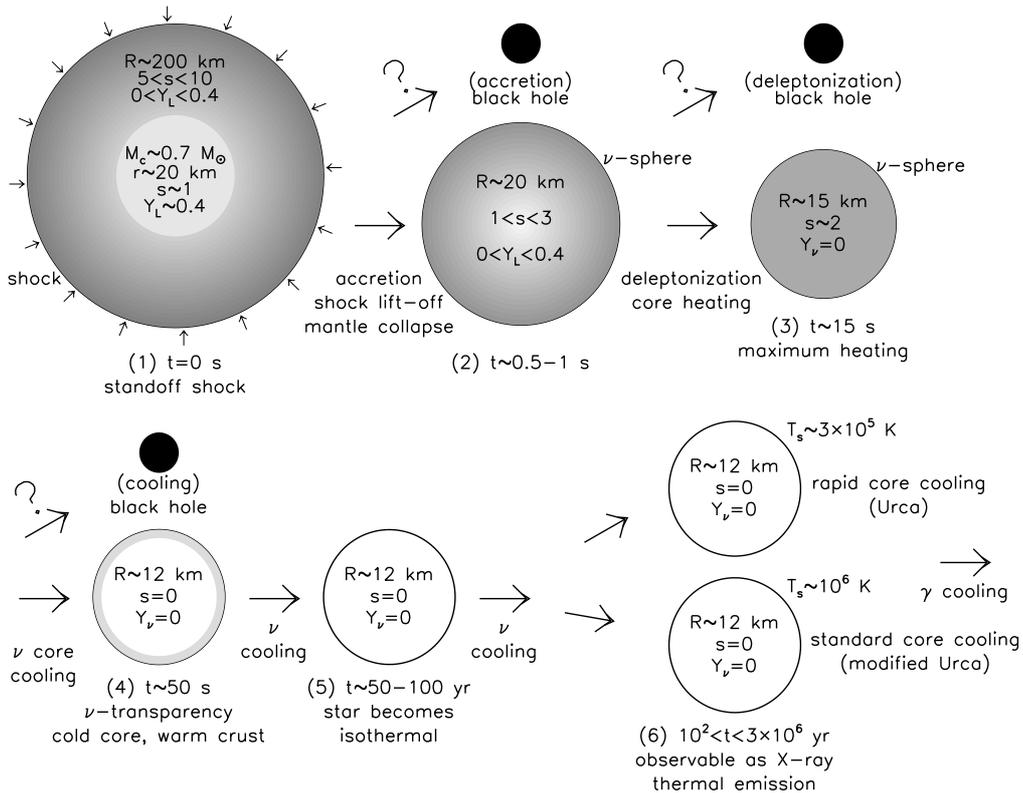}
\caption[The evolution of neutron stars.]{The main stages of the
evolution of neutron stars~\cite{PNS2}. Shading indicates
approximate relative temperatures.}
\label{pns}
\end{center}
\end{figure}
The evolution of protoneutron stars goes through several
stages~\cite{Burrows, PNS1, PNS2} as is shown in Fig.~\ref{pns}.
During the supernova explosion, there goes a shock through the
outer mantle of the protoneutron star. The outer mantle is of
low density and high entropy, accretes matter and loses energy
by $\beta$ decays and neutrino emission. The core has a mass of
about 0.7~$\msun$ in which neutrinos are trapped. The lepton to
baryon fraction is $Y_L \simeq 0.4$. The whole protoneutron star
in this first stage of evolution has an approximate radius of
about 200 kilometers. After approximately five seconds,
accretion becomes less important. The mantle collapses because
of the loss of lepton pressure caused by deleptonization. In
this stage of evolution, the lepton to baryon fraction is in the
range $0 < Y_L < 0.4$, and the radius amounts to approximately
20 kilometers. If a lot of matter accretes onto the protoneutron
star so that it exceeds its maximum mass then the protoneutron
star collapses to a black hole. At approximately 15 seconds
after the supernova explosion, the protoneutron star is
dominated by neutrino diffusion causing deleptonization and
heating of the protoneutron star. In this stage of evolution,
the lepton to baryon fraction is $Y_L = 0$, the radius of the
protoneutron star is approximately 15 kilometers, and its
temperature is heating up to 30~MeV~$\lesssim T
\lesssim$~60~MeV. There is also the possibility of forming a
black hole by deleptonization. Fifty seconds after the supernova
explosion, the protoneutron star becomes transparent to
neutrinos so that the inner part of the star cools down. But the
crust remains warm because of its lower neutrino emissivity, $T
\simeq 3 \cdot 10^6$~K. After 50--100 years, also the crust
cools down by neutrino emission and the star becomes isothermal.
In later stages, the star cools down by direct URCA processes,
\be
\mathrm{n} \longrightarrow \mathrm{p} + \mathrm{e}^- +
\bar\nu_\mathrm{e} \; , \qquad
\mathrm{p} + \mathrm{e}^- \longrightarrow \mathrm{n} +
\nu_\mathrm{e} \; ,
\ee
or modified URCA processes,
\be
\mathrm{n} + \mathrm{n} 
\longrightarrow \mathrm{n} + \mathrm{p} + \mathrm{e}^- + \bar
\nu_\mathrm{e} \; , \qquad
\mathrm{n} + \mathrm{p} + \mathrm{e}^- 
\longrightarrow \mathrm{n} + \mathrm{n} + \nu_\mathrm{e} \; ,
\ee
neutrino and photo emission. A cold neutron star has been
formed.
\subsection{Pulsars}
Before neutron stars were discovered, theorists speculated about
the existence of neutron stars. In 1932, Landau called them
weird stars. In 1934, Baade and Zwicky realized that there is a
connection between supernovae of type~II and neutron stars. The
first neutron star calculations were done by Tolman,
Oppenheimer, and Volkoff in 1939 who created mass-radius
diagrams of neutron stars~\cite{TOV}. Further work has been done
by Wheeler et al.\ (1960--1966) and Pacini (1967).

\begin{figure}[H]
\begin{center}
\includegraphics[width=0.5\textwidth]{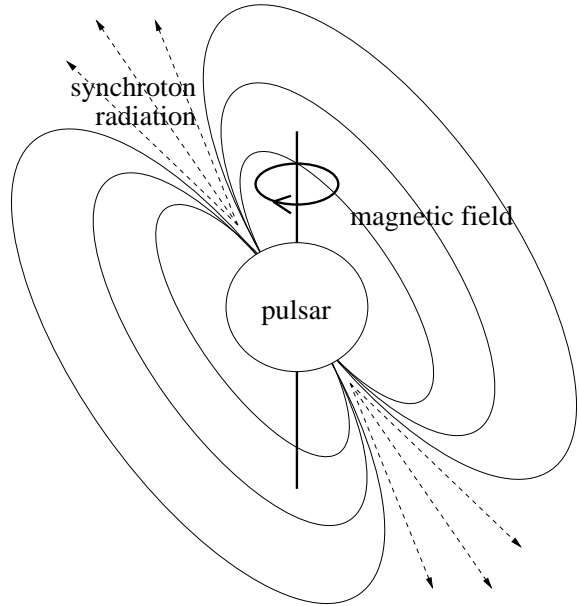}
\caption[A schematic representation of a pulsar.]{A schematic
representation of a pulsar~\cite{diploma_thesis_Ruester}.}
\label{pulsar}
\end{center}
\end{figure}

In the summer of 1967 in Cambridge/England, Jocelyn Bell, a
student of Anthony Hewish who got the Nobel prize in 1974,
detected a neutron star as a pulsar for the first time. Neutron
stars can be observed as pulsars which are \textbf{pulsa}ting
sources of \textbf{r}adiation because they can be identified
by their very precise radio pulses. Pulsars possess a strong
magnetic field of about 10$^{12}$~G in which highly energetic
electrons gyrate and thereby produce synchroton radiation which
is emitted at the magnetic poles of the pulsar, see
Fig.~\ref{pulsar}. Usually, the rotation axis of a pulsar is
inclined to the axis of the magnetic field. Thereby, the cone of
the synchroton radiation can be detected only once in a rotation
period of the pulsar. This is the pulsation phenomenon of
pulsars which is also called the lighthouse effect.

From observations with radio telescopes one knows that pulsars
have periods in the range of 1.6 milliseconds to several
seconds. They rotate so fast because of angular-momentum
conservation. In Sec.~\ref{star_formation}, it was mentioned
that stars (gas balls) rotate. This rotation is kept during the
life of the stars. When red super giants collapse to pulsars,
the rotation velocity increases very much. The pulsar periods
are very stable and therefore increase not much in time. The
Crab pulsar for example has a rotation period of 33 milliseconds
and this changes only about 0.036\% per year. From the present
pulsar period $P$ and its time derivative one is able to
determine the characteristic age $\tau = P / \dot{P}$ of a
pulsar. Because of the rapid rotation, a pulsar has an oblate
shape and therefore is not exactly spherically symmetric.
Observations like gravitational redshift measurements and mass
determinations in binary systems as well as theoretical
calculations show that the masses of neutron stars or pulsars,
respectively, approximately amount to 1.5~$\msun$, and that they
have radii of about ten kilometers. Therefore such compressed,
massive objects have an unbelievable mass density which is of
about 10$^{14}$~g/cm$^3$. A further observation in pulsars are
so-called glitches, sudden small jumps in the rotation period of
pulsars. They are most probably caused by vortices and
rearrangements in the crust of the pulsar. Because of that, it
comes to a decrease of the angular momentum in the superfluid
and an increase of the angular momentum in the
crust~\cite{Alford_Bowers_Rajagopal}. Pulsars show proper
motions which originate in so-called pulsar kicks. A possible
explanation is that they are created if neutrinos emit
asymmetrically during the supernova explosion which leads to a
propulsion of the pulsar. Another observation in pulsars of
binary systems are x-ray emissions and bursts: mass is
transferred from companion stars onto the accretion disc of the
pulsars. Because of the strong magnetic fields of the pulsars,
matter from the accretion disc is diverted to the poles of the
magnetic field. At this places, nuclear fusion processes are
initiated, which emit x-rays.
\subsection{Structure of neutron stars}
\label{Structure_of_neutron_stars}
\begin{figure}[H]
\begin{center}
\includegraphics[width=0.9\textwidth]{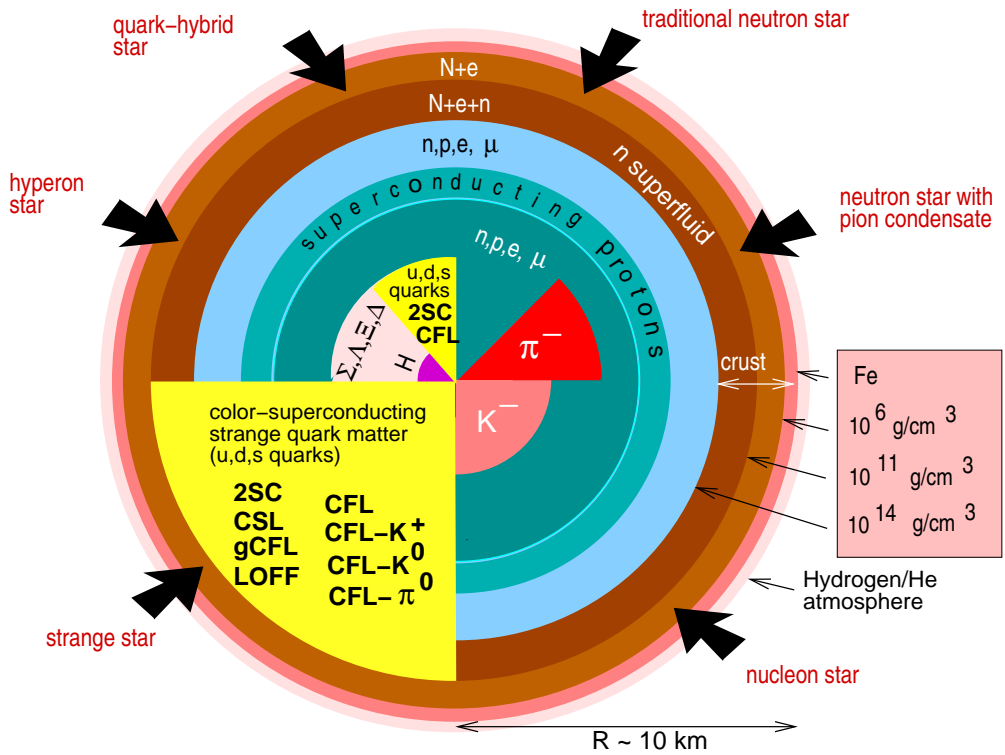}
\caption[Cross-sections of neutron stars.]{Cross-sections of
neutron stars (schematic)~\cite{Weber}. Each sector shows
another possible type of neutron stars.}
\label{ns_structure}
\end{center}
\end{figure}

In Sec.~\ref{Neutron_stars}, it was shown that neutron stars
consist of neutrons because of inverse $\beta$ decays
(\ref{inversebetadecay}). This does not mean that neutron stars
only consist of neutrons. For different mass densities, there
are different phases of matter inside neutron stars, see
Fig.~\ref{ns_structure}.

Neutron stars consist of an atmosphere of electrons, nuclei, and
atoms. Only a fraction of the electrons are bound to nuclei. The
equation of state was calculated by Feynman, Metropolis, and
Teller~\cite{FMT}. They found out that the nuclei in this
regime, where the mass density $\rho\lesssim 10^4$~g/cm$^3$, are
mainly $^{56}$Fe.

In neutron stars with temperatures above typically 100~eV,
between the atmosphere and the solid crust a layer is present,
where nuclei and electrons are in a liquid phase called the
ocean~\cite{ocean}.

One assumes complete ionization of the atoms, when the spacing
between the nuclei becomes small compared to the Thomas-Fermi
radius $r_\mathrm{TF} \simeq a_0 Z^{- 1/3}$ of an isolated
neutral atom. In this equation, $a_0$ is the Bohr radius and $Z$
the charge number. The mass density approximately amounts to
$\rho \simeq A m_u n_N$, where $A$ is the mass number, $m_u$ the
atomic mass unit, and $n_N$ the number density of nuclei, which
depends on the radius of a spherical nucleus whose volume is the
average volume per nucleus~\cite{Ravenhall}, $\frac43 \pi r_c^3
= 1 / n_N$. By combining the last three equations, one finds
that the outer crust of cold neutron stars begins when $\rho
\sim 10^4$~g/cm$^3$~$\gg 3 A Z$~g/cm$^3$. This shell consists of
nuclei and free electrons. The equation of state was originally
calculated by Baym, Pethick, and Sutherland (BPS)~\cite{BPS} and
improved in Refs.~\cite{Haensel, Ruester_crust} using up to date
nuclear data. The BPS model is valid for zero temperature which
is a good approximation for the crust of nonaccreting cold
neutron stars. The outer crust of nonaccreting cold neutron
stars contains nuclei and free electrons. The latter become
relativistic above $\rho \sim 10^7$~g/cm$^3$. The nuclei are
arranged in a \textbf{b}ody-\textbf{c}entered \textbf{c}ubic
(bcc) lattice. The contribution of the lattice has a small
effect on the equation of state but it changes the equilibrium
nucleus to a larger mass number and lowers the total energy of
the system because it will minimize the Coulomb interaction
energy of the nuclei. The latter are stabilized against $\beta$
decay by the filled electron sea. At $\rho \sim 10^4$~g/cm$^3$,
$^{56}$Fe is the true ground state. With increasing mass
density, it is not the true ground state any more because the
nuclei capture electrons, emit neutrinos and become neutron
richer. When the mass density $\rho \simeq 4.3 \cdot
10^{11}$~g/cm$^3$, the so-called neutron drip line is reached.
Neutrons begin to drip out of the nuclei and become free. This
happens because the equilibrium nuclei become more and more
neutron-rich, and finally no more neutrons can be bound to
nuclei. At the neutron drip point, the inner crust of neutron
stars begins. At $\rho \simeq 2 \cdot 10^{14}$~g/cm$^3$ nuclei
do not exist anymore, signalling the end of the neutron star
crust.
The equation of state of the inner crust was calculated
by Baym, Bethe, and Pethick~\cite{BBP} and another equation of
state for this regime was derived by Negele and
Vautherin~\cite{NV}. Also a relativistic mean field model has
been used to describe the density regime of the neutron star
crust within the Thomas-Fermi approximation (see
Ref.~\cite{Shen} and references therein). For higher densities,
the nuclei disintegrate and their constituents, the protons and
neutrons, become superfluid. Muons also begin to appear in these
shells. The equation of state of this regime can be calculated
by using non-relativistic many-body theories~\cite{Akmal98} or
relativistic nuclear field theories~\cite{Walecka, RMF,
Hanauske00, SZ02}. What kind of matter exists in the cores of
neutron stars depends on their central densities. Neutron stars
with lower central densities consist of protons, neutrons,
electrons, and muons while others with larger central densities
can consist of hyperons, a pion or kaon condensate. In neutron
stars with huge central densities, $n \approx 10 n_0$, even the
protons, neutrons, and hyperons can disintegrate into their
constituents: quarks which are deconfined. Such neutron stars
contain a quark core~\cite{Weber} whose true ground state is
strange quark matter~\cite{sqmatter} so that not only the light
up and down quarks but also the strange quarks occur if their
mass is low compared to the strange quark chemical potential.
The remaining three quark flavors are too heavy to participate
in the quark matter of neutron stars. Because of the dominant
attractive interaction in the antitriplet channel, the true
ground state of strange quark matter in the cores of neutron
stars is color-superconducting strange quark
matter~\cite{CSCReviewRischke, lecturesShovkovy,
CSCReviewRajagopal_Wilczek, CSCReviewHong, CSCReviewAlford,
CSCReviewSchaefer, CSCReviewRen, CSCReviewMei}.
\begin{figure}[H]
\begin{center}
\includegraphics[width=0.5\textwidth]{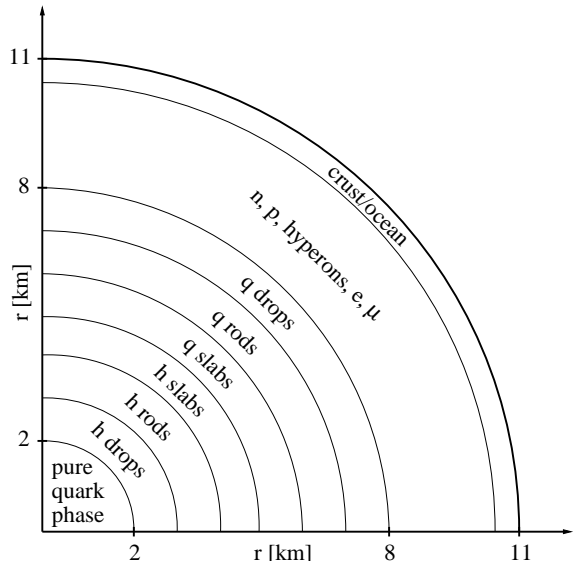}
\caption[The quark-hadron phase transition in neutron
stars.]{A schematic representation of the quark-hadron phase
transition in neutron stars with a mass of about
1.4~$\msun$~\cite{Glendenning}.}
\label{qhpt}
\end{center}
\end{figure}
In nature, phase transitions can take place either through sharp
boundaries between pure phases which are located next to each
other or through mixed-phase regions. Thus, it does not
necessarily mean that the phase transitions in neutron stars
have to take place as sudden as represented in
Fig.~\ref{ns_structure}. In Fig.~\ref{qhpt}, one can see that
the quark-hadron phase transition is not separated into two
distinct shells but that there is a smooth transition from the
hadronic phase into the quark phase. This transition goes in
several stages, and by the effects of Coulomb forces and surface
tensions, some interesting structures are formed. Quarks at
first appear in a structure of drops in a shell under the pure
hadronic phase in which the amount of hadrons exceeds the amount
of quarks. With increasing density, quarks form a structure of
rods, and finally a structure of slabs and are surrounded by
hadronic matter. At higher density, the amount of quarks exceeds
the amount of hadrons which form a structure of slabs. With
increasing density, hadrons form a structure of rods, and
finally a structure of drops that are surrounded by quark
matter. Ultimately, the cores of neutron stars with huge central
densities consist of pure quark matter.
\subsection{Properties of neutron star matter}
The matter in neutron stars is in its ground state. It is in
nuclear equilibrium which means that the energy cannot be
lowered by strong, weak, or electromagnetic interactions. When
matter is in equilibrium concerning weak interactions, one calls
it $\beta$-equilibrated  matter or one says that matter is in
$\beta$ equilibrium. It means that the reactions for (inverse)
$\beta$ decay~(\ref{inversebetadecay}) or the reactions for weak
interactions, respectively, are in equilibrium for all lepton
families,
\be
\label{betaequilibrium}
\mathrm{p} + \beta^- \longleftrightarrow \mathrm{n} +
\nu_\beta \; .
\ee
In this equilibrated reaction, $\beta \equiv \left( e, \mu, \tau
\right)$. The leptons are listed in Table~\ref{leptons}. They
are spin-$\frac12$ particles, and therefore fermions.
\begin{table}[H]
\begin{center}
\begin{tabular}{|l||c|r|r|}
\hline
Lepton & Mass [MeV] & $Q$ [e] & $g$ \\
\hline
\hline
electron (e)                 & $0.51099906 \pm 0.00000015$ &
$-1$ & $2$ \\
electron neutrino ($\nu_e$)  & $< 7.3 \cdot 10^{-6}$       &
$ 0$ & $1$ \\
\hline
muon ($\mu$)                 & $105.658389 \pm 0.000034$   &
$-1$ & $2$ \\
muon neutrino ($\nu_\mu$)    & $< 0.27$                    &
$ 0$ & $1$ \\
\hline
$\tau$ lepton ($\tau$)       & $1776.3 \pm 2.4$            &
$-1$ & $2$ \\
$\tau$ neutrino ($\nu_\tau$) & $< 31$                      &
$ 0$ & $1$ \\
\hline
\end{tabular}
\end{center}
\caption[The masses, electric charges, and degeneracy factors of
the leptons.]{The masses, electric charges, and degeneracy
factors $g$ of the leptons~\cite{Taschenbuch_der_Physik}. The
abbreviation for the respective lepton is shown in brackets.
There also exists the corresponding antiparticle for each lepton
with charge $-Q$. They are not shown here.}
\label{leptons}
\end{table}
Since the quark flavor content of protons and neutrons is
\be
\label{qc_neutron}
\mathrm{p = (u,u,d)} \; , \qquad
\mathrm{n = (u,d,d)} \; ,
\ee
one can express the reaction~(\ref{betaequilibrium}) in terms of
quark flavors,
\be
\label{betaequilibrium2}
\mathrm{u} + \beta^- \longleftrightarrow \mathrm{d} +
\nu_\beta \; .
\ee
By weak interactions, also the transformation from up into
strange quarks is possible,
\be
\label{betaequilibrium3}
\mathrm{u} + \beta^- \longleftrightarrow \mathrm{s} +
\nu_\beta \; .
\ee
Neutrinos carry only lepton number. That is why the chemical
potential of each neutrino family is equal to the chemical
potential of each lepton family,
\be
\label{neutrino_lepton}
\mu_{\nu_\beta} \equiv \mu_{L_\beta} \; .
\ee
From the equilibrated reactions~(\ref{betaequilibrium2})
and~(\ref{betaequilibrium3}), one can directly write down the
equation for the corresponding chemical potentials,
\be
\mu_u + \mu_\beta = \mu_{d,s} + \mu_{L_\beta} \; ,
\ee
where $\mu_{d,s}$ means that $\mu_s = \mu_d$, a fact that
directly comes out of the equilibrated
reactions~(\ref{betaequilibrium2}) and~(\ref{betaequilibrium3}).
The last equation can be rewritten as
\be
\label{mus}
\mu_u = \mu_{d,s} + \mu_Q \; ,
\ee
where
\be
\label{muQ}
\mu_Q \equiv \mu_{L_\beta} - \mu_\beta
\ee
is the chemical potential of electric charge. This means that
electrons, muons, and $\tau$ leptons carry both, lepton number
and electric charge,
\be
\label{mu_beta}
\mu_\beta = \mu_{L_\beta} - \mu_Q \; .
\ee
Because of the quark content of the neutron~(\ref{qc_neutron}),
one can define the baryon chemical potential
\be
\mu_B \equiv 3 \mu \equiv \mu_n = \mu_u + 2 \mu_d \; ,
\ee
where $\mu$ is the quark chemical potential and $\mu_n$ the
chemical potential of neutrons. This relation together with the
fact that $\mu_s = \mu_d$ is used to solve the
equation~(\ref{mus}) for each quark flavor. One obtains
\be
\label{quark_flavor_beta_equilibrium}
\mu_{ff'} = \mu \delta_{ff'} + \mu_Q Q_{ff'} \; ,
\ee
where
\be
\label{matrix_of_electric_charge}
Q = \diag \left(\frac23, -\frac13, -\frac13 \right)
\ee
is the matrix of electric charge in flavor space.

With Eq.~(\ref{quark_flavor_beta_equilibrium}), one
automatically satisfies $\beta$ equilibrium in normal quark
matter where the color symmetry $[SU(3)_c]$ is not broken. This
is not the case in color-superconducting quark matter. There,
one has to know the quark chemical potential of each quark color
and flavor. In order to satisfy $\beta$ equilibrium for each
quark color and flavor, one starts with the equation for $\beta$
equilibrium for each quark
flavor~(\ref{quark_flavor_beta_equilibrium}) and adds to it the
terms for each color which consist of color chemical potentials
and the generators of the $[SU(3)_c]$ group~\cite{colorNJL},
\be
\label{hatmu8}
\mu_{ff'}^{ii'} = \left( \mu \delta_{ff'}
+ \mu_Q Q_{ff'} \right) \delta^{ii'}
+ \delta_{ff'} \sum_{c = 1}^8 \mu_c T_c^{ii'} \; .
\ee
In this equation, the color indices $i$ and $i'$ are
superscripted while the flavor indices $f$ and $f'$ are
subscripted. But not all of the eight color chemical potentials
are nonzero. This can be proven by calculating the tadpoles and
was done for the 2SC phase in Ref.~\cite{Gerhold_Rebhan} where
only a nonzero color chemical potential $\mu_8$ is present. I
extended this calculation for the color-superconducting phases
which I had to investigate for this thesis and got the result
that there are nonzero color chemical potentials $\mu_3$ and
$\mu_8$. Therefore, the chemical potential for each color and
flavor can be simplified to
\be
\label{quark_color_and_flavor_beta_equilibrium}
\mu_{ff'}^{ii'} = \left( \mu \delta_{ff'}
+ \mu_Q Q_{ff'} \right) \delta^{ii'}
+ \, ( \mu_3 T_3^{ii'} + \mu_8 T_8^{ii'} ) \, \delta_{ff'}
\ee
for the purposes of this thesis. Later on, I omit the double
color and flavor indices of the quark chemical potential
matrix~(\ref{quark_color_and_flavor_beta_equilibrium})
and denote $\mu_{ff'}^{ii'}$ by $\mu_f^i$ because the quark
chemical potential
matrix~(\ref{quark_color_and_flavor_beta_equilibrium}) is
diagonal in color-flavor space. Later on, also the double flavor
indices of the matrix of electric
charge~(\ref{matrix_of_electric_charge}) will be omitted because
it is diagonal in flavor space, e.g.\ $Q_{ff'}$ is denoted by
$Q_f$. Also, the double color indices of the generators of the
$[SU(3)_c]$ group, $T_3^{ii'}$ and $T_8^{ii'}$, can be omitted
because they are diagonal in color space, e.g.\ $T_3^{ii'}$ can
be denoted by $T_3^i$, and $T_8^{ii'}$ can be denoted by
$T_8^i$.

Stars are bound by gravity and have to be electrically charge
neutral, otherwise they would be unstable and explode because of
repulsive Coulomb forces. The number density of electrically
charged particles is given by
\be
\label{n_Q_introduction}
n_Q = \langle \psi^\dagger Q \psi \rangle - n_e - n_\mu - n_\tau
\; ,
\ee
where $\psi$ is the quark spinor, and $n_e$, $n_\mu$, and
$n_\tau$ are the number densities of electrons, muons, and
$\tau$ leptons, respectively. The electrical charge neutrality
condition,
\be
\label{electric_neutrality}
n_Q \equiv \frac{\partial p}{\partial \mu_Q} = 0 \; ,
\ee
demands that the number density of electrically charged
particles $n_Q$, which can be calculated by taking the
derivative of the pressure $p$ with respect to the chemical
potential of electric charge, is equal to zero.

Stars without color-superconducting quark matter are
automatically color neutral because the $[SU(3)_c]$ color
symmetry is not broken. This is not the case in
color-superconducting quark matter. If stars consist of
color-superconducting quark matter, then they have to be color
neutral because on the one hand, one is not able to observe
color charges in nature, on the other hand, stars will not be
stable if they are not color neutral. In the following, I show
that the color number densities $n_3$ and $n_8$ have to be equal
to zero in order to fulfill color neutrality.

The spinor of quark colors is defined by
\be
\psi = \left( \psi_r, \psi_g, \psi_b \right)^T \; .
\ee
Herewith, the number densities of quarks read,
\be
n_r \equiv \langle \psi_r^\dagger \psi_r \rangle \; , \qquad
n_g \equiv \langle \psi_g^\dagger \psi_g \rangle \; , \qquad
n_b \equiv \langle \psi_b^\dagger \psi_b \rangle \; ,
\ee
so that
\bsub
\label{n_n3_n8}
\bea
n &\equiv& \langle \psi^\dagger \psi \rangle = n_r + n_g + n_b
\; , \\
n_3 &\equiv& \langle \psi^\dagger T_3 \psi \rangle = \textstyle
\frac12 \displaystyle \left( n_r - n_g \right) \; , \\
n_8 &\equiv& \langle \psi^\dagger T_8 \psi \rangle = \textstyle
\frac{1}{2\sqrt{3}} \displaystyle \left( n_r + n_g -2 n_b
\right) \; .
\eea
\esub
I only need to consider the color number densities $n_3$ and
$n_8$ because there are only nonzero color chemical potentials
$\mu_3$ and $\mu_8$ in the color-superconducting phases which I
had to investigate for this thesis. In order to fulfill color
neutrality, equal number densities of red, green, and blue
quarks are necessary,
\be
n_r = n_g = n_b \; .
\ee
By inserting this into Eqs.~(\ref{n_n3_n8}), one obtains the
conditions for color neutrality,
\be
n_3 = n_8 = 0 \; .
\ee
The color neutrality conditions
\be
\label{color_neutrality}
n_3 \equiv \frac{\partial p}{\partial \mu_3} = 0 \; , \qquad
n_8 \equiv \frac{\partial p}{\partial \mu_8} = 0 \; ,
\ee
demand that the number densities of color charges $n_3$ and
$n_8$, which can be calculated by taking the derivative of the
pressure $p$ with respect to the corresponding color chemical
potential, are equal to zero.

In QCD, color neutrality is realized dynamically due to the
generation of gluon condensates $\langle A_0^3 \rangle \neq 0$
and $\langle A_0^8 \rangle \neq 0$~\cite{Gerhold_Rebhan,
Kryjevski, Dietrich_Rischke}. The appearance of such condensates
is equivalent to nonzero values of the corresponding color
chemical potentials $\mu_3$ and $\mu_8$.
\subsection{Toy models of neutral normal quark matter}
\label{toy_model_NQ}
In this thesis, I present the phase diagram of neutral quark
matter. This will be done in Chapter~\ref{phase_diagram}. For a
better understanding of the properties of neutral quark matter,
it is advantageous to introduce some simple toy models.
Therefore, some formulae of thermodynamics and statistical
mechanics are needed~\cite{Jelitto, Greiner}. The pressure for
non-interacting massive fermions and antifermions at nonzero
temperature $T$ reads~\cite{lecturesRischke, Rischke_Thermo},
\be
\label{pressure1}
p = \frac{g T}{2 \pi^2} \int_0^\infty \ud k \, k^2
\left\{ \ln \left[ 1 + \exp \left( - \frac{ E - \mu }{ T }
\right) \right] + \ln \left[ 1 + \exp \left(- \frac{ E + \mu }{
T } \right) \right] \right\} \; ,
\ee
where $g$ is the degeneracy factor, $k \equiv |\fettu{k}|$ is
the momentum, $E = \sqrt{ k^2 + m^2 }$ is the relativistic total
energy, $m$ is the mass, and $\mu$ is the chemical potential of
the fermions. The first term corresponds to the pressure of
fermions while the second one corresponds to the pressure of
antifermions. By integration by parts, one obtains,
\be
\label{pressure2}
p = \frac{g}{6 \pi^2} \int_0^\infty \ud k \, \frac{k^4}{E}
\left[ n_F \left( \frac{ E - \mu}{ T } \right) + n_F \left(
\frac{E + \mu}{ T } \right) \right] \; ,
\ee
where
\be
\label{Fermi-Dirac}
n_F \left( x \right) \equiv \frac{1}{\e^x + 1} \; ,
\ee
is the Fermi-Dirac distribution function. At zero temperature,
the Fermi-Dirac distribution function becomes a Heaviside
function, cf.\ Eq.~(\ref{distribution_function}),
\be
\label{zeroT}
\lim_{T \rightarrow 0} n_F \left( \frac{x}{T} \right) =
\Heaviside \left( - x \right) \; ,
\ee
so that one has to integrate from $k = 0$ up to the Fermi
momentum,
\be
k_F \equiv \sqrt{ \mu^2 - m^2 } \; .
\ee
Herewith, one obtains from Eq.~(\ref{pressure2}) the pressure of
non-interacting massive fermions for zero temperature,
\be
\label{pressure3}
p = \frac{g}{6 \pi^2} \int_0^{k_F} \ud k \,
\frac{k^4}{E} \; .
\ee
Only the contribution from fermions survives. The pressure of
non-interacting massive fermions for zero temperature can also
be obtained by using the equation,
\be
\label{pressure4}
p = \frac{g}{2 \pi^2} \int_0^{k_F} \ud k \, k^2 \left( \mu - E
\right) \; .
\ee
This result can be calculated directly from the
pressure~(\ref{pressure1}) by using the relation,
\be
\label{zeroT2}
\lim_{T \rightarrow 0} \ln \left[ 1 + \exp \left( - \frac{x}{T}
\right) \right] = - x \, \Heaviside \left( - x \right) \; ,
\ee
or by integration by parts of Eq.~(\ref{pressure3}). The
integrals in Eqs.~(\ref{pressure3}) and~(\ref{pressure4}) have
an analytical solution,
\be
p = \frac{g}{24 \pi^2} \left[ k_F \mu^3 - \frac52 m^2 k_F \mu -
\frac32 m^4 \ln \left( \frac{m}{k_F + \mu} \right) \right] \; ,
\ee
while the integrals in Eqs.~(\ref{pressure1})
and~(\ref{pressure2}) can only be solved numerically. In the
limit of vanishing mass, one obtains from Eq.~(\ref{pressure2})
the pressure of non-interacting massless fermions and
antifermions at nonzero temperature,
\be
p = \frac{g}{24 \pi^2} \left( \mu^4 + 2 \pi^2 \mu^2 T^2 +
\frac{7}{15} \pi^4 T^4 \right) \; .
\ee
Details how to get this result are shown in
Sec.~\ref{massless_fermions_at_nonzero_T} in the Appendix. In
the limit of zero mass and zero temperature, the pressure of
fermions reads,
\be
p = \frac{g}{24 \pi^2} \mu^4 \; .
\ee
The number density of non-interacting fermions can be obtained
by
\be
n = \frac{\partial p}{\partial \mu} \; ,
\ee
which, in the case of massive fermions at nonzero temperature,
leads to the result,
\be
n = \frac{g}{2 \pi^2} \int_0^\infty \ud k \, k^2 \left[ n_F
\left( \frac{E -\mu}{T} \right) - n_F \left( \frac{E + \mu}{T}
\right) \right] \; .
\ee
The first term is the contribution of fermions while the second
one is the contribution of antifermions. The integral can only
be solved numerically, but in the case of zero temperature, one
obtaines an analytical result by using Eq.~(\ref{zeroT}), 
\be
n = \frac{g}{6 \pi^2} k_F^3 \; .
\ee
The number density of massless fermions and antifermions at
nonzero temperature reads
\be
n = \frac{g}{6 \pi^2} \left( \mu^3 + \pi^2 \mu T^2 \right)
\; ,
\ee
so that the number density of massless fermions at zero
temperature is
\be
\label{n_massless_T0}
n = \frac{g}{6 \pi^2} \mu^3 \; .
\ee

From these formulae of thermodynamics and statistical mechanics
one is able to construct simple toy models for quark matter. In
the following, I present toy models of non-interacting normal
quark matter in neutron stars. Toy models for
color-superconducting quark matter will be presented in
Sec.~\ref{CSCinNeutronStars}. If a protoneutron star consists of
normal quark matter, then one has to consider neutral
$\beta$-equilibrated quark matter at nonzero temperature.
Electrons and muons are present in quark matter in protoneutron
stars in order to make them electrically neutral. Charm, bottom,
and top quarks as well as $\tau$ leptons are too heavy so that
they do not exist in the cores of compact stars where the quark
chemical potential $\mu \simeq 500$~MeV. But in protoneutron
stars, neutrinos are trapped. They can be treated as massless in
good approximation. The pressure of a simple toy model of normal
quark matter in protoneutron stars reads
\bea
p_\mathrm{NQ} &=& \frac{3 T}{\pi^2} \sum_{f = u}^s
\int_0^\infty \ud k \, k^2 \left\{ \ln \left[ 1 + \exp \left( -
\frac{E_f - \mu_f}{T} \right) \right] + \ln \left[ 1 + \exp
\left( - \frac{E_f + \mu_f}{T} \right) \right] \right\}
\nonumber \\
&+& \frac{T}{\pi^2} \sum_{\beta = e}^\mu \int_0^\infty \ud k
\, k^2 \left\{ \ln \left[ 1 + \exp \left( - \frac{E_\beta -
\mu_\beta}{T} \right) \right] + \ln \left[ 1 + \exp \left( -
\frac{E_\beta + \mu_\beta}{T} \right) \right] \right\}
\nonumber \\
&+& \frac{1}{24 \pi^2} \sum_{\beta = e}^\mu \left(
\mu_{L_\beta}^4 + 2 \pi^2 \mu_{L_\beta}^2 T^2 + \frac{7}{15}
\pi^4 T^4 \right) - B \; ,
\eea
where the first line in this equation is the contribution of
quarks, the second line is the contribution of electrons and
muons, and the third line is the contribution of massless
electron and muon neutrinos to the pressure. In the third line,
also the bag pressure is subtracted. The relativistic energies
of quarks, electrons, and muons are given by $E_f = ( k^2 +
M_f^2 )^{1/2}$ and $E_\beta = (k^2 + m_\beta^2)^{1/2}$, where
$M_f$ and $m_\beta$ are the masses of the quark flavors ($f =u,
d, s$) and leptons ($\beta = e, \mu$). The chemical potentials
of the quark flavors and the leptons are denoted by
$\mu_f$~(\ref{quark_flavor_beta_equilibrium}) or $\mu_\beta$,
respectively, where
\be
\mu_\beta \equiv \mu_{L_\beta} - \mu_Q \; ,
\ee
cf.\ Eq.~(\ref{muQ}). Because of color symmetry and spin
degeneracy, the degeneracy factor of the quark contribution to
the pressure is $g = 2 N_c = 6$. The degeneracy factor of
electrons and muons is $g = 2$ because of spin degeneracy. For
neutrinos the spin is always opposite the momentum and this is
referred to as left-handed, whereas the antineutrinos are always
right-handed. That is why the degeneracy factor of neutrinos is
$g = 1$. The chemical potential of muon neutrinos can be set
equal to zero which is a good approximation for matter in
protoneutron stars. Electrical neutrality can be achieved by
using Eq.~(\ref{electric_neutrality}),
\be
\label{solve_for_muQ}
n_Q \equiv \frac{\partial p_\mathrm{NQ}}{\partial \mu_Q} =
\sum_{f = u}^s \frac{\partial
p_\mathrm{NQ}}{\partial \mu_f} \frac{\partial
\mu_f}{\partial \mu_Q} + \sum_{\beta = e}^\mu
\frac{\partial p_\mathrm{NQ}}{\partial \mu_\beta} 
\frac{\partial \mu_\beta}{\partial \mu_Q}
= \sum_{f = u}^s Q_f n_f - \sum_{\beta = e}^\mu n_\beta = 0 \; ,
\ee
where
\be
n_f = \frac{3}{\pi^2} \int_0^\infty \ud k \, k^2 \left[ n_F
\left( \frac{E_f - \mu_f}{T} \right) - n_F \left( \frac{E_f +
\mu_f}{T} \right) \right]
\ee
is the net number density of each quark flavor, and
\be
\label{number_density_beta}
n_\beta = \frac{1}{\pi^2} \int_0^\infty \ud k \, k^2 \left[ n_F
\left( \frac{E_\beta - \mu_\beta}{T} \right) - n_F \left(
\frac{E_\beta + \mu_\beta}{T} \right) \right]
\ee
is the number density of electrons or muons, respectively. As
one expects, the condition for electrical neutrality of
non-interacting normal quark matter in $\beta$
equilibrium~(\ref{solve_for_muQ}) is of the form $\sum_i Q_i n_i
= 0$, where $Q_i$ is the electric charge of the particle
species~$i$ and $n_i$ its number density. In order to have
neutral quark matter, for a given quark chemical potential
$\mu$, the chemical potential of electric charge $\mu_Q$ has to
be found by solving for it in Eq.~(\ref{solve_for_muQ}).

A simple toy model for cold normal quark matter in neutron stars
can be obtained by taking the limit $T \rightarrow 0$ in the
above simple toy model for protoneutron stars. Also, neutrinos
can be neglected because in cold neutron stars, they are not
trapped any more. Therefore, the pressure of a simple toy model
of cold normal quark matter in neutron stars reads
\be
p_\mathrm{NQ} = \frac{1}{\pi^2} \sum_{f = u}^s
\int_0^{k_{F_f}}
\ud k \, \frac{k^4}{E_f} + \frac{1}{3 \pi^2}
\sum_{\beta = e}^\mu \int_0^{k_{F_\beta}} \ud k \,
\frac{k^4}{E_\beta} \; ,
\ee
where $k_{F_f} \equiv ( \mu_f^2 - M_f^2 )^{1/2}$ and
$k_{F_\beta} \equiv ( \mu_\beta^2 - m_\beta^2 )^{1/2}$ are the
Fermi momenta of the quark flavors~$f$ and leptons~$\beta$,
respectively. Again, electrical neutrality can be achieved by
solving Eq.~(\ref{solve_for_muQ}) for the chemical potential
$\mu_Q$ for a given quark chemical potential $\mu$. But for $T
\rightarrow 0$, the expressions for the number density of each
quark flavor and for each lepton respectively simplify to
\be
\label{quark_flavor_and_lepton_number_densities}
n_f = \frac{k_{F_f}^3}{\pi^2} \; , \qquad
n_\beta = \frac{k_{F_\beta}^3}{3 \pi^2} \; .
\ee
In the limit of zero temperature, zero up and down quark masses,
and massless electrons, and by neglecting the contribution of
muons, one can solve Eq.~(\ref{solve_for_muQ}) for the chemical
potential of electric charge~\cite{Alford_Rajagopal},
\be
\label{muQsimple_with_s}
\mu_Q \simeq - \frac{M_s^2}{4 \mu} \; ,
\ee
where the Taylor expansion of the strange quark Fermi momentum,
\be
k_{F_s} = \sqrt{ \mu_s^2 - M_s^2 } \simeq
\mu_s - \frac{M_s^2}{2 \mu_s}
\simeq \mu_s - \frac{M_s^2}{2 \mu} \; ,
\ee
is used, and terms which are of higher order are neglected
because their contributions are small. If the contribution of
the strange quarks is neglected, one obtains,
\be
\label{muQsimple_without_s}
\mu_Q \simeq - \frac{\mu}{5} \; .
\ee
Inserting this into Eq.~(\ref{quark_flavor_beta_equilibrium})
and calculating the respective quark flavor number
densities~(\ref{quark_flavor_and_lepton_number_densities}) leads
to the result that there are nearly twice as many down quarks as
up quarks. The simple results for the chemical potential of
charge~(\ref{muQsimple_with_s}) and~(\ref{muQsimple_without_s})
are a good approximation for $\mu \gg M_s$ as one can see by
comparing it with Fig.~5 in Ref.~\cite{RuesterCSCstars}. One
also realizes that strange quarks help neutralizing quark matter
and therefore less electrons are needed. For the ideal but
unrealistic case of zero strange quark mass, no electrons are
present in neutral normal quark matter.
\section{Color superconductivity in neutron stars}
\label{Color_superconductivity_in_neutron stars}
From the statements made in
Sec.~\ref{Structure_of_neutron_stars}, one can expect strange
quark matter in the cores of neutron stars where the density is
so large that deconfined quark matter is able to exist. Because
of the dominant attractive interaction in the antitriplet
channel, quarks form Cooper pairs. The typical temperatures
inside (proto)neutron stars are so low that the diquark
condensate is not melted. That is why one expects not (only)
normal strange quark matter but even color-superconducting
strange quark matter in the cores of neutron stars.

The color-superconducting gap affects the transport properties,
e.g.\ conductivities and viscosities which have an influence on
the cooling rates and on the rotation period of neutron stars.
It also modifies the thermodynamic properties, e.g.\ the
specific heat and the equation of state which have an influence
on the mass-radius relation of color-superconducting neutron
stars~\cite{lecturesShovkovy}. In Ref.~\cite{RuesterCSCstars},
the effect of color superconductivity on the mass and the radius
of compact stars made of pure quark matter is investigated. The
authors confirmed the result of
Ref.~\cite{Blaschke_Fredriksson_Grigorian_Oeztas} that color
superconductivity does not alter the mass and the radius of 
quark stars, if the diquark-coupling constant is chosen to
reproduce vacuum properties such as the pion-decay constant. The
reason is that color superconductivity in neutral quark matter
has a tiny effect on the equation of state, cf.\ Figs.~5.9
and 5.10 in Ref.~\cite{diploma_thesis_Ruester}. The
color-superconducting gap has significant effects on the mass
and radius of quark stars only if the diquark-coupling constant
is artifically increased whereby the value of the gap itself
artificially increases. For gaps on the order of 300~MeV, the
mass and radius of quark stars are approximately twice as large
as for normal-conducting quark stars so that such quark stars
are of the same size and mass as ordinary neutron stars.
Therefore, it is impossible to decide whether a compact star
consists of normal conducting or color-superconducting quark
matter, or simply of hadronic matter.
\begin{figure}[H]
\begin{center}
\makebox[0.99\textwidth][l]{
\includegraphics[width=0.45\textwidth]{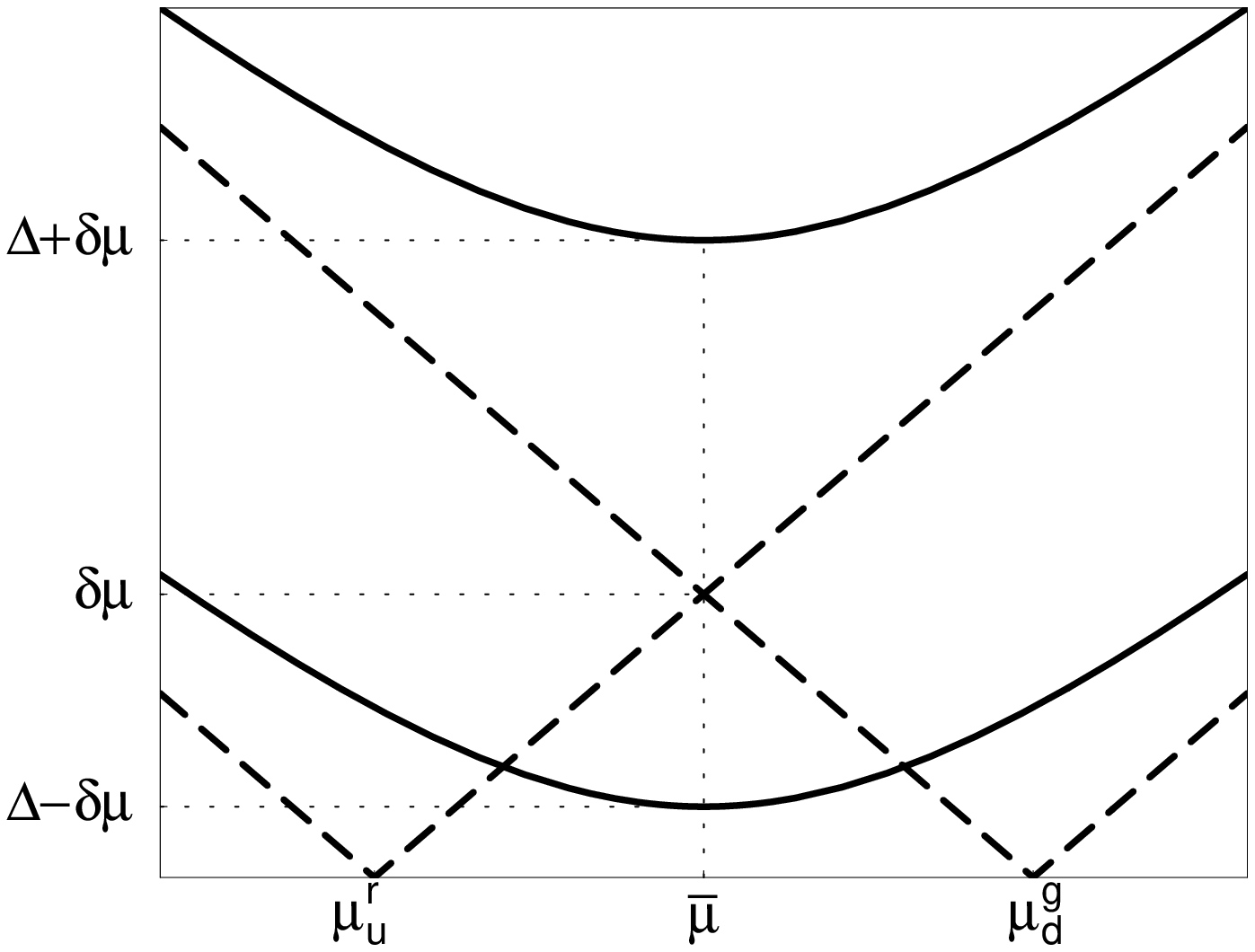}
\hspace{0.035\textwidth}
\includegraphics[width=0.45\textwidth]{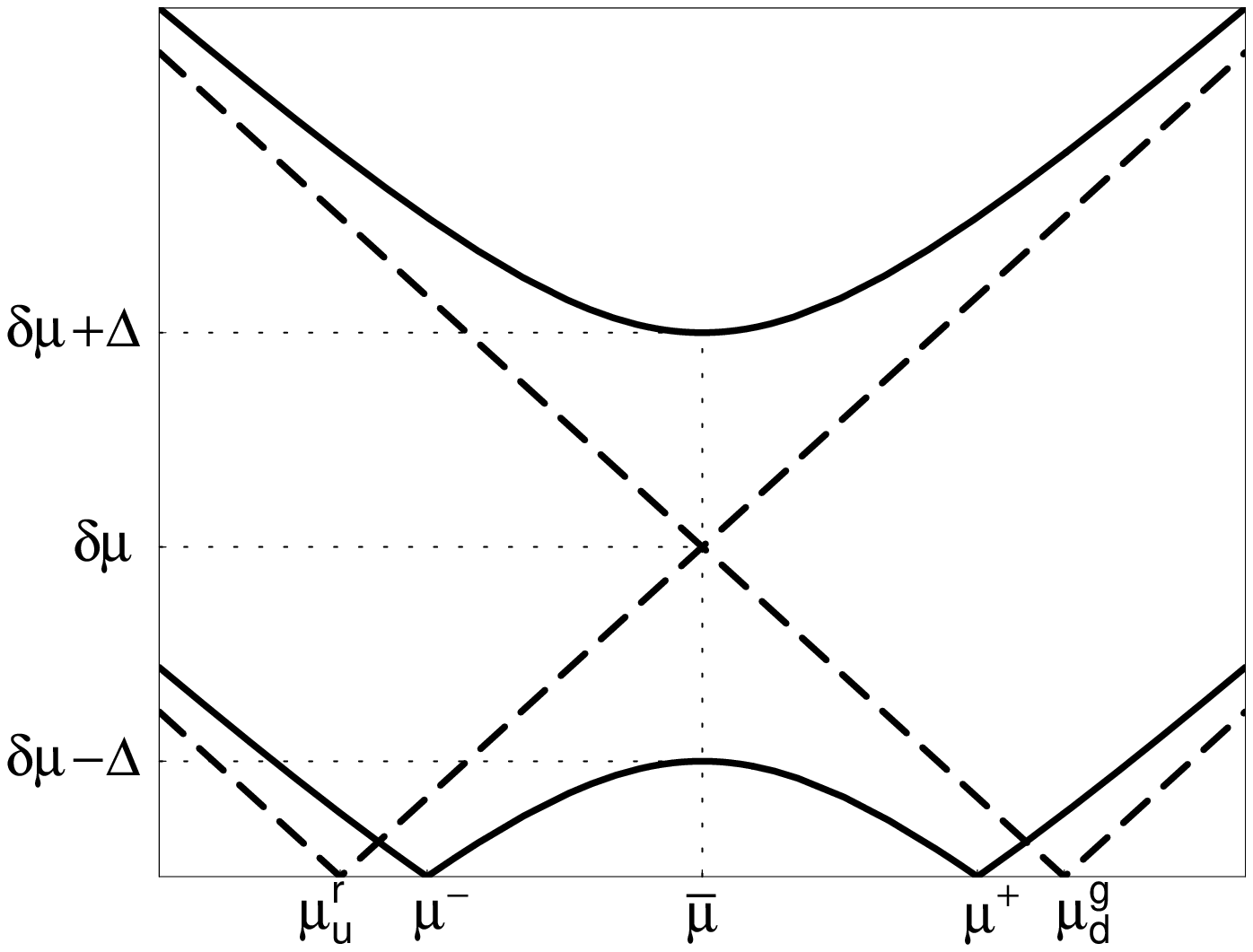}}
\caption[The low-energy part of the dispersion relations.]{The
low-energy part of the dispersion relations, i.e.\ the
quasiparticle energies as a function of the quark momentum. Left
panel: dispersion relations of quark quasiparticles in the 2SC
phase (solid lines) and the normal quark phase (dashed lines).
Right panel: dispersion relations of quark quasiparticles in the
g2SC phase (solid lines) and the normal quark phase (dashed
lines). The dispersion relation with the lowest energy in the
g2SC phase shows a gapless mode~\cite{lecturesShovkovy, g2SC_1,
g2SC_2}. The value of the averaged chemical potential of the up
and down quarks which pair is denoted by $\bar \mu$ and one half
of the difference of their chemical potentials is denoted by
$\delta \mu$. In the massless case, the chemical potentials of
the free up and down quarks are replaced by $\mu^\pm \equiv \bar
\mu \pm [ ( \delta \mu )^2 - \Delta^2 ]^{1/2}$.}
\label{gapless_modes_intro}
\end{center}
\end{figure}
In some cases, color superconductivity is accompanied by baryon
superfluidity or the electromagnetic Meissner effect. Baryon
superfluidity causes rotational vortices while the
electromagnetic Meissner effect entails magnetic flux tubes in
the cores of neutron stars.

At large strange quark masses, neutral two-flavor quark matter
in $\beta$ equilibrium can have another rather unusual ground
state called the gapless two-flavor color superconductor
(g2SC)~\cite{g2SC_1}. While the symmetry in the g2SC ground
state is the same as that in the conventional 2SC phase, the
spectrum of the fermionic quasiparticles is different, see
Fig.~\ref{gapless_modes_intro}. The g2SC phase appears at
intermediate values of the diquark coupling constant while the
2SC phase appears in the regime of strong diquark coupling.
Gapless modes are created if the mismatch between the Fermi
momenta of the quarks which pair becomes large. In the case of
the g2SC phase, $\delta \mu > \Delta$ where $\delta \mu \equiv
\mu_e /2 $. The existence of gapless color-superconducting
phases was confirmed in Refs.~\cite{RuesterCSCstars,
Gubankova_Liu_Wilczek, Mishra_and_Mishra}, and generalized to
nonzero temperatures in Refs.~\cite{g2SC_2, Liao_and_Zhuang}. It
is also shown that a gapless CFL (gCFL) phase appears in neutral
strange quark matter~\cite{Alford_Kouvaris_Rajagopal1,
Alford_Kouvaris_Rajagopal2}. But gapless color-superconducting
quark phases are unstable in some regions of the phase diagram
of neutral quark matter because of chromomagnetic
instabilities~\cite{chromomagnetic_instability} so that another
phase will be the preferred state. Chromomagnetic instabilities
occur even in regular color-superconducting quark phases. The
author of Ref.~\cite{Fukushima_unstable} shows that
chromomagnetic instabilities occur only at low temperatures in
neutral color-superconducting quark matter. The author of
Ref.~\cite{Huang} points out that the instabilities might be
caused by using BCS theory in mean-field approximation, where
phase fluctuations have been neglected. With the increase of the
mismatch between the Fermi surfaces of paired fermions, phase
fluctuations play more and more an important role, and soften
the superconductor. Strong phase fluctuations will eventually
quantum-disorder the superconducting state, and turn the system
into a phase-decoherent pseudogap state. By using an effective
theory of the CFL state, the author of Ref.~\cite{p_wave}
demonstrates that the chromomagnetic instability is resolved by
the formation of an inhomogeneous meson condensate. The authors
of Ref.~\cite{Gorbar} describe a new phase in neutral two-flavor
quark matter within the framework of the Ginzburg-Landau
approach, in which gluonic degrees of freedom play a crucial
role. They call it a gluonic phase. In this phase, gluon
condensates cure a chromomagnetic instability in the 2SC
solution and lead to spontaneous breakdown of the color gauge
symmetry, the $[ U(1)_\mathrm{em} ]$ and the rotational $SO(3)$
group. In other words, the gluonic phase describes an
anisotropic medium in which color and electric superconductivity
coexist. In Ref.~\cite{Giannakis}, it was suggested that the
chromomagnetic instability in gapless color-superconducting
phases indicates the formation of the
Larkin-Ovchinnikov-Fulde-Ferrell (LOFF) phase~\cite{LOFF} which
is discussed in Ref.~\cite{LOFF_quark} in the context of quark
matter. Other possibilities could be the formation of a spin-one
color-superconducting quark phase, a mixed phase, or a
completely new state. The authors of
Ref.~\cite{Shovkovy_Hanauske_Huang} suggest that a mixed phase
composed of the 2SC phase and the normal quark phase may be more
favored if the surface tension is sufficiently
small~\cite{Reddy_Rupak}. The authors of
Ref.~\cite{Alford_Rajagopal_Reddy_Wilczek} suggest a single
first-order phase transition between CFL and nuclear matter.
Such a transition, in space, could take place either through a
mixed phase region or at a single sharp interface with
electron-free CFL and electron-rich nuclear matter in stable
contact. The authors of
Ref.~\cite{Alford_Rajagopal_Reddy_Wilczek} constructed a model
for such an interface.
\subsection{Toy models of neutral color-superconducting quark
matter}
\label{CSCinNeutronStars}
In this subsection, I show some toy models for
color-superconducting quark matter. In these toy models, which
are valid for zero temperature, up and down quarks are treated
as massless, and the strange quark mass is incorporated via a
shift in the strange quark chemical potential which is a good
approximation. Neutrinos are not present in cold quark matter in
neutron stars so that $\mu_e \equiv - \mu_Q$. Also, muons are
not taken into account in the following toy models but electrons
which are treated as massless for simplicity. In the equations
which satisfy the neutrality conditions of these toy models,
terms of higher order will be neglected. The pressure of the toy
model for color-superconducting quark matter in the 2SC phase
reads~\cite{Alford_Rajagopal},
\bea
p_\mathrm{2SC} &=& \frac{1}{\pi^2} \sum_{i = r}^g
\sum_{f = u}^d \int_0^{\left( k_F \right)_f^i}
\ud k \, k^2 \left( \mu_f^i - E_f \right) \nonumber \\ 
&+& \frac{1}{12 \pi^2} \sum_{f = u}^d \left(
\mu_f^b \right)^4 + \frac{1}{3 \pi^2} \sum_{i = r}^b
\int_0^{\left( k_F \right)_s^i} 
\ud k \, \frac{k^4}{E_s} \nonumber \\ 
&+& \frac{\mu_e^4}{12 \pi^2} + \frac{\mu^2 \Delta^2}{\pi^2} - B
\; ,
\eea
where the first line is the contribution of gapped
quasiparticles, the second line is the contribution of free blue
up and free blue down quarks as well as free strange quarks, and
the third line is the contribution of massless electrons, the
pressure correction due to the four gapped quasiparticles, and
the bag pressure. The chemical potential for each quark color
and flavor is given by
Eq.~(\ref{quark_color_and_flavor_beta_equilibrium}), and their
Fermi momenta by
\be
\left( k_F \right)_f^i \equiv \sqrt{ ( \mu_f^i )^2 - M_f^2} \; .
\ee
Up and down quarks are treated as massless in this toy model,
and the strange quark Fermi momenta for each color will be
calculated by the simplified expression,
\be
\label{shift}
\left( k_F \right)_s^i = \sqrt{ ( \mu_s^i )^2 - M_s^2 } \simeq
\mu_s^i - \frac{M_s^2}{2 \mu_s^i} \simeq \mu_s^i -
\frac{M_s^2}{2 \mu} \; ,
\ee
as mentioned above. All quarks which pair have the same common
(averaged) Fermi momenta,
\be
k_F^\mathrm{common} \equiv k_{F_1} \equiv \left( k_F \right)_u^r
= \left( k_F \right)_d^g = k_{F_2} \equiv \left( k_F
\right)_d^r = \left( k_F \right)_u^g = \mu+\frac16 \mu_Q +
\frac{1}{2 \sqrt{3}} \mu_8 \; .
\ee
In the 2SC phase, the Fermi momenta $k_{F_1}$ and $k_{F_2}$ are
equal. Therefore, only one common Fermi momentum
$k_F^\mathrm{common}$ is used. The neutrality
conditions~(\ref{electric_neutrality})
and~(\ref{color_neutrality}) require~\cite{Alford_Rajagopal},
\be
\mu_Q \simeq - \frac{M_s^2}{ 2 \mu } \; , \qquad
\mu_3 = 0 \; , \qquad
\mu_8 \simeq 0 \; .
\ee
In the case without strange quarks, one obtains
\be
\mu_Q \simeq - \frac{\mu}{2} \; , \qquad
\mu_3 = 0 \; , \qquad
\mu_8 \simeq 0 \; .
\ee
These approximations are in good agreement with the exact
results for $\mu \gg M_s$, cf.\ Figs.~5 and 6 in
Ref.~\cite{RuesterCSCstars}. In this toy model, the pressure
difference of the 2SC phase to the normal quark phase
reads~\cite{Alford_Rajagopal}
\be
\delta p_\mathrm{2SC} \simeq \frac{16 \mu^2 \Delta^2 - M_s^4}{16
\pi^2}
\ee
so that the 2SC phase is preferred to the normal quark phase
when
\be
\Delta > \frac{M_s^2}{4 \mu} \; .
\ee

The pressure of the toy model for the CFL phase reads
\be
\label{CFL_toy}
p_\mathrm{CFL} = \frac{1}{\pi^2} \sum_{i = r}^b \sum_{f =
u}^s \int_0^{\left( k_F \right)_f^i} \ud k \, k^2 \left( \mu_f^i
- E_f \right) + \frac {\mu_e^4}{12 \pi^2} + \frac{3 \mu^2
\Delta^2}{\pi^2} - B \; ,
\ee
where the first term is the contribution of the nine gapped
quarks, the second term is the contribution of electrons, the
third term is the correction to the pressure due to the gap, and
the fourth term is the bag pressure. As in the 2SC phase, the
quarks which pair have the same Fermi momenta,
\bsub
\label{common_Fermi_momenta_CFL}
\bea
k_{F_1} &\equiv& \left( k_F \right)_u^r = \left( k_F \right)_d^g
= \left( k_F \right)_s^b = \mu - \frac{M_s^2}{6 \mu} \; ,
\\
k_{F_2} &\equiv& \left( k_F \right)_d^r = \left( k_F \right)_u^g
= \mu + \frac16 \mu_Q + \frac{1}{2 \sqrt{3}} \mu_8 \; , \\
k_{F_3} &\equiv& \left( k_F \right)_s^r = \left( k_F \right)_u^b
= \mu + \frac16 \mu_Q + \frac14 \mu_3 - \frac{1}{4 \sqrt{3}}
\mu_8 - \frac{M_s^2}{4 \mu} \; , \\
k_{F_4} &\equiv& \left( k_F \right)_s^g = \left( k_F \right)_d^b
= \mu - \frac13 \mu_Q - \frac14 \mu_3 - \frac{1}{4 \sqrt{3}}
\mu_8 - \frac{M_s^2}{4 \mu} \; .
\eea
\esub
In the CFL phase, the neutrality condition for $\mu_3$,
Eq.~(\ref{color_neutrality}), requires that
\be
\mu_3 = - \mu_Q \; .
\ee
It is very useful to know that in the CFL phase the following
relation is valid:
\be
\label{CFLrelation}
n_Q - n_3 - \frac{1}{\sqrt{3}} n_8 + n_e = 0 \; ,
\ee
where $n_e$ is the number density of electrons. Because of the
neutrality conditions, $n_Q$, $n_3$, and $n_8$ have to be equal
to zero so that one concludes from Eq.~(\ref{CFLrelation}) that
also $n_e$ and with it $\mu_e$ have to be zero which is in
agreement with the arguments made in
Ref.~\cite{enforced_neutrality}. It means that in the CFL phase
there are neither electrons nor muons nor $\tau$ leptons
allowed, otherwise the neutrality conditions will be violated.
That is why one obtains the simple result
\be
\mu_Q = \mu_3 = 0 \; .
\ee
In order to satisfy the neutrality condition for $\mu_8$,
Eq.~(\ref{color_neutrality}), one gets,
\be
\label{mu8_CFL}
\mu_8 \simeq - \frac{M_s^2}{\sqrt{3} \mu} \; .
\ee
In this toy model, the pressure difference of the CFL phase to
the normal quark phase reads~\cite{Alford_Rajagopal,
Alford_Rajagopal_Reddy_Wilczek},
\be
\delta p_\mathrm{CFL} \simeq \frac{48 \mu^2 \Delta^2 - 3
M_s^4}{16 \pi^2}
\ee
so that the CFL phase is preferred to the normal quark phase
when
\be
\Delta > \frac{M_s^2}{4 \mu} \; .
\ee
This is the same requirement as for the 2SC phase, and because
the pressure difference of the CFL phase to normal quark matter
is three times larger than the pressure difference of the 2SC
phase to normal quark matter, the authors of
Ref.~\cite{Alford_Rajagopal} claim that the 2SC phase is absent
and the CFL phase is the preferred state in compact stars. In
Chapter~\ref{phase_diagram}, I shall show that this statement is
not correct in general so that these simple toy models cannot
replace a thorough analysis of the preferred quark phases by an
NJL model~\cite{NJL, Kunihiro_Hatsuda, Klevansky,
Hatsuda_Kunihiro, Buballa_Habilitationsschrift}
for example. That is why it is so important to construct the
phase diagram of neutral quark matter with a more precise model
which is done in Chapter~\ref{phase_diagram}.
\selectlanguage{american}
\chapter{The phase diagram of neutral quark matter}
\label{phase_diagram}
At sufficiently high densities and sufficiently low temperatures
quark matter is a color superconductor~\cite{Barrois, Frautschi,
Bailin_Love}. This conclusion follows naturally from arguments
similar to those employed in the case of ordinary
low-temperature superconductivity in metals and
alloys~\cite{BCS}. Of course, the case of quark matter is more
complicated because quarks, unlike electrons, come in various
flavors (e.g., up, down, and strange) and carry non-Abelian
color charges. This phenomenon was studied in detail by various
authors~\cite{Alford_Rajagopal_Wilczek, CFL_discoverers,
Shovkovy_Wijewardhana, Pisarski_Rischke1, Pisarski_Rischke2,
spin-1, Schmitt_Wang_Rischke, LOFF_quark, regular_self-energies,
Son, Schaefer2}. Many different phases were discovered, and
recent studies~\cite{g2SC_1, g2SC_2, Gubankova_Liu_Wilczek,
Mishra_and_Mishra, Liao_and_Zhuang, Alford_Kouvaris_Rajagopal1,
Alford_Kouvaris_Rajagopal2, Ruester_Shovkovy_Rischke, Iida, 
Fukushima, Shovkovy_Ruester_Rischke,
Ruester_Werth_Buballa_Shovkovy_Rischke1, Blaschke,
Ruester_Werth_Buballa_Shovkovy_Rischke2,
Abuki_Kitazawa_Kunihiro, Abuki_Kunihiro, Ruester_book} show that
even more new phases exist. (For reviews and lectures on color
superconductivity see Refs.~\cite{CSCReviewRischke,
lecturesShovkovy, CSCReviewRajagopal_Wilczek, CSCReviewHong,
CSCReviewAlford, CSCReviewSchaefer, CSCReviewRen,
CSCReviewMei}.)

In nature, the most likely place where color superconductivity
may occur is the interior of neutron stars. Therefore, it is of
great importance to study the phases of dense matter under the
conditions that are typical for the interior of stars. For
example, one should appreciate that matter in the bulk of a star
is neutral and $\beta$-equilibrated. By making use of rather
general arguments, it was suggested in
Ref.~\cite{Alford_Rajagopal} that such conditions favor the CFL
phase and disfavor the 2SC phase. In trying to refine the
validity of this conclusion, it was recently realized that,
depending on the value of the constituent (medium modified)
strange quark mass, the ground state of neutral and
$\beta$-equilibrated dense quark matter may be different from
the CFL phase. In particular, the g2SC phase~\cite{g2SC_1,
g2SC_2} is likely to be the ground state in the case of a large
strange quark mass. On the other hand, in the case of a
moderately large strange quark mass, the CFL and gCFL
phases~\cite{Alford_Kouvaris_Rajagopal1} are favored. At
nonzero temperature, $T\neq 0$, some other phases were proposed
as well~\cite{Iida}.

I note that the analysis in this thesis is restricted to locally
neutral phases only. This automatically excludes, for example,
crystalline~\cite{LOFF_quark} and mixed~\cite{Reddy_Rupak,
Shovkovy_Hanauske_Huang, Neumann} phases. Also, in the mean
field approximation utilized here, I cannot get any phases with
meson condensates~\cite{Bedaque, Kaplan_Reddy,
Kryjevski_Kaplan_Schaefer}.

In this chapter, I present the phase diagram of neutral quark
matter. In Sec.~\ref{The_phase_diagram_of_massless_quarks}, I
show the phase diagram of massless neutral three-flavor quark
matter. In
Sec.~\ref{phase_diagram_self-consistent}, quark masses are
treated self-consistently within the framework of a three-flavor
NJL model~\cite{Buballa_Habilitationsschrift}, and the phase
diagram with a self-consistent treatment of quark masses is
presented. In Sec.~\ref{phase_diagram_neutrino_trapping}, this
NJL model is extended to nonzero neutrino chemical potentials,
and the influence of neutrinos on the phase diagram is
discussed.

I use the following conventions: I calculate in natural units,
$\hbar = c = k_B = 1$, and utilize the Dirac definition of the
$\gamma$-matrices which are shown in Sec.~\ref{gamma_matrices}
in the Appendix. Latin indices run from one to three while Greek
indices run from zero to three. Four-vectors are denoted by
capital Latin letters while three-vectors are written in the
bold upright font. The space-time vector is defined as $X^\mu =
\left(x^0,x^1,x^2,x^3\right) = \left(t,\fettu{x}\right)$, the
four-momentum vector is denoted as $K^\mu = \left( k^0,
\fettu{k} \right)$, and the metric tensor is given by
$g_{\mu\nu} = \diag \left( 1,-1,-1,-1 \right)$. Absolute values
of vectors are denoted by italic Latin letters, e.g.\ $k =
|\fettu{k}|$. The direction of a vector is indicated by the hat
symbol, e.g.\ $\hat{\fettu{k}} = \fettu{k} / k$. I use the
imaginary-time formalism, i.e., the space-time integration is
defined as $\int_X=\int_0^{1/T} \ud \tau \int_V \ud^3
\fettu{x}$, where $\tau$ is the Euclidean time coordinate and
$V$ the three-volume of the system. The delta function is
defined as $\delta^{\left( 4 \right)} \left( X - Y \right)
\equiv \delta \left( \tau_X - \tau_Y \right) \delta^{\left( 3
\right)} \left( \fettu{x} - \fettu{y} \right)$. Energy-momentum
sums are written as: $T/V\sum_K = T \sum_n \int \ud^3 \fettu{k}
/ ( 2 \pi^3 ) = T/ ( 2 \pi^2 ) \sum_n \int \ud k \, k^2$, where
the sum runs over the Matsubara frequencies $\omega_n = 2 n \pi
T \equiv i k_0$ for bosons and $\omega_n = \left( 2 n + 1
\right) \pi T \equiv i k_0$ for fermions, respectively.
\section{The phase diagram of massless quarks}
\label{The_phase_diagram_of_massless_quarks}
In this section, I study the phase diagram of massless neutral
three-flavor quark matter at zero and nonzero temperature as a
function of the strange quark mass as well as a function of the
quark chemical potential. I start with the QCD Lagrangian
density in order to derive the QCD grand partition function.
Then, the Cornwall-Jackiw-Tomboulis (CJT) formalism~\cite{CJT}
is used to calculate the QCD effective action. The
gluon-exchange interaction between quarks is approximated by a
point-like four-fermion coupling. From the effective action of
quarks, I derive the pressure of color-superconducting quark
matter. In order to allow for the most general ground state, I
employ a nine-parameter ansatz for the gap
matrix~\cite{Ruester_Shovkovy_Rischke}. The effects of the
strange quark mass are incorporated by a shift of the chemical
potential of strange quarks, $\mu^i_s \rightarrow \mu^i_s -
m_s^2/ ( 2 \mu )$, where $i=r,g,b$ is the color index, $m_s$ is
the strange quark mass, and $\mu$ is the quark chemical
potential. This shift reflects the reduction of the Fermi
momenta of strange quarks due to their mass. Such an approach is
certainly reliable at small values of the strange quark mass. I
assume that it is also qualitatively correct at large values of
the strange quark mass. In order to draw the phase diagram of
massless neutral three-flavor quark matter, I solve nine gap
equations together with the conditions of electric and color
charge neutrality.
\subsection{Quantum chromodynamics}
The QCD Lagrangian density is given by
\be
\label{L_QCD}
\mathcal{L} = \bar \psi \left( i \covariant - \hat m \right)
\psi - \frac14 F_{\mu\nu}^a F_a^{\mu\nu} +
\mathcal{L}_\mathrm{gauge} \; .
\ee
For $N_c$ colors and $N_f$ flavors, $\psi$ is the $4 N_c
N_f$-dimensional spinor of quark fields, $\bar\psi \equiv
\psi^\dagger \gamma_0$ is the Dirac conjugate spinor,
$\covariant \equiv \gamma^\mu D_\mu$, where $\gamma^\mu$ are the
Dirac matrices, and $\hat m$ is the quark-mass matrix. The
covariant derivative is defined as
\be
\label{covariant_derivative}
D_\mu \equiv \partial_\mu - i g A_\mu^a T_a \; ,
\ee
where $g$ is the strong-coupling constant, $A_\mu^a$ are the
gluon fields, and $T_a$ are the generators of the $[SU(N_c)_c]$
group. The gluon field-strength tensor is defined as
\be
F_a^{\mu\nu} = \partial^\mu A_a^\nu - \partial^\nu A_a^\mu + g
f_{abc} A_b^\mu A_c^\nu \; ,
\ee
where $f_{abc}$ are the structure constants of the $[SU(N_c)_c]$
group. The first term in the QCD Lagrangian
density~(\ref{L_QCD}) corresponds to interacting quarks while
the second one corresponds to the gluons. The term
$\mathcal{L}_\mathrm{gauge}$ comprises gauge fixing terms and
the contribution from Faddeev-Popov ghosts. Up to irrelevant
constants, the grand partition function of QCD is given
by~\cite{Kapusta, LeBellac}
\be
\mathcal{Z} = \int \D \bar\psi \D \psi \D A_\mu^a \exp \left\{ I
\left[ \bar\psi, \psi, A \right] \right\} \; ,
\ee
where
\be
I \left[ \bar\psi, \psi, A \right] = \int_X \left( \mathcal{L} +
\mu \mathcal{N} + \mu_Q \mathcal{N}_Q + \mu_a \mathcal{N}_a
\right) \; ,
\ee
is the QCD action. The conserved quantities,
\be
\mathcal{N} \equiv \bar\psi \gamma_0 \psi \; , \qquad
\mathcal{N}_Q \equiv \bar\psi \gamma_0 Q \psi \; , \qquad
\mathcal{N}_a \equiv \bar\psi \gamma_0 T_a \psi \; ,
\ee
are the quark number density operator, the operator of electric
charge density of the quarks, and the operators of color charge
densities of the quarks, respectively. The QCD action can be
split into the part of interacting quarks, into the part of
gluons, and into the part of gauge fixing terms and
Faddeev-Popov ghosts,
\be
\label{QCD_action}
I \left[ \bar\psi, \psi, A \right] = I_{\psi_\mathrm{int}}
\left[ \bar\psi, \psi , A \right] + I_A \left[ A \right] +
I_\mathrm{gauge} \left[ A \right] \; ,
\ee
where
\bsub
\bea
I_{\psi_\mathrm{int}} \left[ \bar\psi, \psi, A \right] &=&
\int_{X,Y} \bar\psi \left( X \right) [ \mathcal{G}_0^+ ]^{-1}
\left( X,Y \right) \psi \left( Y \right) \; , \\
I_A \left[ A \right] &=& -\frac14 \int_X F_{\mu\nu}^a \left( X
\right) F_a^{\mu\nu} \left( X \right) \; , \\
I_\mathrm{gauge} \left[ A \right] &=& \int_X
\mathcal{L}_\mathrm{gauge} \; .
\eea
\esub
The inverse tree-level propagator for quarks and
charge-conjugate quarks, respectively, is given by
\bsub
\bea
{[ \mathcal{G}_0^+ ]}{}^{-1} \left( X,Y \right) &\equiv& \, ( i
\covariant_X + \hat\mu \gamma_0 - \hat m ) \, \delta^{\left(
4 \right)} \left( X - Y \right) \; , \\
{[ \mathcal{G}_0^- ]}{}^{-1} \left( X,Y \right) &\equiv& \, ( i
\covariant_X^C - \hat\mu \gamma_0 - \hat m ) \, \delta^{\left( 4
\right)} \left( X - Y \right) \; ,
\eea
\esub
see Sec.~\ref{The_inverse_Dirac_propagator} in the Appendix. The
matrix of quark chemical potentials is denoted by $\hat\mu$ and
is defined by Eq.~(\ref{hatmu8}). The covariant derivative in
the inverse tree-level propagator of quarks is given by
Eq.~(\ref{covariant_derivative}) while the charge-conjugate
covariant derivative in the inverse tree-level propagator of
charge-conjugate quarks reads,
\be
\covariant_\mu^C = \partial_\mu + i g A_\mu^a T_a^T \; .
\ee

By using the charge-conjugate
spinors~(\ref{charge-conjugate_spinors_space-time}) and by
introducing the Nambu-Gorkov basis with the $2 \cdot 4 N_c
N_f$-dimensional quark spinors,
\be
\bar\Psi \equiv
\left( \bar\psi, \bar\psi_C \right) \; , \qquad
\Psi \equiv \left(
\begin{array}{c}
\psi \\
\psi_C
\end{array}
\right) \; ,
\ee
one can rewrite the QCD action of interacting quarks,
\be
I_{\psi_\mathrm{int}} \left[ \bar\Psi, \Psi , A \right] =
\frac12 \int_{X,Y} \bar\Psi \left( X \right) \mathcal{S}_0^{-1}
\left( X,Y \right) \Psi \left( Y \right) \; ,
\ee
where
\be
\label{tree-level}
\mathcal{S}_0^{-1} \equiv \left(
\begin{array}{cc}
[ \mathcal{G}_0^+ ]^{-1} & 0 \\
0 & [ \mathcal{G}_0^- ]^{-1}
\end{array}
\right)
\ee
is the inverse tree-level propagator for Nambu-Gorkov quarks.

By adding a bilocal source term to the
QCD action~(\ref{QCD_action}), one
obtains~\cite{CSCReviewRischke, Schmitt_PhD},
\be
I \left[ \bar\Psi, \Psi, A, \mathcal{K} \right] \equiv I \left[
\bar\Psi, \Psi, A \right] + \frac12 \int_{X,Y} \bar\Psi \left( X
\right) \mathcal{K} \left( X,Y \right) \Psi \left( Y \right) \;
,
\ee
where
\be
\label{bilocal_source}
\mathcal{K} \equiv \left(
\begin{array}{cc}
\sigma^+ & \varphi^- \\
\varphi^+ & \sigma^-
\end{array}
\right) \; .
\ee
The four entries of $\mathcal{K}$ are not independent. Due to
charge-conjugation invariance of the action, $\sigma^- \equiv C
\left( \sigma^+ \right)^T C^{-1}$, and since the action has to
be real-valued, $\varphi^- \equiv \gamma_0 \left( \varphi^+
\right)^\dagger \gamma_0$. In the presence of an external source
$\mathcal{K}$, the grand partition function of QCD, up to
irrelevant constants, reads,
\be
\label{Z_QCD_bilocal}
\mathcal{Z} \left[ \mathcal{K} \right] = \int
\D \bar\Psi \D \Psi \D A_\mu^a \exp \left\{ I
\left[ \bar\Psi, \Psi, A, \mathcal{K} \right] \right\} \; .
\ee

The QCD effective action can be derived from the grand partition
function of QCD~(\ref{Z_QCD_bilocal}) by using the
CJT formalism~\cite{CSCReviewRischke, RuesterCSCstars,
Schmitt_PhD, CJT, Miransky_Shovkovy_Wijewardhana, Takagi,
Abuki},
\bea
\label{QCD_effective_action}
\Gamma \left[ \bar\Psi, \Psi, A, \mathcal{S},
\mathcal{D} \right]
&=& I \left[ \bar\Psi, \Psi, A \right] - \frac12 \Tr \ln
\mathcal{D}^{-1} - \frac12 \Tr \left( \mathcal{D}_0^{-1}
\mathcal{D} - 1 \right) \nonumber \\
&+& \frac12 \Tr \ln \mathcal{S}^{-1} + \frac12 \Tr \left(
\mathcal{S}_0^{-1} \mathcal{S} - 1 \right) + \Gamma_2 \left[
\bar\Psi, \Psi, A, \mathcal{S}, \mathcal{D} \right] \; .
\eea
The quantities $\mathcal{D}$ and $\mathcal{S}$ are the
gluon and quark propagator, respectively. The inverse
tree-level quark propagator $\mathcal{S}_0^{-1}$ was introduced
in Eq.~(\ref{tree-level}). Correspondingly, $\mathcal{D}_0^{-1}$
is the inverse tree-level gluon propagator. The traces run over
space-time, Nambu-Gorkov, color-flavor, and Dirac indices. The
factor $\frac12$ in front of the fermionic one-loop terms
compensates the doubling of degrees of freedom in the
Nambu-Gorkov basis. The functional $\Gamma_2$ is the sum of all
two-particle irreducible (2PI) diagrams. It is impossible to
evaluate all 2PI diagrams exactly. However, the advantage of the
QCD effective action~(\ref{QCD_effective_action}) is that
truncating the sum $\Gamma_2$ after a finite number of terms
still provides a well-defined many-body approximation. Later on,
the gluon-exchange interaction between quarks is approximated by
a point-like four-fermion coupling. This effectively removes
dynamical gluon degrees of freedom, such that one does not need
to worry about gauge fixing or possible ghost degrees of
freedom. Therefore, the latter are already omitted in
Eq.~(\ref{QCD_effective_action}).
The stationary points of the effective
action~(\ref{QCD_effective_action}) determine the expectation
values of the one- and two-point functions,
\be
\label{stationary}
\frac{\delta \Gamma}{\delta \bar\Psi} = 0 \; , \qquad
\frac{\delta \Gamma}{\delta \Psi} = 0 \; , \qquad
\frac{\delta \Gamma}{\delta A_\mu^a} = 0 \; , \qquad
\frac{\delta \Gamma}{\delta \mathcal{D}} = 0 \; , \qquad
\frac{\delta \Gamma}{\delta \mathcal{S}} = 0 \; .
\ee

The first two equations yield the Dirac equation for the
quark fields $\Psi$ and $\bar\Psi$ in the presence of the
gluon field $A_\mu^a$. The third equation is the Yang-Mills
equation for the gluon field,
\be
D_\nu^{ab} F^{\nu\mu}_b \left( X \right) =
\frac{\delta}{\delta A_\mu^a \left (X \right)} \left[ \frac12
\Tr \left( \mathcal{D}_0^{-1} \mathcal{D} - \mathcal{S}_0^{-1}
\mathcal{S} \right) - \Gamma_2 \right] \; ,
\ee
where $D_\nu^{ab}=\partial_\nu \delta^{ab}-g f^{abc}
A_\nu^c(X)$ is the covariant derivative in the adjoint
representation. The first two terms on the right-hand side are
the contributions from gluon and fermion
tadpoles~\cite{Gerhold_Rebhan}.

The functional derivative with respect to $A_\mu^a$ acting on
the trace is nontrivial because of the dependence of the inverse
tree-level propagators $\mathcal{D}_0^{-1}$ and
$\mathcal{S}_0^{-1}$ on the gluon field,
cf.\ Eq.~(\ref{tree-level}). The last term is nonzero if
$\Gamma_2$ contains 2PI diagrams with an explicit dependence on
$A_\mu^a$. As shown in Ref.~\cite{Gerhold_Rebhan}, the solution
of the Yang-Mills equation in the 2SC phase is a constant
background field $A_\mu^a \sim g_{\mu 0} \delta^{a8}$. This
background field acts like a color chemical potential $\mu_8$
and provides the color-charge neutrality of the 2SC
phase~\cite{Gerhold_Rebhan}. In this manner, also the color
chemical potential $\mu_3$ is present in three-flavor color
superconductors where the color chemical potentials $\mu_3$ and
$\mu_8$ ensure color neutrality. Later on, I shall remove the
gluon degrees of freedom by approximating the non-local gluon
exchange with a point-like four-fermion coupling. The constant
background field $A_\mu^a$ then disappears from the treatment,
and the color chemical potentials $\mu_3$ and $\mu_8$ assume the
role of the background field to ensure color neutrality.

The fourth equation~(\ref{stationary}) is the Dyson-Schwinger
equation for the gluon propagator,
\be
{ \mathcal{D}^{-1} }_{ab}^{\mu\nu} \left( X,Y \right) = {
\mathcal{D}_0^{-1} }_{ab}^{\mu\nu} \left( X,Y \right) +
\Pi_{ab}^{\mu\nu} \left( X,Y \right) \; ,
\ee
where
\be
\label{gluon_self-energy}
\Pi_{ab}^{\mu\nu} \left( X,Y \right) = -2 \, \frac{\delta
\Gamma_2}{\delta \mathcal{D}_{ba}^{\nu\mu} \left( Y,X \right)}
\ee
is the gluon self-energy. Since I shall approximate the gluon
exchange by a four-fermion coupling, I do not need to solve
the Dyson-Schwinger equation for the gluon propagator.

The fifth equation~(\ref{stationary}) is the Dyson-Schwinger
equation for the quark propagator,
\be
\label{Dyson-Schwinger}
\mathcal{S}^{-1} \left( X,Y \right) = \mathcal{S}_0^{-1} \left(
X,Y \right) + \Sigma \left( X,Y \right) \; ,
\ee
where
\be
\label{quark_self-energy}
\Sigma \left( X,Y \right) = 2 \, \frac{\delta \Gamma_2}{\delta
\mathcal{S} \left( Y,X \right)}
\ee
is the quark self-energy. According to their definition, the
self-energies~(\ref{gluon_self-energy})
and~(\ref{quark_self-energy}) are obtained from the set of 2PI
diagrams by opening one internal line. The quark self-energy is
a $2 \times 2$-matrix in Nambu-Gorkov space,
\be
\label{Sigma}
\Sigma = \left(
\begin{array}{cc}
\Sigma^+ & \Phi^- \\
\Phi^+ & \Sigma^-
\end{array}
\right) \; ,
\ee
where the quantities $\Sigma^\pm$ are the regular self-energies,
and $\Phi^\pm$ are the anomalous self-energies. For $\Phi^\pm$,
also the term gap matrices is used. The gap matrices in
connection with the quark self-energy~(\ref{quark_self-energy})
yield the so-called gap equations. By solving these gap
equations, one obtains the gap parameters. In
space-time, $\Sigma^+ \left( X,Y \right)$ $\left[ \Sigma^-
\left( X,Y \right) \right]$ has a quark [charge-conjugate
quark] entering at $X$ and another quark [charge-conjugate
quark] emerging at $Y$. The anomalous self-energies have to be
interpreted as follows: a quark [charge-conjugate quark]
enters $\Phi^+ \left( X,Y \right)$ $\left[ \Phi^- \left( X,Y
\right) \right]$ at $X$ and, in contrast to the regular
self-energies, here, a charge-conjugate quark [an ordinary
quark] emerges at $Y$. This is typical for systems with a
fermion-fermion condensate in the ground
state~\cite{Fetter_Walecka}. The self-energies $\Phi^\pm$
symbolize this condensate. This is why the crucial quantities
regarding color superconductivity are the gap matrices
$\Phi^\pm$. A nonzero value of $\Phi^\pm$ is equivalent to
Cooper pairing, or, in other words, to a nonvanishing diquark
expectation value. The self-energies in Eq.~(\ref{Sigma}) are
related in the same way as the bilocal sources in
Eq.~(\ref{bilocal_source}),
\be
\Sigma^- \equiv C \left( \Sigma^+ \right)^T C^{-1} \; , \qquad
\Phi^- \equiv \gamma_0 \left( \Phi^+ \right)^\dagger \gamma_0 \;
.
\ee
The quark propagator in Nambu-Gorkov space can be
determined from the Dyson-Schwinger
equation~(\ref{Dyson-Schwinger}), see
Sec.~\ref{The_tree-level_quark_propagator} in the Appendix,
\be
\mathcal{S} = \left(
\begin{array}{cc}
\mathcal{G}^+ & \Xi^- \\
\Xi^+ & \mathcal{G}^-
\end{array}
\right) \; ,
\ee
where
\be
\mathcal{G}^\pm = \left\{ [ \mathcal{G}_0^\pm ]^{-1} +
\Sigma^\pm - \Phi^\mp \left( [ \mathcal{G}_0^\mp ]^{-1} +
\Sigma^\mp \right)^{-1} \Phi^\pm \right\}^{-1} \; ,
\ee
is the propagator for quasiquarks or charge-conjugate
quasiquarks, respectively, and
\be
\Xi^\pm = - \left( [ \mathcal{G}_0^\mp ]^{-1} + \Sigma^\mp
\right)^{-1} \Phi^\pm \mathcal{G}^\pm =  - \, \mathcal{G}^\mp
\Phi^\pm \left( [ \mathcal{G}_0^\pm ]^{-1} + \Sigma^\pm
\right)^{-1} \; ,
\ee
are the anomalous propagators. These anomalous propagators are
typical for superconducting systems~\cite{Fetter_Walecka} and
account for the possibility that in the presence of a
Cooper-pair condensate, symbolized by $\Phi^\pm$, a fermion can
always be absorbed in the condensate, while its
charge-conjugate counterpart is emitted from the condensate and
continues to propagate. 

The QCD pressure is, up to a prefactor $T/V$, equal to the
QCD effective action~(\ref{QCD_effective_action}) determined at
the stationary points~(\ref{stationary}) which is denoted by
$\Gamma^*$,
\be
p = \frac{T}{V} \Gamma^* \; ,
\ee
where $T$ is the temperature and $V$ the volume of the system.
\subsection{The effective action of quarks}
In this section, I investigate the phase diagram of massless
neutral three-flavor quark matter. The quark spinor has the
following color-flavor structure:
\be
\psi = \left( \psi_u^r, \psi_d^r, \psi_s^r, \psi_u^g, \psi_d^g,
\psi_s^g, \psi_u^b, \psi_d^b, \psi_s^b \right)^T \; .
\ee
Since I approximate the gluon-exchange interaction between
quarks by a point-like four-fermion coupling, I only need to
consider the contributions of quarks to the QCD effective
action~(\ref{QCD_effective_action}),
\be
\label{effective_action_of_quarks}
\Gamma \left[ S \right] = \frac12 \Tr \ln S^{-1} + \frac12 \Tr
\left( S_0^{-1} S - 1 \right) + \Gamma_2 \left[ S \right] \; .
\ee
Here, the tree-level quark propagator
$\mathcal{S}_0^{-1}$~(\ref{tree-level}), which occurs in the
QCD effective action~(\ref{QCD_effective_action}), is replaced
by the quark propagator,
\be
S_0^{-1} \equiv \left(
\begin{array}{cc}
[ G_0^+ ]^{-1} & 0 \\
0 & [ G_0^- ]^{-1}
\end{array}
\right) \; ,
\ee
where
\be
\label{G0pmhm1}
{[ G_0^\pm ]}{}^{-1} \left( X,Y \right) \equiv \, ( i \dirac_X
\pm \hat\mu \gamma_0 ) \, \delta^{\left( 4 \right)} \left( X - Y
\right) \; ,
\ee
is the massless inverse Dirac propagator for quarks and
charge-conjugate quarks, respectively, in which the constant
background field $A_\mu^a$ disappears from the treatment, and
the color chemical potentials $\mu_3$ and $\mu_8$ assume the
role of the background field to ensure color neutrality. The
quark chemical potential matrix in color-flavor space is defined
as
\be
\hat\mu = \diag \left( \mu_u^r, \mu_d^r, \mu_s^r, \mu_u^g,
\mu_d^g, \mu_s^g, \mu_u^b, \mu_d^b, \mu_s^b \right) \; ,
\ee
where the chemical potential of each quark is given by
Eq.~(\ref{quark_color_and_flavor_beta_equilibrium}) because
quark matter inside neutron stars is in $\beta$ equilibrium.

Since I shall present the phase diagram of massless neutral
three-flavor quark matter in this section, the mass term of the
inverse Dirac propagators~(\ref{G0pmhm1}) is omitted. At
sufficiently large quark chemical potential, there is no need to
take into account the small up and down quark masses because the
dynamical effect of such masses around the quark Fermi surfaces
is negligible. Of course, the situation with the strange quark
is different because its mass is not very small as compared to
the quark chemical potential $\mu$. The most important effect of
a nonzero strange quark mass is, however, a shift of the strange
quark chemical potential due to the reduction of the Fermi
momentum,
\be
\label{shift2}
\left( k_F \right)_s^i = \sqrt{ ( \mu_s^i )^2 - m_s^2 } \simeq
\mu_s^i - \frac{m_s^2}{2 \mu_s^i} \simeq \mu_s^i -
\frac{m_s^2}{2 \mu} \; ,
\ee
cf.\ Eq.~(\ref{shift}). Here, I have approximated $\mu_s^i$ by
$\mu$ in the denominator. Quantitatively, this does not make a
big difference.
\begin{figure}[H]
\begin{center}
\makebox[0.9\textwidth][l]{
\includegraphics[width=0.255\textwidth]{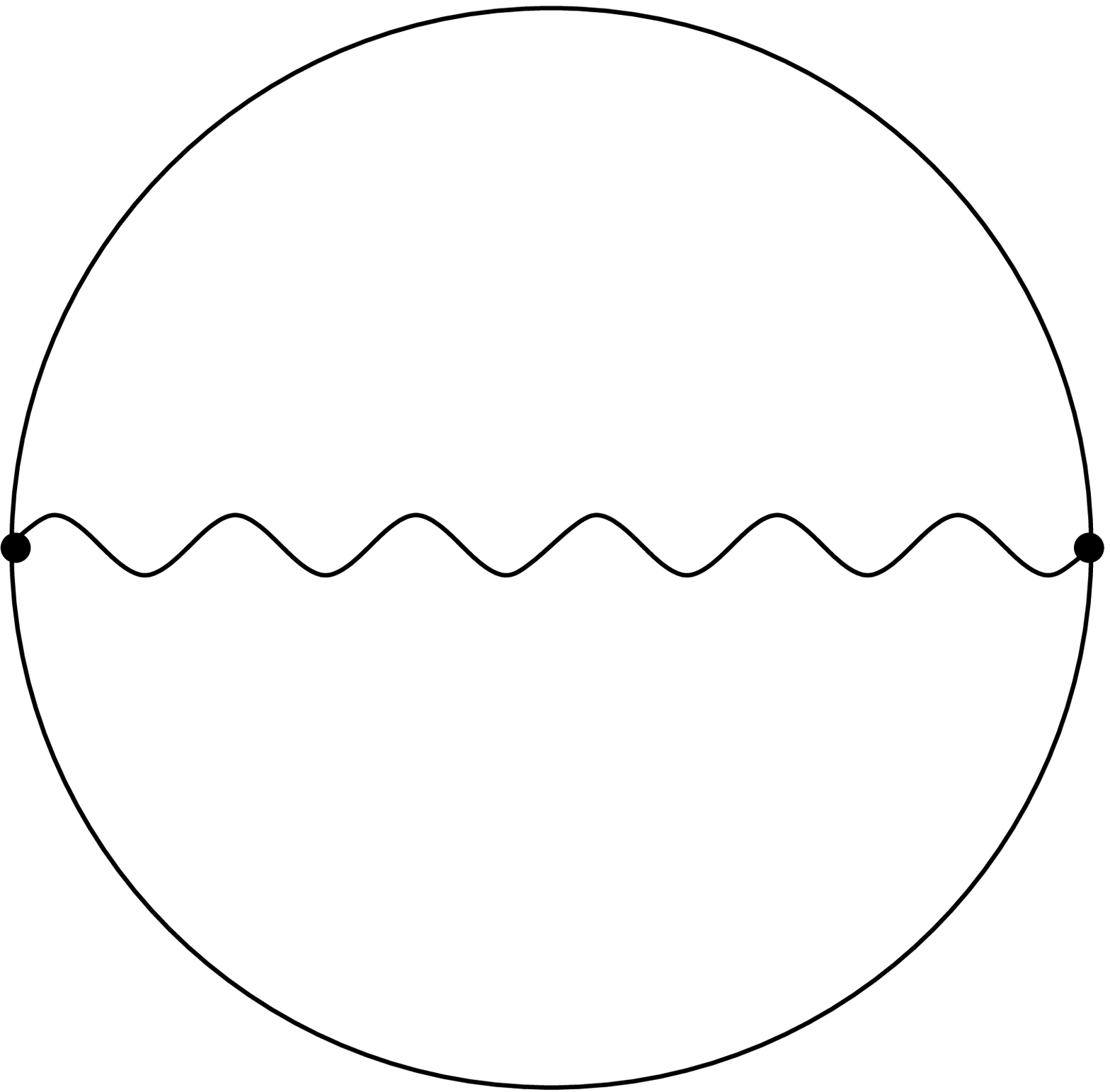}
\hspace{0.09\textwidth}
\includegraphics[width=0.5\textwidth]{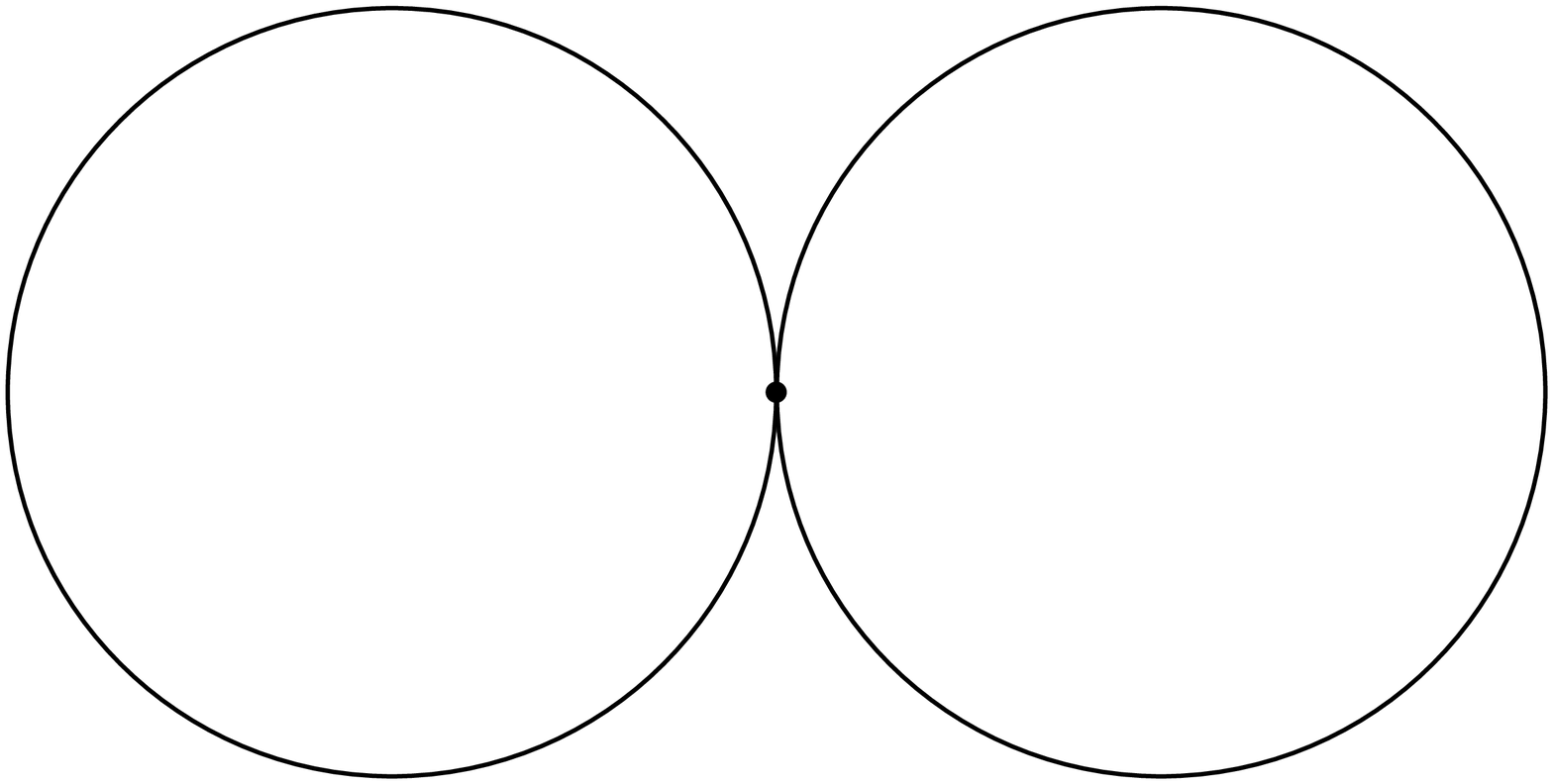}}
\caption[The sunset-type diagram, and the double-bubble
diagram.]{Left panel: the sunset-type diagram. Right panel: the
double-bubble diagram.}
\label{sunset_double-bubble}
\end{center}
\end{figure}
For the sum of all 2PI diagrams I only include the sunset-type
diagram, which is shown in the left panel of
Fig.~\ref{sunset_double-bubble},
\be
\Gamma_2 \left[ S \right] = - \frac{g^2}{4} \int_{X,Y} \Tr
\left[ \Gamma_a^\mu S \left( X,Y \right) \Gamma_b^\nu S \left(
Y,X \right) \right] D_{\mu\nu}^{ab} \left( X,Y \right) \; .
\ee
The trace runs over Nambu-Gorkov, color-flavor, and Dirac
indices. The Nambu-Gorkov vertex is defined as
\be
\Gamma_a^\mu = \left(
\begin{array}{cc}
\gamma^\mu T_a & 0 \\
0 & - \gamma^\mu T_a^T
\end{array}
\right) \; .
\ee

The stationary point of the effective action of
quarks~(\ref{effective_action_of_quarks}),
\be
\frac{\delta \Gamma}{\delta S} = 0 \; ,
\ee
is the Dyson-Schwinger equation for the quark propagator,
\be
\label{Dyson-Schwinger2}
S^{-1} \left( X,Y \right) = S_0^{-1} \left( X,Y \right) + \Sigma
\left( X,Y \right) \; ,
\ee
where
\be
\Sigma \left( X,Y \right) = 2 \, \frac{\delta \Gamma_2}{\delta S
\left( Y,X \right)} = - g^2 \Gamma_a^\mu S \left( X,Y \right)
\Gamma_b^\nu D_{\mu\nu}^{ab} \left( Y,X \right)
\ee
is the quark self-energy. The effective action at the stationary
point which is determined by the Dyson-Schwinger equation for
the quark propagator~(\ref{Dyson-Schwinger2}) reads,
\be
\label{Gamma_star}
\Gamma^* = \frac12 \Tr \ln S^{-1} - \frac14 \Tr \left( \Sigma S
\right) \; .
\ee

For translationally invariant systems, it is advantageous to
work in energy-momentum space instead of in space-time,
\bsub
\label{FT}
\bea
S^{-1} \left( X,Y \right) &=& \frac{T}{V} \sum_K \e^{-i K
\left(X - Y\right)} S^{-1} \left( K \right) \; , \\
S \left( X,Y \right) &=& \frac{T}{V} \sum_K \e^{-i K \left(X -
Y\right)} S \left( K \right) \; , \\
D_{\mu\nu}^{ab} \left( X,Y \right) &=& \frac{T}{V} \sum_K \e^{-i
K \left(X - Y\right)} D_{\mu\nu}^{ab} \left( K \right) \; , \\
\Sigma \left( K \right) &=& \int_Z \e^{i K Z} \Sigma \left( Z
\right) \; ,
\eea
\esub
where I assumed translational invariance of propagators and
self-energies, $Z \equiv X - Y$. In energy-momentum space, the
Dyson-Schwinger equation for the quarks reads,
\be
S^{-1} \left( K \right) = S_0^{-1} \left( K \right) + \Sigma
\left( K \right) \; .
\ee
The inverse free quark propagator in energy-momentum space is
given by
\be
S_0^{-1} \left( K \right) = \left( 
\begin{array}{cc}
[ G_0^+ ]^{-1} & 0 \\
0 & [ G_0^- ]^{-1}
\end{array}
\right) \; ,
\ee
where the massless inverse Dirac propagator for quarks and
charge-conjugate quarks, respectively, reads,
\be
\label{G0pmhm1K}
[ G_0^\pm ]^{-1} \left( K \right) = \gamma_0 \left( k_0 \pm
\hat\mu \right) - \gamma \cdot \fettu{k} \; .
\ee
The quark self-energy is obtained by
\be
\label{Sigma2}
\Sigma = \left( 
\begin{array}{ll}
\, 0 & \Phi^- \\
\Phi^+ & \, 0
\end{array}
\right)
= - g^2 \frac{T}{V} \sum_Q \Gamma_a^\mu S \left( Q \right)
\Gamma_b^\nu D_{\mu\nu}^{ab} \; ,
\ee
where
\be
\label{gap_equation}
\Phi^+ = g^2 \frac{T}{V} \sum_Q \gamma^\mu T_a^T \Xi^+ \left(
Q \right) \gamma^\nu T_b D_{\mu\nu}^{ab} \; , \qquad
\Phi^- = g^2 \frac{T}{V} \sum_Q \gamma^\mu T_a \Xi^- \left( Q
\right) \gamma^\nu T_b^T D_{\mu\nu}^{ab} \; ,
\ee
are the gap equations which are independent of $K$ because I
approximate the gluon-exchange interaction between quarks by a
point-like four-fermion coupling. The inverse quark propagator
is
\be
\label{Shm1}
S^{-1} \left( K \right) = \left( 
\begin{array}{cc}
[ G_0^+ ]^{-1} & \Phi^- \\
\Phi^+ & [ G_0^- ]^{-1}
\end{array}
\right) \; .
\ee
The regular self-energies play an important role in the dynamics
of chiral symmetry breaking, but they are of less
importance in color-superconducting quark matter. (The effect
of the regular self-energies was studied in
Ref.~\cite{regular_self-energies}.) Therefore, the regular 
self-energies in Eq.~(\ref{Sigma2}) are omitted. The quark
self-energy~(\ref{Sigma2}) contains the Feynman gauged gluon
propagator, see Sec.~\ref{The_Feynman_gauged_gluon_propagator}
in the Appendix,
\be
D_{\mu\nu}^{ab} = - \delta^{ab} \frac{g_{\mu\nu}}{\Lambda^2} \;
,
\ee
which represents the gluon-exchange interaction between quarks
by a point-like four-fermion coupling. In this approximation,
the sunset-type diagram becomes a double-bubble diagram, which
is shown in the right panel of Fig.~\ref{sunset_double-bubble}.

The quark propagator may be obtained by inverting
Eq.~(\ref{Shm1}),
\be
S \left( K \right) = \left(
\begin{array}{cc}
G^+ & \Xi^- \\
\Xi^+ & G^-
\end{array}
\right) \; ,
\ee
where
\be
\label{Gpm}
G^\pm \left( K \right) = \left\{ [ G_0^\pm ]^{-1} - \Phi^\mp
G_0^\mp \Phi^\pm \right\}^{-1}
\ee
is the propagator for quasiquarks or charge-conjugate
quasiquarks, respectively, and
\be
\label{anomalous}
\Xi^\pm \left( K \right) = - G_0^\mp \Phi^\pm G^\pm =  - G^\mp
\Phi^\pm G_0^\pm \; ,
\ee
are the anomalous propagators.

I want to study the most general ansatz of the gap matrix for
the CFL phase in color-flavor space. Therefore, I use the
following nine-parameter ansatz~\cite{Ruester_Shovkovy_Rischke}:
\be
\label{most_general}
\Phi^\pm = \left(
\begin{array}{ccccccccc}
{[ \Phi_{uu}^{rr} ]}{}^\pm & 0 & 0 & 0 & {[ \Phi_{ud}^{rg}
]}{}^\pm & 0 & 0 & 0 & {[ \Phi_{us}^{rb} ]}{}^\pm \\
0 & 0 & 0 & {[ \Phi_{du}^{rg} ]}{}^\pm & 0 & 0 & 0 & 0 & 0 \\
0 & 0 & 0 & 0 & 0 & 0 & {[ \Phi_{su}^{rb} ]}{}^\pm & 0 & 0 \\
0 & {[ \Phi_{du}^{rg} ]}{}^\pm & 0 & 0 & 0 & 0 & 0 & 0 & 0 \\
{[ \Phi_{ud}^{rg} ]}{}^\pm & 0 & 0 & 0 & {[
\Phi_{dd}^{gg}]}{}^\pm & 0 & 0 & 0 & {[ \Phi_{ds}^{gb} ]}{}^\pm
\\ 
0 & 0 & 0 & 0 & 0 & 0 & 0 & {[ \Phi_{sd}^{gb} ]}{}^\pm & 0 \\
0 & 0 & {[ \Phi_{su}^{rb} ]}{}^\pm & 0 & 0 & 0 & 0 & 0 & 0 \\
0 & 0 & 0 & 0 & 0 & {[ \Phi_{sd}^{gb} ]}{}^\pm & 0 & 0 & 0 \\
{[ \Phi_{us}^{rb} ]}{}^\pm & 0 & 0 & 0 & {[ \Phi_{ds}^{gb}
]}{}^\pm & 0 & 0 & 0 & {[ \Phi_{ss}^{bb} ]}{}^\pm
\end{array}
\right) \; .
\ee
This ansatz for the gap matrix is indeed the most general one
for the CFL phase because it can be obtained with some
modifications from the antitriplet and sextet gap
ansatz~(\ref{CFL_33_66}). The difference between the gap
ansatz~(\ref{CFL_33_66}) and the gap matrix~(\ref{most_general})
is that the gap ansatz~(\ref{most_general}) has different
entries for all nonzero color-flavor elements. The wavefunction
requirements are still fulfilled because the gap
matrix~(\ref{most_general}) is symmetric in color-flavor space.

The effective action of quarks at the stationary
point~(\ref{Gamma_star}) in energy-momentum space reads,
\be
\label{Gamma*}
\Gamma^* = \frac12 \sum_K \ln \det \left( \frac{S^{-1}}{T}
\right) - \frac14 \sum_K \Tr \left( \Sigma S \right) \; ,
\ee
where I used Eqs.~(\ref{FT}), the derivation of the
Fourier transformed kinetic part of the QCD grand partition
function which is shown in
Sec.~\ref{The_determinant_of_the_inverse_quark_propagator}
in the Appendix, and the relation $\ln \det A = \Tr \ln A$
which is proven in Sec.~\ref{The_logarithm_of_the_determinant}
in the Appendix.
\subsection{Propagators and self-energies in projector
representation}
\label{projector_representation}
I only represented the color-flavor structure in the gap
ansatz~(\ref{most_general}). Together with the Dirac part, the
complete gap matrices read~\cite{Pisarski_Rischke3},
\be
\Phi^+ = \sum_{c,e} \phi_c^e \mathcal{P}_c^e (\fettu{k}) \; ,
\qquad
\Phi^- = \sum_{c,e} \phi_c^e \mathcal{P}_{-c}^{-e} (\fettu{k})
\; ,
\ee
where $\phi_c^e$ is only the color-flavor part of the gap
ansatz~(\ref{most_general}), and
\be
\mathcal{P}_c^e ( \fettu{k}) = \frac14 \, ( 1 + c \gamma_5 )( 1
+ e \gamma_0 \fett{\gamma} \cdot \hat{\fettu{k}} )
\ee
are the energy-chirality projectors, see
Eq.~(\ref{energy-chirality_projectors}). The gap parameters
$\phi_c^e{}_{ff'}^{ii'}$ are real-valued numbers.
It will be very advantageous in later calculations if
the (inverse) massless Dirac propagators are also written in
terms of the energy-chirality projectors,
\bsub
\label{GPce}
\bea
{[ G_0^\pm ]}{}^{-1} &=& \gamma_0 \sum_{c,e} \, [ \hat{G}_0^\pm
]^{-1} \mathcal{P}_c^e \left( \fettu{k} \right) = \sum_{c,e} \,
[ \check{G}_0^\pm ]^{-1} \mathcal{P}_c^e \left( \fettu{k}
\right) \gamma_0 \; , \\
G_0^\pm &=& \sum_{c,e} \hat{G}_0^\pm \mathcal{P}_c^e \left(
\fettu{k} \right) \gamma_0 = \gamma_0 \sum_{c,e} \check{G}_0^\pm
\mathcal{P}_c^e \left( \fettu{k} \right) \; .
\eea
\esub
Let the matrix of quark chemical potentials $\tilde\mu$ and the
color-flavor part of the gap matrix $\tilde\phi$ be
\textit{arbitrary} for the moment. I separated the Dirac part
and the color-flavor part,
\be
\label{hat_check_G}
[ \hat{G}_0^\pm ]^{-1} \equiv k_0 \pm \tilde\mu - e k \; ,
\qquad
[ \check{G}_0^\pm ]^{-1} \equiv k_0 \pm \tilde\mu + e k \; ,
\ee
of the (inverse) massless Dirac propagators in Eq.~(\ref{GPce}).
Note that the color-flavor matrices in Eq.~(\ref{hat_check_G})
depend on $e$. I did not show an extra index $e$ in the above
equations for simplicity. I shall omit the indices $c$ and $e$
at the color-flavor matrices of propagators and self-energies.

The (inverse) regular propagator reads,
\be
[ G^+ ]^{-1} = \gamma_0 \sum_{c,e} \, [ \hat{G}^+ ]^{-1}
\mathcal{P}_c^e \left( \fettu{k} \right) \; , \quad
G^+ = \sum_{c,e} \hat{G}^+ \mathcal{P}_c^e \left( \fettu{k}
\right) \gamma_0 \; , \quad [ \hat{G}^+ ]^{-1} \equiv [
\hat{G}_0^+ ]^{-1} - \tilde\phi \check{G}_0^- \tilde\phi \; .
\ee
The anomalous propagator can be written as
\be
\label{anomalous2}
\Xi^+ = \sum_{c,e} \hat\Xi^+ \mathcal{P}_{-c}^{-e} \left(
\fettu{k} \right) \; , \qquad \hat\Xi^+ \equiv - \check{G}_0^-
\tilde\phi \hat{G}^+ \; .
\ee
\subsection{The potential part of the effective action of
quarks}
The potential part of the effective action of
quarks~(\ref{Gamma*}),
\be
\Gamma_\mathrm{pot}^* \equiv - \frac14 \sum_K \Tr \left( \Sigma
S \right) \; ,
\ee
can be simplified. After performing the Nambu-Gorkov trace, one
obtains for the potential part of the effective action of
quarks,
\be
\Gamma_\mathrm{pot}^* = - \frac14 \sum_K \Tr \left( \Phi^- \Xi^+
+ \Phi^+ \Xi^- \right) \; .
\ee
With the anomalous propagators~(\ref{anomalous}) and the
relation $\Tr \left( A B \right) = \Tr \left( B A \right)$, one
can easily check that
\be
\Tr \left( \Phi^- \Xi^+ \right) \equiv - \Tr \left( \Phi^- G_0^-
\Phi^+ G^+ \right) = - \Tr \left( \Phi^+ G^+ \Phi^- G_0^-
\right) \equiv \Tr \left( \Phi^+ \Xi^- \right) \; .
\ee
Therefore, the potential part of the effective action of quarks
can be simplified to
\be
\label{Gamma*pot}
\Gamma_\mathrm{pot}^* = - \frac12 \sum_K \Tr \left( \Phi^- \Xi^+
\right) = - \frac12 \sum_K \sum_{c,e} \Tr \left[ \phi \,
\hat\Xi^+ \, \mathcal{P}_{-c}^{-e} \left( \fettu{k} \right)
\right] = - \frac12 \sum_K \sum_{c,e} \Tr \, ( \phi \, \hat\Xi^+
) \; ,
\ee
where the Dirac trace has been performed in the last step. In
the following, I shall further simplify the potential part of
the effective action of quarks~(\ref{Gamma*pot}): I insert the
anomalous propagator~(\ref{anomalous2}) into the gap
equation~(\ref{gap_equation}) and multiply it from the left side
with the energy-chirality projectors in order to project onto
$\phi_c^e$. Thus, the sums over $c$ and $e$ vanish. By
performing the Dirac trace one obtains,
\bea
\phi_c^e &=& g^2 \frac{T}{V} \sum_Q \sum_{c',e'} T_a^T \hat\Xi^+
\left( Q \right) T_b \Tr \left[ \mathcal{P}_c^e \left( \fettu{k}
\right) \gamma^\mu \mathcal{P}_{-c'}^{-e'} \left( \fettu{q}
\right) \gamma^\nu \right] D_{\mu\nu}^{ab} \nonumber \\
&=& g^2 \frac{T}{V} \sum_Q \sum_{e'} T_a^T \hat\Xi^+ \left( Q
\right) T_a \Tr \left[ \mathcal{P}_c \Lambda^{e} \left(
\fettu{k} \right) \gamma^\mu \Lambda^{e'} \left( \fettu{q}
\right) \gamma^\nu \right] D_{\mu\nu} \; .
\eea
In this expression, I explicitly show that the color-flavor part
of the gap matrix carries the indices $c$ and $e$ which arise
from the energy and chirality projections. In the second line,
the color-flavor part of the Feynman gauged gluon propagator has
been performed. The Dirac trace is computed in
Sec.~\ref{The_Dirac_trace} in the Appendix. One obtains
\be
\phi_{ff'}^{ii'} = - \frac{2 g^2}{ \Lambda^2 } \frac{T}{V}
\sum_Q \sum_{e'} T_a^{i''i} ( \hat\Xi^+ ){}_{ff'}^{i''i'''}
\left( Q \right) T_a^{i'''i'} \; ,
\ee
where $T_a^{i''i} \equiv ( T_a^{ii''} )^T$. As one can see, the
gap equations do not depend on the indices $c$ and $e$ because
they do not appear any longer on the right-hand side of this
equation. After summation over $a$ one
obtains~\cite{Shovkovy_Wijewardhana}
\be
\phi_{ff'}^{ii'} = - \frac{2 g^2}{ \Lambda^2 } \frac{T}{V}
\sum_Q \sum_{e'} \left[ \frac12 ( \hat\Xi^+ ){}_{ff'}^{i'i}
\left( Q \right) - \frac16 ( \hat\Xi^+ ){}_{ff'}^{ii'} \left( Q
\right) \right] \; ,
\ee
which can be rewritten as
\be
\label{simplified_gap_eqns}
\phi_{ff'}^{ii'} + 3 \phi_{ff'}^{i'i} = - \frac83
\frac{g^2}{\Lambda^2} \frac{T}{V} \sum_Q \sum_{e'} \, (
\hat\Xi^+ ){}_{ff'}^{ii'} \left( Q \right) \; ,
\ee
where only one term is left on the right-hand side. This
equation can be transformed into
\be
\label{Xi38}
\sum_K \sum_e \, ( \hat\Xi^+ ){}_{ff'}^{ii'} \left( K \right)
= - \frac38 \frac{\Lambda^2}{g^2} \frac{V}{T} \left(
\phi_{ff'}^{ii'} + 3 \phi_{ff'}^{i'i} \right) \; ,
\ee
where I renamed $Q$ in $K$ and $e'$ in $e$. I now want to write
the right-hand side of this equation in a nicer form. Therefore,
I use the relation,
\be
\sum_a T_a^{i''i} \phi_{ff'}^{i''i'''} T_a^{i'''i'} =
\frac12 \phi_{ff'}^{i'i} - \frac16 \phi_{ff'}^{ii'} \; .
\ee
Multiplying by six and rearranging yields,
\be
\phi_{ff'}^{ii'} + 6 \sum_a T_a^{i''i} \phi_{ff'}^{i''i'''}
T_a^{i'''i'} = 3 \phi_{ff'}^{i'i} \; ,
\ee
This expression can be inserted into Eq.~(\ref{Xi38}),
\be
\sum_K \sum_e \, ( \hat\Xi^+ ){}_{ff'}^{ii'} \left( K \right)
= - \frac34 \frac{\Lambda^2}{g^2} \frac{V}{T} \left( \phi +
3 \sum_a T_a^T \phi T_a \right)_{ff'}^{ii'} \; .
\ee
By inserting this result into the potential part of the
effective action of quarks~(\ref{Gamma*pot}) one obtains
\be
\label{Gamma*pot2}
\Gamma_\mathrm{pot}^* = \frac34 \frac{\Lambda^2}{g^2}
\frac{V}{T} \Tr \left[ \phi \left( \phi + 3 \sum_a T_a^T
\phi T_a \right) \right] \; ,
\ee
where the sum over $c$ has been performed which yields an extra
factor of two.
\subsection{The kinetic part of the effective action of quarks}
The inverse quark propagator $S^{-1}$~(\ref{Shm1}) is a $72
\times 72$-matrix and consists of the following subspace
structure: it is a $2 \times 2$-matrix in Nambu-Gorkov space, a
$9 \times 9$-matrix in color-flavor space, and a $4 \times
4$-matrix in Dirac space. In order to calculate the
kinetic (first) part of the effective action in
Eq.~(\ref{Gamma*}), I transform the inverse quark propagator
into block-diagonal form. This is easily achieved by changing
the order of rows and columns in color-flavor and Nambu-Gorkov
space. By such a transformation, the absolute value of a
determinant does not change. Only the sign of the determinant is
changed if there is an odd number of exchanges of rows and
columns. In the case mentioned above, no corrections have to be
made to the result of the determinant because there is an even
number of exchanges of rows and columns. There are six $2 \times
2$-blocks,
\be
\begin{aligned}
S_1^{-1} &= \left(
\begin{array}{cc}
{[ G_0^+{}_s^g ]}{}^{-1} & {[ \Phi_{sd}^{gb} ]}{}^- \\
{[ \Phi_{sd}^{gb} ]}{}^+ & {[ G_0^-{}_d^b ]}{}^{-1}
\end{array}
\right) \; , &
S_2^{-1} &= \left(
\begin{array}{cc}
{[ G_0^+{}_d^b ]}{}^{-1} & {[ \Phi_{sd}^{gb} ]}{}^- \\
{[ \Phi_{sd}^{gb} ]}{}^+ & {[ G_0^-{}_s^g ]}{}^{-1}
\end{array}
\right) \; , \\
S_3^{-1} &= \left(
\begin{array}{cc}
{[ G_0^+{}_s^r ]}{}^{-1} & {[ \Phi_{su}^{rb} ]}{}^- \\
{[ \Phi_{su}^{rb} ]}{}^+ & {[ G_0^-{}_u^b ]}{}^{-1}
\end{array}
\right) \; , &
S_4^{-1} &= \left(
\begin{array}{cc}
{[ G_0^+{}_u^b ]}{}^{-1} & {[ \Phi_{su}^{rb} ]}{}^- \\
{[ \Phi_{su}^{rb} ]}{}^+ & {[ G_0^-{}_s^r ]}{}^{-1}
\end{array}
\right) \; , \\
S_5^{-1} &= \left(
\begin{array}{cc}
{[ G_0^+{}_d^r ]}{}^{-1} & {[ \Phi_{du}^{rg} ]}{}^- \\
{[ \Phi_{du}^{rg} ]}{}^+ & {[ G_0^-{}_u^g ]}{}^{-1}
\end{array}
\right) \; , &
S_6^{-1} &= \left(
\begin{array}{cc}
{[ G_0^+{}_u^g ]}{}^{-1} & {[ \Phi_{du}^{rg} ]}{}^- \\
{[ \Phi_{du}^{rg} ]}{}^+ & {[ G_0^-{}_d^r ]}{}^{-1}
\end{array}
\right) \; ,
\end{aligned}
\ee
and one $6 \times 6$-block in the block-diagonal structure of
the inverse quark propagator~(\ref{Shm1}) in color-flavor and
Nambu-Gorkov space. This $6 \times 6$-block is a $2 \times
2$-matrix in Nambu-Gorkov space,
\be
S_7^{-1} = \left(
\begin{array}{cc}
[ G_0^+ ]_{\left( 3 \times 3 \right)}^{-1} & \Phi_{\left( 3
\times 3 \right)}^- \\
\Phi_{\left( 3 \times 3 \right)}^+ & [ G_0^- ]_{\left( 3 \times
3 \right)}^{-1}
\end{array}
\right) \; ,
\ee
where the Dirac propagators as well as the gap matrices,
\be
[ G_0^\pm ]_{\left( 3 \times 3 \right)}^{-1} = \diag \left(
{[ G_0^\pm{}_u^r ]}{}^{-1}, {[ G_0^\pm{}_d^g ]}{}^{-1}, {[
G_0^\pm{}_s^b ]}{}^{-1} \right) \; , \quad
\Phi_{\left( 3 \times 3 \right)}^\pm = \left(
\begin{array}{ccc}
{[ \Phi_{uu}^{rr} ]}{}^\pm & {[ \Phi_{ud}^{rg} ]}{}^\pm & {[
\Phi_{us}^{rb} ]}{}^\pm \\
{[ \Phi_{ud}^{rg} ]}{}^\pm & {[ \Phi_{dd}^{gg} ]}{}^\pm & {[
\Phi_{ds}^{gb} ]}{}^\pm \\
{[ \Phi_{us}^{rb} ]}{}^\pm & {[ \Phi_{ds}^{gb} ]}{}^\pm & {[
\Phi_{ss}^{bb} ]}{}^\pm
\end{array}
\right) \; ,
\ee
are $3 \times 3$-matrices in color-flavor space.

Since $\det \left[ \diag \left( A, B \right) \right] = \det A
\cdot \det B$~\cite{Fischer}, where $A$ and $B$ are quadratic
matrices, the kinetic part of the effective action of
quarks~(\ref{Gamma*}) is,
\be
\Gamma_\mathrm{kin}^* \equiv \frac12 \sum_K \ln \det \left(
\frac{S^{-1}}{T} \right) = \frac12 \sum_K \sum_{i = 1}^7 \ln
\det \left( \frac{S_i^{-1}}{T} \right) \; .
\ee
In order to simplify the determinants, I make use of the
Gauss-elimination procedure~\cite{Fischer},
\be
\det \left(
\begin{array}{cc}
A & B \\
C & D
\end{array}
\right)
= \det \left[ D \left( A - B D^{-1} C \right) \right] \; ,
\ee
where $A$, $B$, $C$, and $D$ are quadratic matrices and $D$ is
invertible. Hereby, all $2 \times 2$-blocks in color-flavor and
Nambu-Gorkov space become $1 \times 1$-blocks in color-flavor
space, while the original $6 \times 6$-block in color-flavor and
Nambu-Gorkov space becomes a $3 \times 3$-block in color-flavor
space so that the kinetic part of the effective action reads,
\bea
\label{Gamma_mathrm_kin}
\Gamma_\mathrm{kin}^* &=& \frac12 \sum_K \left[ \ln \det \left(
\frac{ [ G_0^-{}_d^b ]^{-1} \, [ G^+{}_s^g ]^{-1} }{T^2} \right)
+ \ln \det \left( \frac{ [ G_0^-{}_s^g ]^{-1} \, [ G^+{}_d^b
]^{-1} }{T^2} \right) \right. \nonumber \\
&& \hspace{7.6mm} + \ln \det \left( \frac{ [ G_0^-{}_u^b ]^{-1}
\, [ G^+{}_s^r ]^{-1} }{T^2} \right) + \ln \det \left( \frac{ [
G_0^-{}_s^r ]^{-1} \, [ G^+{}_u^b ]^{-1} }{T^2} \right)
\nonumber \\
&& \hspace{7.6mm} + \ln \det \left( \frac{ [ G_0^-{}_u^g ]^{-1}
\, [ G^+{}_d^r ]^{-1} }{T^2} \right) + \ln \det \left( \frac{ [
G_0^-{}_d^r ]^{-1} \, [ G^+{}_u^g ]^{-1} }{T^2} \right)
\nonumber \\
&& \hspace{7.1mm} \left. + \ln \det \left( \frac{ [ G_0^-
]_{\left( 3 \times 3 \right)}^{-1} \, [ G^+ ]_{\left( 3 \times 3
\right)}^{-1} }{T^2} \right) \right] \; ,
\eea
where
\be
[ G^+ ]_{\left( 3 \times 3 \right)}^{-1} \equiv [ G_0^+
]_{\left( 3 \times 3 \right)}^{-1} - \Phi_{\left( 3 \times 3
\right)}^- {[ G_0^- ]}_{\left( 3 \times 3
\right)} \Phi_{\left( 3 \times 3 \right)}^+ \; .
\ee
The color and flavor indices at the propagators in
Eq.~(\ref{Gamma_mathrm_kin}) denote the respective elements of
the propagators which are given by Eqs.~(\ref{G0pmhm1K})
and~(\ref{Gpm}). The kinetic part of the effective action of
quarks can be simplified by separating the color-flavor part and
the Dirac part in the propagators which can be achieved by using
the energy-chirality projectors, see
Sec.~\ref{projector_representation}. With the relations $\ln
\det A = \Tr \ln A$ and $\Tr \ln \sum_i a_i \mathcal{P}_i =
\sum_i \ln a_i \Tr \, \mathcal{P}_i$ (see
Secs.~\ref{The_determinant_of_the_inverse_quark_propagator}
and~\ref{The_trace_of_the_logarithm} in the Appendix), only the
color-flavor part is remaining in the kinetic part of the
effective action of quarks,
\bea
\label{Gamma*kin}
\Gamma_\mathrm{kin}^* &=& \sum_K \sum_e \left[ \ln \left( \frac{
[ \check{G}_0^-{}_d^b ]^{-1} \, [ \hat{G}^+{}_s^g ]^{-1} }{T^2}
\right) + \ln \left( \frac{ [ \check{G}_0^-{}_s^g ]^{-1} \, [
\hat{G}^+{}_d^b ]^{-1} }{T^2} \right) + \ln \left( \frac{ [
\check{G}_0^-{}_u^b ]^{-1} \, [ \hat{G}^+{}_s^r ]^{-1} }{T^2}
\right) \right. \nonumber \\
&& \hspace{1cm} + \ln \left( \frac{ [ \check{G}_0^-{}_s^r ]^{-1}
\, [ \hat{G}^+{}_u^b ]^{-1} }{T^2} \right) + \ln \left( \frac{ [
\check{G}_0^-{}_u^g ]^{-1} \, [ \hat{G}^+{}_d^r ]^{-1} }{T^2}
\right) + \ln \left( \frac{ [ \check{G}_0^-{}_d^r ]^{-1} \, [
\hat{G}^+{}_u^g ]^{-1} }{T^2} \right) \nonumber \\
&& \hspace{9.6mm} \left. + \ln \det \left( \frac{ [
\check{G}_0^- ]_{\left( 3 \times 3 \right)}^{-1} \, [ \hat{G}^+
]_{\left( 3 \times 3 \right)}^{-1} }{T^2} \right) \right] \; .
\eea
In this expression, also the sum over $c$ has been performed
which gives an extra factor of two. One can simplify the
effective action of quarks,
\bea
\label{Gamma*kin2}
\Gamma_\mathrm{kin}^* &=& \sum_K \sum_{i = 1}^3 \sum_e \left[
\ln \left( \frac{ k_0^2 - \left( \tilde\epsilon_i^e \right)^2
}{ T^2 } \right) \right. \nonumber \\
&+& \left. \! \ln \left( \frac{ \left( k_0 - \delta\mu_i
\right)^2 - \left[ \epsilon_\fettu{k}^e \left( \bar\mu_i, \phi_i
\right) \right]^2 }{ T^2 } \right) + \ln \left( \frac{ \left(
k_0 + \delta\mu_i \right)^2 - \left[ \epsilon_\fettu{k}^e \left(
\bar\mu_i, \phi_i \right) \right]^2 }{ T^2 } \right) \right] \;
.
\eea
Some computations and definitions have been made in order to get
this result. These are explained in the following:
\be
\label{barmu_deltamu}
\begin{aligned}
\bar\mu_1 &\equiv \frac12 \left( \mu_s^g + \mu_d^b \right) \; ,
& 
\bar\mu_2 &\equiv \frac12 \left( \mu_s^r + \mu_u^b \right) \; ,
& 
\bar\mu_3 &\equiv \frac12 \left( \mu_d^r + \mu_u^g \right) \; ,
\\
\delta\mu_1 &\equiv \frac12 \left( \mu_s^g - \mu_d^b \right) \;
, &
\delta\mu_2 &\equiv \frac12 \left( \mu_s^r - \mu_u^b \right) \;
, &
\delta\mu_3 &\equiv \frac12 \left( \mu_d^r - \mu_u^g \right) \;
,
\end{aligned}
\ee
are the averaged values and half of the differences of various
pairs of quark chemical potentials which come from the six $2
\times 2$-blocks in color-flavor and Nambu-Gorkov space, and
\be
\begin{aligned}
\phi_1 &\equiv \phi_{sd}^{gb} \; , &
\phi_2 &\equiv \phi_{su}^{rb} \; , &
\phi_3 &\equiv \phi_{du}^{rg} \; , \\
\varphi_1 &\equiv \phi_{ds}^{gb} \; , &
\varphi_2 &\equiv \phi_{us}^{rb} \; , &
\varphi_3 &\equiv \phi_{ud}^{rg} \; , \\
\sigma_1 &\equiv \phi_{uu}^{rr} \; , &
\sigma_2 &\equiv \phi_{dd}^{gg} \; , &
\sigma_3 &\equiv \phi_{ss}^{bb} \; ,
\end{aligned}
\ee
are useful definitions for the gap parameters in order to write
Eq.~(\ref{Gamma*kin2}) in a more compact form. The gap
parameters in the first two lines correspond to the attractive
antitriplet channel, while the gap parameters in the last line
correspond to the repulsive sextet channel. The quasiparticle
energies which come from the six $2 \times 2$-blocks in
color-flavor and Nambu-Gorkov space are defined by the following
equation:
\be
\epsilon_\fettu{k}^e \left( \mu, \phi \right) \equiv \sqrt{
\left( k - e \mu \right)^2 + \left| \phi \right|^2 } \; ,
\ee
cf.\ Eqs.~(\ref{epsilon_2SC_intro})
and~(\ref{epsilon_CFL_intro}). The dispersion relations (energy
eigenvalues) of the quasiquarks which come out of the six $2
\times 2$-blocks in color-flavor and Nambu-Gorkov space read,
\be
{( k_0^\pm )}_i = \epsilon_\fettu{k}^e \left( \bar\mu_i, \phi_i
\right) \pm \delta\mu_i \; .
\ee
The most complicated expression arises from the determinant of
the $3 \times 3$-matrix in color-flavor space in
Eq.~(\ref{Gamma*kin}) which can be calculated analytically. For
doing this, one can use computer software~\cite{Maple,
Mathematica}. The determinant has the following form:
\be
\label{determinant}
\det \left( 3 \times 3 \right)_e = k_0^6 + b k_0^4 + c k_0^2 + d
\; .
\ee
The coefficients $b$, $c$, and $d$ are rather complicated
functions of the quark momentum $k$, the three chemical
potentials $\mu_u^r$, $\mu_d^g$, and $\mu_s^b$, and the six gap
parameters $\varphi_i$ and $\sigma_i$, where $i = 1, 2, 3$, and
$e = \pm$ which stands for quasiquarks and quasiantiquarks,
respectively. The determinant~(\ref{determinant}) can always be
factorized as follows:
\be
\det \left( 3 \times 3\right )_e \equiv \prod_{i = 1}^3 \left[
k_0^2 - \left( \tilde\epsilon_i^e \right)^2 \right] \; .
\ee
The functions $\tilde\epsilon^e_i$ are the quasiparticle
energies, dispersion relations, and energy eigenvalues of the
three quasiquarks which come from the $6 \times 6$-block of the
inverse Nambu-Gorkov quark propagator. In order to get their
explicit expressions, one has to solve the cubic equation
\be
\label{cubic_in_xi}
\det \left( 3 \times 3 \right)_e = \xi^3 + b \xi^2 + c \xi + d =
0 \; ,
\ee
where $\xi \equiv k_0^2$. By making use of Cardano's formulae,
see Sec.~\ref{Cubic_equations} in the Appendix, the solutions
can be presented in an analytical form. Because of their very
complicated nature I refrain from presenting them explicitly.

The sum over all fermionic Matsubara frequencies has to be
performed in the kinetic term of the effective action of
quarks~(\ref{Gamma*kin2}). This is also done with the grand
partition function in
Sec.~\ref{Summation_over_the_fermionic_Matsubara_frequencies} in
the Appendix which is of the same form as the kinetic term of
the effective action of quarks. Therefore, the result in
Sec.~\ref{Summation_over_the_fermionic_Matsubara_frequencies}
can be applied to the kinetic term of the effective
action of quarks,
\bea
\Gamma_\mathrm{kin}^* &=& \sum_\fettu{k} \sum_{i = 1}^3 \sum_e
\left\{ \frac{ \tilde\epsilon_i^e - k }{ T } + 2 \ln \left[ 1 +
\exp \left( - \frac{ \tilde\epsilon_i^e }{ T } \right) \right]
\right\} \nonumber \\
&+& 2 \sum_\fettu{k} \sum_{i = 1}^3 \sum_e \left\{ \frac{
\epsilon_\fettu{k}^e \left( \bar\mu_i, \phi_i \right) - k }{ T }
+ \ln \left[ 1 + \exp \left( - \frac{ \epsilon_\fettu{k}^e
\left( \bar\mu_i, \phi_i \right) - \delta\mu_i }{ T } \right)
\right] \right. \nonumber \\
&& \hspace{4.43cm} \left. + \ln \left[ 1 + \exp \left( - \frac{
\epsilon_\fettu{k}^e \left( \bar\mu_i, \phi_i \right) +
\delta\mu_i }{ T } \right) \right] \right\} \; ,
\eea
where irrelevant constants are neglected and the vacuum
contribution is subtracted.
\subsection{The pressure of color-superconducting quark matter}
Finally, by combining all contributions to the effective action
of quarks~(\ref{Gamma*}), I derive the result for the pressure
$p \equiv \frac{T}{V} \Gamma^*$ of color-superconducting quark
matter,
\bea
\label{pressure_massless}
p &=& \frac{T}{\pi^2} \sum_{\beta = e}^\mu \int_0^\infty \ud k
\, k^2 \left\{ \ln \left[ 1 + \exp \left( - \frac{ E_\beta -
\mu_\beta }{ T } \right) \right] + \ln \left[ 1 + \exp \left( -
\frac{ E_\beta + \mu_\beta }{ T } \right) \right] \right\}
\nonumber \\
&+& \frac34 \frac{\Lambda^2}{g^2} \sum_{i=1}^3 \left( \phi_{i}^2
+ \varphi_{i}^2 + 6 \phi_{i} \varphi_{i} + 2 \sigma_{i}^2
\right) \nonumber \\
&+& \frac{1}{ 2 \pi^2 } \sum_{i = 1}^3 \sum_e \int_0^\kappa \ud
k \, k^2 \left\{ \tilde\epsilon_i^e - k + 2 T \ln \left[ 1 +
\exp \left( - \frac{ \tilde\epsilon_i^e }{ T } \right) \right]
\right\} \nonumber \\
&+& \frac{1}{\pi^2} \sum_{i=1}^3 \sum_e \int_0^\kappa \ud k \,
k^2 \left\{ \epsilon_\fettu{k}^e \left( \bar\mu_i, \phi_i
\right) - k + T \ln \left[ 1 + \exp \left( - \, \frac{
\epsilon_\fettu{k}^e \left( \bar\mu_i, \phi_i \right) -
\delta\mu_i }{ T } \right) \right] \right. \nonumber \\
&& \hspace{5.52cm} \left. + \, T \ln \left[ 1 + \exp \left( - \,
\frac{ \epsilon_\fettu{k}^e \left( \bar\mu_i, \phi_i \right) +
\delta\mu_i }{ T } \right) \right] \right\} \; ,
\eea
where I have performed the trace in the potential part of the
effective action of quarks~(\ref{Gamma*pot2}). I also converted
the sums over all $\fettu{k}$ into integrals by using the
relation,
\be
\label{sum_int}
\frac{1}{V} \sum_\fettu{k} \longrightarrow
\int_{-\infty}^{+\infty} \frac{\ud^3 \fettu{k}}{\left( 2 \pi
\right)^3} \longrightarrow \frac{1}{2 \pi^2} \int_0^\infty \ud k
\, k^2 \; .
\ee
The first term in the pressure~(\ref{pressure_massless}) is the
contribution of leptons, i.e.\ electrons and muons. They are
added in order to make color-superconducting quark matter
electrically neutral. In principle, the contribution of
neutrinos should be added as well. In this section, however,
their contribution is neglected which is a good approximation
for neutron stars after deleptonization. The dispersion
relations of the leptons are given by $E_\beta = ( k^2 +
m_\beta^2 )^{1/2}$, where $m_e = 0.51099906$~MeV is the electron
mass and $m_\mu = 105.658389$~MeV is the muon
mass~\cite{Taschenbuch_der_Physik}. In various applications, a
bag constant could be added to the pressure if necessary. In
order to render the integrals in the expression for the pressure
finite, I introduced a three-momentum cutoff $\kappa$. In QCD
with dynamical gluons, of course, such a cutoff would not be
necessary. Here, however, I use a model in which the
gluon-exchange interaction between quarks is approximated by a
point-like four-fermion coupling. Such a model is
nonrenormalizable.

The stationary conditions of the
pressure~(\ref{pressure_massless}) with respect to the nine gap
parameters read,
\be
\label{nine_gap_eqns}
\frac{\partial p}{\partial \phi_j} = 0 \; , \qquad
\frac{\partial p}{\partial \varphi_j} = 0 \; , \qquad
\frac{\partial p}{\partial \sigma_j} = 0 \; ,
\ee
cf.\ Eq.~(\ref{gap_eqn_intro}). In order to find the values of
the gap parameters, one can solve for them in
Eq.~(\ref{simplified_gap_eqns}). But a much simpler and
equivalent method is to solve the gap
equations~(\ref{nine_gap_eqns}) for the nine gap parameters,
\bsub
\label{gap_eqns}
\bea
\frac32 \frac{\Lambda^2}{g^2} \left( \phi_j + 3 \varphi_j
\right) + \frac{\phi_j}{2 \pi^2} \sum_e \int_0^\kappa \frac{\ud
k \, k^2}{\epsilon_\fettu{k}^e \left( \bar\mu_j, \phi_j \right)}
\left[ \tanh \left( \frac{\epsilon_\fettu{k}^e \left( \bar\mu_j,
\phi_j \right) - \delta\mu_j}{2 T} \right) \right.
\hspace{0.9mm} \nonumber \\
\left. + \tanh \left( \frac{\epsilon_\fettu{k}^e
\left( \bar\mu_j, \phi_j \right) + \delta\mu_j}{2 T} \right)
\right] &=& 0 \; , \\
\frac32 \frac{\Lambda^2}{g^2} \left( \varphi_j + 3 \phi_j
\right) + \frac{1}{2 \pi^2} \sum_{i = 1}^3 \sum_e \int_0^\kappa
\ud k \, k^2 \frac{\partial \tilde\epsilon_i^e}{\partial
\varphi_j} \tanh \left( \frac{ \tilde\epsilon_i^e}{2 T} \right)
&=& 0 \; , \\
3 \frac{\Lambda^2}{g^2} \sigma_j + \frac{1}{2 \pi^2} \sum_{i =
1}^3 \sum_e \int_0^\kappa \ud k \, k^2 \frac{\partial
\tilde\epsilon_i^e}{\partial \sigma_j} \tanh \left( \frac{
\tilde\epsilon_i^e}{2 T} \right) &=& 0 \; ,
\eea
\esub
which has to be done numerically. In order to obtain these
results, I made use of the relation,
\be
\tanh \left( \frac{x}{2} \right) = 1 - 2 n_F \left( x \right) \;
,
\ee
where $n_F$ is the Fermi-Dirac distribution function which was
defined by Eq.~(\ref{Fermi-Dirac}).

Matter in neutron stars has to satisfy the conditions of charge
neutrality. The condition for electric charge
neutrality~(\ref{electric_neutrality}) reads,
\bea
\label{nQ}
n_Q \! &=& - \sum_{\beta = e}^\mu n_\beta + \frac{1}{2 \pi^2}
\sum_{i = 1}^3 \sum_e \int_0^\kappa \ud k \, k^2 \frac{\partial
\tilde\epsilon_i^e}{\partial \mu_Q} \tanh \left(
\frac{\tilde\epsilon_i^e}{2 T} \right) \nonumber \\
&+& \frac{1}{2 \pi^2} \sum_{i = 1}^3 \sum_e \int_0^\kappa \ud k
\, k^2 \frac{\partial \bar\mu_i}{\partial \mu_Q} \,
\frac{\bar\mu_i - e k}{\epsilon_\fettu{k}^e \left( \bar\mu_i,
\phi_i \right)} \! \left[ \tanh \left(
\frac{\epsilon_\fettu{k}^e \left( \bar\mu_i, \phi_i \right) -
\delta\mu_i}{2 T} \right) + \tanh \left(
\frac{\epsilon_\fettu{k}^e \left( \bar\mu_i, \phi_i \right) +
\delta\mu_i}{2 T} \right) \right] \nonumber \\
&+& \frac{1}{\pi^2} \sum_{i = 1}^3 \sum_e \int_0^\kappa \ud k \,
k^2 \frac{\partial \left( \delta\mu_i \right)}{\partial \mu_Q}
\left[ n_F \left( \frac{\epsilon_\fettu{k}^e \left( \bar\mu_i,
\phi_i \right) - \delta\mu_i}{T} \right) - n_F \left(
\frac{\epsilon_\fettu{k}^e \left( \bar\mu_i, \phi_i \right) +
\delta\mu_i}{T} \right) \right] = 0 \; ,
\eea
where $n_\beta$ is the number density of electrons and muons,
respectively, see Eq.~(\ref{number_density_beta}). The
conditions for color neutrality~(\ref{color_neutrality}) are,
\bsub
\label{n3n8}
\bea
n_3 &=& \frac{1}{2 \pi^2} \sum_{i = 1}^3 \sum_e \int_0^\kappa
\ud k \, k^2 \frac{\partial \tilde\epsilon_i^e}{\partial \mu_3}
\tanh \left( \frac{\tilde\epsilon_i^e}{2 T} \right) \nonumber \\
&+& \frac{1}{2 \pi^2} \sum_{i = 1}^3 \sum_e \int_0^\kappa \ud k
\, k^2 \frac{\partial \bar\mu_i}{\partial \mu_3} \,
\frac{\bar\mu_i - e k}{\epsilon_\fettu{k}^e \left( \bar\mu_i,
\phi_i \right)} \left[ \tanh \left( \frac{\epsilon_\fettu{k}^e
\left( \bar\mu_i, \phi_i \right) - \delta\mu_i}{2 T} \right) +
\tanh \left( \frac{\epsilon_\fettu{k}^e \left( \bar\mu_i, \phi_i
\right) + \delta\mu_i}{2 T} \right) \right] \nonumber \\
&+& \frac{1}{\pi^2} \sum_{i = 1}^3 \sum_e \int_0^\kappa \ud k \,
k^2 \frac{\partial \left( \delta\mu_i \right)}{\partial \mu_3}
\left[ n_F \left( \frac{\epsilon_\fettu{k}^e \left( \bar\mu_i,
\phi_i \right) - \delta\mu_i}{T} \right) - n_F \left(
\frac{\epsilon_\fettu{k}^e \left( \bar\mu_i, \phi_i \right) +
\delta\mu_i}{T} \right) \right] = 0 \; , \\
n_8 &=& \frac{1}{2 \pi^2} \sum_{i = 1}^3 \sum_e \int_0^\kappa
\ud k \, k^2 \frac{\partial \tilde\epsilon_i^e}{\partial \mu_8}
\tanh \left( \frac{\tilde\epsilon_i^e}{2 T} \right) \nonumber \\
&+& \frac{1}{2 \pi^2} \sum_{i = 1}^3 \sum_e \int_0^\kappa \ud k
\, k^2 \frac{\partial \bar\mu_i}{\partial \mu_8} \,
\frac{\bar\mu_i - e k}{\epsilon_\fettu{k}^e \left( \bar\mu_i,
\phi_i \right)} \left[ \tanh \left( \frac{\epsilon_\fettu{k}^e
\left( \bar\mu_i, \phi_i \right) - \delta\mu_i}{2 T} \right) +
\tanh \left( \frac{\epsilon_\fettu{k}^e \left( \bar\mu_i, \phi_i
\right) + \delta\mu_i}{2 T} \right) \right] \nonumber \\
&+& \frac{1}{\pi^2} \sum_{i = 1}^3 \sum_e \int_0^\kappa \ud k \,
k^2 \frac{\partial \left( \delta\mu_i \right)}{\partial \mu_8}
\left[ n_F \left( \frac{\epsilon_\fettu{k}^e \left( \bar\mu_i,
\phi_i \right) - \delta\mu_i}{T} \right) - n_F \left(
\frac{\epsilon_\fettu{k}^e \left( \bar\mu_i, \phi_i \right) +
\delta\mu_i}{T} \right) \right] = 0 \; .
\eea
\esub
The derivatives of the averaged values of various pairs of quark
chemical potentials which come out of the $2 \times 2$-blocks in
color-flavor and Nambu-Gorkov space with respect to the chemical
potential of electric charge $\mu_Q$ and the color chemical
potentials $\mu_3$ and $\mu_8$, respectively, are,
\be
\begin{aligned}
\frac{\partial \bar\mu_1}{\partial \mu_Q} &= -\frac13 \; , &
\frac{\partial \bar\mu_2}{\partial \mu_Q} &= \frac16 \; , &
\frac{\partial \bar\mu_3}{\partial \mu_Q} &= \frac16 \; , \\
\frac{\partial \bar\mu_1}{\partial \mu_3} &= -\frac14 \; , &
\frac{\partial \bar\mu_2}{\partial \mu_3} &= \frac14 \; , &
\frac{\partial \bar\mu_3}{\partial \mu_3} &= 0 \; , \\
\frac{\partial \bar\mu_1}{\partial \mu_8} &=
-\frac{1}{4\sqrt{3}} \; , &
\frac{\partial \bar\mu_2}{\partial \mu_8} &=
-\frac{1}{4\sqrt{3}} \; , &
\frac{\partial \bar\mu_3}{\partial \mu_8} &= \frac{1}{2\sqrt{3}}
\; .
\end{aligned}
\ee
The derivatives of half of the differences of various pairs of
quark chemical potentials which come out of the $2 \times
2$-blocks in color-flavor and Nambu-Gorkov space with respect to
the chemical potential of electric charge $\mu_Q$ and the color
chemical potentials $\mu_3$ and $\mu_8$, respectively, read,
\be
\begin{aligned}
\frac{\partial \left( \delta\mu_1 \right)}{\partial \mu_Q} &= 0
\; , &
\frac{\partial \left( \delta\mu_2 \right)}{\partial \mu_Q} &=
-\frac12 \; , &
\frac{\partial \left( \delta\mu_3 \right)}{\partial \mu_Q} &=
-\frac12 \; , \\
\frac{\partial \left( \delta\mu_1 \right)}{\partial \mu_3} &=
-\frac14 \; , &
\frac{\partial \left( \delta\mu_2 \right)}{\partial \mu_3} &=
\frac14 \; , & 
\frac{\partial \left( \delta\mu_3 \right)}{\partial \mu_3} &=
\frac12 \; , \\
\frac{\partial \left( \delta\mu_1 \right)}{\partial \mu_8} &=
\frac{\sqrt{3}}{4} \; , &
\frac{\partial \left( \delta\mu_2 \right)}{\partial \mu_8} &=
\frac{\sqrt{3}}{4} \; , &
\frac{\partial \left( \delta\mu_3 \right)}{\partial \mu_8} &= 0
\; .
\end{aligned}
\ee
The number density of quarks with color $i$ and flavor $f$ can
be obtained by
\be
n_f^i = \frac{\partial p}{\partial \mu_f^i} \; .
\ee
By using this relation, the number density of red up, green
down, and blue strange quarks respectively reads,
\be
\label{rugdbs}
n_f^i = \frac{1}{2 \pi^2} \sum_{j = 1}^3 \sum_e \int_0^\kappa
\ud k \, k^2 \frac{\partial \tilde\epsilon_j^e}{\partial
\mu_f^i} \tanh \left( \frac{\tilde\epsilon_j^e}{2 T} \right) \;
.
\ee
The number density of green strange, red strange, and red down
quarks (upper sign) as well as the number density of blue down,
blue up, and green up quarks (lower sign) is,
\bea
\label{gsrsrd_bdbugu}
n_f^i &=& \frac{1}{4 \pi^2} \sum_e \int_0^\kappa \ud k \, k^2
\frac{ \bar\mu_j - e k }{ \epsilon_\fettu{k}^e \left( \bar\mu_j,
\phi_j \right) } \left[ \tanh \left( \frac{\epsilon_\fettu{k}^e
\left( \bar\mu_j, \phi_j \right) - \delta\mu_j}{2 T} \right) +
\tanh \left( \frac{\epsilon_\fettu{k}^e \left( \bar\mu_j, \phi_j
\right) + \delta\mu_j}{2 T} \right) \right] \nonumber \\
&\pm& \frac{1}{2 \pi^2} \sum_e \int_0^\kappa \ud k \, k^2 \left[
n_F \left( \frac{\epsilon_\fettu{k}^e \left( \bar\mu_j, \phi_j
\right) - \delta\mu_j}{T} \right) - n_F \left(
\frac{\epsilon_\fettu{k}^e \left( \bar\mu_j, \phi_j \right) +
\delta\mu_j}{T} \right) \right] \; .
\eea
In this equation for the quark number densities, those
$\bar\mu_j$ and $\delta\mu_j$~(\ref{barmu_deltamu}) have
to be taken which contain the respective quark chemical
potential that corresponds to the respective quark number
density which one wants to calculate. Since the quasiparticle
distribution relations $\tilde\epsilon_i^e$ are obtained
analytically by solving the cubic equation~(\ref{cubic_in_xi}),
the derivatives of $\tilde\epsilon_i^e$ with respect to the gap
parameters and the chemical potentials can also be computed
analytically. Because of their very complicated nature, I
refrain from presenting them explicitly.

All these thermodynamic quantities are valid for nonzero
temperature. I also want to show the results in the special case
of zero temperature. Therefore, Eqs.~(\ref{zeroT})
and~(\ref{zeroT2}), and the following relations are useful:
\be
\lim_{T \rightarrow 0} \tanh \left( \frac{x}{T} \right) = 1 - 2
\Heaviside \left( - x \right) \; ,
\ee
and
\bsub
\bea
\Heaviside \left[ - \epsilon_\fettu{k}^e \left( \bar\mu_i,
\phi_i \right) + \delta\mu_i \right] + \Heaviside \left[ -
\epsilon_\fettu{k}^e \left( \bar\mu_i, \phi_i \right) -
\delta\mu_i \right] &=& \Heaviside \left[ - \epsilon_\fettu{k}^e
\left( \bar\mu_i, \phi_i \right) + \left| \delta\mu_i \right|
\right] \; , \\
\Heaviside \left[ - \epsilon_\fettu{k}^e \left( \bar\mu_i,
\phi_i \right) + \delta\mu_i \right] - \Heaviside \left[ -
\epsilon_\fettu{k}^e \left( \bar\mu_i, \phi_i \right) -
\delta\mu_i \right] &=& \Heaviside \left[ - \epsilon_\fettu{k}^e
\left( \bar\mu_i, \phi_i \right) + \left| \delta\mu_i \right|
\right] \sgn \left( \delta\mu_i \right) \; .
\eea
\esub
In this context, it is important to mention that the
quasiparticle energies $\epsilon_\fettu{k}^e \left( \bar\mu_i,
\phi_i \right)$ and $\tilde\epsilon_i^e$ are positive numbers.

In the limit of zero temperature, the pressure of
color-superconducting quark matter~(\ref{pressure_massless})
reads,
\bea
p &=& \frac{1}{3 \pi^2} \sum_{\beta = e}^\mu
\int_0^{k_{F_\beta}} \ud k \, \frac{k^4}{E_\beta} + \frac34
\frac{\Lambda^2}{g^2} \sum_{i = 1}^3 \left( \phi_i^2 +
\varphi_i^2 + 6 \phi_i \varphi_i + \sigma_i^2 \right) +
\frac{1}{2 \pi^2} \sum_{i = 1}^3 \sum_e \int_0^\kappa \ud k \,
k^2 \left( \tilde\epsilon_i^e - k \right) \nonumber \\
&+& \frac{1}{\pi^2} \sum_{i = 1}^3 \sum_e \int_0^\kappa \ud k \,
k^2 \left[ \epsilon_\fettu{k}^e \left( \bar\mu_i, \phi_i \right)
- k \right] + \frac{1}{\pi^2} \sum_{i = 1}^3
\int_{\mu_i^-}^{\mu_i^+} \ud k \, k^2 \left[ \left| \delta\mu_i
\right| - \epsilon_\fettu{k}^e \left( \bar\mu_i, \phi_i \right)
\right] \; ,
\eea
where
\be
\mu_i^\pm \equiv \bar\mu_i \pm \sqrt{ \left| \delta\mu_i
\right|^2 - \left| \phi_i \right|^2 } \; .
\ee
At zero temperature, the gap equations~(\ref{gap_eqns}) are,
\bsub
\label{gap_eqns_T0}
\bea
\frac32 \frac{\Lambda^2}{g^2} \left( \phi_j + 3 \varphi_j
\right) + \frac{\phi_j}{\pi^2} \sum_e \left( \int_0^{\mu_j^-}
\frac{\ud k \, k^2}{\epsilon_\fettu{k}^e \left( \bar\mu_j,
\phi_j \right)} + \int_{\mu_j^+}^\kappa \frac{\ud k \,
k^2}{\epsilon_\fettu{k}^e \left( \bar\mu_j, \phi_j \right)}
\right) &=& 0 \; , \\
\frac32 \frac{\Lambda^2}{g^2} \left( \varphi_j + 3 \phi_j
\right) + \frac{1}{2 \pi^2} \sum_{i = 1}^3 \sum_e \int_0^\kappa
\ud k \, k^2 \frac{\partial \tilde\epsilon_i^e}{\partial
\varphi_j} &=& 0 \; , \\
3 \frac{\Lambda^2}{g^2} \sigma_j + \frac{1}{2 \pi^2} \sum_{i =
1}^3 \sum_e \int_0^\kappa \ud k \, k^2 \frac{\partial
\tilde\epsilon_i^e}{\partial \sigma_j} &=& 0 \; .
\eea
\esub
From Eq.~(\ref{nQ}), the condition for electric charge
neutrality at zero temperature can be easily derived,
\bea
\label{neutrality_nQ_T0}
n_Q &=& - \sum_{\beta = e}^\mu n_\beta + \frac{1}{2 \pi^2}
\sum_{i = 1}^3 \sum_e \int_0^\kappa \ud k \, k^2 \frac{\partial
\tilde\epsilon_i^e}{\partial \mu_Q} + \frac{1}{\pi^2} \sum_{i =
1}^3 \sgn \left( \delta\mu_i \right) \int_{\mu_i^-}^{\mu_i^+}
\ud k \, k^2 \frac{\partial \left( \delta\mu_i \right)}{\partial
\mu_Q} \nonumber \\
&+& \frac{1}{\pi^2} \sum_{i = 1}^3 \sum_e \left(
\int_0^{\mu_i^-} \ud k \, k^2 \frac{\partial \bar\mu_i}{\partial
\mu_Q} \, \frac{\bar\mu_i - e k}{\epsilon_\fettu{k}^e \left(
\bar\mu_i, \phi_i \right)} + \int_{\mu_i^+}^\kappa \ud k \, k^2
\frac{\partial \bar\mu_i}{\partial \mu_Q} \, \frac{\bar\mu_i - e
k}{\epsilon_\fettu{k}^e \left( \bar\mu_i, \phi_i \right)}
\right) = 0 \; ,
\eea
where the number density of electrons and muons respectively
is given by
Eq.~(\ref{quark_flavor_and_lepton_number_densities}). The color
charge neutrality conditions~(\ref{n3n8}) at zero temperature
read,
\bsub
\label{neutrality_n3n8_T0}
\bea
n_3 &=& \frac{1}{\pi^2} \sum_{i = 1}^3 \sum_e \left(
\int_0^{\mu_i^-} \ud k \, k^2 \frac{\partial \bar\mu_i}{\partial
\mu_3} \, \frac{\bar\mu_i - e k}{\epsilon_\fettu{k}^e \left(
\bar\mu_i, \phi_i \right)} + \int_{\mu_i^+}^\kappa \ud k \, k^2
\frac{\partial \bar\mu_i}{\partial \mu_3} \, \frac{\bar\mu_i - e
k}{\epsilon_\fettu{k}^e \left( \bar\mu_i, \phi_i \right)}
\right) \nonumber \\
&+& \frac{1}{\pi^2} \sum_{i = 1}^3 \sgn \left( \delta\mu_i
\right) \int_{\mu_i^-}^{\mu_i^+} \ud k \, k^2 \frac{\partial
\left( \delta\mu_i \right)}{\partial \mu_3} + \frac{1}{2 \pi^2}
\sum_{i = 1}^3 \sum_e \int_0^\kappa \ud k \, k^2 \frac{\partial
\tilde\epsilon_i^e}{\partial \mu_3} = 0 \; , \\
n_8 &=& \frac{1}{\pi^2} \sum_{i = 1}^3 \sum_e \left(
\int_0^{\mu_i^-} \ud k \, k^2 \frac{\partial \bar\mu_i}{\partial
\mu_8} \, \frac{\bar\mu_i - e k}{\epsilon_\fettu{k}^e \left(
\bar\mu_i, \phi_i \right)} + \int_{\mu_i^+}^\kappa \ud k \, k^2
\frac{\partial \bar\mu_i}{\partial \mu_8} \, \frac{\bar\mu_i - e
k}{\epsilon_\fettu{k}^e \left( \bar\mu_i, \phi_i \right)}
\right) \nonumber \\
&+& \frac{1}{\pi^2} \sum_{i = 1}^3 \sgn \left( \delta\mu_i
\right) \int_{\mu_i^-}^{\mu_i^+} \ud k \, k^2 \frac{\partial
\left( \delta\mu_i \right)}{\partial \mu_8} + \frac{1}{2 \pi^2}
\sum_{i = 1}^3 \sum_e \int_0^\kappa \ud k \, k^2 \frac{\partial
\tilde\epsilon_i^e}{\partial \mu_8} = 0 \; .
\eea
\esub
The quark number densities of red up, green down, and blue
strange quarks at zero temperature can be easiliy derived from
Eq.~(\ref{rugdbs}),
\be
n_f^i = \frac{1}{2 \pi^2} \sum_{j = 1}^3 \sum_e \int_0^\kappa
\ud k \, k^2 \frac{\partial \tilde\epsilon_j^e}{\partial
\mu_f^i} \; .
\ee
The quark number densities of green strange, red strange, and
red down quarks (upper sign) as well as the number densities of
blue down, blue up, and green up quarks (lower sign) at zero
temperature can be obtained by Eq.~(\ref{gsrsrd_bdbugu}),
\be
n_f^i = \frac{1}{2 \pi^2} \sum_e \left( \int_0^{\mu_j^-} \ud k
\, k^2 \frac{\bar\mu_j - e k}{\epsilon_\fettu{k}^e \left(
\bar\mu_j, \phi_j \right)} + \int_{\mu_j^+}^\kappa \ud k \, k^2
\frac{\bar\mu_j - e k}{\epsilon_\fettu{k}^e \left( \bar\mu_j,
\phi_j \right)} \right) \pm \frac{\sgn \left( \delta\mu_j
\right)}{6 \pi^2} \left[ \left( \mu_j^+ \right)^3 - \left(
\mu_j^- \right)^3 \right] \; .
\ee

I use the following model parameters: the strength of the
diquark coupling and the value of the cutoff are fixed as,
\be
\frac{g^2}{\Lambda^2} = 45.1467~\mathrm{GeV}^{-2} \; , \qquad
\kappa = 0.6533~\mathrm{GeV} \; .
\ee
The model parameters are chosen to reproduce several key
observables of vacuum QCD such as the pion decay constant.

For given temperature $T$, strange quark mass $m_s$, and quark
chemical potential $\mu$, I solved a coupled system of twelve
non-linear equations, i.e.\ nine gap equations and three
neutrality conditions, in order to obtain the values of the nine
gap parameters and of the chemical potentials of electric and
color charge. This was done numerically~\cite{numrec}.
\subsection{Results at zero temperature}
\label{Results_at_zero_temperature}
In this subsection, I focus on three-flavor quark matter at zero
temperature. It is clear that, for small and moderate values of
the strange quark mass, the ground state of neutral quark matter
should correspond to either the regular (gapped) CFL phase
or the gCFL phase. At very large strange quark mass and/or
relatively weak coupling, the ground state can also be either a
regular (gapped) or gapless 2SC color superconductor.
\begin{figure}[H]
\begin{center}
\includegraphics[width=0.95\textwidth]{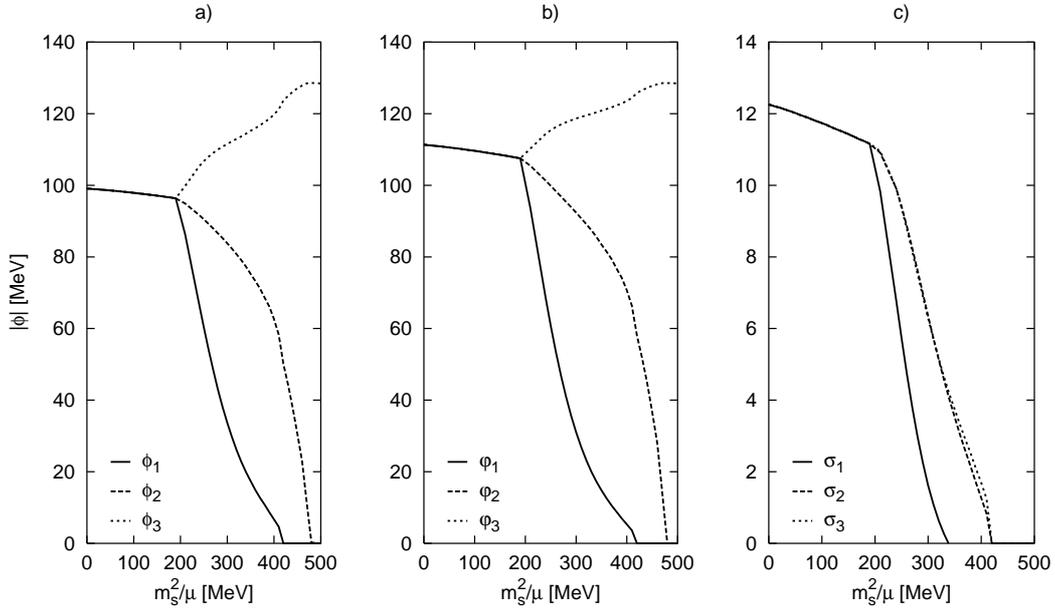}
\caption[The gap parameters as a function of $m_s^2/\mu$.]{The
absolute values of the gap parameters as a function of
$m_s^2/\mu$ for electrical and color neutral
color-superconducting quark matter at $T = 0$ and $\mu =
500$~MeV. The actual values of the gap parameters shown in
panel~a) are negative.}
\label{phi_ms2mu}
\end{center}
\end{figure}
In order to see how the phase structure of neutral three-flavor 
quark matter changes with the strange quark mass $m_s$, I solve
a coupled set of twelve equations, i.e., nine gap
equations~(\ref{gap_eqns_T0}) and three neutrality
conditions~(\ref{neutrality_nQ_T0})
and~(\ref{neutrality_n3n8_T0}), for various values of $m_s$,
keeping the quark chemical potential fixed. In this calculation
I take $\mu=500$~MeV. The results for the absolute values of the
gap parameters and the chemical potentials $\mu_Q$, $\mu_3$, and
$\mu_8$ are shown in Figs.~\ref{phi_ms2mu} and~\ref{chp_ms2mu},
respectively. Note that, strictly speaking, the gap parameters
do not coincide with the actual values of the gaps in the
quasiparticle spectra. In the case of the CFL phase, for
example, there is a degenerate octet of quasiparticles with a
gap $\phi_\mathrm{octet}= \left| \phi_{1} \right|$ and a singlet
state with a gap $\phi_\mathrm{singlet} = 3 \varphi_{1} - \left|
\phi_{1} \right|$. In the CFL phase, $\phi_{1} = \phi_{2} =
\phi_{3} < 0$, $\varphi_{1} = \varphi_{2} = \varphi_{3} > 0$,
and $\sigma_{1} = \sigma_{2} = \sigma_{3} > 0$. Also, in the CFL
phase, the following relation between the gap parameters is
satisfied: $\sigma_{i} = \varphi_{i} - \left| \phi_{i} \right|
\equiv 2 \phi_{(6,6)}$, $i=1,2,3$, where $\phi_{(6,6)}$ is the
sextet gap in the notation of Ref.~\cite{Shovkovy_Wijewardhana}.
\begin{figure}[H]
\begin{center}
\includegraphics[width=0.95\textwidth]{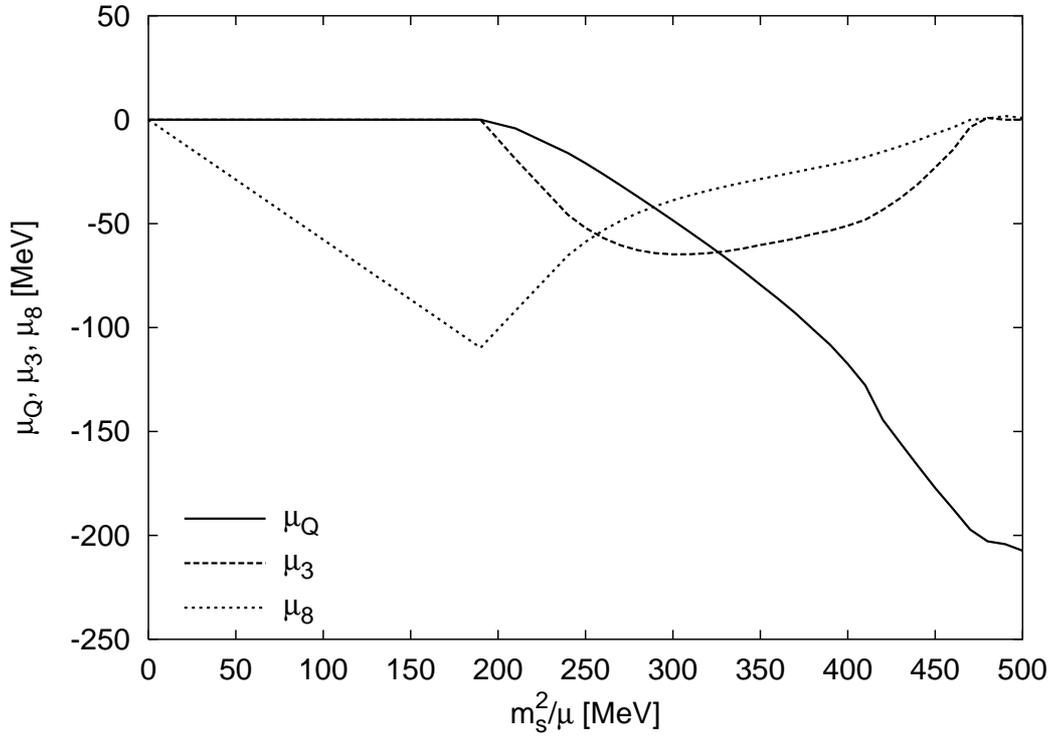}
\caption[The chemical potentials of electric and color charge as
a function of $m_s^2/\mu$.]{The electrical and color chemical
potentials as a function of $m_s^2/\mu$ of electrical and color
neutral color-superconducting quark matter at $T = 0$ and $\mu =
500$~MeV.}
\label{chp_ms2mu}
\end{center}
\end{figure}
\begin{figure}[H]
\begin{center}
\includegraphics[width=0.95\textwidth]{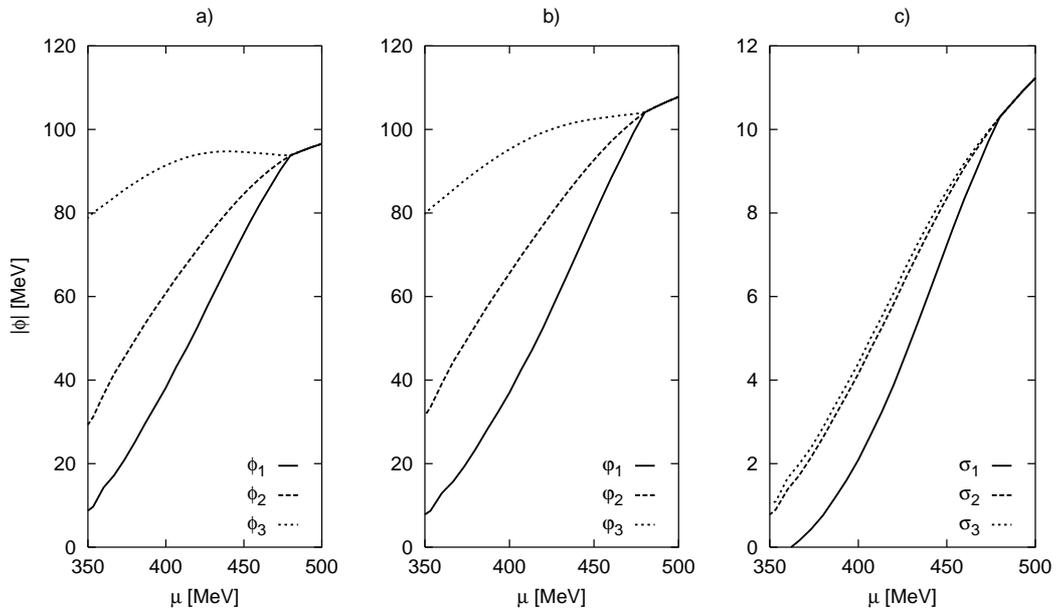}
\caption[The gap parameters as a function of the quark chemical
potential.]{The absolute values of the gap parameters as a
function of $\mu$ of electrical and color neutral
color-superconducting quark matter at $T = 0$ and $m_s =
300$~MeV. The actual values of the gap parameters shown in
panel~a) are negative.}
\label{phi_mu}
\end{center}
\end{figure}
The results in Figs.~\ref{phi_ms2mu} and~\ref{chp_ms2mu} extend
the results of Refs.~\cite{Alford_Kouvaris_Rajagopal1,
Alford_Kouvaris_Rajagopal2} by considering a more general
ansatz for the gap matrix that takes into account, in
particular, the pairing in the symmetric sextet channel. The
effect of including pairing in the symmetric channel is a
splitting between the pairs of gaps ($|\phi_{1}|$,
$\varphi_{1}$), ($|\phi_{2}|$, $\varphi_{2}$), and
($|\phi_{3}|$, $\varphi_{3}$) that is also reflected in the
change of the quasiparticle spectra. In agreement with the
general arguments of Refs.~\cite{CFL_discoverers,
Shovkovy_Wijewardhana, Schaefer2}, the symmetric sextet gaps are
rather small, see Fig.~\ref{phi_ms2mu}~c). This explains the
fact why the splittings between the above mentioned pairs of gap
parameters are not very large [compare the results in
Figs.~\ref{phi_ms2mu}~a) and b)].
\begin{figure}[H]
\begin{center}
\includegraphics[width=0.95\textwidth]{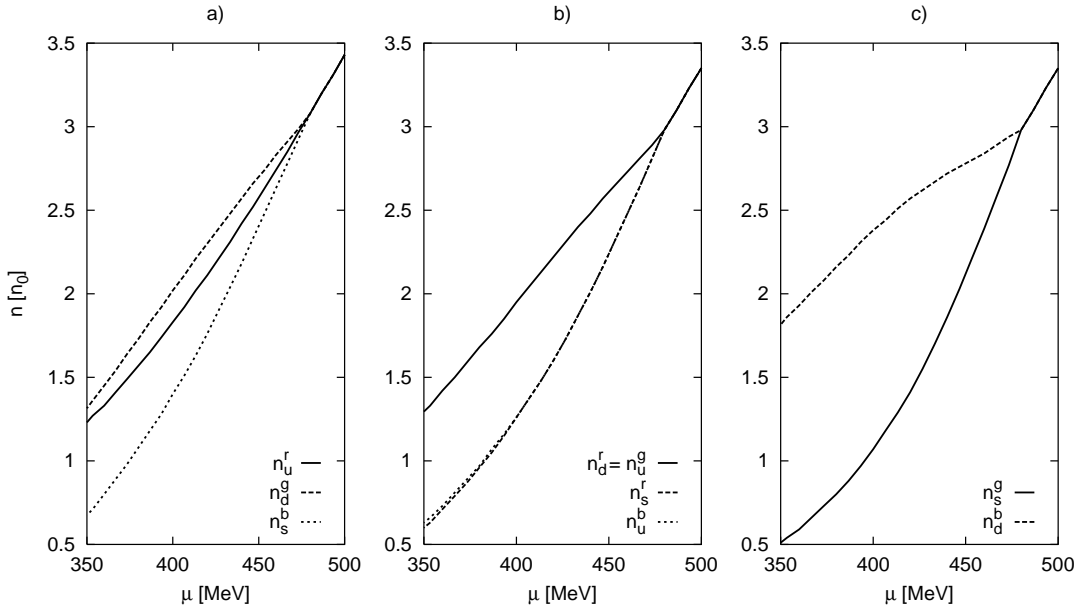}
\caption[The quark number densities as a function of the quark
chemical potential.]{The number densities of each quark color
and flavor as a function of $\mu$ for electrical and color
neutral color-superconducting quark matter at $T = 0$ and $m_s =
300$~MeV. The densities are given in units of the saturation
density of nuclear matter, $n_0 = 0.15$ fm$^{-3}$.}
\label{n_mu}
\end{center}
\end{figure}
The effects of the strange quark mass are incorporated by a
shift of the strange quark chemical potential~(\ref{shift2}).
Such an approach is certainly reliable at small values of the
strange quark mass. However, I assumed that it is also
qualitatively correct at large values of the strange quark mass.
By comparing the results for the gap parameters shown in
panel~a) of Fig.~\ref{phi_ms2mu} with the results for the gap
parameters obtained by taking the strange quark mass properly
into account shown in the left panel of Fig.~1 in
Ref.~\cite{Shovkovy_Ruester_Rischke}, one can see that both
results are qualitatively in good agreement.

I confirm that the phase transition from the CFL phase to the
gCFL phase happens at a critical value of the parameter
$m_s^2/\mu$ that is in good agreement with the simple estimate
of Refs.~\cite{Alford_Kouvaris_Rajagopal1,
Alford_Kouvaris_Rajagopal2},
\be
\frac{m_s^2}{\mu} \simeq 2 \Delta \simeq 190~\mathrm{MeV} \;
.
\ee
The qualitative results for the chemical potentials $\mu_Q$,
$\mu_3$, and $\mu_8$ in Fig.~\ref{chp_ms2mu} are in agreement
with the corresponding results obtained in
Refs.~\cite{Alford_Kouvaris_Rajagopal1,
Alford_Kouvaris_Rajagopal2} as well. In my notation, the color
chemical potential $\mu_8$ fulfills the identity,
\be
\mu_8 = - \frac{m_s^2}{\sqrt{3} \mu} \; ,
\ee
in the CFL phase, cf.\ Eq.~(\ref{mu8_CFL}). While the CFL phase
requires no electrons to remain neutral, the pairing in the gCFL
phase is distorted and a nonzero density of electrons appears.
This is seen directly from the dependence of the chemical
potential of electric charge $\mu_Q$ in Fig.~\ref{chp_ms2mu}
which becomes nonzero only in the gCFL phase. This observation
led the authors of Refs.~\cite{Alford_Kouvaris_Rajagopal1,
Alford_Kouvaris_Rajagopal2} to the conclusion that the phase
transition between the CFL and gCFL phase is an insulator-metal
phase transition, and that the value of the electron density is
a convenient order parameter in the description of such a
transition. In fact, one could also choose one of the
differences between number densities of mutually paired quarks
as an alternative choice for the order parameter~\cite{g2SC_1, 
g2SC_2}. In either case, there does not seem to exist any
continuous symmetry that is associated with such an order
parameter. To complete the discussion of the chemical
potentials, I add that the other color chemical potential
$\mu_3$ is zero only in the CFL phase at $T=0$.

The effects of a nonzero strange quark mass on the phase
structure of neutral strange quark matter could be viewed from
a different standpoint that, in application to stars, may look 
more natural. This is the case where the dependence on the quark
chemical potential is studied at a fixed value of $m_s$. The
corresponding numerical results are shown in Fig.~\ref{phi_mu}.
(Note once again that $\phi_{i}$, where $i=1,2,3$, have negative
values, and I always plotted their absolute values.) In this
particular calculation I chose $m_s=300$~MeV.
\begin{figure}[H]
\begin{center}
\includegraphics[width=0.95\textwidth]{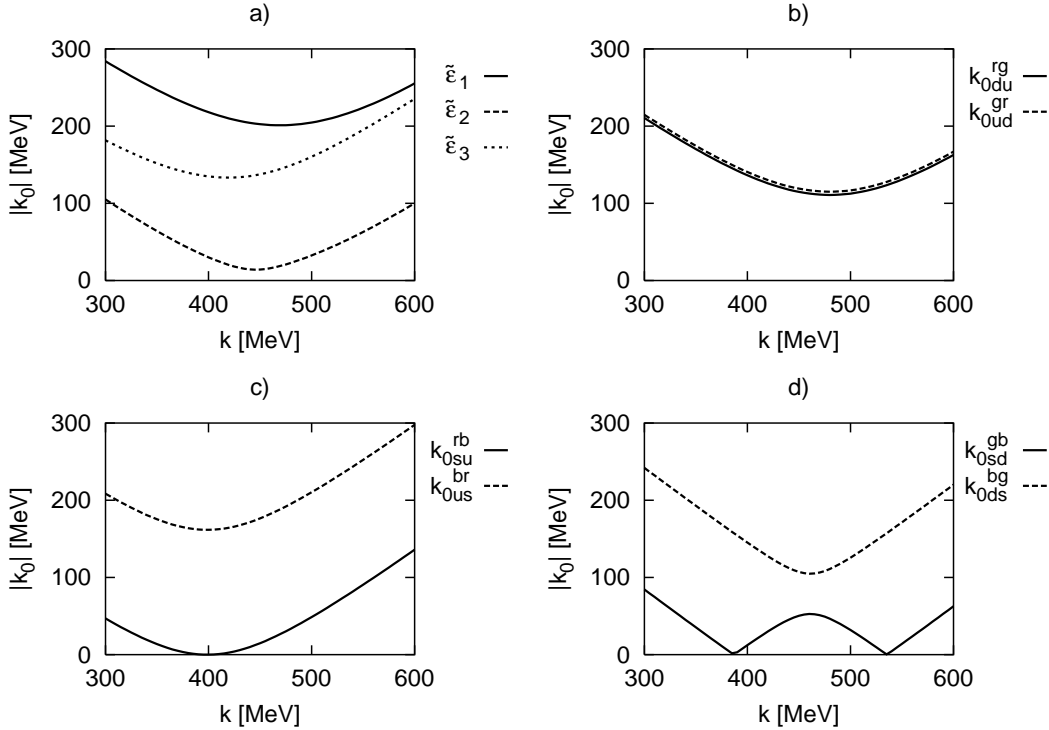}
\caption[The quasiparticle dispersion relations.]{The
quasiparticle dispersion relations for electrical and color
neutral color-superconducting quark matter at $T = 0$, $\mu =
500$~MeV, and $m_s = 400$~MeV.}
\label{epsilon}
\end{center}
\end{figure}
At large values of the quark chemical potential [$\mu \gtrsim
m_s^2/(2\Delta) \sim 475$~MeV which is similar to the small
strange quark mass limit considered before], the ground state of
quark matter is the CFL phase. At smaller values of the chemical
potential, the ground state of dense matter is the gCFL phase.
In this case, there are nine gap parameters all of which are
different from each other. One could also check that the density
of quarks that pair are not equal in the gCFL phase. This can be 
seen from Fig.~\ref{n_mu} where all nine quark number densities
are plotted for the same value of the strange quark mass,
$m_s = 300$~MeV. Only $n_d^r = n_u^g$ and $n_s^r \approx n_u^b$,
all other quark number densities are different from each other.
This agrees with the general criterion of the appearance of
gapless phases at $T=0$ that was proposed in Refs.~\cite{g2SC_1,
g2SC_2} in the case of two-flavor quark matter. In the ordinary
CFL phase, in contrast, one finds that $n_u^r = n_d^g = n_s^b$,
and $n_d^r = n_s^r = n_u^g = n_s^g = n_u^b = n_d^b$.

In order to see that the gCFL phase indeed describes a gapless
superconductor, it is necessary to show that the dispersion
relations of quasiparticles contain gapless excitations. In
Fig.~\ref{epsilon}, the dispersion relations of all nine
quasiparticles are plotted. The dispersion relations for the
corresponding antiparticles are not shown. From
Fig.~\ref{epsilon}, one can see that there is indeed a gapless
mode in the green-strange--blue-down sector. This is the same
that was found in Refs.~\cite{Alford_Kouvaris_Rajagopal1,
Alford_Kouvaris_Rajagopal2}. Note also that, in agreement with 
Refs.~\cite{Alford_Kouvaris_Rajagopal1,
Alford_Kouvaris_Rajagopal2}, the red-strange--blue-up
quasiparticle has a dispersion relation that is nearly
quadratic, $k_{0su}^{ \phantom{0} rb } \simeq | k - k^* |^2$
with $k^* \approx 400$~MeV for the given choice of parameters,
see Fig.~\ref{epsilon}~c). The nearly quadratic dispersion
relation resembles the situation at the transition between the
2SC phase, where $n_u^r = n_d^g$, and the gapless 2SC phase,
where $n_u^r$ and $n_d^g$ are different. This explains the
approximate equality $n_s^r \approx n_u^b$ mentioned above.
\subsection{Results at nonzero temperature}
In this subsection, I present the results for the phase
structure of dense neutral three-flavor quark matter in the
plane of temperature and $m_s^2 / \mu$, as well as in the plane
of temperature and quark chemical potential.

I discuss the temperature dependence of the gap parameters in
the two qualitatively different cases of small and large values
of the strange quark mass. As one could see in
Sec.~\ref{Results_at_zero_temperature}, the zero-temperature
properties of neutral quark matter were very different in these
two limits.
\begin{figure}[H]
\begin{center}
\includegraphics[width=0.95\textwidth]{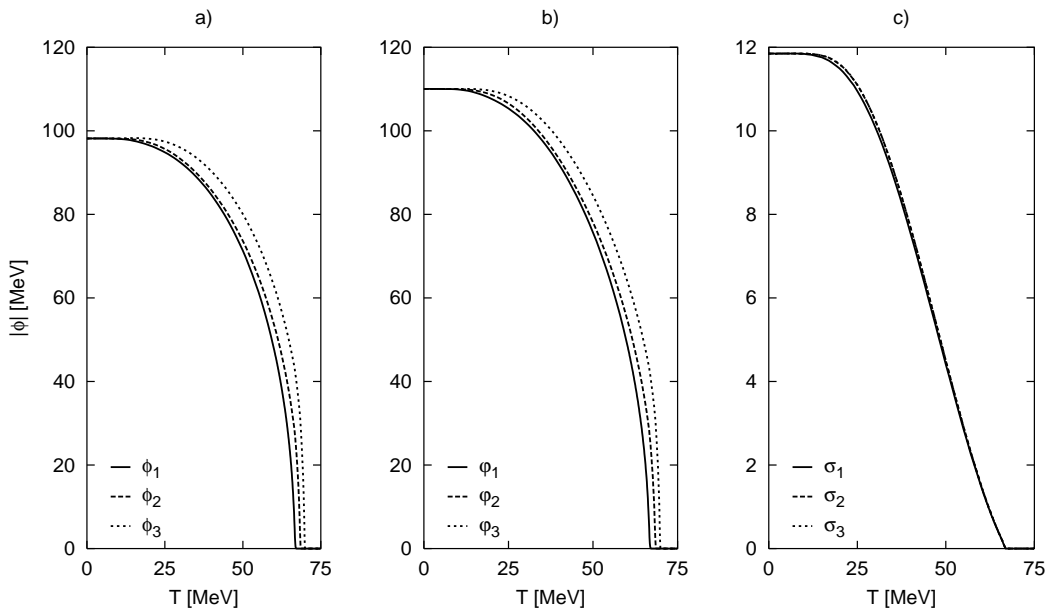}
\caption[The temperature dependence of the gaps for a small
strange quark mass.]{The temperature dependence of the gaps in
the case of a small strange quark mass, $m_s^2 / \mu = 80$~MeV.
Note, that the actual values of the gap parameters shown in
panel~a) are negative. The results are plotted for $\mu =
500$~MeV.}
\label{phit80}
\end{center}
\end{figure}
The results for the temperature dependence of the gap parameters 
are shown in Figs.~\ref{phit80},~\ref{phit80zoom},
and~\ref{phit320} for two different values of the strange quark
mass that represent the two qualitatively different regimes. In
the case of a small strange quark mass (i.e., the case of
$m_s^2/\mu = 80$~MeV which is shown in Figs.~\ref{phit80}
and~\ref{phit80zoom}), the zero-temperature limit corresponds to
the CFL phase. This is seen from the fact that the three
different gaps shown in every panel of Fig.~\ref{phit80} merge
as $T \rightarrow 0$. At nonzero temperature, on the other hand,
the gap parameters are not the same. This suggests that, similar
to the zero-temperature case of Figs.~\ref{phi_ms2mu}
and~\ref{phi_mu}, a phase transition to the gCFL phase happens
at some nonzero temperature. However, I shall show below that
there is no phase transition between the CFL and gCFL phases at
\textit{any} nonzero temperature. Instead, there is an
insulator-metal crossover transition between the CFL phase and a
so-called \textit{metallic} CFL (mCFL) phase. At this point, all
quasiparticles are still gapped. At some higher temperature, the
mCFL phase is replaced by the gCFL phase.
\begin{figure}[H]
\begin{center}
\includegraphics[width=0.95\textwidth]{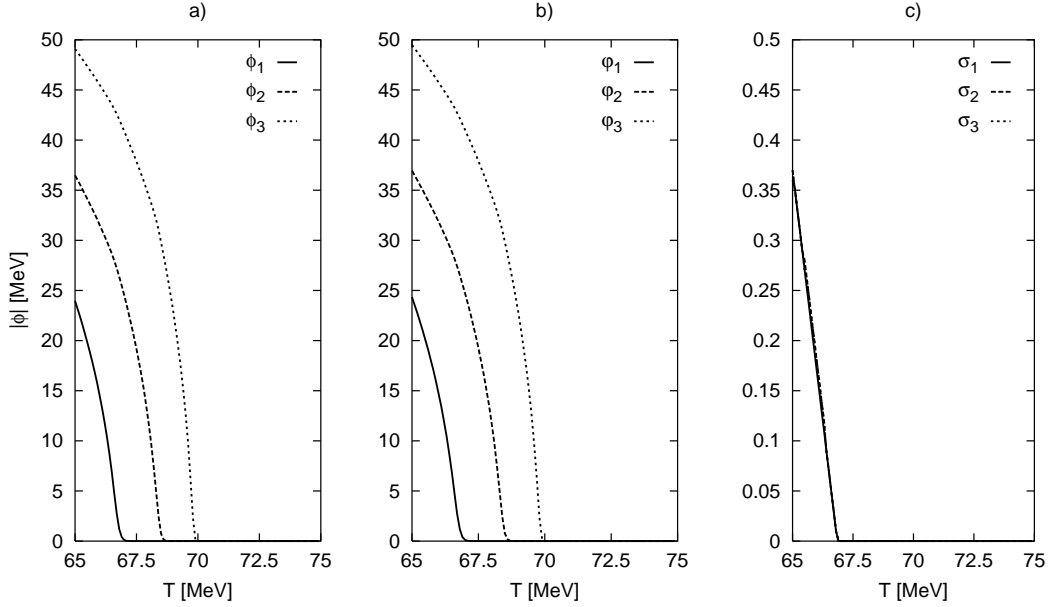}
\caption[The near-critical temperature dependence of the gaps
for a small $m_s$.]{The near-critical temperature dependence of
the gaps in the case of a small strange quark mass, $m_s^2/\mu =
80$~MeV. Note, that the actual values of the gap parameters
shown in panel~a) are negative. The results are plotted for $\mu
= 500$~MeV.}
\label{phit80zoom}
\end{center}
\end{figure}
\begin{figure}[H]
\begin{center}
\includegraphics[width=0.95\textwidth]{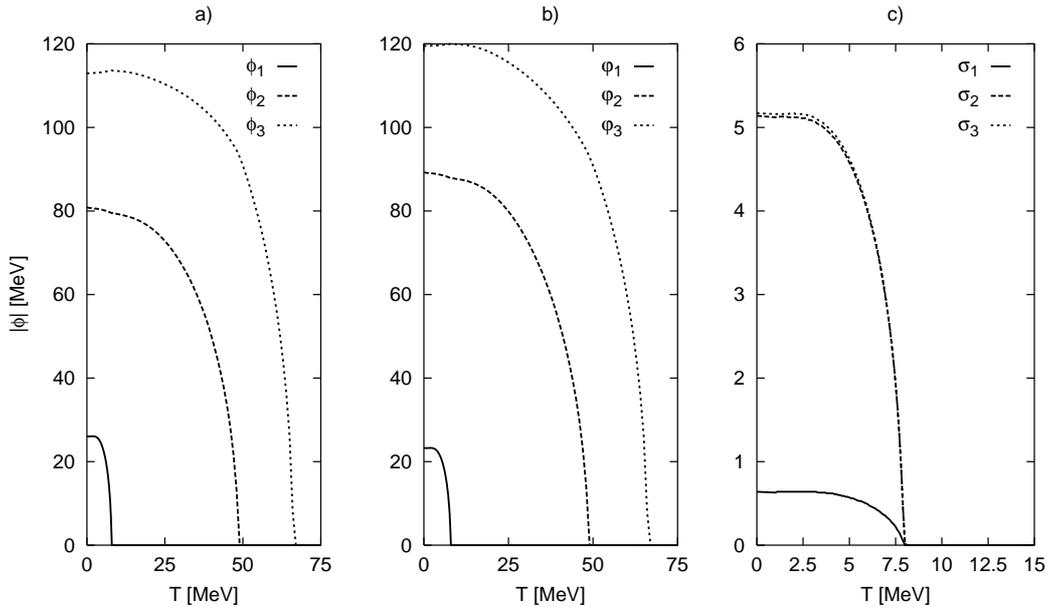}
\caption[The temperature dependence of the gaps for a large
strange quark mass.]{The temperature dependence of the gaps in
the case of a large strange quark mass, $m_s^2/\mu = 320$~MeV.
Note, that the actual values of the gap parameters shown in
panel~a) are negative. The results are plotted for $\mu =
500$~MeV.}
\label{phit320}
\end{center}
\end{figure}
If the temperature is increased even further, there are three
consecutive phase transitions. These correspond to the three
phase transitions predicted in Ref.~\cite{Iida} in the limit
of a small strange quark mass. In order to resolve these, I
show a close-up of the near-critical region of Fig.~\ref{phit80}
in Fig.~\ref{phit80zoom}. The three transitions that one
observes are the following:
\begin{enumerate}
\item transition from the gCFL phase to the so-called uSC
phase,
\item transition from the uSC phase to the 2SC phase,
\item transition from the 2SC phase to the normal quark
phase.
\end{enumerate}
Here, the notation uSC (dSC) stands for
\textbf{s}uper\textbf{c}onductivity in which all three colors of
the \textbf{u}p (\textbf{d}own) quark flavor participate in
diquark pairing~\cite{Iida}. My results differ from those of
Ref.~\cite{Iida} in that the dSC phase is replaced by the uSC
phase. The reason is that, in my case, the first gaps which
vanish with increasing temperature are $\phi_1$ and $\varphi_1$,
see Fig.~\ref{phit80zoom}, while in their case $\Delta_2$
(corresponding to my $\phi_2$ and $\varphi_2$) disappears first.
The authors of Ref.~\cite{Fukushima} also obtain the dSC phase
instead of the uSC phase at temperatures close to the critical
temperature and at small values of the strange quark mass. I
also obtain the dSC phase with my numerical calculations when I
use the model parameters of Ref.~\cite{Fukushima}. In fact, the
main difference between these two studies is the value of the
cutoff parameter in the model. From this, I conclude that the
size of the uSC or dSC region, respectively, in the phase
diagram is particularly sensitive to the choice of the cutoff
parameter in the model~\cite{Shovkovy_Ruester_Rischke}. Although
I use the same terms for the phases that were introduced in
Ref.~\cite{Iida}, I distinguish between the gapped phases (e.g.,
CFL and mCFL phase) and the gapless phases (e.g., gCFL). Also,
in order to reflect the physical properties of the mCFL phase, I
prefer to use the term \textit{metallic} CFL, instead of
\textit{modified} CFL as in Ref.~\cite{Iida}. (Note that, in
that work, the mCFL phase also encompasses the gCFL phase.)
\begin{figure}[H]
\begin{center}
\includegraphics[width=0.95\textwidth]{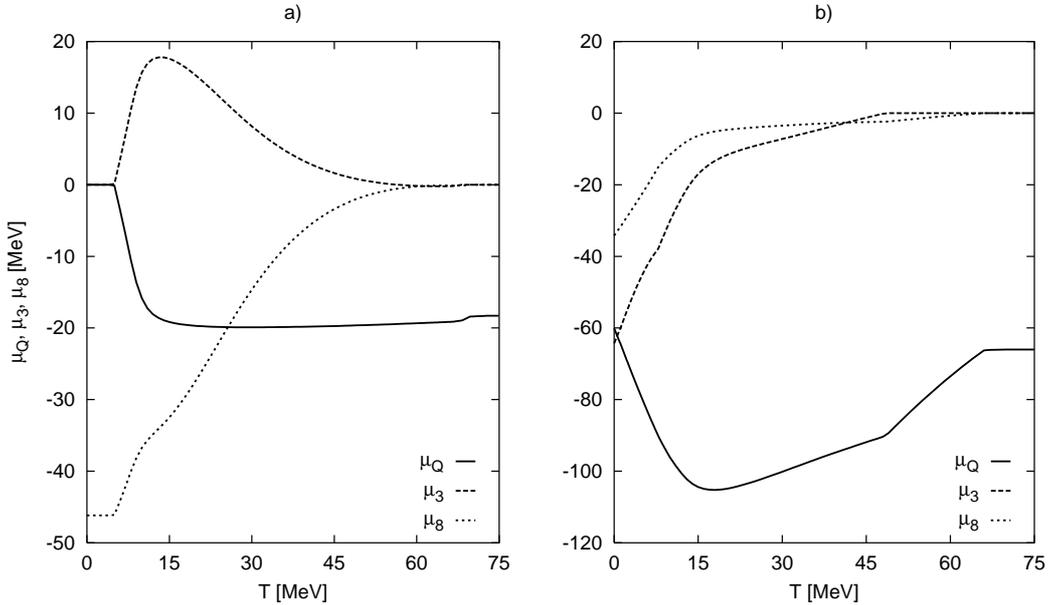}
\caption[The temperature dependence of the electrical and color
chemical potentials.]{The temperature dependence of the
electrical and color chemical potentials for $m_s^2/\mu =
80$~MeV (left panel) and for $m_s^2/\mu = 320$~MeV (right
panel). The quark chemical potential is taken to be $\mu =
500$~MeV.}
\label{mut}
\end{center}
\end{figure}
In the case of a large strange quark mass (i.e., the case of
$m_s^2/\mu = 320$~MeV shown in Fig.~\ref{phit320}), the
zero-temperature limit corresponds to the gCFL phase. By looking
at the corresponding temperature dependence of the gap
parameters, one can see that this case is a natural
generalization of the previous limit of a small strange quark
mass. There are also three consecutive phase transitions. It is
noticeable, however, that the separation between the different
transitions becomes much wider at large $m_s$.
\begin{figure}[H]
\begin{center}
\includegraphics[width=0.95\textwidth]{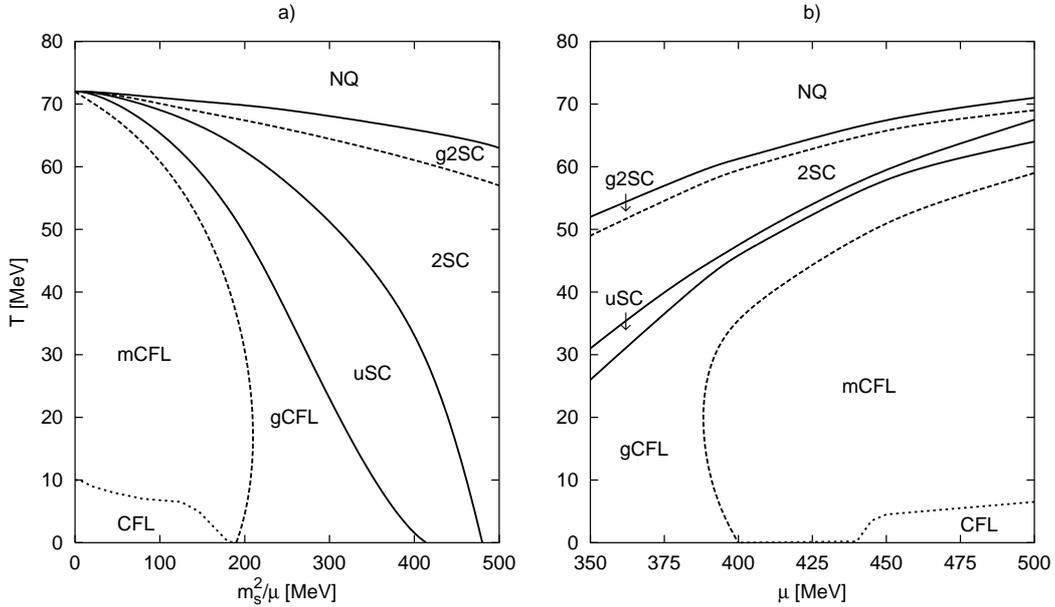}
\caption[The phase diagram of massless neutral three-flavor
quark matter.]{The phase diagram of massless neutral
three-flavor quark matter in the plane of temperature and
$m_s^2/\mu$ (left panel) and in the plane of temperature and
quark chemical potential (right panel). The results in the left
panel are for a fixed value of the quark chemical potential,
$\mu = 500$~MeV. The results in the right panel are for a fixed
value of the strange quark mass, $m_s = 250$~MeV. The dashed
lines are associated with the appearance of additional gapless
modes in the spectra. The dotted lines indicate the
insulator-metal crossover.}
\label{phasediagram_massless}
\end{center}
\end{figure}
I now take a closer look at the transition between the CFL, the
mCFL, and the gCFL phase. At zero temperature, there was no
symmetry connected with the order parameter, i.e., the number
density of electrons, that is associated with the CFL
$\rightarrow$ gCFL phase transition. At nonzero temperature, the
electron density is not strictly zero in the CFL phase as soon
as $m_s \neq 0$. Indeed, the arguments of
Ref.~\cite{enforced_neutrality} regarding the enforced
neutrality of the CFL phase do not apply at $T \neq 0$. This is
why I conclude that the insulator-metal transition between the
CFL and the mCFL phase is just a smooth crossover at $T \neq 0$.
Of course, in principle, one can never exclude the existence of
a first-order phase transition. My numerical analysis, however,
reveals a crossover. The transition can only be identified by a
rapid increase of the chemical potential of electric charge or
the electron density in a relatively narrow window of
temperatures, see panel~a) in Fig.~\ref{mut}. The location of
the maximum of the corresponding susceptibility (i.e., $\ud n_Q/
\ud T$) is then associated with the transition point.

The transition between the mCFL and the gCFL phase corresponds
to the appearance of gapless quasiparticle modes in the
spectrum. There is no way of telling from the temperature
dependences in Figs.~\ref{phit80},~\ref{phit80zoom}
and~\ref{phit320}, whether the corresponding CFL and/or 2SC
phases are gapless or not. This additional piece of information
can only be extracted from the behavior of the quasiparticle
spectra. I also investigated them, but I do not show them
explicitly.

My results for the phase structure of dense neutral
three-flavor quark matter are summarized in
Fig.~\ref{phasediagram_massless}. I show the phase diagram in
the plane of temperature and $m_s^2/\mu$ at a fixed value of the
quark chemical potential, $\mu = 500$~MeV, and in the plane of
temperature and quark chemical potential at a fixed value of the
strange quark mass, $m_s = 250$~MeV. The three solid lines
denote the three phase transitions discussed above. The two
dashed lines mark the appearance of gapless modes in the mCFL
and 2SC phases. I term these as the mCFL $\rightarrow$ gCFL and
2SC $\rightarrow$ g2SC crossover transitions. In addition, as I
mentioned above, there is also an insulator-metal type
transition between the CFL and mCFL phase. This is marked by the
dotted lines in Fig.~\ref{phasediagram_massless}.

The effects of the strange quark mass are incorporated only by a
shift of the strange quark chemical potential~(\ref{shift2}).
By comparing the phase diagram shown in panel~a) of
Fig.~\ref{phasediagram_massless} with that one shown in the
right panel of Fig.~1 in Ref.~\cite{Shovkovy_Ruester_Rischke},
where the strange quark mass is properly taken into account, one
can see that both phase diagrams are quantitatively in good
agreement for low strange quark masses. For large strange quark
masses, the phase diagrams are still qualitatively in good
agreement.
\section{The phase diagram with a self-consistent treatment of
quark masses}
\label{phase_diagram_self-consistent}
The first attempt to obtain the phase diagram of dense, locally
neutral three-flavor quark matter as a function of the strange
quark mass, the quark chemical potential, and the temperature
was made in Ref.~\cite{Ruester_Shovkovy_Rischke}. I presented
this work in Sec.~\ref{The_phase_diagram_of_massless_quarks}. It
was shown that, at zero temperature and small values of the
strange quark mass, the ground state of quark matter corresponds
to the CFL phase~\cite{CFL_discoverers, Shovkovy_Wijewardhana,
Schaefer2}. At some critical value of the strange quark mass,
this is replaced by the gCFL
phase~\cite{Alford_Kouvaris_Rajagopal1,
Alford_Kouvaris_Rajagopal2}. In addition, several other phases
were found at nonzero temperature. For instance, it was shown
that there should exist a metallic CFL (mCFL) phase, a so-called
uSC phase, as well as the standard 2SC
phase~\cite{Alford_Rajagopal_Wilczek, Pisarski_Rischke1,
regular_self-energies, Son} and the g2SC phase~\cite{g2SC_1,
g2SC_2}.

In Ref.~\cite{Ruester_Shovkovy_Rischke}, the effect of the
strange quark mass was incorporated only approximately by a
shift of the strange quark chemical potential, $\mu_s
\rightarrow \mu_s - m_s^2/( 2 \mu )$. Such an approach is
reliable at small values of the strange quark mass but for
larger values of the strange quark mass, the results are only
qualitatively in good agreement by comparing them to the case,
where the strange quark mass is properly taken into
account~\cite{Shovkovy_Ruester_Rischke}. The phase diagram of
Ref.~\cite{Ruester_Shovkovy_Rischke} was further developed in
Refs.~\cite{Fukushima, Shovkovy_Ruester_Rischke} where the
shift-approximation in dealing with the strange quark was not
employed any more. So far, however, quark masses were treated as
free parameters, rather than dynamically generated quantities.

In this section, I study the phase diagram of dense, locally
neutral three-flavor quark matter within the framework of a
NJL model~\cite{NJL, Kunihiro_Hatsuda, Klevansky,
Hatsuda_Kunihiro, Buballa_Habilitationsschrift},
treating dynamically generated quark masses self-consistently. I
introduce the model and, within this model, I derive a complete
set of gap equations and charge neutrality conditions. The
results are summarized in the phase diagram in the plane of
temperature $T$ and quark chemical potential $\mu$. Some results
within this approach were also obtained in Refs.~\cite{Blaschke,
Abuki_Kitazawa_Kunihiro, Abuki_Kunihiro, SRP}.
\subsection{Model}
In this section, I use a three-flavor NJL
model~\cite{Kunihiro_Hatsuda, Klevansky, Hatsuda_Kunihiro, 
Buballa_Habilitationsschrift}. The Lagrangian density is given
by~\cite{Buballa_Habilitationsschrift,
Ruester_Werth_Buballa_Shovkovy_Rischke1,
Ruester_Werth_Buballa_Shovkovy_Rischke2},
\bea
\label{L_NJL}
\mathcal{L} &=& \bar\psi \left( i \dirac - \hat m \right) \psi
\nonumber \\
&+& G_S \sum_{a = 0}^8 \left[ ( \bar\psi \lambda_a \psi )^2 + (
\bar\psi i \gamma_5 \lambda_a \psi )^2 \right] \nonumber \\
&+& G_D \sum_{\gamma, c} \left[ \bar\psi_\alpha^a i \gamma_5
\epsilon^{\alpha\beta\gamma} \epsilon_{abc} ( \psi_C )_\beta^b
\right] \left[ ( \bar\psi_C )_\rho^r i \gamma_5
\epsilon^{\rho\sigma\gamma} \epsilon_{rsc} \psi_\sigma^s \right]
\nonumber \\
&-& K \left\{ {\det}_f \left[ \bar\psi \left( 1 + \gamma_5
\right) \psi \right] + {\det}_f \left[ \bar\psi \left( 1 -
\gamma_5 \right) \psi \right] \right\} \; ,
\eea
where the quark spinor field,
\be
\psi = \left( \psi_u^r, \psi_d^r, \psi_s^r, \psi_u^g, \psi_d^g,
\psi_s^g, \psi_u^b, \psi_d^b, \psi_s^b \right)^T \; ,
\ee
carries color ($a = r,g,b$) and flavor ($\alpha = u,d,s$)
indices. The matrix of quark current masses is given by
\be
\hat m = \diag \left( m_u, m_d, m_s, m_u, m_d, m_s, m_u, m_d,
m_s \right) \; ,
\ee
and $\lambda_a$ with $a = 1, \ldots, 8$ are the Gell-Mann
matrices in flavor space~(\ref{Gell-Mann}), $\lambda_0 \equiv
\sqrt{2/3}$. In the Lagrangian density, I use the
charge-conjugate quark
spinors~(\ref{charge-conjugate_spinors_space-time}). The
Dirac conjugate quark spinor is given by $\bar\psi \equiv
\psi^\dagger \gamma_0$.

Throughout this work, I shall assume isospin symmetry on the
Lagrangian level, $m_{u,d} \equiv m_u = m_d$, whereas $m_s$ will
be different, thus explicitly breaking $[SU(3)_f]$ flavor
symmetry. The terms in the Lagrangian density~(\ref{L_NJL}) can
straightforwardly be generalized to any number of flavors $N_f$.
I only deal with the lighter up, down, and strange quarks
because charm, bottom, and top quarks are too heavy to occur in
the cores of neutron stars.

The term in the first line in the Lagrangian
density~(\ref{L_NJL}), $\mathcal{L}_\mathrm{Dirac}$, is the
Dirac Lagrangian density of non-interacting massive quarks. The
terms in the second line correspond to a $U(N_f)_\ell \times
U(N_f)_r$ symmetric four-point interaction with the scalar
coupling constant $G_S$. Therefore, this part of the Lagrangian
density is termed symmetric Lagrangian density,
$\mathcal{L}_\mathrm{sym}$. The term in the third line of the
Lagrangian density, $\mathcal{L}_\mathrm{diquark}$, describes a
scalar diquark interaction in the color-antitriplet and
flavor-antitriplet channel. For symmetry reasons there also
should be a pseudoscalar diquark interaction with the same
diquark coupling constant $G_D$. This term would be important to
describe Goldstone boson condensation in the CFL
phase~\cite{Buballa_Forbes}. In this study, however, I neglect
this possibility, and therefore drop the pseudoscalar diquark
term. The terms in the last line of the Lagrangian density,
$\mathcal{L}_\mathrm{det}$, are called the 't~Hooft interaction,
where $K$ is the coupling constant. The determinants in these
terms have to be taken in flavor space which means that
$\mathcal{L}_\mathrm{det}$ is a maximally flavor-mixing $2
N_f$-point interaction, involving an incoming and an outgoing
quark of each flavor. Consequently, for three flavors, the
't~Hooft term is a six-point interaction term of the form,
\be
{\det}_f \, ( \bar\psi \mathcal{O} \psi ) \equiv \sum_{i,j,k}
\epsilon_{ijk} \, 
( \bar\psi_u \mathcal{O} \psi_i ) \, 
( \bar\psi_d \mathcal{O} \psi_j ) \, 
( \bar\psi_s \mathcal{O} \psi_k ) \; ,
\ee
where $i$, $j$, and $k$ are flavor indices.

The 't~Hooft interaction is $SU(N_f)_\ell \times SU(N_f)_r$
symmetric, but it breaks the $U_A(1)$ axial
symmetry~\cite{tHooft}, which was left unbroken by
$\mathcal{L}_\mathrm{sym}$. It thus translates the $U_A(1)$
anomaly, which in QCD arises at quantum level from the gluon
sector, to a tree-level interaction in a pure quark model. The
't~Hooft term is phenomenologically important to get the correct
mass splitting of the $\eta$ and $\eta'$ mesons. In the chiral
limit ($m_u = m_d = m_s = 0$), the $\eta'$ mass is lifted to a
nonzero value by $\mathcal{L}_\mathrm{det}$, while the other
pseudoscalar mesons, including the $\eta$, remain massless.
There are many other terms which are consistent with the
symmetries and which could be added to the Lagrangian. But for
simplicity, however, I shall restrict myself to the Lagrangian
density~(\ref{L_NJL}). This Lagrangian density should be viewed
as an effective model of strongly interacting matter that
captures at least some key features of QCD dynamics. The
Lagrangian density contains three interaction terms which are
chosen to respect the symmetries of QCD.

In the Lagrangian density~(\ref{L_NJL}), there are six
parameters: the quark current masses $m_{u,d}$ and $m_s$, the
coupling constants $G_S$, $G_D$, and $K$, and the cutoff
$\Lambda$. I use the following set of model
parameters~\cite{RKH}:
\be
\label{NJL_model_parameters}
\begin{aligned}
m_{u,d} &= 5.5~\mathrm{MeV} \; , \\
m_s &= 140.7~\mathrm{MeV} \; , \\
G_S \Lambda^2 &= 1.835 \; , \\
K \Lambda^5 &= 12.36 \; , \\
\Lambda &= 602.3~\mathrm{MeV} \; .
\end{aligned}
\ee
After fixing the current masses of the up and down quarks at
equal values, $m_{u,d} = 5.5$~MeV, the other four parameters in
Eq.~(\ref{NJL_model_parameters}) are chosen to reproduce the
following four observables of vacuum QCD~\cite{RKH}: $m_\pi =
135.0$~MeV, $m_K = 497.7$~MeV, $m_{\eta'} = 957.8$~MeV, and
$f_\pi = 92.4$~MeV. This parameter set gives $m_\eta =
514.8$~MeV~\cite{RKH}. In Ref.~\cite{RKH}, the diquark coupling
constant $G_D$ was not fixed by the fit of the meson spectrum in
vacuum. In general, it is expected to be of the same order as
the quark-antiquark coupling constant $G_S$. In this section, I
study the following two cases in detail:
\begin{enumerate}
\item the regime of intermediate diquark coupling strength with
$G_D = \frac34 G_S$,
\item the regime of strong diquark coupling strength with $G_D =
G_S$.
\end{enumerate}
In fact, the value of the diquark coupling constant for the
regime of intermediate coupling strength is used in many
studies because this value for the diquark coupling constant is
predicted by the Fierz transformation in vacuum. Therefore, it
is a standard value, and, if not mentioned otherwise, throughout
this thesis, the standard value of the diquark coupling
constant,
\be
\label{G_D}
G_D = \frac34 G_S \; ,
\ee
is used.

In order to obtain the Lagrangian density~(\ref{L_NJL}) in
mean-field approximation, one can adopt the
rules~\cite{Hatsuda_Kunihiro},
\bsub
\bea
\bar\psi_i \psi_i \bar\psi_j \psi_j &\longrightarrow&
\langle \bar\psi_i \psi_i \rangle \bar\psi_j \psi_j
+ \langle \bar\psi_j \psi_j \rangle \bar\psi_i \psi_i
- \langle \bar\psi_i \psi_i \rangle \langle \bar\psi_j \psi_j
\rangle \; , \\
\bar\psi_i \psi_i \bar\psi_j \psi_j \bar\psi_k \psi_k
&\longrightarrow&
\sum_{\substack{i,j,k \\ \mathrm{(cyclic)}}}
\langle \bar\psi_i \psi_i \rangle
\langle \bar\psi_j \psi_j \rangle \bar\psi_k \psi_k
- 2 \, \langle \bar\psi_i \psi_i \rangle \langle \bar\psi_j
\psi_j \rangle \langle \bar\psi_k \psi_k \rangle \; ,
\eea
\esub
where the last terms on the right-hand side are indispensable to
avoid a double counting of the interaction energy. This is why
there is also a factor of two in the six-fermion interaction.
One should note that in this approach, the condensates are
treated as variational parameters which are determined so as to
give a maximum point of the pressure.

The diquark condensates $\Delta_c$ are analogously defined as in
Eq.~(\ref{diquark_condensate}). The quark-antiquark condensates
are defined as,
\be
\sigma_\alpha \propto \langle \bar\psi_\alpha \psi_\alpha
\rangle \; .
\ee
Herewith, I arrive at the following Langangian density in
mean-field approximation:
\be
\label{L_mfa}
\mathcal{L} = \bar\psi \, ( i \dirac - \hat M ) \, \psi
- 2 G_S \sum_{\alpha = 1}^3 \sigma_\alpha^2
+ \frac12 \bar\psi \Phi^- \psi_C + \frac12 \bar\psi_C \Phi^+
\psi
- \frac{1}{4 G_D} \sum_{c = 1}^3 \left| \Delta_c \right|^2
+ 4 K \sigma_u \sigma _d \sigma_s \; ,
\ee
where the constituent quark mass matrix is defined as,
\be
\hat M = \diag \left( M_u, M_d, M_s, M_u, M_d, M_s, M_u, M_d,
M_s \right) \; .
\ee
The constituent mass of each quark flavor can be obtained by
\be
\label{M_alpha}
M_\alpha = m_\alpha - 4 G_S \sigma_\alpha + 2 K \sigma_\beta
\sigma_\gamma \; ,
\ee
where the set of flavor indices ($\alpha, \beta, \gamma$) is a
permutation of ($u,d,s$). The Lagrangian density in mean-field
approximation contains the gap matrices in color-flavor and
Dirac space,
\be
\label{gap_ansatz}
\Phi^\pm = \pm \gamma_5 \hat\Phi \; ,
\ee
which fulfill the relation $\Phi^- = \gamma_0 ( \Phi^+ )^\dagger
\gamma_0$. The color-flavor part of the gap matrices reads,
\be
\label{gap_ansatz_color-flavor}
\hat\Phi_{\alpha \beta}^{ab}
= \sum_c \epsilon^{abc} \epsilon_{\alpha \beta c} \Delta_c
= \left(
\begin{array}{ccccccccc}
0 & 0 & 0 & 0 & \phantom{-} \Delta_3 & 0 & 0 & 0 & \phantom{-}
\Delta_2 \\
0 & 0 & 0 & - \Delta_3 & 0 & 0 & 0 & 0 & 0 \\
0 & 0 & 0 & 0 & 0 & 0 & - \Delta_2 & 0 & 0 \\
0 & - \Delta_3 & 0 & 0 & 0 & 0 & 0 & 0 & 0 \\
\phantom{-} \Delta_3 & 0 & 0 & 0 & 0 & 0 & 0 & 0 & \phantom{-}
\Delta_1 \\
0 & 0 & 0 & 0 & 0 & 0 & 0 & - \Delta_1 & 0 \\
0 & 0 & - \Delta_2 & 0 & 0 & 0 & 0 & 0 & 0 \\
0 & 0 & 0 & 0 & 0 & - \Delta_1 & 0 & 0 & 0 \\
\phantom{-} \Delta_2 & 0 & 0 & 0 & \phantom{-} \Delta_1 & 0 & 0
& 0 & 0
\end{array}
\right) \; ,
\ee
where $\Delta_c$ are real-valued gap parameters. Here, as
before, $a$ and $b$ refer to the color components, and $\alpha$
and $\beta$ refer to the flavor components. The gap parameters
$\Delta_1$, $\Delta_2$, and $\Delta_3$ correspond to the
down-strange, the up-strange, and the up-down diquark
condensates, respectively. All three of them originate from the
color-antitriplet, flavor-antitriplet diquark pairing channel.
For simplicity, the color and flavor symmetric condensates are
neglected in this study. They were shown to be small [see
Figs.~\ref{phi_ms2mu}~c),~\ref{phi_mu}~c),~\ref{phit80}~c),
and~\ref{phit320}~c)] and not crucial for the qualitative
understanding of the phase
diagram~\cite{Ruester_Shovkovy_Rischke}.

I should note that I have restricted myself to field
contractions corresponding to the Hartree approximation in
Eq.~(\ref{L_mfa}). In a more complete treatment, among others,
the 't~Hooft interaction term gives also rise to mixed
contributions containing both diquark and quark-antiquark
condensates, i.e., $\propto \sum_{\alpha = 1}^3 \sigma_\alpha
|\Delta_\alpha|^2$~\cite{Rapp_Schaefer_Shuryak_Velkovsky}. In
this study, as in Refs.~\cite{Buballa_Habilitationsschrift,
SRP}, I neglected such terms for simplicity. While their
presence may change the results quantitatively, one does not
expect them to modify the qualitative structure of the phase
diagram.

Up to irrelevant constants, the grand partition function is
given by
\be
\label{Z_NJL}
\mathcal{Z} = \int \D \bar\psi \D \psi \exp \left\{ I \left[
\bar\psi, \psi \right] \right\} \; ,
\ee
where
\be
\label{action}
I \left[ \bar\psi, \psi \right] = \int_X \left( \mathcal{L} +
\mu \mathcal{N} + \mu_Q \mathcal{N}_Q + \mu_a \mathcal{N}_a
\right) \; ,
\ee
is the action. The conserved quantities,
\be
\mathcal{N} \equiv \bar\psi \gamma_0 \psi \; , \qquad
\mathcal{N}_Q \equiv \bar\psi \gamma_0 Q \psi \; , \qquad
\mathcal{N}_a \equiv \bar\psi \gamma_0 T_a \psi \; ,
\ee
are the quark number density operator, the operator of electric
charge density of the quarks, and the operators of color charge
densities of the quarks, respectively. The action can be split
into a kinetic and a potential part,
\be
I \left[ \bar\psi, \psi \right] = I_\mathrm{kin} \left[
\bar\psi, \psi \right] + I_\mathrm{pot} \; .
\ee
By inserting the Lagrangian density in mean-field
approximation~(\ref{L_mfa}) into the action~(\ref{action}), one
obtains for the kinetic part of the action,
\be
I_\mathrm{kin} \left[ \bar\psi, \psi \right] \int_X \left(
\bar\psi [ G_0^+ ]^{-1} \psi + \frac12 \bar\psi \Phi^- \psi_C +
\frac12 \bar\psi_C \Phi^+ \psi \right) \; ,
\ee
and for the potential part of the action,
\be
I_\mathrm{pot} = \int_X \left( - 2 G_S \sum_{\alpha = 1}^3
\sigma_\alpha^2 - \frac{1}{4 G_D} \sum_{c = 1}^3 \left| \Delta_c
\right|^2 + 4 K \sigma_u \sigma_d \sigma_s \right) \; .
\ee
The massive inverse Dirac propagator for quarks and
charge-conjugate quarks, respectively, is given by
\be
[ G_0^\pm ]^{-1} = i \dirac \pm \hat\mu \gamma_0 - \hat M \; ,
\ee
where
\be
\hat\mu = \diag \left( \mu_u^r, \mu_d^r, \mu_s^r, \mu_u^g,
\mu_d^g, \mu_s^g, \mu_u^b, \mu_d^b, \mu_s^b \right) \; ,
\ee
is the matrix of quark chemical potentials. The chemical
potentials for each quark color and flavor are defined by
Eq.~(\ref{quark_color_and_flavor_beta_equilibrium}) because
quark matter inside neutron stars is in $\beta$ equilibrium.

By introducing the Nambu-Gorkov basis,
\be
\bar\Psi \equiv
\left( \bar\psi, \bar\psi_C \right) \; , \qquad
\Psi \equiv \left(
\begin{array}{c}
\psi \\
\psi_C
\end{array}
\right) \; ,
\ee
the kinetic part of the action can be rewritten as
\be
I_\mathrm{kin} \left[ \bar\Psi, \Psi \right] = \frac12 \int_X
\bar\Psi S^{-1} \Psi \; ,
\ee
where I have used the inverse quark propagator in Nambu-Gorkov
space,
\be
\label{invS}
S^{-1} = \left( 
\begin{array}{cc}
[ G_0^+ ]^{-1} & \Phi^- \\
\Phi^+ & [ G_0^- ]^{-1}
\end{array}
\right) \; .
\ee

In momentum space, the massive inverse Dirac propagator for
quarks and charge-conjugate quarks, respectively, reads,
\be
[ G_0^\pm ]^{-1} = \gamma_0 \left( k_0 \pm \hat\mu \right) -
\fett{\gamma} \cdot \fettu{k} - \hat M \; .
\ee
The kinetic part of the grand partition function~(\ref{Z_NJL})
is Fourier transformed in
Sec.~\ref{The_determinant_of_the_inverse_quark_propagator} in
the Appendix. With that result, one obtains for the pressure of
color-superconducting quark matter, $p = \frac{T}{V} \ln
\mathcal{Z}$,
\be
p = \frac{T}{2 V} \sum_K \ln \det \left( \frac{S^{-1}}{T}
\right) - 2 G_S \sum_{\alpha = 1}^3 \sigma_\alpha^2 - \frac{1}{4
G_D} \sum_{c = 1}^3 \left| \Delta_c \right|^2 + 4 K \sigma_u
\sigma_d \sigma_s + p_\beta \; ,
\ee
where I also added the contribution of leptons, $p_\beta$, which
will be specified later.

The most complicated expression in the pressure is the
determinant because the inverse quark propagator~(\ref{invS}) is
a $72 \times 72$-matrix with the following substructure: it is a
$2 \times 2$-matrix in Nambu-Gorkov space, a $9 \times 9$-matrix
in color-flavor space and a $4 \times 4$-matrix in Dirac space.
By using the relation $\det \left( c A \right) = c^d \det
A$~\cite{Fischer}, where $c$ is a factor and $A$ is a quadratic
matrix with dimension $d$, one obtains for the kinetic part of
the pressure,
\be
\label{p_kin}
p_\mathrm{kin} \equiv \frac{T}{2 V} \sum_K \ln \det \left(
\frac{S^{-1}}{T} \right) = \frac{T}{2 V} \sum_K \ln \left(
\frac{\det S^{-1}}{T^{72}} \right) \; ,
\ee
where it is useful to rewrite the determinant of the inverse
quark propagator with the relation $\det \left( A B \right) =
\det A \cdot \det B$, where the matrices $A$ and $B$ are
quadratic~\cite{Fischer},
\be
\det S^{-1} = \det \left( \gamma_0 \gamma_0 S^{-1} \right) =
\det \gamma_0 \det \left( \gamma_0 S^{-1} \right) = \det \left(
\gamma_0 S^{-1} \right) \; .
\ee
By using the spin projectors~(\ref{spin_projectors}),
\begin{equation}
\mathcal{P}_s = \frac12 \, ( 1 + s \, \fett{\sigma} \cdot
\hat{\fettu{k}} ) \; ,
\end{equation}
where $s=\pm$ stands for projections onto states with spin up or
spin down, respectively, one can rewrite the matrices $\gamma_0
[G_0^\pm]^{-1}$ and $\gamma_0 \Phi^\pm$
as~\cite{Ruester_Werth_Buballa_Shovkovy_Rischke1},
\bea
\label{invG0-Ps}
\gamma_0 [ G_0^\pm ]^{-1} &=&
\sum_s \left(
\begin{array}{cc}
k_0 \pm \hat\mu - \hat M & - s k \\
- s k & k_0 \pm \hat\mu + \hat M
\end{array}
\right) \mathcal{P}_s \; , \\
\label{Phi-Ps}
\gamma_0 \Phi^\pm &=&
\pm \sum_s \left(
\begin{array}{cc}
\ \ \, 0 \ & \, \hat\Phi \, \\
- \hat\Phi \ & \ 0 \,
\end{array}
\right) \mathcal{P}_s \; .
\eea
The spin projectors are $2 \times 2$-matrices in spin space.
Thus, I have separated the spin space from the color-flavor
space and particle-antiparticle space. The subdivision of the
Dirac space into the spin space and the particle-antiparticle
space is necessary in order to take the different quark masses
properly into account.

By making use of the definition in Eq.~(\ref{invS}), as well as
Eqs.~(\ref{invG0-Ps}) and~(\ref{Phi-Ps}), I obtain the following
representation:
\be
\gamma_0 S^{-1} = \sum_s \hat S_s^{-1}
\mathcal{P}_s \; ,
\ee
where 
\be
\hat S_s^{-1} = k_0 - \mathcal{M}_s \; ,
\ee
and 
\be
\label{M_s}
\mathcal{M}_s = \left(
\begin{array}{cccc}
- \hat\mu + \hat M & s k & 0 & \hat\Phi \\
s k & - \hat\mu - \hat M & - \hat\Phi & 0 \\
0 & - \hat\Phi & \hat\mu + \hat M & s k \\
\hat\Phi & 0 & s k & \hat\mu - \hat M
\end{array}
\right) \; ,
\ee
(with $s=\pm$) is real and symmetric. By using the spin
projectors, the spin space is separated from the Nambu-Gorkov
space, color-flavor space, and particle-antiparticle space.
Therefore, the matrices $\hat S_s^{-1}$ and $\mathcal{M}_s$ are
$36 \times 36$-matrices and have the following substructure:
they are $2 \times 2$-matrices in Nambu-Gorkov space, $9 \times
9$-matrices in color-flavor space, and $2 \times 2$-matrices in
particle-antiparticle space. Each element of the matrices which
are shown in Eqs.~(\ref{invG0-Ps}),~(\ref{Phi-Ps}),
and~(\ref{M_s}) are $9 \times 9$-matrices in color-flavor space.

Since there is no explicit energy dependence in $\mathcal{M}_s$,
its eigenvalues $\epsilon_{i}$ determine the quasiparticle
dispersion relations, $k_0 = \epsilon_{i}(k)$. By using the
properties of the projectors $\mathcal{P}_s$ as well as the
matrix relations, $\ln \det A = \Tr \ln A$, and $\Tr \ln
\sum_i a_i \mathcal{P}_i = \sum_i \ln a_i \Tr \, \mathcal{P}_i$,
see Secs.~\ref{The_logarithm_of_the_determinant}
and~\ref{The_trace_of_the_logarithm} in the Appendix, I derive,
\bea
\ln \det \left( \gamma_0 S^{-1} \right)
&=& \Tr \ln \left( \gamma_0 S^{-1} \right) = \Tr \ln \sum_s \hat
S_s^{-1} \mathcal{P}_s \nonumber \\
&=& \sum_s \Tr \ln \hat S_s^{-1} = \sum_s \ln \det
\hat S_s^{-1} = \ln \left( \det \hat S_+^{-1} \cdot \det \hat
S_-^{-1} \right) \; ,
\eea
where the traces in the first line run over Nambu-Gorkov,
color-flavor, particle-antiparticle, and spin indices, while the
trace in the second line only runs over Nambu-Gorkov,
color-flavor, and particle-antiparticle indices. It turns out
that the two determinants appearing on the right-hand side of
this equation are equal, i.e., $\det \hat S_-^{-1} = \det \hat
S_+^{-1}$. From the physics viewpoint, this identity reflects
the twofold spin degeneracy of the spectrum of quark
quasiparticles. The formal proof of this degeneracy is
straightforward after noticing that the following matrix
relation is satisfied:
\be
\hat S_{-s}^{-1} = \mathcal{R} \hat S_s^{-1} \mathcal{R}^{-1} \;
,
\ee
where
\be
\mathcal{R} = \diag \left( 1, -1, -1, 1 \right)
\ee
is a matrix with the properties $\mathcal{R} = \mathcal{R}^T =
\mathcal{R}^\dagger = \mathcal{R}^{-1}$ and $\det \mathcal{R} =
1$.

Another observation which turns out to be helpful in the
calculation is that the determinant $\det \hat S_s^{-1} ( k_0
)$ is an even function of $k_0$, i.e., $\det \hat S_s^{-1} ( -
k_0 ) = \det \hat S_s^{-1} ( k_0 )$. This is a formal
consequence of the following matrix relation:
\be
\hat S_s^{-1} \left( - k_0 \right) = - \mathcal{B} \hat S_s^{-1}
\left( k_0 \right) \mathcal{B}^{-1} \; ,
\ee
where
\be
\mathcal{B} = \left(
\begin{array}{rrrr}
0  & \ \ 0 & 0  & \ \ i \\
0  & \ \ 0 & -i & \ \ 0 \\
0  & \ \ i & 0  & \ \ 0 \\
-i & \ \ 0 & 0  & \ \ 0
\end{array}
\right)
\ee
is a matrix with the properties $\mathcal{B} =
\mathcal{B}^\dagger = \mathcal{B}^{-1}$ and $\det \mathcal{B} =
1$. The invariance of the determinant $\det \hat S_s^{-1} ( k_0
)$ with respect to the change of the energy sign, $k_0
\rightarrow - k_0$, is directly related to the use of the
Nambu-Gorkov basis for quark fields. In this basis, for each
quasiparticle excitation with a positive energy $k_0 = \epsilon(
k )$, there exists a corresponding excitation with a negative
energy $k_0 = - \epsilon ( k )$. Therefore, the result for the
determinant should read,
\be
\label{det_invS}
\det S^{-1} = \prod_{i = 1}^{18} \left( k_0^2 - \epsilon_i^2
\right)^2 \; .
\ee
The expression $k_0^2 - \epsilon_i^2$ in Eq.~(\ref{det_invS})
has to be squared because of the spin degeneracy. Because of the
artificial Nambu-Gorkov degeneracy, $ k_0^2 - \epsilon_i^2
\equiv ( k_0 - \epsilon_i ) \, ( k_0 + \epsilon_i )$, where
$\epsilon_i$ are the 18 positive energy eigenvalues of
Eq.~(\ref{M_s}).

In order to simplify the numerical calculation of the
eigenvalues of the matrix $\mathcal{M}_+$, defined in
Eq.~(\ref{M_s}), I first write it in a block-diagonal form. With
a proper ordering of its 36 rows and 36 columns, it decomposes
into six diagonal blocks of dimension $4 \times 4$ and one
diagonal block of dimension $12 \times 12$. The explicit form of
these blocks reads,
\bsub
\label{blocks}
\bea
\mathcal{M}_+^{(1)} &=& \left(
\begin{array}{cccc}
- \mu_d^r + M_d & k & 0 & - \Delta_3 \\
k &  - \mu_d^r - M_d & \Delta_3 & 0 \\
0 & \Delta_3 & \mu_u^g + M_u & k \\
- \Delta_3 & 0 & k & \mu_u^g - M_u
\end{array}
\right) \; , \\
\mathcal{M}_+^{(2)} &=& \left(
\begin{array}{cccc}
\mu_d^r - M_d & k & 0 & - \Delta_3 \\
k &  \mu_d^r + M_d & \Delta_3 & 0 \\
0 & \Delta_3 & -\mu_u^g - M_u & k \\
- \Delta_3 & 0 & k & -\mu_u^g + M_u
\end{array}
\right) \; , \\
\mathcal{M}_+^{(3)} &=& \left(
\begin{array}{cccc}
- \mu_s^r + M_s & k & 0 & - \Delta_2 \\
k &  - \mu_s^r - M_s & \Delta_2 & 0 \\
0 & \Delta_2 & \mu_u^b + M_u & k \\
- \Delta_2 & 0 & k & \mu_u^b - M_u
\end{array}
\right) \; , \\
\mathcal{M}_+^{(4)} &=& \left(
\begin{array}{cccc}
\mu_s^r - M_s & k & 0 & - \Delta_2 \\
k &  \mu_s^r + M_s & \Delta_2 & 0 \\
0 & \Delta_2 & -\mu_u^b - M_u & k \\
- \Delta_2 & 0 & k & -\mu_u^b + M_u
\end{array}
\right) \; , \\
\mathcal{M}_+^{(5)} &=& \left(
\begin{array}{cccc}
- \mu_s^g + M_s & k & 0 & - \Delta_1 \\
k &  - \mu_s^g - M_s & \Delta_1 & 0 \\
0 & \Delta_1 & \mu_d^b + M_d & k \\
- \Delta_1 & 0 & k & \mu_d^b - M_d
\end{array}
\right) \; , \\
\mathcal{M}_+^{(6)} &=& \left(
\begin{array}{cccc}
\mu_s^g - M_s & k & 0 & - \Delta_1 \\
k &  \mu_s^g + M_s & \Delta_1 & 0 \\
0 & \Delta_1 & -\mu_d^b - M_d & k \\
- \Delta_1 & 0 & k & -\mu_d^b + M_d
\end{array}
\right) \; , \\
\mathcal{M}_+^{(7)} &=& \left(
\begin{array}{@{\extracolsep{1mm}}ccc}
\tilde{\mathcal{M}}_u^r & \tilde\Delta_3 & \tilde\Delta_2 \\
\tilde\Delta_3 & \tilde{\mathcal{M}}_d^g & \tilde\Delta_1 \\
\tilde\Delta_2 & \tilde\Delta_1 & \tilde{\mathcal{M}}_s^b
\end{array}
\right) \; .
\eea
\esub
Here, I have only shown the color-flavor structure of the matrix
$\mathcal{M}_+^{(7)}$ because the repesentation of the total
structure of the $12 \times 12$-matrix $\mathcal{M}_+^{(7)}$
would be too large. Each color-flavor element in
$\mathcal{M}_+^{(7)}$ is a $4 \times 4$-matrix in Nambu-Gorkov
and particle-antiparticle space,
\be
\begin{aligned}
\tilde{\mathcal{M}}_u^r &= \left(
\begin{array}{cccc}
-\mu_u^r -M_u &       k      &       0      &       0      \\
       k      & -\mu_u^r+M_u &       0      &       0      \\
       0      &       0      & \mu_u^r -M_u &       k      \\
       0      &       0      &       k      & \mu_u^r +M_u
\end{array}
\right) \; , &
\tilde\Delta_1 &= \left(
\begin{array}{@{\extracolsep{-2mm}}cccc}
   0      &     0                &     0                &
   -\Delta_1 \\
   0      &     0                & \phantom{-} \Delta_1 &   0   
      \\
   0      & \phantom{-} \Delta_1 &     0                &   0   
      \\
-\Delta_1 &     0                &     0                &   0
\end{array}
\right) \; , \\
\tilde{\mathcal{M}}_d^g &= \left(
\begin{array}{cccc}
-\mu_d^g -M_d &       k      &       0      &       0      \\
       k      & -\mu_d^g+M_d &       0      &       0      \\
       0      &       0      & \mu_d^g -M_d &       k      \\
       0      &       0      &       k      & \mu_d^g +M_d
\end{array}
\right) \; , &
\tilde\Delta_2 &= \left(
\begin{array}{@{\extracolsep{-2mm}}cccc}
   0      &     0                &     0                &
   -\Delta_2 \\
   0      &     0                & \phantom{-} \Delta_2 &   0   
      \\
   0      & \phantom{-} \Delta_2 &     0                &   0   
      \\
-\Delta_2 &     0                &     0                &   0
\end{array}
\right) \; , \\
\tilde{\mathcal{M}}_s^b &= \left(
\begin{array}{cccc}
-\mu_s^b -M_s &       k      &       0      &       0      \\
       k      & -\mu_s^b+M_s &       0      &       0      \\
       0      &       0      & \mu_s^b -M_s &       k      \\       
       0      &       0      &       k      & \mu_s^b +M_s
\end{array}
\right) \; , &
\tilde\Delta_3 &= \left(
\begin{array}{@{\extracolsep{-2mm}}cccc}
   0      &     0                &     0                &
   -\Delta_3 \\
   0      &     0                & \phantom{-} \Delta_3 &   0   
      \\
   0      & \phantom{-} \Delta_3 &     0                &   0   
      \\
-\Delta_3 &     0                &     0                &   0
\end{array}
\right) \; .
\end{aligned}
\ee
The inverse quark propagator is a $72 \times 72$-matrix, and
therefore, it has 72 energy eigenvalues. Because of spin
degeneracy half of them are equal. Out of 36 eigenvalues from
all seven blocks~(\ref{blocks}), there are 18 positive and 18
negative eigenvalues because of the Nambu-Gorkov degeneracy. Out
of total 18 positive eigenvalues, nine of them correspond to the
nine quasiquarks and the other nine correspond to the nine
quasiantiquarks. I obtained the 36 eigenvalues of
Eq.~(\ref{blocks}) numerically~\cite{numrec}.

Here, it might be interesting to note that the eigenvalues of
the first six $4\times 4$ matrices in Eq.~(\ref{blocks}) can be
calculated analytically in the limit when two quark masses
appearing in each of them are equal. For example, when
$M_d = M_u$, the four eigenvalues of matrix
$\mathcal{M}_+^{(1)}$ are given by
\be
\lambda^{(1)}_i = \pm \sqrt{ \left( \frac{ \mu_d^r + \mu_u^g }{
2 } \pm \sqrt{ M_u^2 + k^2 } \right)^2 + \Delta_3^2 } - \frac{
\mu_d^r - \mu_u^g }{ 2 } \; ,
\ee
while the eigenvalues of $\mathcal{M}_+^{(2)}$ differ only by
the sign in front of the second term,
\be
\lambda^{(2)}_i = \pm \sqrt{ \left( \frac{ \mu_d^r + \mu_u^g }{
2 } \pm \sqrt{ M_u^2 + k^2 } \right)^2 + \Delta_3^2 } + \frac{
\mu_d^r - \mu_u^g }{ 2 } \; .
\ee
When the value of $\delta M \equiv M_d - M_u$ is nonzero but
small, the corrections to the above eigenvalues are $\pm M_u
\delta M / (2 \sqrt{ M_u^2 + k^2} )$ with the plus sign in the
case of antiparticle modes, and the minus sign in the case of
particle modes. The eigenvalues of $\mathcal{M}_+^{(3)}$ and
$\mathcal{M}_+^{(4)}$ in the limit $M_s = M_u$, as well as the
eigenvalues of $\mathcal{M}_+^{(5)}$ and $\mathcal{M}_+^{(6)}$
in the limit $M_s = M_d$, are similar.

It turns out that in regular color-superconducting quark phases,
there are two positive-energy eigenvalues and two
negative-energy eigenvalues in each of the first six matrices in
Eq.~(\ref{blocks}). The two negative-energy eigenvalues are a
consequence of the artificial Nambu-Gorkov degeneracy. These
negative-energy eigenvalues do not have the same absolute value
as the positive-energy eigenvalues in the same matrix. That it
why these negative-energy eigenvalues have their positive
counterparts in another of these first six matrices. It is easy
to extract the 18 positive-energy eigenvalues out of the total
36 energy eigenvalues. This is done by sorting the energy
eigenvalues in descending order~\cite{numrec} in each of the
first six matrices. After that, I take the first two energy
eigenvalues from each of the first six matrices. The lower one
of these positive energy eigenvalues corresponds to a quasiquark
and the larger one corresponds to a quasiantiquark. It is
necessary to remember to which of the first six matrices they
belong to in order to determine from which of the matrices a
gapless mode comes from. In gapless color-superconducting quark
phases, an energy eigenvalue of at least one of the first six
matrices becomes positive (negative) which has been
negative (positive) in regular color-superconducting quark
phases. This means in the case of one gapless mode that one of
the first six matrices has three positive but only one negative
energy eigenvalues, and another one has only one positive but
three negative energy eigenvalues. The matrix from which arise
three positive-energy eigenvalues shows a gapless mode.
Therefore, if one sorts the energy eigenvalues in each of the
first six matrices in descending order and then extracts from
each of these matrices the first two energy eigenvalues, one
obtains at least one negative-energy eigenvalue in a gapless
color-superconducting quark phase by this method. Since I do not
need such negative energy eigenvalues, I take the absolute value
of them. Because of the Nambu-Gorkov degeneracy, they can be
found by using the following relation:
\be
\label{relation_epsilon}
\epsilon_i \left( \mathcal{M}_+^{( 2 n - 1 )} \right) = -
\epsilon_i \left( \mathcal{M}_+^{( 2 n )} \right) \; ,
\ee
where $n = 1, 2, 3$. This relation means that for example the
matrices $\mathcal{M}_+^{(1)}$ and $\mathcal{M}_+^{(2)}$ have
the same four absolute values of energy eigenvalues but in each
of these two matrices the four energy eigenvalues have different
signs. So, if one finds a negative-energy eigenvalue in matrix
$\mathcal{M}_+^{(1)}$ which has been positive in a regular phase
before, then the gapless mode arises from $\mathcal{M}_+^{(2)}$.

The relation~(\ref{relation_epsilon}) is easily proven by
introducing the matrix,
\be
\mathcal{U} = \diag \left( 1, -1, 1, -1 \right)
\ee
with the properties $\mathcal{U} = \mathcal{U}^T =
\mathcal{U}^\dagger = \mathcal{U}^{-1}$ and $\det \mathcal{U} =
1$,
\be
- k_0 - \mathcal{M}_+^{( 2 n )} = - \mathcal{U} \left( k_0 -
\mathcal{M}_+^{( 2 n - 1 )} \right) \mathcal{U}^{-1} \; ,
\ee
where $n = 1, 2, 3$.

The $12 \times 12$-matrix in Eq.~(\ref{blocks}),
$\mathcal{M}_+^{(7)}$, always has six positive and six negative
eigenvalues. Therefore, it is easy to extract the six
positive-energy eigenvalues out of $\mathcal{M}_+^{(7)}$. I sort
the energy eigenvalues in descending order~\cite{numrec} and then
take the first six one. The first three out of the six
positive-energy eigenvalues correspond to the three
quasiantiquarks while the second three out of the six
positive-energy eigenvalues correspond to three quasiparticles.
The remaining six negative-energy eigenvalues arise because of
the artificial Nambu-Gorkov degeneracy. Their absolute values
are equal to the positive-energy eigenvalues of
$\mathcal{M}_+^{(7)}$.

By using Eq.~(\ref{det_invS}), the kinetic part of the
pressure~(\ref{p_kin}) can be rewritten as
\be
p_\mathrm{kin} = \frac{T}{V} \sum_{i = 1}^{18} \sum_K \ln \left(
\frac{k_0^2 - \epsilon_i^2}{ T^2 } \right) \; .
\ee
After performing the sum over all fermionic Matsubara
frequencies, see
Sec.~\ref{Summation_over_the_fermionic_Matsubara_frequencies} in
the Appendix, and by noting that $\ln \mathcal{Z}_\mathrm{kin}
\equiv \frac{V}{T} p_\mathrm{kin}$ and $\delta\mu \equiv 0$, one
obtains,
\be
p_\mathrm{kin} = \frac{1}{V} \sum_{i = 1}^{18} \sum_\fettu{k}
\left\{ \epsilon_i + 2 T \ln \left[ 1 + \exp \left( - \frac{
\epsilon_i }{ T } \right) \right] \right\} \; ,
\ee
which can be converted into an integral representation by using
Eq.~(\ref{sum_int}). Therefore, the pressure of
color-superconducting quark matter reads,
\bea
\label{p_NJL}
p &=& \frac{1}{2 \pi^2} \sum_{i = 1}^{18} \int_0^\Lambda \ud k \,
k^2 \left\{ \epsilon_i + 2 T \ln \left[ 1 + \exp \left( - \frac{
\epsilon_i }{ T } \right) \right] \right\} - 2 G_S \sum_{\alpha
= 1}^3 \sigma_\alpha^2 - \frac{1}{4 G_D} \sum_{c = 1}^3 \left|
\Delta_c \right|^2 + 4 K \sigma_u \sigma_d \sigma_s \nonumber \\
&+& \frac{T}{\pi^2} \sum_{\beta = e}^\mu \int_0^\infty \ud k \,
k^2 \left\{ \ln \left[ 1 + \exp \left( - \frac{ E_\beta -
\mu_\beta }{ T } \right) \right] + \ln \left[ 1 + \exp \left( -
\frac{ E_\beta + \mu_\beta }{ T } \right) \right] \right\} \; ,
\eea
where the term in the last line is the contribution of leptons,
$p_\beta$, i.e.\ electrons and muons. They are added in order to
make color-superconducting quark matter electrically neutral. In
principle, the contribution of neutrinos should be added as
well. In this section, however, their contribution is neglected
which is a good approximation for neutron stars after
deleptonization. The effect of neutrino trapping is included
in Ref.~\cite{Ruester_Werth_Buballa_Shovkovy_Rischke2} and is
the subject in Sec.~\ref{phase_diagram_neutrino_trapping} of my
thesis. The dispersion relations of the leptons are given
by $E_\beta = ( k^2 + m_\beta^2 )^{1/2}$, where $m_e =
0.51099906$~MeV is the electron mass and $m_\mu =
105.658389$~MeV is the muon mass~\cite{Taschenbuch_der_Physik}.
In various applications, a bag constant could be added to the
pressure, and also the vacuum contribution could be substracted
from the pressure if necessary. In order to render the integrals
in the expression for the pressure finite, I used the
three-momentum cutoff $\Lambda$.

In the limit of zero temperature, the pressure of
color-superconducting quark matter reads,
\be
\label{p_NJL_T0}
p = \frac{1}{2 \pi^2} \sum_{i = 1}^{18} \int_0^\Lambda \ud k \,
k^2 \epsilon_i - 2 G_S \sum_{\alpha = 1}^3 \sigma_\alpha^2 -
\frac{1}{4 G_D} \sum_{c = 1}^3 \left| \Delta_c \right|^2 + 4 K
\sigma_u \sigma_d \sigma_s + \frac{1}{3 \pi^2} \sum_{\beta =
e}^\mu \int_0^{k_{F_\beta}} \ud k \, \frac{k^4}{E_\beta} \; .
\ee
In order to obtain the values for the quark-antiquark
condensates $\sigma_\alpha$ and the color-superconducting gap
parameters $\Delta_c$, I solve the following six stationary
conditions:
\be
\label{NJL_gap_eqns}
\frac{\partial p}{\partial \sigma_\alpha} = 0 \; , \qquad
\frac{\partial p}{\partial \Delta_c} = 0 \; .
\ee
After the quark-antiquark condensates are known, the quark
constituent masses can be determined by Eq.~(\ref{M_alpha}). In
order to enforce the conditions of local electric and color
charge neutrality in color-superconducting quark matter, it is
necessary to require the three Eqs.~(\ref{electric_neutrality})
and~(\ref{color_neutrality}) to be satisfied. These fix the
values of the three corresponding chemical potentials, $\mu_Q$,
$\mu_3$ and $\mu_8$. After these are fixed, only the quark
chemical potential $\mu$ and the temperature $T$ are left as
free parameters.

The number density of each quark is given by
\be
n_f^i = \frac{\partial p}{\partial \mu_f^i} \; ,
\ee
and the number density of each quark flavor reads,
\be
n_f = \sum_{i=r}^b n_f^i \; .
\ee
Since the energy eigenvalues are computed numerically, the
derivatives, i.e.\ the six gap equations, the three neutrality
conditions, and the quark number densities have to be computed
numerically. The number densities of the leptons can be
calculated using Eq.~(\ref{number_density_beta}) in the case
of nonzero temperature and
Eq.~(\ref{quark_flavor_and_lepton_number_densities}) in the case
of zero temperature.
\subsection{Results}
In order to obtain the phase diagram, I have to find the ground
state of quark matter for each given set of the parameters in
the model. In the case of locally neutral quark matter, there
are two parameters that should be specified: temperature $T$ and
quark chemical potential $\mu$. After these are fixed, one has
to compare the values of the pressure in all competing neutral
phases of quark matter. The ground state corresponds to the
phase with the highest pressure.

Before calculating the pressure, given by Eq.~(\ref{p_NJL}) for
nonzero temperature and by Eq.~(\ref{p_NJL_T0}) for zero
temperature, one has to find the values of the chiral and the
color-superconducting order parameters, $\sigma_\alpha$ and
$\Delta_c$, as well as the values of the three charge chemical
potentials, $\mu_Q$, $\mu_3$ and $\mu_8$. These are obtained by
solving the coupled set of six gap
equations~(\ref{NJL_gap_eqns}) together with the three
neutrality conditions~(\ref{electric_neutrality})
and~(\ref{color_neutrality}). By using standard numerical
recipes~\cite{numrec}, it is not extremely difficult to find a
solution to the given set of nine nonlinear equations.
Complications arise, however, due to the fact that often the
solution is not unique.

The existence of different solutions to the same set of
equations,~(\ref{electric_neutrality}),
(\ref{color_neutrality}), and~(\ref{NJL_gap_eqns}), reflects the
physical fact that there could exist several competing neutral
phases with different physical properties. Among these phases,
all but one are unstable or metastable. In order to take this
into account in my study, I look for the solutions of the eight
types which are listed in Table~\ref{csc_quark_phases}.
\begin{table}[H]
\begin{center}
\begin{tabular}{|l||c|c|c|}
\hline
Phase   & $\Delta_1$   & $\Delta_2$   & $\Delta_3$ \\
\hline
\hline
NQ      & $-$          & $-$          & $-$ \\
2SC     & $-$          & $-$          & $\checkmark$ \\
2SC$us$ & $-$          & $\checkmark$ & $-$ \\
2SC$ds$ & $\checkmark$ & $-$          & $-$ \\
uSC     & $-$          & $\checkmark$ & $\checkmark$ \\
dSC     & $\checkmark$ & $-$          & $\checkmark$ \\
sSC     & $\checkmark$ & $\checkmark$ & $-$ \\
CFL     & $\checkmark$ & $\checkmark$ & $\checkmark$ \\
\hline
\end{tabular}
\end{center}
\caption[The classification of eight color-superconducting quark
phases.]{The classification of eight color-superconducting quark
phases. The unmarked gap parameters ($-$) are zero while the
checkmarked gap parameters ($\checkmark$) are nonzero in the
respective color-superconducting quark phases. Note, that I do
not distinguish here between the CFL and the mCFL
phase~\cite{Ruester_Shovkovy_Rischke}.}
\label{csc_quark_phases}
\end{table}
\begin{figure}[H]
\begin{center}
\includegraphics[width=0.83\textwidth]{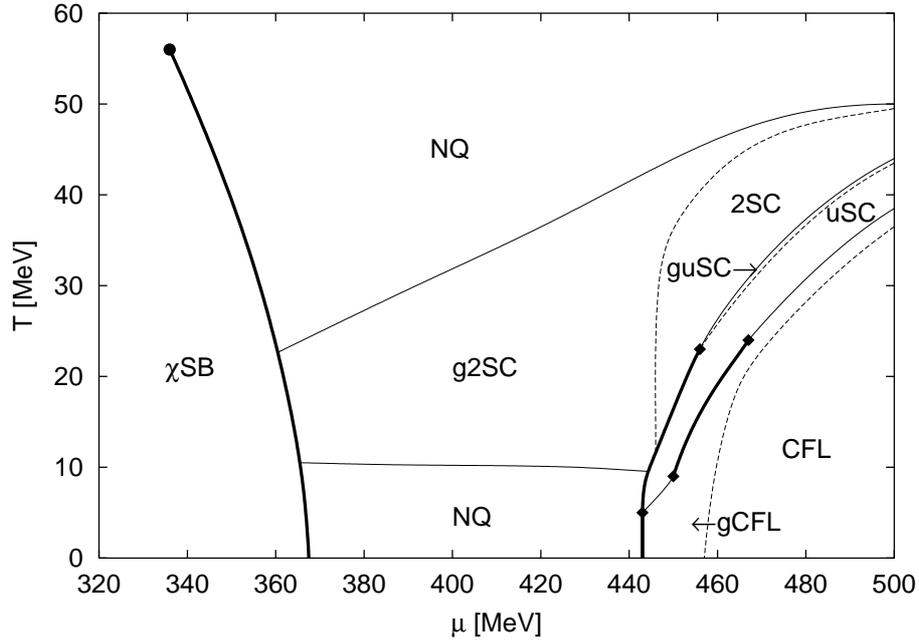}
\caption[The phase diagram of neutral quark matter without
neutrinos for $G_{D} = \frac34 G_{S}$.]{The phase diagram of
neutral quark matter in the regime of intermediate diquark
coupling strength, $G_{D} = \frac34 G_{S}$. First-order phase
boundaries are indicated by bold solid lines, whereas the thin
solid lines mark second-order phase boundaries between two
phases which differ by one or more nonzero diquark condensates.
The dashed lines indicate the \mbox{(dis-)appearance} of gapless
modes in different phases, and they do not correspond to phase
transitions.}
\label{phasediagram}
\end{center}
\end{figure}
\begin{figure}[H]
\begin{center}
\includegraphics[width=0.83\textwidth]{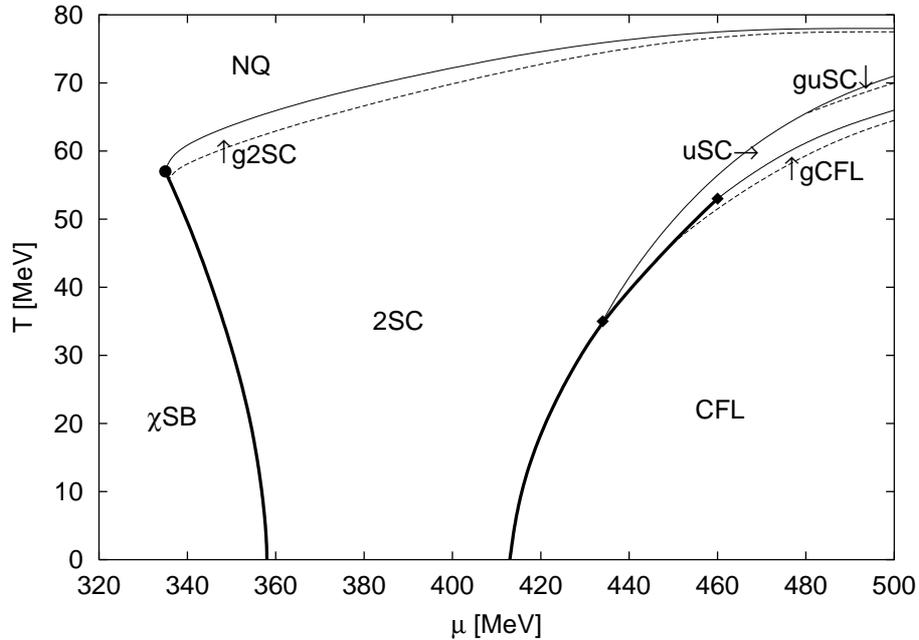}
\caption[The phase diagram of neutral quark matter without
neutrinos for $G_{D} = G_{S}$.]{The phase diagram of neutral
quark matter in the regime of strong diquark coupling, $G_{D} =
G_{S}$. The meaning of the various line types is the same as in
Fig.~\ref{phasediagram}.}
\label{phasediagram_strong}
\end{center}
\end{figure}
\begin{figure}[H]
\begin{center}
\includegraphics[width=0.98\textwidth]{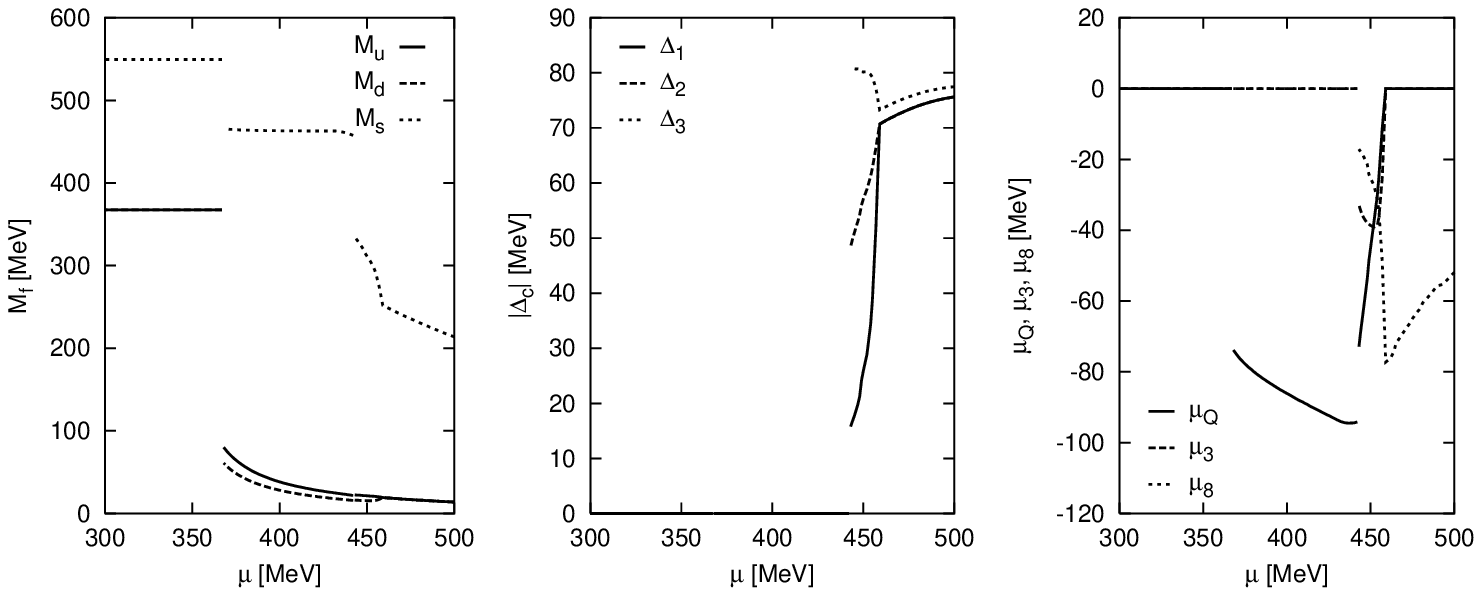}
\includegraphics[width=0.98\textwidth]{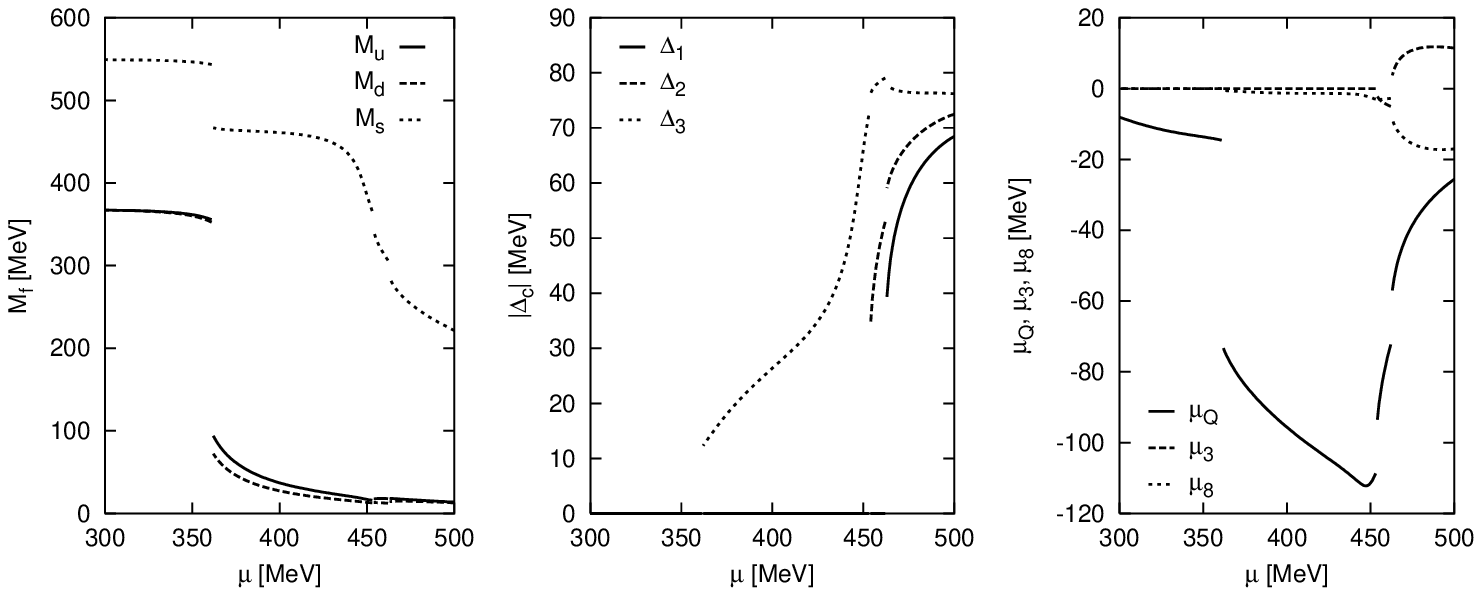}
\includegraphics[width=0.98\textwidth]{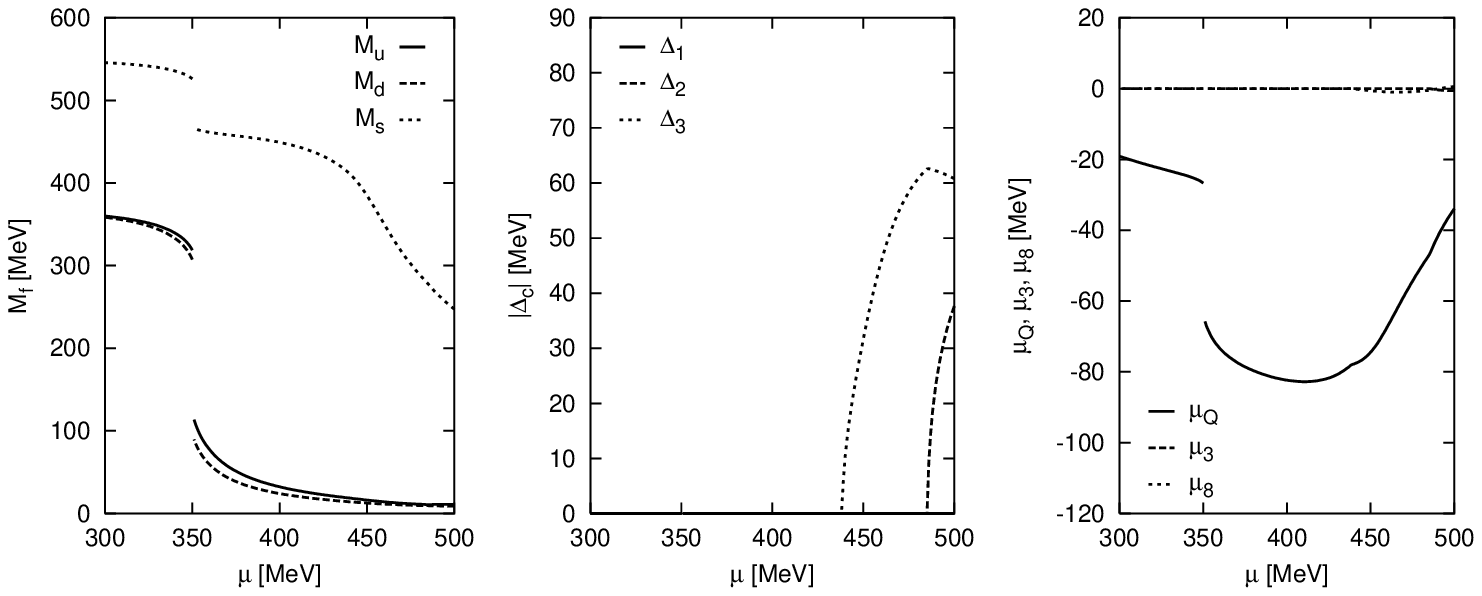}
\caption[Quark masses, gap parameters, and chemical potentials
at $G_D = \frac34 G_S$.]{The dependence of the quark masses, of
the gap parameters, and of the electric and color charge
chemical potentials on the quark chemical potential at a fixed
temperature, $T = 0$~MeV (three upper panels), $T =
20$~MeV (three middle panels), and $T = 40$~MeV (three lower
panels). The diquark coupling strength is $G_D = \frac34 G_S$.}
\label{plot0_20_40}
\end{center}
\end{figure}
Then, I calculate the values of the pressure in all
nonequivalent phases, and determine the ground state as the
phase with the highest pressure. After this is done, I
additionally study the spectrum of low-energy quasiparticles in
search for the existence of gapless modes. This allows me to
refine the specific nature of the ground state.

In the definition of the eight phases in terms of $\Delta_c$ in
Table~\ref{csc_quark_phases}, I have ignored the
quark-antiquark condensates $\sigma_\alpha$. In fact, in the
chiral limit ($m_\alpha = 0$), the quantities $\sigma_\alpha$
are good order parameters and I could define additional
sub-phases characterized by nonvanishing values of one or more
$\sigma_\alpha$. With the model parameters at hand, however,
chiral symmetry is broken explicitly by the nonzero current
quark masses, and the values of $\sigma_\alpha$ never vanish.
Hence, in a strict sense it is impossible to define any new
phases in terms of $\sigma_\alpha$.

Of course, this does not exclude the possibility of
discontinuous changes in $\sigma_\alpha$ at some line in the
plane of temperature and quark chemical potential, thereby
constituting a first-order phase transition line. It is
generally expected that the chiral phase transition remains
first order at low temperatures, even for nonzero quark masses.
Above some critical temperature, however, this line could end in
a critical endpoint, and there is only a smooth crossover at
higher temperatures. Among others, this picture emerges from
NJL-model studies, both, without~\cite{cr_point_NJL} and
with~\cite{Buballa_Oertel} diquark pairing (see also
Ref.~\cite{Buballa_Habilitationsschrift}). Therefore, I expect a
similar behavior in my analysis.

\begin{figure}[H]
\begin{center}
\includegraphics[width=0.98\textwidth]{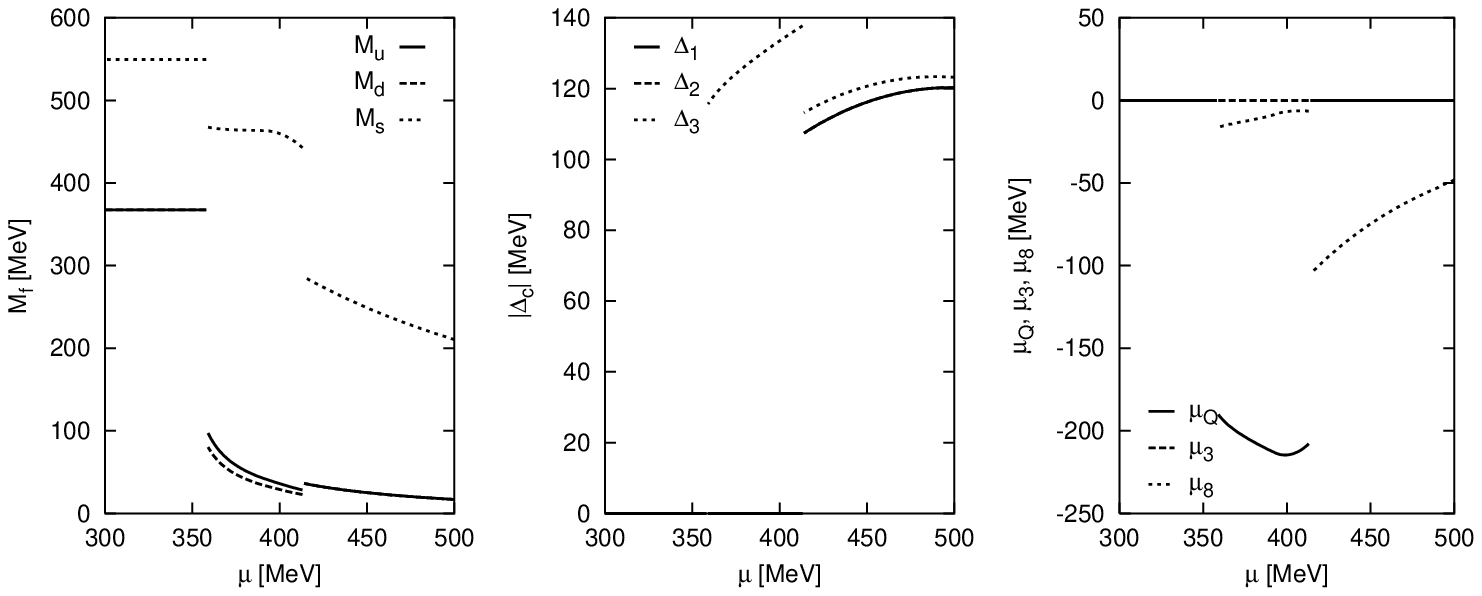}
\includegraphics[width=0.98\textwidth]{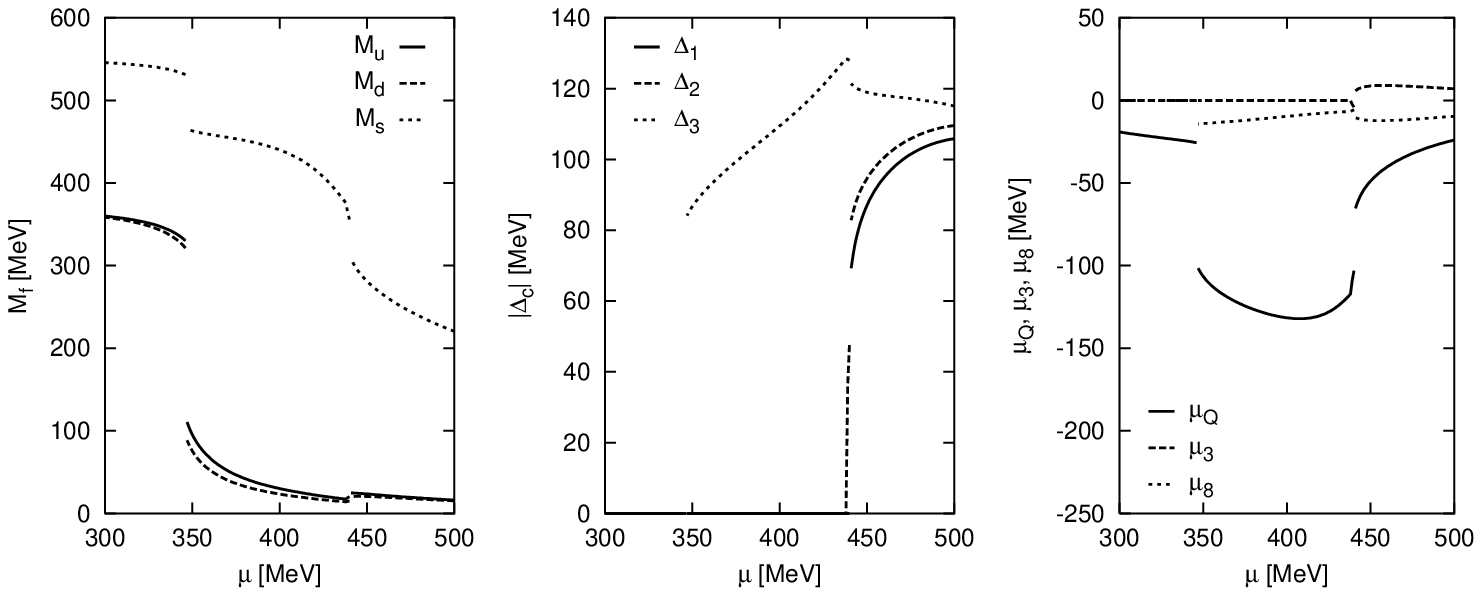}
\includegraphics[width=0.98\textwidth]{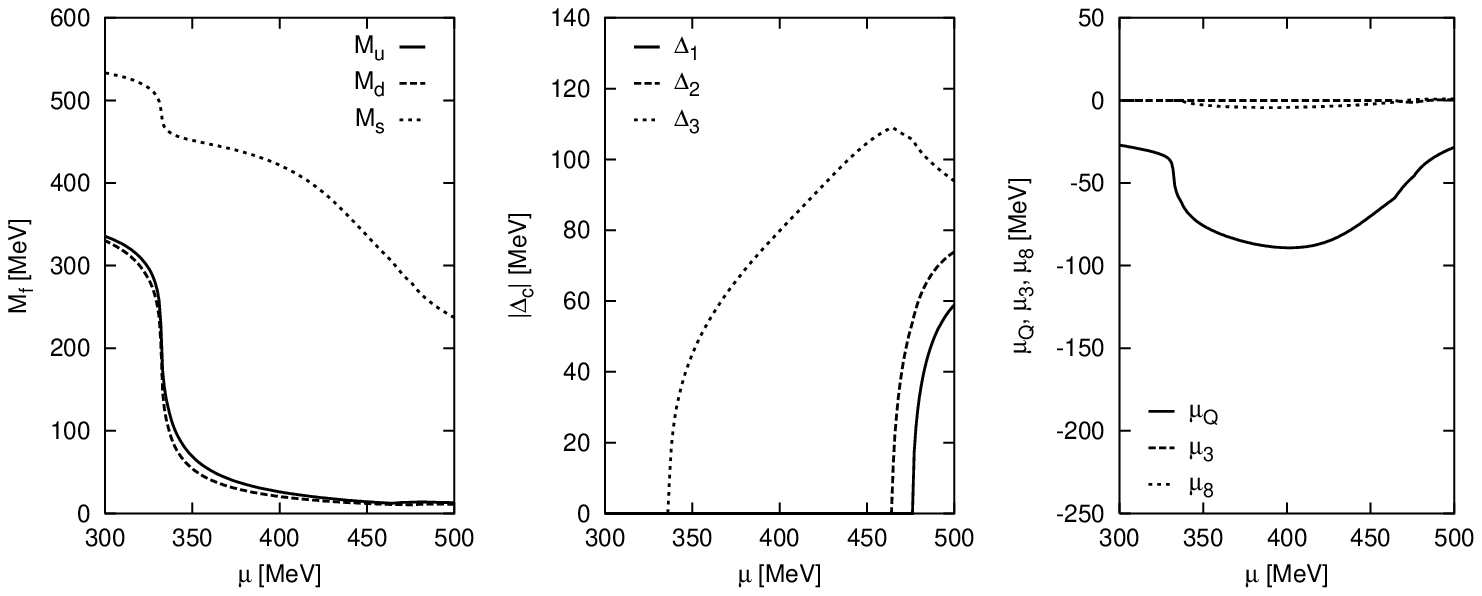}
\caption[Quark masses, gap parameters, and chemical potentials
at $G_D = G_S$.]{The dependence of the quark masses, of the gap
parameters, and of the electric and color charge chemical
potentials on the quark chemical potential at a fixed
temperature, $T = 0$~MeV (three upper panels), $T = 40$~MeV
(three middle panels), and $T = 60$~MeV (three lower panels).
The diquark coupling strength is $G_D = G_S$.}
\label{plot0_40_60_strong}
\end{center}
\end{figure}

My numerical results for neutral quark matter are summarized in
Figs.~\ref{phasediagram} and~\ref{phasediagram_strong}. These
are the phase diagrams in the plane of temperature $T$ and quark
chemical potential $\mu$ in the case of an intermediate diquark
coupling strength, $G_D = \frac34 G_S$, and in the case of a
strong coupling, $G_D = G_S$, respectively. The corresponding
dynamical quark masses, gap parameters, and three charge
chemical potentials are displayed in Figs.~\ref{plot0_20_40}
and~\ref{plot0_40_60_strong}, respectively. All quantities are
plotted as functions of $\mu$ for three different fixed values
of the temperature: $T = 0, 20, 40$~MeV in the case of $G_D =
\frac34 G_S$ (see Fig.~\ref{plot0_20_40}) and $T = 0, 40,
60$~MeV in the case of $G_D = G_S$ (see
Fig.~\ref{plot0_40_60_strong}).

Let me begin with the results in the case of the diquark
coupling being $G_D = \frac34 G_S$. In the region of small quark
chemical potentials and low temperatures, the phase diagram is
dominated by the normal phase in which the approximate chiral
symmetry is broken, and in which quarks have relatively large
constituent masses. This is denoted by $\chi$SB in
Fig.~\ref{phasediagram}. With increasing the temperature, this
phase changes smoothly into a normal quark phase (NQ) in which
quark masses are relatively small. Because of explicit breaking
of the chiral symmetry in the model at hand, there is no need
for a phase transition between the two regimes.

However, as pointed out above, the symmetry argument does not
exclude the possibility of a first-order chiral phase
transition. As expected, at lower temperatures I find a line of
first-order chiral phase transitions. It is located within a
relatively narrow window of the quark chemical
potentials (336~MeV~$\lesssim \mu \lesssim$~368~MeV) which are
of the order of the vacuum values of the light-quark constituent
masses. (For the parameters used in my calculations one obtains
$M_u = M_d = 367.7$~MeV and $M_s = 549.5$~MeV in
vacuum~\cite{RKH}.) At this critical line, the quark chiral
condensates, as well as the quark constituent masses, change
discontinuously. With increasing temperature, the size of the
discontinuity decreases, and the line terminates at the endpoint
located at $(T_\mathrm{cr}, \mu_\mathrm{cr}) \approx
(56, 336)$~MeV, see Fig.~\ref{phasediagram}.

The location of the critical endpoint is consistent with other
mean-field studies of NJL models with similar sets of
parameters~\cite{Buballa_Habilitationsschrift, cr_point_NJL,
Buballa_Oertel}. This agreement does not need to be exact
because, in contrast to the studies in
Refs.~\cite{Buballa_Habilitationsschrift, cr_point_NJL,
Buballa_Oertel}, here I imposed the condition of electric charge
neutrality in quark matter. (Note that the color neutrality is
satisfied automatically in the normal quark phase.) One may
argue, however, that the additional constraint of neutrality is
unlikely to play a big role in the vicinity of the endpoint.

It is appropriate to mention here that the location of the
critical endpoint might be affected very much by fluctuations of
the composite chiral fields. These are not included in the
mean-field studies of the NJL model. In fact, this is probably
the main reason for their inability to pin down the location of
the critical endpoint consistent with lattice QCD
calculations~\cite{lattice}. It is fair to mention that the
current lattice QCD calculations are not very reliable at
nonzero $\mu$ either. Therefore, the predictions of this study,
as well as of those in Refs.~\cite{Buballa_Habilitationsschrift,
cr_point_NJL, Buballa_Oertel}, regarding the critical endpoint
cannot be considered as very reliable.

\begin{figure}[H]
\begin{center}
\includegraphics[width=0.9\textwidth]{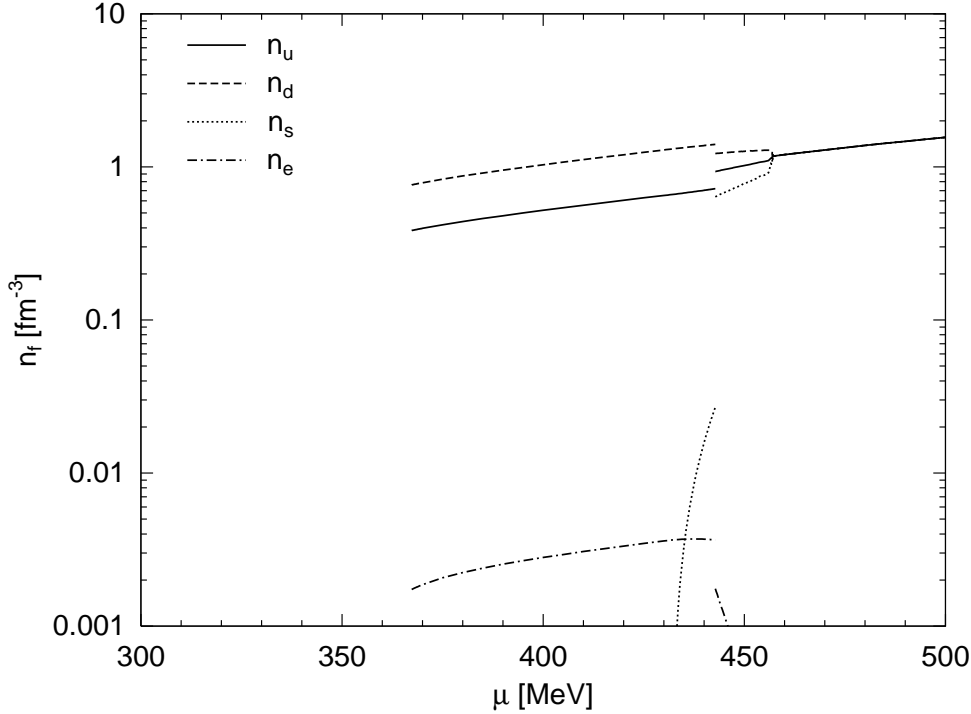}
\caption[The number densities of quarks and electrons at
$T = 0$ and $G_D = \frac34 G_S$.]{The dependence of the number
densities of quarks and electrons on the quark chemical
potential at $T = 0$~MeV for the diquark coupling strength $G_D
= \frac34 G_S$. Note that the densities of all three quark
flavors coincide above $\mu = 457$~MeV. The density of muons
vanishes for all values of $\mu$.}
\label{densT0log}
\end{center}
\end{figure}

When the quark chemical potential exceeds some critical value
and the temperature is not too large, a Cooper instability with
respect to diquark condensation should develop in the system.
Without enforcing neutrality, i.e., if the chemical potentials
of up and down quarks are equal, this happens immediately after
the chiral phase transition, when the density becomes
nonzero~\cite{Buballa_Oertel}. In the present model, this is
not the case at low temperatures.
In order to understand this, let me inspect the various
quantities at $T=0$ which are displayed in the upper three
panels of Fig.~\ref{plot0_20_40}. At the chiral phase boundary,
the up and down quark masses become relatively small, whereas
the strange quark mass experiences only a moderate drop of about
84~MeV induced by the 't~Hooft interaction. This is not
sufficient to populate any strange quark states at the given
chemical potential, and the system mainly consists of up and
down quarks together with a small fraction of electrons, see
Fig.~\ref{densT0log}. The electric charge chemical
potential which is needed to maintain neutrality in this regime
is between about $-73$~MeV and $-94$~MeV. It turns out that the
resulting splitting of the up and down quark Fermi momenta is
too large for the given diquark coupling strength to enable
diquark pairing and the system stays in the normal quark phase.
At $\mu \approx 432$~MeV, the chemical potential felt by the
strange quarks, $\mu - \mu_Q / 3$, reaches the strange quark
mass and the density of strange quarks becomes nonzero. At
first, this density is too small to play a sizeable role in
neutralizing the quark matter, or in enabling strange-nonstrange
cross-flavor diquark pairing, see Fig.~\ref{densT0log}.
The first instability happens at $\mu_\mathrm{gCFL} \approx
443$~MeV, where a first-order phase transition from the NQ phase
to the gCFL phase takes place. This is directly related to a
drop of the strange quark mass by about 121~MeV. As a
consequence, strange quarks become more abundant and pairing
gets easier. Yet, in the gCFL phase, the strange quark mass is
still relatively large, and the standard BCS pairing between
strange and light (i.e., up and down) quarks is not possible. In
contrast to the regular CFL phase, the gCFL phase requires a
nonzero density of electrons to stay electrically neutral. At $T
= 0$, therefore, one could use the value of the electron density
as a formal order parameter that distinguishes these two
phases~\cite{Alford_Kouvaris_Rajagopal1,
Alford_Kouvaris_Rajagopal2}.

\begin{figure}[H]
\begin{center}
\includegraphics[width=0.9\textwidth]{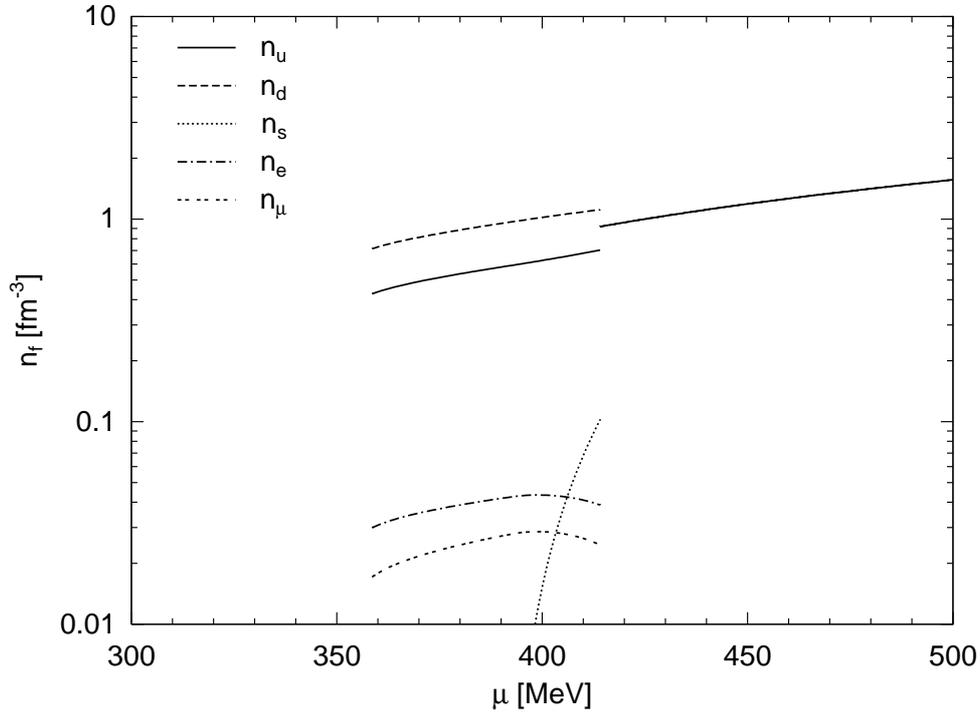}
\caption[The number densities of quarks, electrons, and muons at
$T = 0$ and $G_D = G_S$.]{The dependence of the number
densities of quarks, electrons, and muons on the quark chemical
potential at $T = 0$~MeV for the diquark coupling strength $G_D
= G_S$. Note, that the densities of all three quark flavors
coincide above $\mu = 414$~MeV.}
\label{densT0log_strong}
\end{center}
\end{figure}

With increasing the chemical potential further (still at $T=0$),
the strange quark mass decreases and the cross-flavor Cooper
pairing gets stronger. Thus, the gCFL phase eventually turns
into the regular CFL phase at $\mu_\mathrm{CFL} \approx
457$~MeV. The electron density goes to zero at this point, as it
should. This is indicated by the vanishing value of $\mu_Q$ in
the CFL phase, see the upper right panel in
Fig.~\ref{plot0_20_40}. I remind that the CFL phase is neutral
because of having equal number densities of all three quark
flavors, $n_u = n_d = n_s$, see Figs.~\ref{densT0log}
and~\ref{densT0log_strong}. This equality is enforced by the
pairing mechanism, and this is true even when the quark masses
are not exactly equal~\cite{enforced_neutrality}.
The same NJL model at zero temperature was studied previously in
Ref.~\cite{SRP}. My results agree qualitatively with those of
Ref.~\cite{SRP} only when the quark chemical potential is larger
than the critical value for the transition to the CFL phase at
457~MeV. The appearance of the gCFL phase for 443~MeV~$\lesssim
\mu \lesssim$~457~MeV was not recognized in Ref.~\cite{SRP}.
Instead it was suggested that there exists a narrow (about
12~MeV wide) window of values of the quark chemical potential
around $\mu \approx 450$~MeV in which the 2SC phase is the
ground state. By carefully checking the same region, I find
that the 2SC phase does not appear there.

\begin{figure}[H]
\begin{center}
\includegraphics[width=0.9\textwidth]{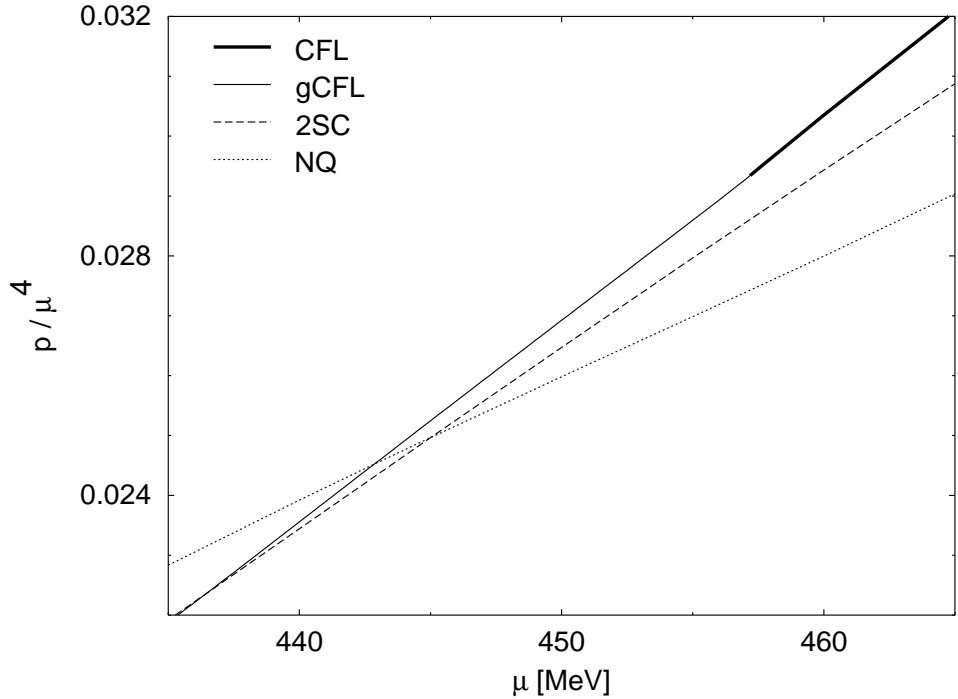}
\caption[The pressures of different phases of neutral
color-superconducting quark matter.]{The pressures of different
phases of neutral color-superconducting quark matter divided by
$\mu^4$ as a function of the quark chemical potential $\mu$ at
$T = 0$: regular CFL phase (bold solid line), gapless CFL phase
(thin solid line), 2SC phase (dashed line), normal quark
phase (dotted line). The diquark coupling strength is
$G_D=\frac34 G_S$.}
\label{pmu0}
\end{center}
\end{figure}

This is illustrated in Fig.~\ref{pmu0} where the pressure of
three different solutions is displayed. Had I ignored the gCFL
solution (thin solid line), the 2SC solution (dashed line) would
indeed be the most favored one in the interval between $\mu
\approx 445$~MeV and $\mu \approx 457$~MeV. After including the
gCFL phase in the analysis, this is no longer the case.

Now let me turn to the case of nonzero temperature. One might
suggest that this should be analogous to the zero temperature
case, except that Cooper pairing is somewhat suppressed by
thermal effects. In contrast to this na\"{\i}ve expectation, the
thermal distributions of quasiparticles together with the local
neutrality conditions open qualitatively new possibilities that
were absent at $T = 0$. As in the case of the two-flavor model
of Refs.~\cite{g2SC_1, g2SC_2}, a moderate thermal smearing of
mismatched Fermi surfaces could increase the probability of
creating zero-momentum Cooper pairs without running into a
conflict with Pauli blocking. This leads to the appearance of
several stable color-superconducting phases that could not exist
at zero temperatures.

With increasing the temperature, the first qualitatively new
feature in the phase diagram appears when 5~MeV~$\lesssim T
\lesssim$~10~MeV. In this temperature interval, the NQ phase is
replaced by the uSC phase when the quark chemical potential
exceeds the critical value of about 444~MeV. The corresponding
transition is a first-order phase transition, see
Fig.~\ref{phasediagram}. Increasing the chemical potential
further by several MeV, the uSC phase is then replaced by the
gCFL phase, and the gCFL phase later turns gradually into the
(m)CFL phase. (In this study, I do not distinguish between the
CFL phase and the mCFL phase~\cite{Ruester_Shovkovy_Rischke}.)
Note that, in the model at hand, the transition between the uSC
and the gCFL phase is of second order in the following two
temperature intervals: 5~MeV~$\lesssim T \lesssim$~9~MeV and $T
\lesssim 24$~MeV. On the other hand, it is a first-order phase
transition when 9~MeV~$\lesssim T \lesssim$~24~MeV. Leaving
aside its unusual appearance, this is likely to be an accidental
property in the model for a given set of parameters. For a
larger value of the diquark coupling, in particular, such a
feature does not appear, see Fig.~\ref{phasediagram_strong}.
The transition from the gCFL to the CFL phase is a smooth
crossover at $T > 0$. The reason is that the electron density is
not a good order parameter that could be used to distinguish the
gCFL from the CFL phase when the temperature is nonzero. This is
also confirmed by my numerical results for the electric charge
chemical potential $\mu_Q$ in Fig.~\ref{plot0_20_40}. While at
zero temperature the value of $\mu_Q$ vanishes identically in
the CFL phase, this is not the case at nonzero temperatures.

Another new feature in the phase diagram appears when the
temperature is above about 11~MeV. In this case, with increasing
the quark chemical potential, the Cooper instability happens
immediately after the $\chi$SB phase. The corresponding critical
value of the quark chemical potential is rather low, about
365~MeV. The first color-superconducting phase is the g2SC
phase~\cite{g2SC_1, g2SC_2}. This phase is replaced with the 2SC
phase in a crossover transition only when $\mu \gtrsim 445$~MeV.
The 2SC phase is then followed by the gapless uSC (guSC) phase,
by the uSC phase, by the gCFL phase and, eventually, by the CFL
phase (see Fig.~\ref{phasediagram}).

In the NJL model at hand, determined by the parameters in
Eq.~(\ref{NJL_model_parameters}), I do not find the dSC phase as
the ground state anywhere in the phase diagram. This is similar
to the conclusion of Refs.~\cite{Ruester_Shovkovy_Rischke,
Shovkovy_Ruester_Rischke}, but differs from that of
Refs.~\cite{Iida, Fukushima}. This should not be surprising
because, as was noted earlier~\cite{Shovkovy_Ruester_Rischke},
the appearence of the dSC phase is rather sensitive to a
specific choice of parameters in the NJL model.

The phase diagram in Fig.~\ref{phasediagram} has a very specific
ordering of quark phases. One might ask if this ordering is
robust against the modification of the parameters of the model
at hand. Below, I argue that some features are indeed quite
robust, while others are not. 

It should be clear that the appearance of color-superconducting
phases under the stress of neutrality constraints is very
sensitive to the strength of diquark coupling. In the case of
two-flavor quark matter, this was demonstrated very clearly in
Refs.~\cite{g2SC_1, g2SC_2} at zero as well as at nonzero
temperatures. Similar conclusions were also reached in the study
of three-flavor quark matter at zero
temperature~\cite{Abuki_Kitazawa_Kunihiro}.

In the model at hand, it is instructive to study the phase
diagram in the regime of strong diquark coupling, $G_D = G_S$.
The corresponding results are summarized in the diagram in
Fig.~\ref{phasediagram_strong}. As one can see, the main
qualitative difference between the diagrams in
Figs.~\ref{phasediagram} and~\ref{phasediagram_strong} occurs at
intermediate values of the quark chemical potential. While at
$G_D = \frac34 G_S$, there is a large region of the g2SC phase
sandwiched between the low-temperature and high-temperature NQ
phases, this is not the case at stronger coupling, $G_D = G_S$.
The regions of the gapless phases shrink with increasing diquark
coupling constant. Above some value of the diquark coupling
constant in the regime of (very) strong coupling, all gapless
phases disappear~\cite{Abuki_Kunihiro}.

The last observation can easily be explained by the fact that
with increasing diquark coupling strength, the condensation
energy also increases and therefore Cooper pairing is favorable,
even if there is a larger mismatch of the Fermi surfaces due to
charge neutrality constraints. Moreover, in the presence of
large gaps, the Fermi surfaces are smeared over a region which
is of the order of the values of the gaps. Therefore additional
thermal smearing is of no further help, and it is not surprising
that the thermal effects in a model with sufficiently strong
coupling are qualitatively the same as in models without
neutrality constraints imposed: thermal fluctuations can only
destroy the pairing. In the model with a not very strong
coupling, on the other hand, the interplay of the charge
neutrality and thermal fluctuations is more subtle. The normal
phase of cold quark matter develops a Cooper instability and
becomes a color superconductor only after a moderate thermal
smearing of the quark Fermi surfaces is introduced~\cite{g2SC_1,
g2SC_2}.

Other than this, the qualitative features of the phase diagrams
in Figs.~\ref{phasediagram} and~\ref{phasediagram_strong} are
similar. Of course, in the case of the stronger coupling, the
critical lines lie systematically at higher values of the
temperature and at lower values of the quark chemical potential.
In this context one should note that the first-order phase
boundary between the two normal regimes $\chi$SB and NQ
is insensitive to the diquark coupling. Therefore, upon
increasing $G_D$ it stays at its place until it is eventually
displaced by the expanding 2SC phase. As a result, there is
no longer a critical endpoint in Fig.~\ref{phasediagram_strong},
but only a critical point where the first-order
normal($\chi$SB)-2SC phase boundary changes into second order.
\section{The phase diagram with the effect of neutrino trapping}
\label{phase_diagram_neutrino_trapping}
Quark matter is expected to be color-superconducting. At
extremely large densities, namely when the quark chemical
potential $\mu$ is much larger than the constituent,
medium-modified quark masses, the ground state of matter is
given by the CFL phase~\cite{CFL_discoverers} (for studies of
QCD at large densities, see also
Refs.~\cite{Shovkovy_Wijewardhana, Schaefer2}). At the highest
densities existing in stars, however, the chemical potential is
unlikely to be much larger than 500~MeV, while the constituent
mass of the strange quarks is not smaller than the current mass,
which is about 100~MeV. In stellar matter, therefore, the
heavier strange quarks may not be able to participate in diquark
Cooper pairing as easily as the light up and down quarks. Then,
the pairing of light quarks can lead to the two-flavor
color-superconducting ground
state~\cite{Alford_Rajagopal_Wilczek, Pisarski_Rischke1,
regular_self-energies, Son}. It should be pointed out, however,
that the 2SC phase is subject to large penalties after imposing
the charge neutrality and $\beta$-equilibrium
conditions~\cite{Alford_Rajagopal}. Indeed, when the fraction of
strange quarks and leptons is small, the electric neutrality
requires roughly twice as many down quarks as up quarks. In this
case, Cooper pairing of up quarks with down quarks of opposite
momenta becomes rather difficult. Then, depending on the details
of the interaction, the gCFL
phase~\cite{Alford_Kouvaris_Rajagopal1,
Alford_Kouvaris_Rajagopal2}, the g2SC phase~\cite{g2SC_1,
g2SC_2}, or even the normal quark matter phase (NQ) may be more
favored. Also, there exist other reasonable
possibilities~\cite{spin-1, Schmitt_Wang_Rischke, Huang, Gorbar,
LOFF_quark, Reddy_Rupak, Shovkovy_Hanauske_Huang, Neumann,
Schmitt, Aguilera, Hong_2} which, in view of the known
instabilities in the gapless
phases~\cite{chromomagnetic_instability, Huang, Giannakis}, are
considered to be very promising.

In Ref.~\cite{Ruester_Werth_Buballa_Shovkovy_Rischke1}, the
phase diagram of neutral quark matter was obtained which was
described by a NJL-type model with the parameter set from
Ref.~\cite{RKH}. This work is presented in
Sec.~\ref{phase_diagram_self-consistent}. In contrast to
previous studies in Refs.~\cite{Ruester_Shovkovy_Rischke,
Fukushima, Shovkovy_Ruester_Rischke}, dynamically generated
quark masses were treated self-consistently in
Ref.~\cite{Ruester_Werth_Buballa_Shovkovy_Rischke1}.
(For earlier studies on color superconductivity, treating quark
masses as dynamical quantities, see also
Refs.~\cite{Buballa_Oertel, HZC}.)

In this section, I follow the same approach to study the effect
of a nonzero neutrino (or, more precisely, lepton-number)
chemical potential on the structure of the phase diagram
\cite{Ruester_Werth_Buballa_Shovkovy_Rischke2}. This is expected
to have a potential relevance for the physics of protoneutron
stars where neutrinos are trapped during the first few seconds
of the stellar evolution. In application to protoneutron stars,
it is of interest to cover a range of parameters that could
provide a total lepton fraction in quark matter of up to about
0.4. This is the value of the lepton-to-baryon charge ratio in
iron cores of progenitor stars. Because of the conservation of
both, lepton and baryon charges, this value is also close to the
lepton fraction in protoneutron stars at early times, when the
leptons do not have a chance to diffuse through dense matter
and escape from the star.

The effect of neutrino trapping on color-superconducting quark
matter has been previously discussed in Ref.~\cite{SRP}. It was
found that a nonzero neutrino chemical potential favors the 2SC
phase and disfavors the CFL phase. This is not unexpected
because the neutrino chemical potential is related to the
conserved lepton number in the system and therefore it also
favors the presence of (negatively) charged leptons. This helps
2SC-type pairing because electrical neutrality in quark matter
can be achieved without inducing a very large mismatch between
the Fermi surfaces of up and down quarks. The CFL phase, on the
other hand, is electrically and color neutral \textit{in the
absence} of charged leptons when $T
=0$~\cite{enforced_neutrality}. A nonzero neutrino chemical
potential can only spoil CFL-type pairing.

In this section, I extend the analysis of Ref.~\cite{SRP} by
performing a more systematic survey of the phase diagram in the
space of temperature, quark, and lepton-number chemical
potentials. This also includes the possibility of gapless phases
which have not been taken into account in Ref.~\cite{SRP}.
\subsection{Model}
\label{model+neutrinos}
I consider a system of up, down, and strange quarks, in weak
equilibrium with charged leptons and the corresponding
neutrinos. I assume that the lepton sector of the model is given
by an ideal gas of massive electrons ($m_e = 0.51099906$~MeV)
and muons ($m_\mu =
105.658389$~MeV)~\cite{Taschenbuch_der_Physik}, as well as
massless electron and muon neutrinos. I do not take into account
the $\tau$ lepton, which is too heavy to play any role. Also, I
neglect the possibility of neutrino mixing, and therefore I do
not take into account the $\tau$ neutrino either. In the quark
sector, I use the same three-flavor NJL model as in
Sec.~\ref{phase_diagram_self-consistent}, cf.\ 
Eq.~(\ref{L_NJL}).

In the model at hand, there are six mutually commuting conserved
charge densities. They split naturally into the following three
classes:
\begin{description}
\item[electric charge:] This is related to the $U(1)$ symmetry
of electromagnetism. The corresponding charge density is given
by
\be
\label{n_Q}
n_Q = \langle \psi^\dagger Q \psi \rangle - n_e - n_\mu \; ,
\ee
where $Q$ is the electric charge matrix of the
quarks~(\ref{matrix_of_electric_charge}), and $n_e$ and $n_\mu$
denote the number densities of electrons and muons,
respectively, cf.\ Eq.~(\ref{n_Q_introduction}).
\item[two lepton charges:] As long as the neutrinos are trapped
and their oscillations are neglected, the lepton family numbers
are conserved. The corresponding densities read,
\be
n_{L_e} = n_e + n_{\nu_e} \; , \qquad
n_{L_\mu} = n_\mu + n_{\nu_\mu} \; ,
\ee
\item[baryon number and two color charges:] The $SU(3)$-color
symmetry and $U(1)$ baryon number symmetry imply the
conservation of three independent charge densities in the quark
sector,
\be
n = \langle \psi^\dagger \psi \rangle \; , \qquad
n_3 = \langle \psi^\dagger T_3 \psi \rangle \; , \qquad
n_8 = \langle \psi^\dagger T_8 \psi \rangle \; ,
\ee
where $T_3$ and $T_8$ are the matrices associated with the two
mutually commuting color charges of the $[SU(3)_c]$ gauge group,
cf.\ Eq.~(\ref{n_n3_n8}). Note that $n$ is the quark number
density that is related to the baryon number density as follows:
$n_B = n / 3$. An alternative choice of the three conserved
charges is given by the number densities of red, green and blue
quarks, i.e.,
\be
n_r = n_B + n_3 + \frac{1}{\sqrt{3}} n_8 \; , \qquad n_g = n_B -
n_3 + \frac{1}{\sqrt{3}} n_8 \; , \qquad n_b = n_B -
\frac{2}{\sqrt{3}} n_8 \; .
\ee
\end{description}
The six conserved charge densities defined above are related to
the six chemical potentials of the model. These are the quark
chemical potential $\mu = \mu_B / 3$, the two color chemical
potentials $\mu_3$ and $\mu_8$, the electric charge chemical
potential $\mu_Q$, and the two lepton-number chemical potentials
$\mu_{L_e}$ and $\mu_{L_\mu}$.

In chemical equilibrium, the chemical potentials of all
individual quark and lepton species can be expressed in terms of
these six chemical potentials according to their content of
conserved charges. For the quarks, which carry quark number,
color and electric charge, this is related to the matrix of the
quark chemical
potentials~(\ref{quark_color_and_flavor_beta_equilibrium}). The
neutrinos, on the other hand, carry lepton number only,
cf.\ Eq.~(\ref{neutrino_lepton}). Finally, electrons and muons
carry both, lepton number and electric charge, see
Eq.~(\ref{mu_beta}). As in
Sec.~\ref{phase_diagram_self-consistent}, the quark part of the
model is treated in the mean-field (Hartree) approximation,
allowing for the presence of both, quark-antiquark condensates
and scalar diquark condensates. By using the same gap ansatz
(\ref{gap_ansatz})--(\ref{gap_ansatz_color-flavor}) and by
following the same steps in the derivation as in
Sec.~\ref{phase_diagram_self-consistent} and by including the
ideal-gas contribution for the leptons, I arrive at the
following expressions for the pressure at nonzero and zero
temperature, respectively:
\bsub
\label{p_NJL_neutrinos}
\bea
p &=& \frac{1}{2 \pi^2} \sum_{i = 1}^{18} \int_0^\Lambda \ud k
\, k^2 \left\{ \epsilon_i + 2 T \ln \left[ 1 + \exp \left( -
\frac{ \epsilon_i }{ T } \right) \right] \right\} - 2 G_S
\sum_{\alpha = 1}^3 \sigma_\alpha^2 - \frac{1}{4 G_D} \sum_{c =
1}^3 \left| \Delta_c \right|^2 + 4 K \sigma_u \sigma_d \sigma_s
\nonumber \\ &+& \frac{T}{\pi^2} \sum_{\beta = e}^\mu
\int_0^\infty \ud k \, k^2 \left\{ \ln \left[ 1 + \exp \left( -
\frac{ E_\beta - \mu_\beta }{ T } \right) \right] + \ln \left[ 1
+ \exp \left( - \frac{ E_\beta + \mu_\beta }{ T } \right)
\right] \right\} \nonumber \\
&+& \frac{1}{24 \pi^2} \sum_{\beta = e}^\mu \left(
\mu_{\nu_\beta}^4 + 2 \pi^2 \mu_{\nu_\beta}^2 T^2 + \frac{7}{15}
\pi^4 T^4 \right) \; , \\
p &=& \frac{1}{2 \pi^2} \sum_{i = 1}^{18} \int_0^\Lambda \ud k
\, k^2 \epsilon_i - 2 G_S \sum_{\alpha = 1}^3 \sigma_\alpha^2 -
\frac{1}{4 G_D} \sum_{c = 1}^3 \left| \Delta_c \right|^2 + 4 K
\sigma_u \sigma_d \sigma_s \nonumber \\
&+& \frac{1}{3 \pi^2} \sum_{\beta = e}^\mu \int_0^{k_{F_\beta}}
\ud k \, \frac{k^4}{E_\beta} + \frac{1}{24 \pi^2} \sum_{\beta =
e}^\mu \mu_{\nu_\beta}^4 \; ,
\eea
\esub
where $\epsilon_i$ are eighteen independent positive-energy
eigenvalues, see
Sec.~\ref{phase_diagram_self-consistent} for details. In order
to obtain the values for the quark-antiquark condensates
$\sigma_\alpha$ and the color-superconducting gap parameters
$\Delta_c$, I solve the following six stationary conditions:
\be
\label{gapeqns_neutrinos}
\frac{\partial p}{\partial \sigma_\alpha} = 0 \; , \qquad
\frac{\partial p}{\partial \Delta_c} = 0 \; .
\ee
In order to enforce the conditions of local charge neutrality in
quark matter, one also requires
Eqs.~(\ref{electric_neutrality}) and~(\ref{color_neutrality}) to
be satisfied. By solving these, I determine the values of the
three corresponding chemical potentials $\mu_Q$, $\mu_3$, and
$\mu_8$ for a given set of the other chemical potentials, $\mu$,
$\mu_{L_e}$, $\mu_{L_\mu}$, and for a given temperature $T$. In
general, therefore, the phase diagram of dense quark matter with
neutrino trapping should span a four-dimensional parameter
space.

Note that instead of using the chemical potentials, $\mu$,
$\mu_{L_e}$, and $\mu_{L_\mu}$ as free parameters in the study
of the phase diagram, one may also try to utilize the quark
number density and the two lepton fractions,
\be
n \equiv \frac{ \partial p }{\partial \mu} \; , \qquad
Y_{L_e} \equiv 3 \frac{n_{L_e}}{n} \; , \qquad
Y_{L_\mu} \equiv 3 \frac{n_{L_\mu}}{n} \; ,
\ee
where the two lepton densities $n_{L_\beta}$ are defined by
\be
n_{L_e} \equiv \frac{ \partial p }{\partial \mu_{L_e}} \; ,
\qquad
n_{L_\mu} \equiv \frac{ \partial p }{\partial \mu_{L_\mu}} \; .
\ee
In some cases, the choice of $n$, $Y_{L_e}$, and $Y_{L_\mu}$ as
free parameters is indeed very useful. For instance, this is
helpful in order to determine the initial state of matter inside
protoneutron stars at very early times, when the lepton
fractions are approximately the same as in the progenitor stars
(i.e., $Y_{L_e} \approx 0.4$ and $Y_{L_\mu} = 0$). The problem
is, however, that such an approach becomes ambiguous in the
vicinity of first-order phase transitions, where the baryon
number density as well as the lepton fractions are in general
discontinuous. For this reason, it is more appropriate to study
the phase structure of (dense) QCD at given fixed values of the
chemical potentials $\mu$, $\mu_{L_e}$, and $\mu_{L_\mu}$.
Unlike densities, the chemical potentials change continuously
when the system crosses a boundary of a first-order phase
transition. (It should be noted that the chemical potentials
$\mu_Q$, $\mu_3$, and $\mu_8$ may change discontinuously at a
boundary of a first-order phase transition because of the
long-range Coulomb interaction enforcing the constraints $n_Q =
0$, $n_3 = 0$, and $n_8 = 0$.)
\subsection{Simplified considerations}
\label{Simplified_considerations}
As mentioned before, neutrino trapping favors the 2SC phase and
strongly disfavors the CFL phase~\cite{SRP}. This is a
consequence of the modified $\beta$-equilibrium condition in the
system. In this section, I would like to emphasize that this is
a model-independent effect. In order to understand the physics
behind it, it is instructive to start my consideration from a
very simple toy model. Later, many of its qualitative features
will be also observed in my self-consistent numerical analysis
of the NJL model.

Let me first assume that strange quarks are very heavy and
consider a gas of non-interacting massless up and down
quarks in the normal quark phase at $T = 0$. As required by
$\beta$ equilibrium, electrons and electron neutrinos are also
present in the system. (Note that in this section I neglect
muons and muon neutrinos for simplicity.)

In the absence of Cooper pairing, the densities of quarks and 
leptons are given by
\be
n_u = \frac{\mu_u^3}{\pi^2} \; , \qquad
n_d = \frac{\mu_d^3}{\pi^2} \; , \qquad
n_e = \frac{\mu_e^3}{3 \pi^2} \; , \qquad
n_{\nu_e} = \frac{\mu_{\nu_e}^3}{6 \pi^2} \; ,
\ee
cf.\ Eq.~(\ref{n_massless_T0}). Expressing the chemical
potentials through $\mu$, $\mu_Q$ and $\mu_{L_e}$ by using
Eq.~(\ref{quark_flavor_beta_equilibrium}), and imposing
electric charge neutrality by setting Eq.~(\ref{n_Q}) equal to
zero, one arrives at the following relation:
\be
\label{toy1}
2 \, ( 1 + \frac{2}{3}y )^3 - \, (1 - \frac{1}{3} y)^3 - \, (x -
y)^3 = 0 \; ,
\ee
where I have introduced the chemical potential ratios $x =
\mu_{L_e} / \mu$ and $y = \mu_Q / \mu$. The above cubic equation
can be solved for $y$ (electric chemical potential) at any given
$x$ (lepton-number chemical potential) with Cardano's formulae
which are shown in Sec.~\ref{Cubic_equations} in the Appendix.
The result can be used to calculate the ratio of quark chemical
potentials, $\mu_d / \mu_u = (3 - y) / (3 + 2 y)$.

\begin{figure}[H]
\begin{center}
\includegraphics[width=0.9\textwidth]{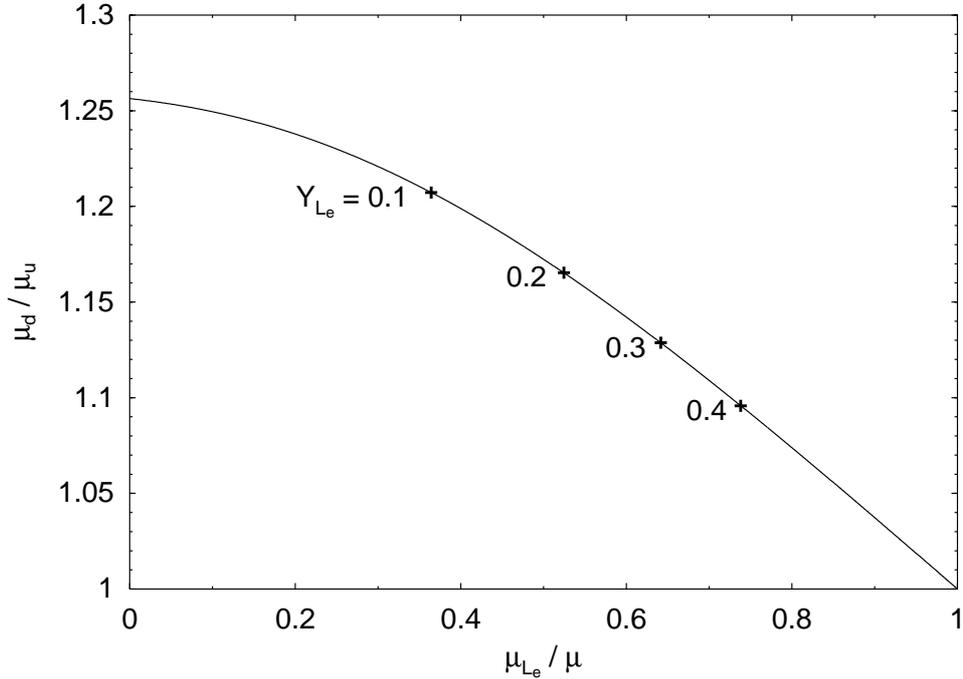}
\caption[Ratio of down and up quark chemical potentials as a
function of $\mu_{L_e} / \mu$.]{Ratio of down and up quark
chemical potentials as a function of $\mu_{L_e} / \mu$ in the
toy model. The crosses mark the solutions at several values of
the lepton fraction.}
\label{toy}
\end{center}
\end{figure}

The ratio $\mu_d / \mu_u$ as a function of $\mu_{L_e} / \mu$ is
shown in Fig.~\ref{toy}. At vanishing $\mu_{L_e}$, one finds
$y \approx -0.219$ and, thus, $\mu_d / \mu_u \approx 1.256$
(note that this value is very close to $2^{1/3} \approx 1.26$).
This result corresponds to the following ratios of the number
densities in the system: $n_u / n_d \approx 0.504$ and $n_e /
n_d \approx 0.003$, reflecting that the density of electrons is
tiny and the charge of the up quarks has to be balanced by
approximately twice as many down quarks,
cf.\ Sec.~\ref{toy_model_NQ}.
At $\mu_{L_e} = \mu$, on the other hand, the real solution to
Eq.~(\ref{toy1}) is $y = 0$, i.e., the up and down Fermi momenta
become equal. This can be seen most easily if one inverts the
problem and solves Eq.~(\ref{toy1}) for $x$ at given $y$. When 
$y = 0$ one finds $x = 1$, meaning that $\mu_d = \mu_u$ and, in
turn, suggesting that pairing between up and down quarks is
unobstructed at $\mu_{L_e} = \mu$. This is in contrast to the
case of vanishing $\mu_{L_e}$, when the two Fermi surfaces are
split by about 25\%, and pairing is difficult.

It is appropriate to mention that many features of the above
considerations would not change much even when Cooper pairing
is taken into account. The reason is that the corresponding
corrections to the quark densities are parametrically suppressed
by a factor of order $( \Delta / \mu )^2$, where $\Delta$ is the
gap parameter.

In order to estimate the magnitude of the effect in the case of
quark matter in protoneutron stars, I indicate several typical
values of the lepton fractions $Y_{L_e}$ in Fig.~\ref{toy}. As
mentioned earlier, $Y_{L_e}$ is expected to be of order $0.4$
right after the collapse of the iron core of the progenitor
star. According to Fig.~\ref{toy}, this corresponds to $\mu_d /
\mu_u \approx 1.1$, i.e., while the splitting between the up and
down Fermi surfaces does not disappear completely, it gets
reduced considerably compared to its value in the absence of
trapped neutrinos. This reduction substantially facilitates the
cross-flavor pairing of up and down quarks. The effect is
gradually washed out during about a dozen of seconds of the
deleptonization period when the value of $Y_{L_e}$ decreases to
zero.

The toy model is easily modified to the opposite extreme of
three massless quark flavors, where the number density of
strange quarks reads,
\be
n_s = \frac{\mu_s^3}{\pi^2} \; .
\ee
Basically, this corresponds to replacing Eq.~(\ref{toy1}) by
\be
2 \, ( 1 + \frac23 y )^3 - 2 \, ( 1 - \frac13 y )^3 - \, (x -
y)^3 = 0 \; .
\ee
In the absence of neutrino trapping, $x = 0$, the only real
solution to this equation is $y = 0$, indicating that the
chemical potentials (which also coincide with the Fermi momenta)
of up, down and strange quarks are equal,
cf.\ Sec.~\ref{toy_model_NQ}. This reflects the fact that the
system with equal densities of up, down and strange quarks is
neutral by itself, without electrons. With increasing $x \propto
\mu_{L_e}$, the solution requires a nonzero $y \propto \mu_Q$,
suggesting that up-down and up-strange pairing becomes more
difficult. To see this more clearly, I can go one step further
in the analysis of the toy model.

Let me assume that the quarks are paired in a regular, i.e.,
fully gapped, CFL phase at $T = 0$. Then, as shown in
Ref.~\cite{enforced_neutrality}, the quark part of the matter is
automatically electrically neutral. Hence, if I want to keep the
whole system electrically and color neutral, there must be no
electrons. Obviously, this is easily realized without trapped
neutrinos by setting $\mu_Q$ equal to zero. At non-vanishing
$\mu_{L_e}$ the situation is more complicated. The quark part is
still neutral by itself and therefore no electrons are admitted.
Hence, the electron chemical potential $\mu_e = \mu_{L_e} -
\mu_Q$ must vanish, and consequently $\mu_Q$ should be nonzero
and equal $\mu_{L_e}$. It is natural to ask what should be the
values of the color chemical potentials $\mu_3$ and $\mu_8$ in
the CFL phase when $\mu_{L_e} \neq 0$.

In order to analyze the stress on the CFL phase due to nonzero
$\mu_{L_e}$, I follow the same approach as in
Sec.~\ref{CSCinNeutronStars}. In this analytical consideration,
I also account for the effect of the strange quark mass simply
by shifting the strange quark chemical potential by $-M_s^2 / (2
\mu)$. In my notation, CFL-type pairing requires the Fermi
momenta defined by Eq.~(\ref{common_Fermi_momenta_CFL}). These
are used to calculate the pressure in the toy model,
\be
\label{p_toy}
p_\mathrm{toy} = \frac{1}{\pi^2} \sum_{a = r}^b \sum_{\alpha =
u}^s \int_0^{\left( k_F \right)_\alpha^a} \ud k \, k^2 \left(
\mu_\alpha^a - k \right) + \frac{\mu_e^4}{12 \pi^2} +
\frac{\mu_{L_e}^4}{24 \pi^2}  + \frac{3 \mu^2 \Delta^2}{\pi^2}
\; ,
\ee
where in contrast to the toy model for the CFL phase defined by
Eq.~(\ref{CFL_toy}), the contributions of electrons and electron
neutrinos are included and the bag pressure is neglected. With
Eqs.~(\ref{CFLrelation}) and~(\ref{p_toy}), one easily derives
the neutrality conditions~(\ref{electric_neutrality})
and~(\ref{color_neutrality}). Thus, it becomes obvious that
charge neutrality requires $\mu_Q = \mu_{L_e}$. The neutrality
condition, $n_3 = 0$, requires that $\mu_3=-\mu_Q$ which means
that $\mu_3 = - \mu_{L_e}$. Finally, one can check that the
third neutrality condition, $n_8 = 0$, requires,
\be
\mu_{8} = -\frac{\mu_{L_e}}{\sqrt{3}} - \frac{M_s^2}{\sqrt{3}
\mu} \; .
\ee
The results for the charge chemical potentials $\mu_Q$, $\mu_3$,
and $\mu_{8}$ imply the following magnitude of stress on pairing
in the CFL phase:
\bsub
\label{mismatch}
\bea
\delta\mu_{(rd,gu)} &=& \frac{\mu_u^g - \mu_d^r}{2} = \mu_{L_e}
\; , \\
\delta\mu_{(rs,bu)} &=& \frac{\mu_u^b - \mu_s^r}{2} = \mu_{L_e}
+ \frac{M_s^2}{2 \mu} \; , \\
\delta\mu_{(gs,bd)} &=& \frac{\mu_d^b - \mu_s^g}{2} =
\frac{M_s^2}{2 \mu} \; .
\eea
\esub
Note that there is no mismatch between the values of the
chemical potentials of the other three quarks, $\mu_u^r =
\mu_d^g = \mu_s^b = \mu - M_s^2 / ( 6 \mu )$.

From Eq.~(\ref{mismatch}) one can see that the largest mismatch
occurs in the $(rs, bu)$ pair (for positive $\mu_{L_e}$). The
CFL phase can withstand the stress only if the value of
$\delta\mu_{(rs,bu)}$ is less than $\Delta_2$. A larger mismatch
should drive a transition to a gapless phase exactly as in
Refs.~\cite{g2SC_1, g2SC_2, Alford_Kouvaris_Rajagopal1,
Alford_Kouvaris_Rajagopal2}. Thus, the critical value of the
lepton-number chemical potential is,
\be
\label{mu_L_cr}
\mu_{L_e}^\mathrm{(cr)} \approx \Delta_2 - \frac{M_s^2}{2 \mu}
\; .
\ee
When $\mu_{L_e} > \mu_{L_e}^\mathrm{(cr)}$, the CFL phase turns
into the gCFL$^\prime$ phase, which is a variant of the gCFL
phase~\cite{Alford_Kouvaris_Rajagopal1,
Alford_Kouvaris_Rajagopal2}. By definition, the gapless mode
with a linear dispersion relation in the gCFL$^\prime$ phase is
$rs$--$bu$ instead of $gs$--$bd$ as in the standard gCFL phase.
(Let me remind that the mode $a \alpha$--$b \beta$ is defined
by its dispersion relation which interpolates between the
dispersion relations of hole-type excitations of $a
\alpha$-quarks at small momenta, $k \ll \mu_\alpha^a$, and
particle-type excitations of $b \beta$-quarks at large momenta,
$k \gg \mu_\beta^b$.)

In order to see what this means for the physics of protoneutron
stars, I should again try to relate the value of $\mu_{L_e}$ to
the lepton fraction. There are no electrons in the (regular) CFL
phase at $T = 0$. Therefore the entire lepton number is carried
by neutrinos. For the baryon density I may neglect the pairing
effects to first approximation and employ the ideal-gas
relations. This yields,
\be
Y_{L_e} \approx \frac16 \left( \frac{\mu_{L_e}}{\mu} \right)^3
\; .
\ee
Inserting typical numbers, $\mu = 500$~MeV and $\mu_{L_e}
\lesssim \Delta \approx 50$~MeV--100~MeV, one finds $Y_{L_e}
\lesssim 10^{-4}$--$10^{-3}$. Thus, there is practically no
chance to find a sizeable amount of leptons in the CFL phase.
The constraint gets relaxed slightly at nonzero temperatures
and/or in the gCFL phase, but the lepton fraction remains rather
small even then (my numerical results indicate that, in general,
$Y_{L_e} \lesssim 0.05$ in the CFL phase).
\subsection{Results}
The simple toy-model considerations in
Sec.~\ref{Simplified_considerations} give a qualitative
understanding of the effect of neutrino trapping on the mismatch
of the quark Fermi momenta and, thus, on the pairing properties
of two- and three-flavor quark matter. Now I turn to a more
detailed numerical analysis of the phase diagram in the
framework of the NJL model defined in
Sec.~\ref{model+neutrinos}.

In the numerical calculations, I use the same set of model
parameters as in Sec.~\ref{phase_diagram_self-consistent} of my
thesis, see Eq.~(\ref{NJL_model_parameters}). The parameters are
chosen to reproduce several key observables of
vacuum QCD~\cite{RKH}. In this section, I choose a diquark
coupling constant $G_D = \frac34 G_S$.

In order to obtain the phase diagram, one has to determine the
ground state of matter for each given set of the parameters. As
discussed in Sec.~\ref{model+neutrinos}, in the case of locally
neutral matter with trapped neutrinos, there are four parameters
that should be specified: the temperature $T$, the quark
chemical potential $\mu$ as well as the two lepton family
chemical potentials $\mu_{L_e}$ and $\mu_{L_\mu}$. After these
are fixed, the values of the pressure in all competing neutral
phases of quark matter should be compared. This is determined by
using the same algorithm as in
Sec.~\ref{phase_diagram_self-consistent}. The complete set of
equations~(\ref{electric_neutrality}),~(\ref{color_neutrality}),
and~(\ref{gapeqns_neutrinos}) is solved for each of the eight
phases allowed by symmetries. Then, the corresponding values of
the pressure are determined from Eq.~(\ref{p_NJL_neutrinos}).
The phase with the largest pressure is the ground state.

In this section, I always assume that the muon lepton-number
chemical potential vanishes, i.e., $\mu_{L_\mu} = 0$. This is
expected to be a good approximation for matter inside
protoneutron stars. My analysis can thus be interpreted as an
extension of the $T$--$\mu$ phase diagram which is discussed in
Sec.~\ref{phase_diagram_self-consistent} into the $\mu_{L_e}$
direction. Consequently, the complete phase structure requires a
three-dimensional presentation.

\subsubsection{The three-dimensional phase diagram}
\begin{figure}[H]
\begin{center}
\includegraphics[width=0.97\textwidth]{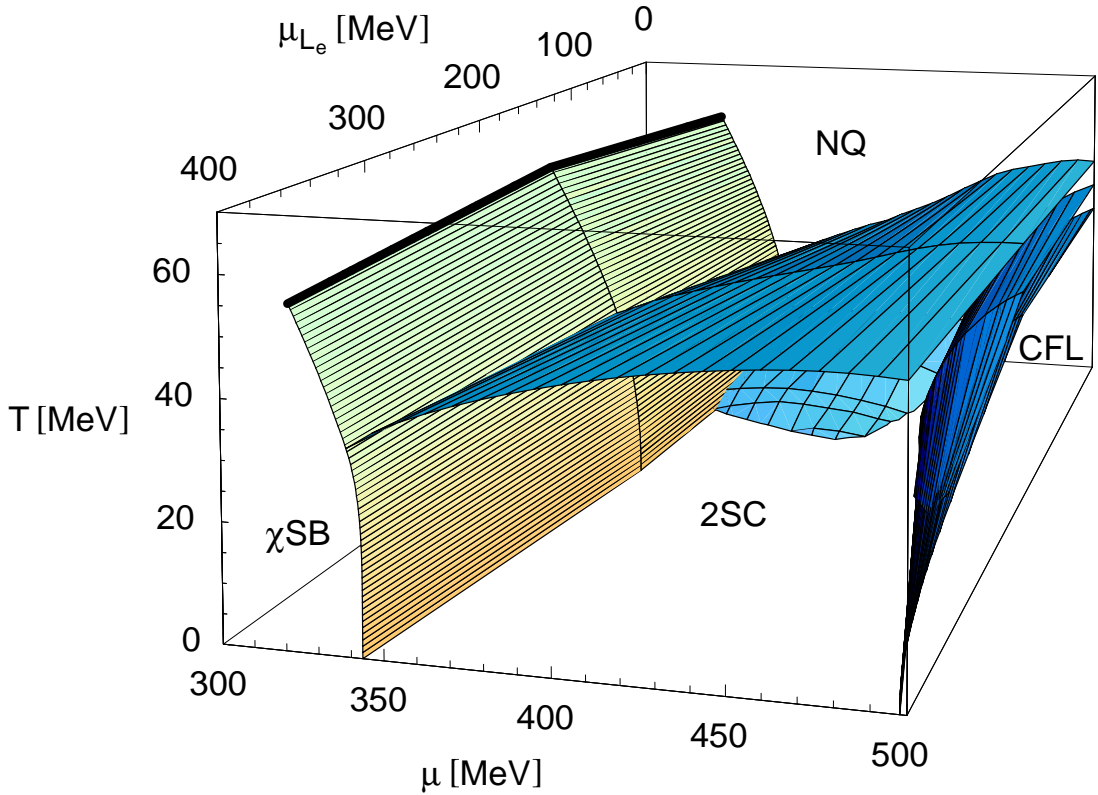}
\caption[The three-dimensional phase diagram of neutral
three-flavor quark matter.]{The phase diagram of neutral
three-flavor quark matter in the three-dimensional space spanned
by the quark chemical potential $\mu$, the lepton-number
chemical potential $\mu_{L_e}$, and the temperature $T$.}
\label{phase3d}
\end{center}
\end{figure}

The general features of the phase diagram in the
three-dimensional space, spanned by the quark chemical potential
$\mu$, the lepton-number chemical potential $\mu_{L_e}$, and the
temperature $T$ are depicted in Fig.~\ref{phase3d}. Because of
the rather complicated structure of the diagram, only the four
main phases ($\chi$SB, NQ, 2SC, and CFL) are shown explicitly.
Although it is not labeled, a thin slice of a fifth phase, the
uSC phase, squeezed in between the 2SC and CFL phases, can also
be seen. In addition, at small values of the lepton-number
chemical potential and small temperatures, another unmarked
region between $\mu \simeq 370$~MeV and $\mu \simeq 440$~MeV
exists. In Fig.~\ref{phase3d}, however, only a small part of the
surface at its boundary with the 2SC phase can be seen. This
region corresponds to the second piece of the normal quark
matter phase which is disconnected from the main region (the
appearance of a small disconnected region can also be deduced
from the $T$--$\mu$ phase diagram in Fig.~\ref{phasediagram}).
While lacking detailed information, the phase diagram in
Fig.~\ref{phase3d} gives a clear overall picture. Among other
things, one sees, for example, that the CFL phase becomes
strongly disfavored with increasing $\mu_{L_e}$ and gets
gradually replaced by the 2SC phase.
In order to discuss the structure of the phase diagram in more
details I proceed by showing several two-dimensional slices of
it. These are obtained by keeping one of the chemical
potentials, $\mu$ or $\mu_{L_e}$, fixed and varying the other
two parameters.

\subsubsection{The $\fett{T}$--$\fett{\mu}$ phase diagram}
\begin{figure}[H]
\begin{center}
\includegraphics[width=0.83\textwidth]{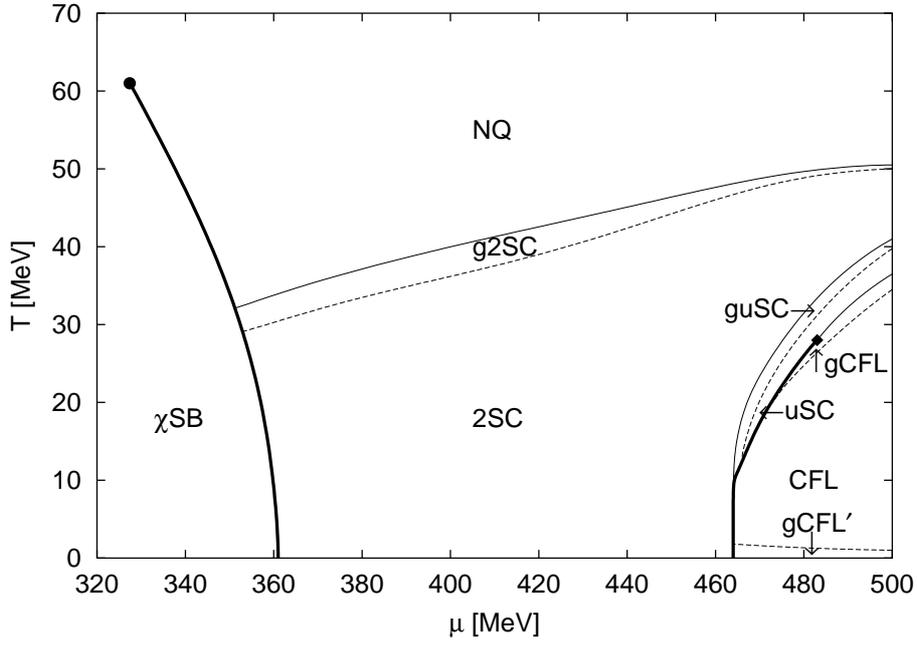}
\caption[The phase diagram of neutral quark matter at
$\mu_{L_e} = 200$~MeV.]{The phase diagram of neutral quark
matter at a fixed lepton-number chemical potential $\mu_{L_e} =
200$~MeV.}
\label{phase200}
\end{center}
\end{figure}

\begin{figure}[H]
\begin{center}
\includegraphics[width=0.83\textwidth]{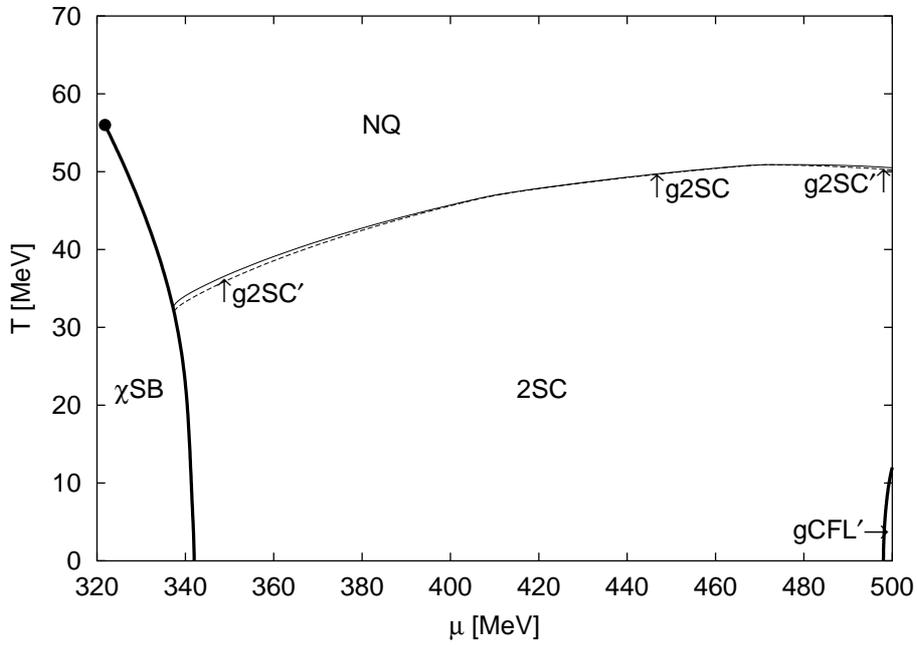}
\caption[The phase diagram of neutral quark matter at
$\mu_{L_e} = 400$~MeV.]{The phase diagram of neutral quark
matter at a fixed lepton-number chemical potential $\mu_{L_e} =
400$~MeV.}
\label{phase400}
\end{center}
\end{figure}

The phase diagrams at two fixed values of the lepton-number
chemical potential, $\mu_{L_e} = 200$~MeV and $\mu_{L_e} =
400$~MeV are presented in Figs.~\ref{phase200}
and~\ref{phase400}. The phase diagram at $\mu_{L_e} = 0$~MeV,
see Fig.~\ref{phasediagram}, has already been discussed in
Sec.~\ref{phase_diagram_self-consistent}. The general effects of
neutrino trapping can be understood by analyzing the
similarities and differences between these three phase diagrams.
In this section, I use the same convention for line styles as in
Sec.~\ref{phase_diagram_self-consistent}: thick and thin solid
lines denote first-order and second-order phase transitions,
respectively; dashed lines indicate the (dis-)appearance
of gapless modes in different phases.

\begin{figure}[H]
\begin{center}
\includegraphics[width=0.98\textwidth]{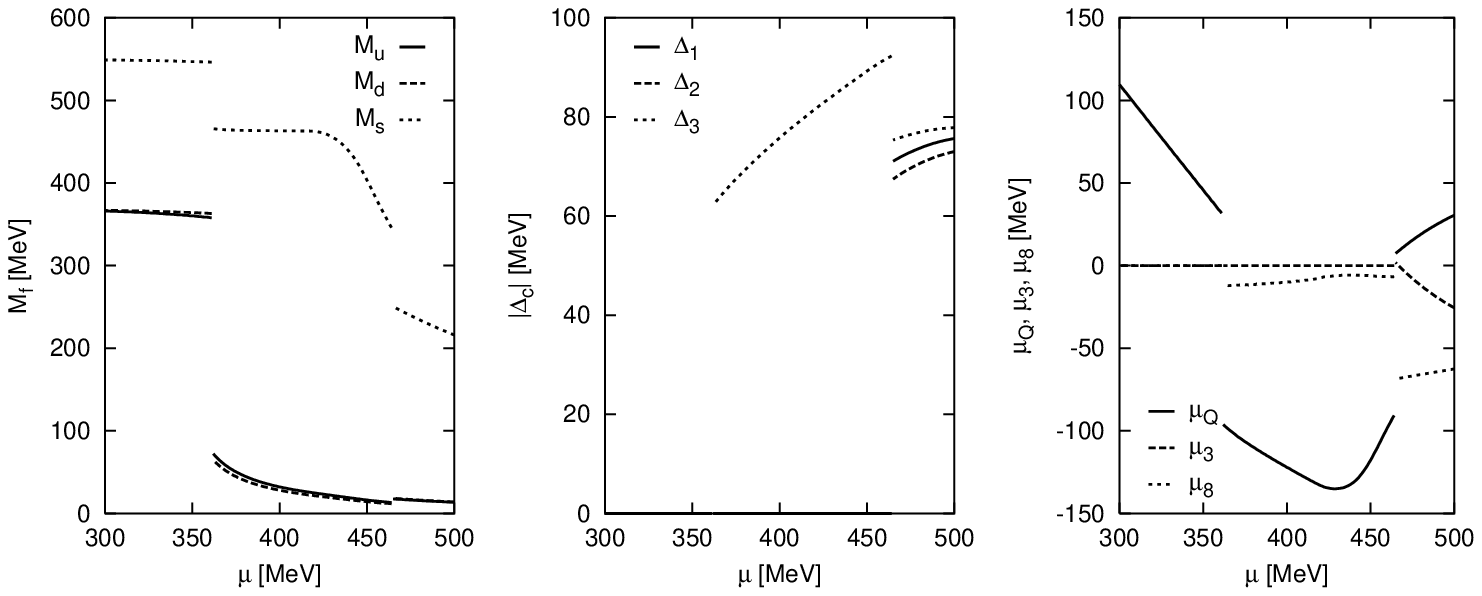}
\includegraphics[width=0.98\textwidth]{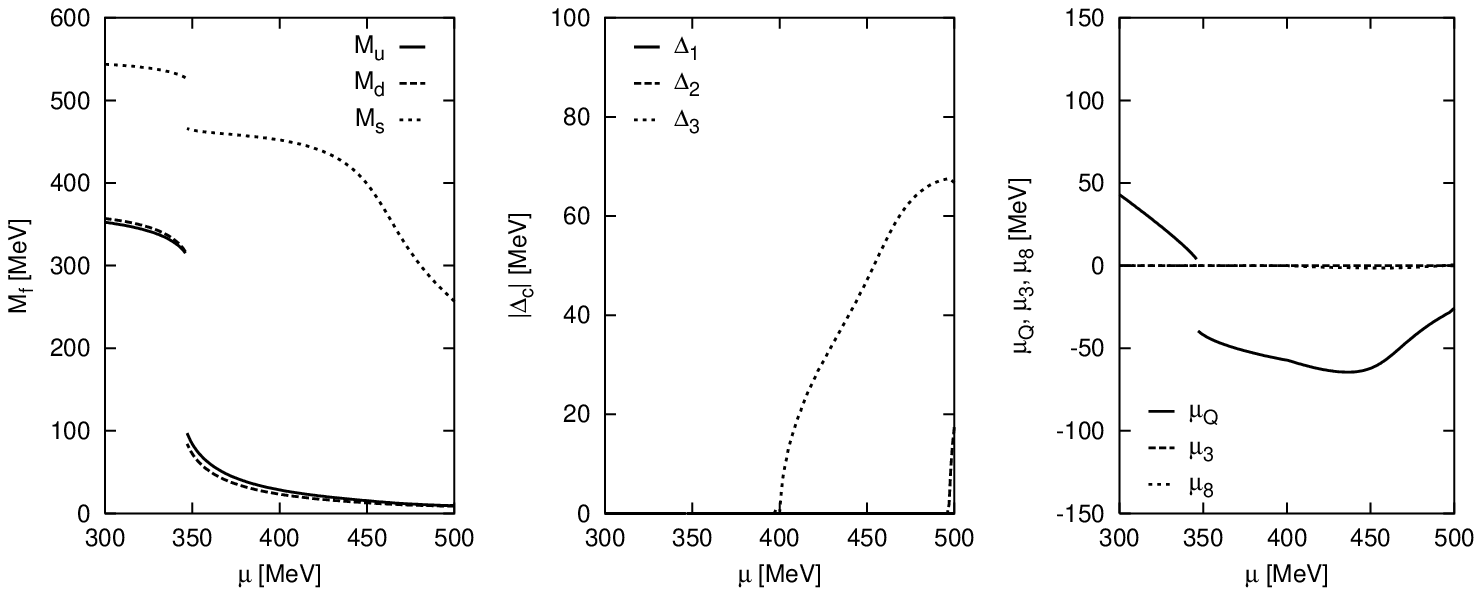}
\caption[Quark masses, gap parameters, and chemical potentials
at $\mu_{L_e} = 200$~MeV.]{Dependence of the quark masses, of
the gap parameters, and of the electric and color charge
chemical potentials on the quark chemical potential at a fixed
temperature, $T = 0$~MeV  (three upper panels) and $T = 40$~MeV
(three lower panels). The lepton-number chemical potential is
kept fixed at $\mu_{L_e} = 200$~MeV.}
\label{plot0-40_200}
\end{center}
\end{figure}

\begin{figure}[H]
\begin{center}
\includegraphics[width=0.98\textwidth]{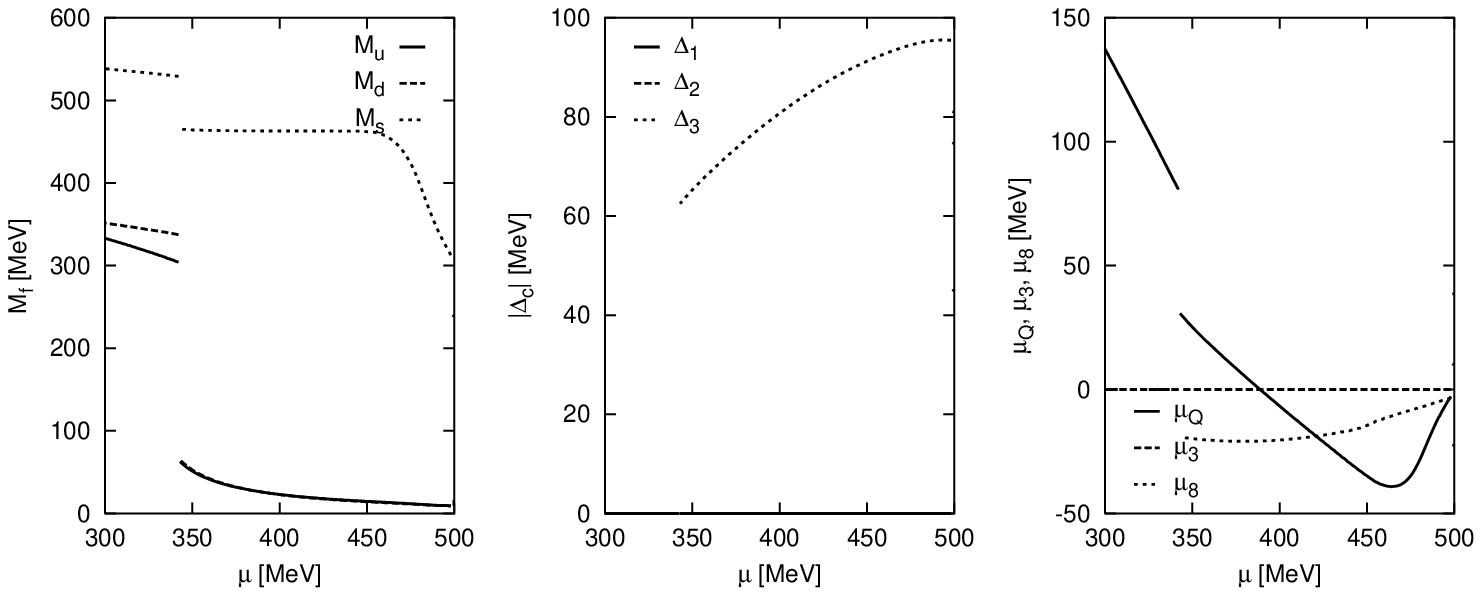}
\includegraphics[width=0.98\textwidth]{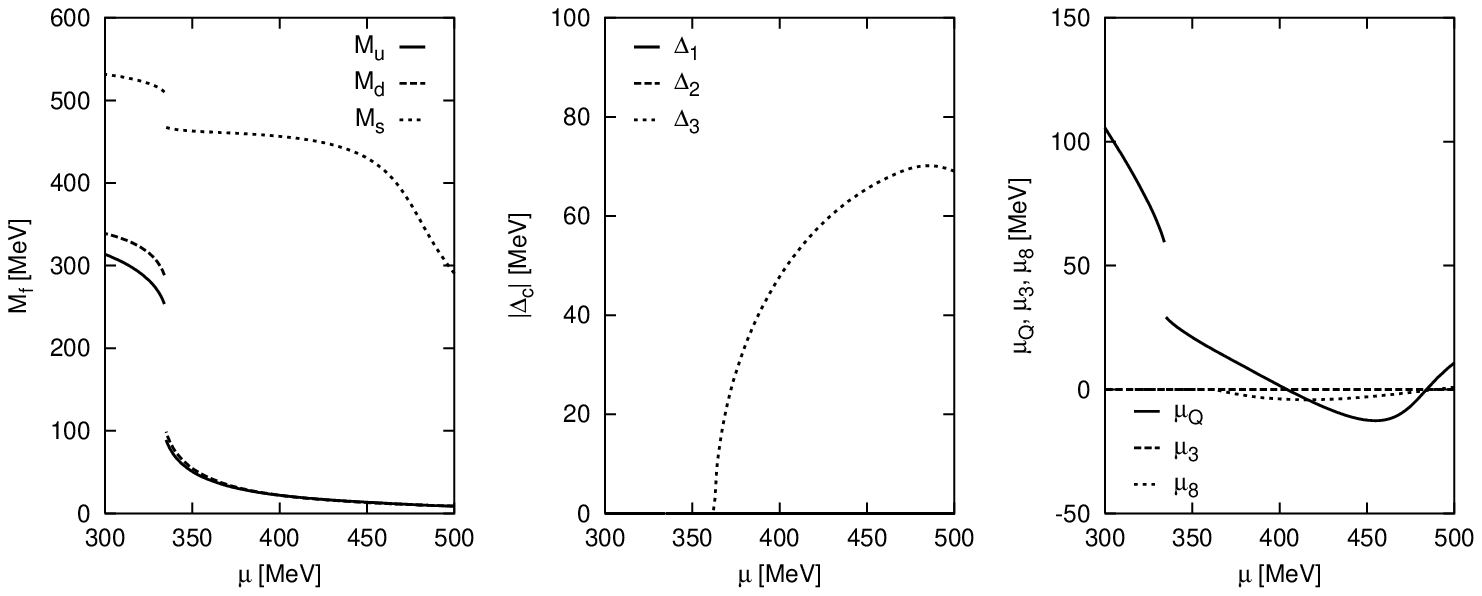}
\caption[Quark masses, gap parameters, and chemical potentials
at $\mu_{L_e} = 400$~MeV.]{Dependence of the quark masses, of
the gap parameters, and of the electric and color charge
chemical potentials on the quark chemical potential at a fixed
temperature,  $T = 0$~MeV (three upper panels) and $T = 40$~MeV
(three lower panels). The lepton-number chemical potential is
kept fixed at $\mu_{L_e} = 400$~MeV.}
\label{plot0-40_400}
\end{center}
\end{figure}

Here it is appropriate to note that, in the same model, a
schematic version of the $T$--$\mu$ phase diagram at $\mu_{L_e}
= 200$~MeV was first presented in Ref.~\cite{SRP}, see the right
panel of Fig.~4 there. If one ignores the complications due to
the presence of the uSC phase and various gapless phases, the
results of Ref.~\cite{SRP} are in qualitative agreement with my
findings.
In order to understand the basic characteristics of different
phases in the phase diagrams in Figs.~\ref{phase200}
and~\ref{phase400}, I also present the results for the dynamical
quark masses, the gap parameters, and the charge chemical
potentials. These are plotted as functions of the quark chemical
potential in Figs.~\ref{plot0-40_200} and~\ref{plot0-40_400},
for two different values of the temperature in the case of
$\mu_{L_e} = 200$~MeV and $\mu_{L_e} = 400$~MeV, respectively.
In each of the diagrams, there are roughly four distinct
regimes. At low temperature and low quark chemical potential,
there is a region in which the approximate chiral symmetry is
spontaneously broken by large $\bar\psi \psi$-condensates. The
corresponding phase is denoted by $\chi$SB. In this regime,
quarks have relatively large constituent masses which are close
to the vacuum values, see Figs.~\ref{plot0-40_200}
and~\ref{plot0-40_400}. Here, the density of all quark flavors
is very low and even vanishes at $T = 0$. There is no diquark
pairing in this phase. The $\chi$SB phase is rather insensitive
to the presence of a nonzero lepton-number chemical potential.
With increasing $\mu_{L_e}$ the phase boundary is only slightly
shifted to lower values of $\mu$. This is just another
manifestation of the strengthening of the 2SC phase due to
neutrino trapping.

With increasing temperature, the $\bar\psi \psi$-condensates
melt and the $\chi$SB phase turns into the NQ phase where the
quark masses are relatively small. Because of the explicit
breaking of the chiral symmetry by the current quark masses,
there is no need for a phase transition between the two regimes.
In fact, at low chemical potentials, I find only a smooth
crossover, whereas there is a first-order phase transition in a
limited region, 320~MeV~$\lesssim \mu \lesssim$~360~MeV. In
contrast to the $\chi$SB regime, the high-temperature NQ phase
extends to arbitrary large values of $\mu$. All these
qualitative features are little affected by the lepton-number
chemical potential.

The third regime is located at relatively low temperatures but
at quark chemical potentials higher than in the $\chi$SB phase.
In this region, the masses of the up and down quarks have
already dropped to values well below their respective chemical
potentials while the strange quark mass is still large, see left
columns of panels in Figs.~\ref{plot0-40_200}
and~\ref{plot0-40_400}. As a consequence, up and down quarks
are quite abundant but strange quarks are essentially absent.
It turns out that the detailed phase structure in this region is
very sensitive to the lepton-number chemical potential. At
$\mu_{L_e} = 0$, as discussed in
Sec.~\ref{phase_diagram_self-consistent}, the pairing between up
and down quarks is strongly hampered by the constraints of
neutrality and $\beta$ equilibrium. As a consequence, there is
no pairing at low temperatures, $T \lesssim 10$~MeV, and
g2SC-type pairing appears at moderate temperatures,
10~MeV~$\lesssim T < T_c$ with the value of $T_c$ in the range
of several dozen MeV, see Fig.~\ref{phasediagram}. The situation
changes dramatically with increasing the value of the
lepton-number chemical potential. Eventually, the
low-temperature region of the normal phase of quark matter is
replaced by the (g)2SC phase (e.g., at $\mu = 400$~MeV, this
happens at $\mu_{L_e} \simeq 110$~MeV). With $\mu_{L_e}$
increasing further, no qualitative changes happen in this part
of the phase diagram, except that the area of the (g)2SC phase
expands slightly.

Finally, the region in the phase diagram at low temperatures and
large quark chemical potentials corresponds to phases in which
the cross-flavor strange-nonstrange Cooper pairing becomes
possible. In general, as the strength of pairing increases with
the quark chemical potential, the system passes through regions
of the gapless uSC (guSC), uSC, and gCFL phases and finally
reaches the CFL phase. (Of course, the intermediate phases may
not always be realized.) The effect of neutrino trapping, which
grows with increasing lepton-number chemical potential, is to
push out the location of the strange-nonstrange pairing region
to larger values of $\mu$. Of course, this is in agreement with
the general arguments in Sec.~\ref{Simplified_considerations}.
\begin{table}[H]
\begin{center}
\begin{tabular}{|l||l|c|c|c|}
\hline
Phase                 & Gapless modes          & $\Delta_1$   &
$\Delta_2$   & $\Delta_3$ \\
\hline
\hline
g2SC                  & $ru$--$gd$, $gu$--$rd$ & $-$          &
$-$          & $\checkmark$ \\
g2SC$^\prime$         & $rd$--$gu$, $gd$--$ru$ & $-$          &
$-$          & $\checkmark$ \\
guSC                  & $rs$--$bu$             & $-$          &
$\checkmark$ & $\checkmark$ \\
gCFL                  & $gs$--$bd$             & $\checkmark$ &
$\checkmark$ & $\checkmark$ \\
gCFL$^\prime$         & $rs$--$bu$             & $\checkmark$ &
$\checkmark$ & $\checkmark$ \\
gCFL$^{\prime\prime}$ & $gs$--$bd$, $rs$--$bu$ & $\checkmark$ &
$\checkmark$ & $\checkmark$ \\
\hline
\end{tabular}
\end{center}
\caption[The classification of gapless phases in
color-superconducting quark matter.]{The classification of
gapless phases in color-superconducting quark matter. The
unmarked gap parameters ($-$) are zero while the checkmarked gap
parameters ($\checkmark$) are nonzero in the respective gapless
color-superconducting quark phases. The gapless modes $a
\alpha$--$b \beta$ are defined by their dispersion relations
which interpolate between the dispersion relations of hole-type
excitations of $a \alpha$-quarks at small momenta, $k \ll
\mu_\alpha^a$, and particle-type excitations of $b \beta$-quarks
at large momenta, $k \gg \mu_\beta^b$.}
\label{gapless_csc_quark_phases}
\end{table}
\begin{figure}[H]
\begin{center}
\includegraphics[angle=270, width=0.83\textwidth]{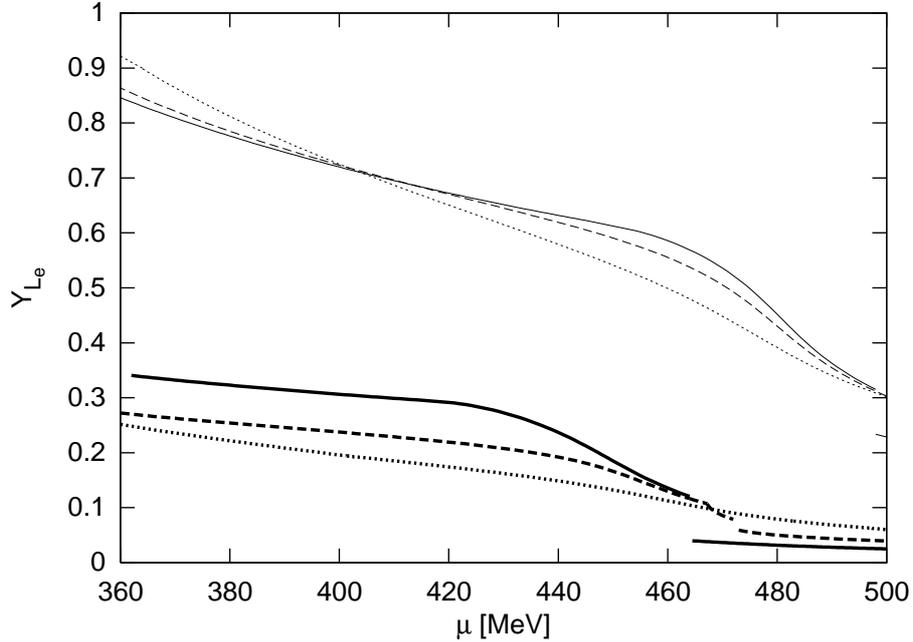}
\caption[The dependence of the electron family lepton fraction
$Y_{L_e}$ on $\mu$.]{Dependence of the electron family lepton
fraction $Y_{L_e}$ for $\mu_{L_e} = 200$~MeV (thick lines) and
$\mu_{L_e} = 400$~MeV (thin lines) on the quark chemical
potential at a fixed temperature, $T = 0$~MeV (solid lines), $T
= 20$~MeV (dashed lines), and $T = 40$~MeV (dotted lines).}
\label{plot_Y_0-20-40}
\end{center}
\end{figure}
It is interesting to note that the growth of the strangeness
content with increasing quark chemical potential could
indirectly be deduced from the behavior of the electric charge
chemical potential $\mu_Q$ at $T = 0$, see the solid lines in
the right panels of Figs.~\ref{plot0-40_200}
and~\ref{plot0-40_400}. The value of $\mu_Q$ reaches its minimum
somewhere in a range of values of the quark chemical potential
around $\mu \simeq 440$~MeV. This corresponds to the point where
the strange quark chemical potential, \mbox{$\mu_s \simeq \mu -
\mu_Q / 3$}, reaches the value of the strange quark mass (see
left panels). Hence, there are essentially no strange quarks at
lower values of $\mu$, and a rapidly increasing amount of
strange quarks at higher values of $\mu$. Since the latter
contribute to the electric neutralization, this is a natural
explanation for the drop of $|\mu_Q|$ above this point.
\begin{figure}[H]
\begin{center}
\includegraphics[width=0.83\textwidth]{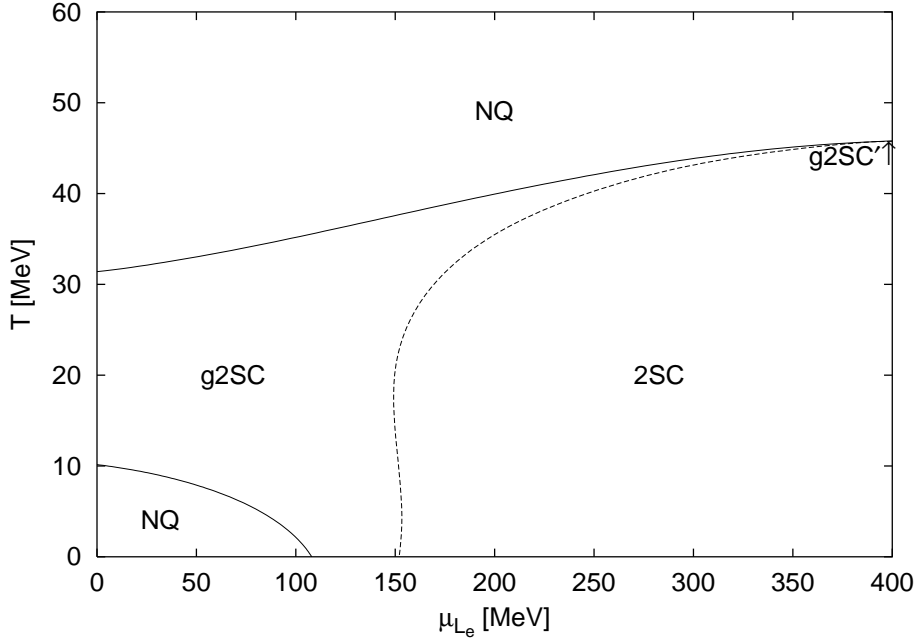}
\caption[The phase diagram of the outer stellar core.]{The
phase diagram of neutral quark matter in the plane of
temperature and lepton-number chemical potential at a fixed
value of the quark chemical potential, $\mu = 400$~MeV. This
phase diagram corresponds to the \textit{outer stellar core}.}
\label{phase_mu400}
\end{center}
\end{figure}
As I mentioned earlier, the presence of the lepton-number
chemical potential $\mu_{L_e}$ leads to a change of the quark
Fermi momenta. This change, in turn, affects Cooper pairing of
quarks, facilitating the appearance of some phases and
suppressing others. As it turns out, there is also another
qualitative effect due to a nonzero value of $\mu_{L_e}$. In
particular, I find several new variants of gapless phases which
do not exist at vanishing $\mu_{L_e}$. In
Figs.~\ref{phase200},~\ref{phase400},~\ref{phase_mu400},
and~\ref{phase_mu500}, these are denoted by the same names,
g2SC or gCFL, but with one or two primes added.

I define the g2SC$^\prime$ as the gapless two-flavor
color-superconducting phase in which the gapless excitations
correspond to $rd$--$gu$ and $gd$--$ru$ modes instead of the
usual $ru$--$gd$ and $gu$--$rd$ ones, i.e., $u$ and $d$
flavors are exchanged as compared to the usual g2SC phase.
The g2SC$^\prime$ phase becomes possible only when the value
of $( \mu_u^r - \mu_d^g ) / 2 \equiv ( \mu_Q + \mu_3 ) / 2$ is
positive and larger than $\Delta_3$. The other phases are
defined in a similar manner. In particular, the gCFL$^\prime$
phase, which was already introduced in
Sec.~\ref{Simplified_considerations}, is indicated by the
gapless $rs$--$bu$ mode, while the gCFL$^{\prime\prime}$ phase
has both, $gs$--$bd$ (as in the gCFL phase) and $rs$--$bu$
gapless modes. The definitions of all gapless phases are
summarized in Table~\ref{gapless_csc_quark_phases}.

\subsubsection{The lepton fraction $\fett{Y_{L_e}}$}
The numerical results for the lepton fraction $Y_{L_e}$ are shown
in Fig.~\ref{plot_Y_0-20-40}. The thick and thin lines
correspond to two different fixed values of the lepton-number
chemical potential, $\mu_{L_e} = 200$~MeV and $\mu_{L_e} =
400$~MeV, respectively. For a fixed value of $\mu_{L_e}$, I find
that the lepton fraction changes only slightly with temperature.
This is concluded from the comparison of the results at $T =
0$~MeV (solid lines), $T = 20$~MeV (dashed lines), and $T =
40$~MeV (dotted lines) in Fig.~\ref{plot_Y_0-20-40}.
\begin{figure}[H]
\begin{center}
\includegraphics[width=0.83\textwidth]{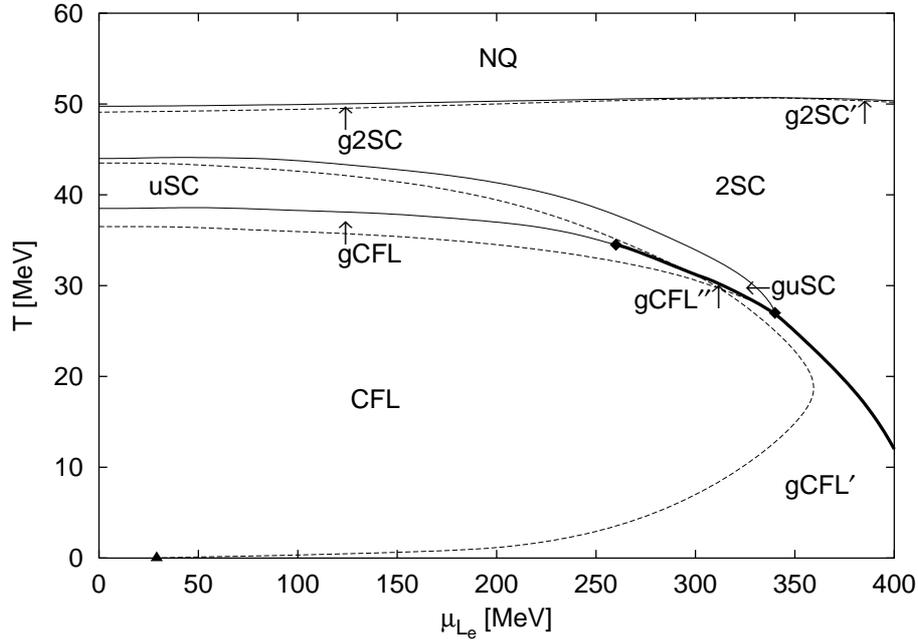}
\caption[The phase diagram of the inner stellar core.]{The
phase diagram of neutral quark matter in the plane of
temperature and lepton-number chemical potential at a fixed
value of the quark chemical potential, $\mu = 500$~MeV. This
phase diagram corresponds to the \textit{inner stellar core}.
The triangle denotes the transition point from the CFL phase to
the gCFL$^\prime$ phase at $T = 0$.}
\label{phase_mu500}
\end{center}
\end{figure}
As is easy to check, at $T = 40$~MeV, i.e., when Cooper pairing
is not so strong, the $\mu$ dependence of $Y_{L_e}$ does not
differ very much from the prediction in the simple two-flavor
model in Sec.~\ref{Simplified_considerations}. By saying this,
of course, one should not undermine the fact that the lepton
fraction in Fig.~\ref{plot_Y_0-20-40} has a visible structure in
the dependence on $\mu$ at $T = 0$~MeV and $T = 20$~MeV. This
indicates that quark Cooper pairing plays a nontrivial role in
determining the value of $Y_{L_e}$.

My numerical study shows that it is hard to achieve values of
the lepton fraction more than about $0.05$ in the CFL phase.
Gapless versions of the CFL phases, on the other hand, could
accommodate the lepton fraction up to about $0.2$ or so,
provided the quark and lepton-number chemical potentials are
sufficiently large.

From Fig.~\ref{plot_Y_0-20-40}, one can also see that the value
of the lepton fraction $Y_{L_e} \approx 0.4$, i.e., the value
expected at the center of the protoneutron star right after its
creation, requires a lepton-number chemical potential
$\mu_{L_e}$ in the range somewhere between $200$~MeV and
$400$~MeV, or slightly higher. The larger the quark chemical
potential~$\mu$, the larger a lepton-number chemical
potential~$\mu_{L_e}$ is needed. Then, in a realistic
construction of a protoneutron star, this is likely to result in
a noticeable gradient of the lepton-number chemical potential at
the initial time. This gradient may play an important role in
the subsequent deleptonization due to neutrino diffusion through
dense matter.

\subsubsection{The $\fett{T}$--$\fett{\mu_{L_e}}$ phase diagram}
Now let me explore the phase diagrams in the plane of
temperature and lepton-number chemical potential, keeping the
quark chemical potential fixed. Two such slices of the phase
diagram are presented in Figs.~\ref{phase_mu400}
and~\ref{phase_mu500}. The first one corresponds to a not 
very large value of the quark chemical potential, $\mu =
400$~MeV. This could be loosely termed as the \textit{outer
stellar core} phase diagram. The second one corresponds to $\mu
= 500$~MeV, and one could associate it with the \textit{inner
stellar core} case. Note, however, that the terms \textit{outer
stellar core} and \textit{inner stellar core} should not be
interpreted literally here. The central densities of
(proto-)neutron stars are subject to large theoretical
uncertainties and, thus, are not known very well. In the model
at hand, the \textit{outer stellar core} case corresponds to a
range of densities around $4 n_0$, while the \textit{inner
stellar core} case corresponds to a range of densities around
$10 n_0$. These values are of the same order of magnitude that
one typically obtains in models (see, e.g., Ref.~\cite{PNS1}).
\begin{figure}[H]
\begin{center}
\includegraphics[width=0.98\textwidth]{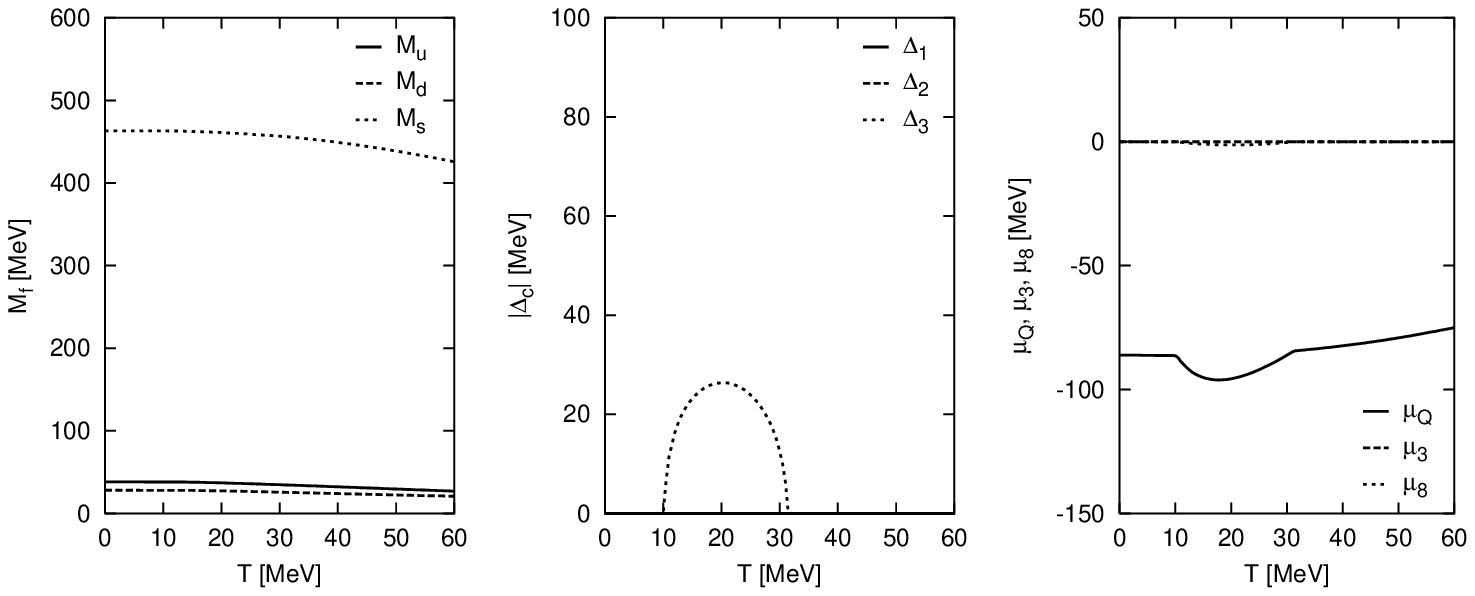}
\includegraphics[width=0.98\textwidth]{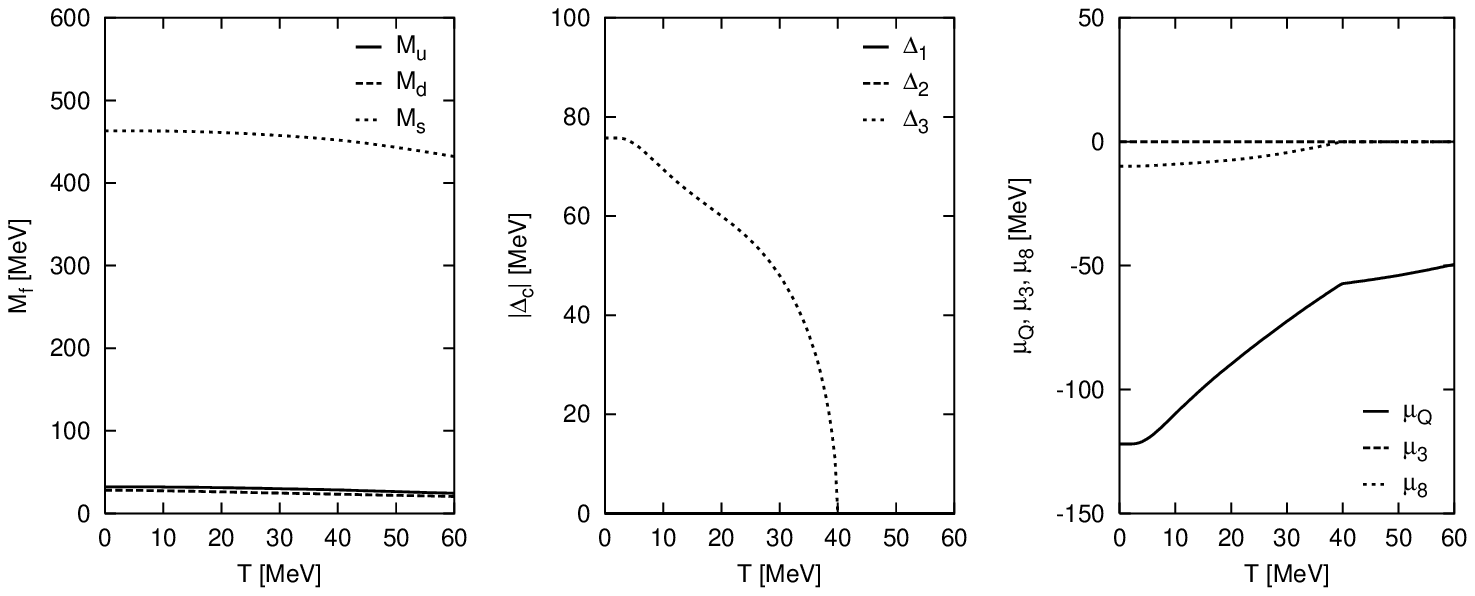}
\includegraphics[width=0.98\textwidth]{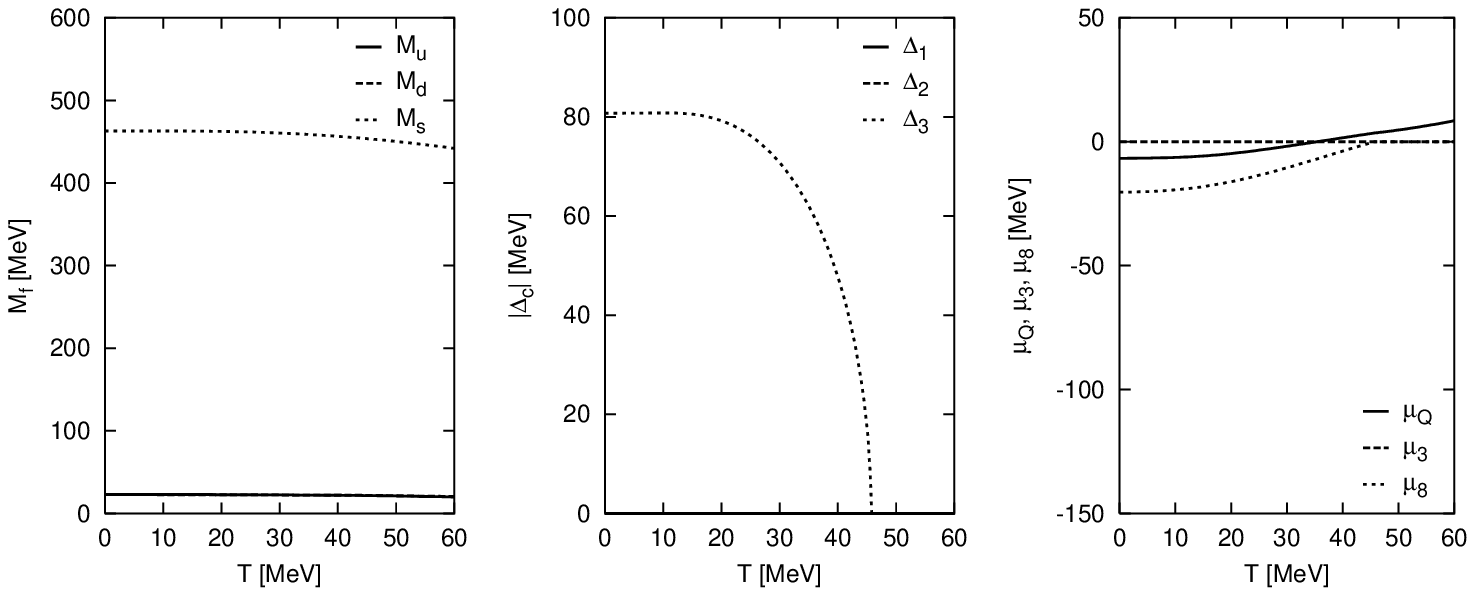}
\caption[Quark masses, gap parameters, and chemical potentials
in the outer stellar core.]{Dependence of the quark masses, of
the gap parameters, and of the electric and color charge
chemical potentials on the temperature at a fixed value of the
lepton-number chemical potential, $\mu_{L_e} = 0$~MeV (three
upper panels), $\mu_{L_e} = 200$~MeV (three middle panels), and
$\mu_{L_e} = 400$~MeV (three lower panels). The quark chemical
potential is $\mu = 400$~MeV (\textit{outer stellar core}
case).}
\label{plot_nu0-200-400_mu400}
\end{center}
\end{figure}
\begin{figure}[H]
\begin{center}
\includegraphics[width=0.98\textwidth]{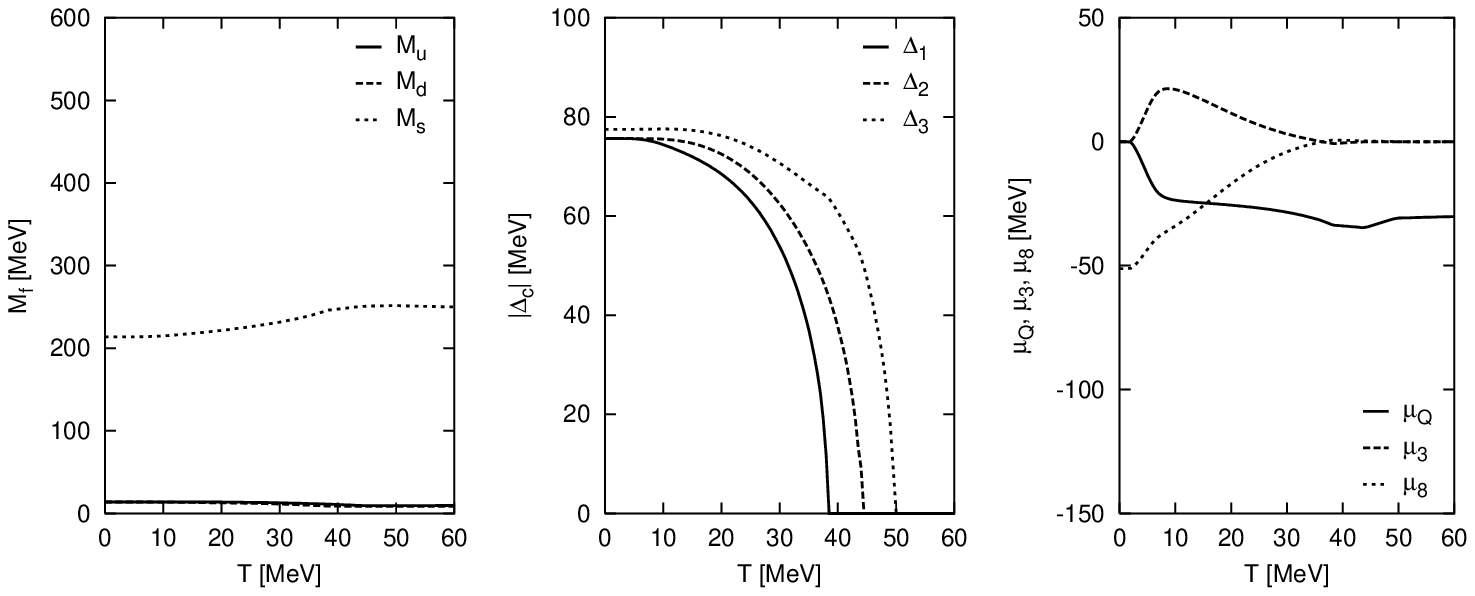}
\includegraphics[width=0.98\textwidth]{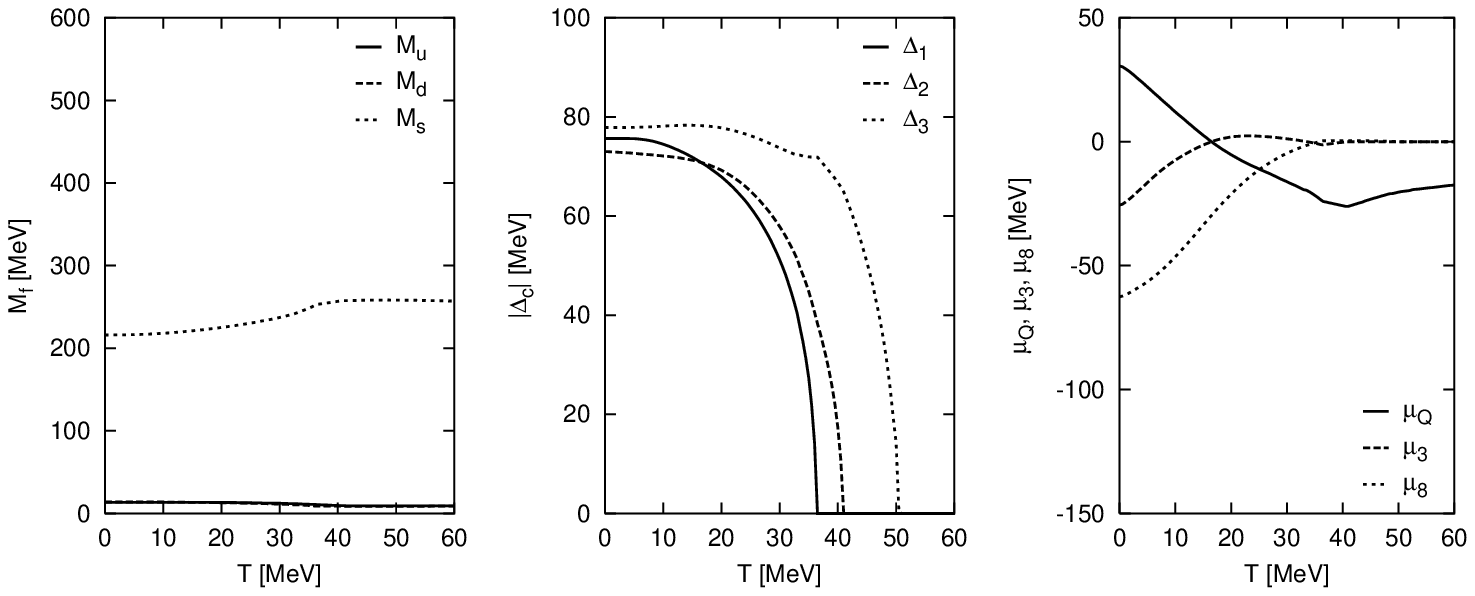}
\includegraphics[width=0.98\textwidth]{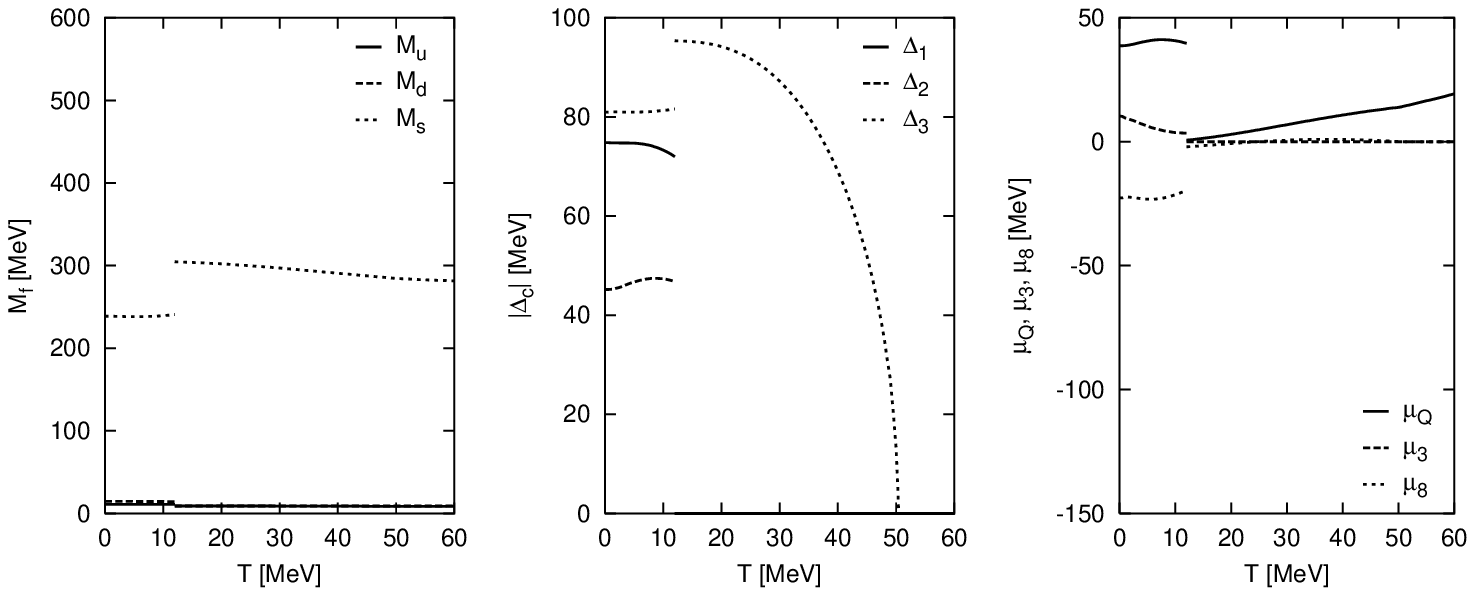}
\caption[Quark masses, gap parameters, and chemical potentials
in the inner stellar core.]{Dependence of the quark masses, of
the gap parameters, and of the electric and color charge
chemical potentials on the temperature at a fixed value of the
lepton-number chemical potential, $\mu_{L_e} = 0$~MeV (three
upper panels), $\mu_{L_e} = 200$~MeV (three middle panels), and
$\mu_{L_e} = 400$~MeV (three lower panels). The quark chemical
potential is $\mu = 500$~MeV (\textit{inner stellar core}
case).}
\label{plot_nu0-200-400_mu500}
\end{center}
\end{figure}
In addition to the phase diagrams, I also show the results for
the dynamical quark masses, the gap parameters, and the charge
chemical potentials. These are plotted as functions of
temperature in Fig.~\ref{plot_nu0-200-400_mu400} ($\mu =
400$~MeV, \textit{outer stellar core} case) and in
Fig.~\ref{plot_nu0-200-400_mu500} ($\mu = 500$~MeV,
\textit{inner stellar core} case), for three different values of
the lepton-number chemical potential, $\mu_{L_e} = 0$~MeV (upper
panels), $\mu_{L_e} = 200$~MeV (middle panels), and $\mu_{L_e} =
400$~MeV (lower panels).

At first sight, the two phase diagrams in
Figs.~\ref{phase_mu400} and~\ref{phase_mu500} look so different
that no obvious connection between them could be made. It is
natural to ask, therefore, how such a dramatic change could
happen with increasing the value of the quark chemical potential
from $\mu = 400$~MeV to $\mu = 500$~MeV. In order to understand
this, it is useful to place the corresponding slices of the
phase diagram in the three-dimensional phase diagram in
Fig.~\ref{phase3d}.

The $\mu=500$~MeV phase diagram corresponds to the right-hand
side surface of the bounding box in Fig.~\ref{phase3d}. This
contains almost all complicated phases with strange-nonstrange
cross-flavor pairing. The $\mu = 400$~MeV phase diagram, on the
other hand, is obtained by cutting the three-dimensional phase
diagram with a plane parallel to the bounding surface, but going
through the middle of the phase diagram. This part of the phase
diagram is dominated by the 2SC and NQ phases. Keeping in mind
the general structure of the three-dimensional phase diagram, it
is also not difficult to understand how the two phase diagrams
in Figs.~\ref{phase_mu400} and~\ref{phase_mu500} transform into
each other.

Several comments are in order regarding the zero-temperature
phase transition from the CFL to gCFL$^\prime$ phase, shown by
a small black triangle in the phase diagram at $\mu = 500$~MeV,  
see Fig.~\ref{phase_mu500}. The appearance of this transition is
in agreement with the analytical result in
Sec.~\ref{Simplified_considerations}. Moreover, the critical
value of the lepton-number chemical potential also turns out to
be very close to the estimate in Eq.~(\ref{mu_L_cr}). Indeed, by
taking into account that $M_s \approx 214$~MeV and $\Delta_2
\approx 76$~MeV, I obtain $\mu_{L_e}^\mathrm{(cr)} = \Delta_2 -
M_s^2 / ( 2 \mu ) \approx 30$~MeV which agrees well with the
numerical value.

In order to estimate how the critical value of $\mu_{L_e}$
changes with decreasing the quark chemical potential below $\mu
= 500$~MeV, one can use the zero-temperature numerical results
for $M_s$ and $\Delta_2$ from
Sec.~\ref{phase_diagram_self-consistent}. Then, one arrives at
the following power-law fit for the $\mu$-dependence of the
critical value:
\be
\label{mu_L_cr-fit}
\mu_{L_e}^\mathrm{(cr)} \approx 0.00575~\mathrm{MeV}^{-1} \left(
\mu - 457.4~\mathrm{MeV} \right) \left( 622.1~\mathrm{MeV} - \mu
\right) \; ,
\ee
for 457.4~MeV~$\leq \mu \leq$~500~MeV. Note that the CFL phase
does not appear at $T = 0$ when $\mu < 457.4$~MeV,
cf.\ Fig.~\ref{phasediagram}. With increasing the values of the
quark chemical potential above $\mu = 500$~MeV, one expects that
the critical value of $\mu_{L_e}$ should continue to increase
for a while, and then decrease when the effects of the cutoff
start to suppress the size of the gap $\Delta_2$. However, the
validity of the fit in Eq.~(\ref{mu_L_cr-fit}) is questionable
there because no numerical data for $\mu > 500$~MeV was used in
its derivation.

A schematic version of the phase diagram in the $T$--$\mu_{L_e}$
plane was presented earlier in Ref.~\cite{SRP}, see the left
panel in Fig.~4 there. In Ref.~\cite{SRP}, the value of the
quark chemical potential was $\mu = 460$~MeV, and therefore a
direct comparison with my results is not easy. One can see,
however, that the diagram of Ref.~\cite{SRP} fits naturally into
the three-dimensional diagram in Fig.~\ref{phase3d}. Also, the
diagram of Ref.~\cite{SRP} is topologically close to my version
of the phase diagram at $\mu = 500$~MeV which is shown in
Fig.~\ref{phase_mu500}. The quantitative difference is not
surprising: the region of the (g)CFL phase is considerably
larger at $\mu = 500$~MeV than at $\mu = 460$~MeV.
\selectlanguage{american}
\chapter{Conclusions}
\label{Conclusions}
The phase diagram of neutral quark matter was poorly understood
as I began with the research on this topic in 2003, see
Fig.~\ref{phase_diagram_schematic}. The task of my thesis is to
illuminate the phase structure of neutral quark matter. From the
phase diagram of neutral quark matter, one can predict in which
state is neutral quark matter in the cores of neutron stars
which are most probably the candidates for neutral
color-superconducting quark matter in nature.
\section{Summary and discussion}
In Chapter~\ref{Introduction}, I gave an introduction to quark
matter, color superconductivity, and stellar evolution. I
presented the most important color-superconducting quark phases
and explained how stars evolve and become neutron stars. At
extremely large densities which are probably present in the
cores of neutron stars, there is a dominant attractive
interaction between the quarks that causes the formation of
quark Cooper pairs~\cite{Barrois, Frautschi, Bailin_Love}. This
causes color superconductivity. Since quarks are spin-$\frac12$
fermions which appear in various colors and flavors, there are
many possibilities of forming quark Cooper pairs. This is why
there exist many different color-superconducting quark phases.

In order to predict which of these phases are energetically
preferred in nature, it is necessary to compute the phase
diagram of neutral quark matter and to determine which phases
are dominant at a given temperature and quark chemical
potential. The phase diagram of neutral quark matter consisting
of up, down, and strange quarks is presented in
Chapter~\ref{phase_diagram}. In this context, it is important to
consider \textit{neutral} quark matter because otherwise there
would occur repulsive Coulomb forces which lead to the explosion
of the neutron stars. Quark matter has to be color neutral
because color-charged objects have never been seen in nature.
Besides, color-charged neutron stars would be unstable. Matter
inside neutron stars is in $\beta$ equilibrium which means that
neutron-star matter is in equilibrium with respect to weak
interactions.

In Sec.~\ref{The_phase_diagram_of_massless_quarks}, I studied
massless neutral three-flavor quark matter at large baryon
densities within an NJL model. The effects of the strange quark
mass were incorporated by a shift of the chemical potential of
strange quarks, see Eq.~(\ref{shift2}). This shift reflects the
reduction of the Fermi momenta of strange quarks due to their
mass. Such an approach is certainly reliable at small values of
the strange quark mass. In Ref.~\cite{Shovkovy_Ruester_Rischke},
where the strange quark mass was properly taken into account,
I confirmed that it is also qualitatively correct at large
values of the strange quark mass. I obtained a very rich phase
structure by varying the strange quark mass, the quark chemical
potential, and the temperature.

At $T = 0$, there are two main possibilities for the strange
quark matter ground state: the CFL and gCFL phases. These
findings confirm the results of
Refs.~\cite{Alford_Kouvaris_Rajagopal1,
Alford_Kouvaris_Rajagopal2} concerning the existence of the gCFL
phase, the estimate of the critical value of the strange quark
mass $m_s$, and the dependence of the chemical potentials on
$m_s$. I also confirmed that it is the color-neutrality
condition, controlled by the color chemical potential $\mu_8$
which drives the transition from the CFL to the gCFL
phase~\cite{Alford_Kouvaris_Rajagopal1,
Alford_Kouvaris_Rajagopal2}. This is in contrast to gapless
two-flavor color superconductivity which results from electrical
neutrality~\cite{g2SC_1, g2SC_2}.

Because I use a nine-parameter ansatz for the gap matrix, see
Eq.~(\ref{most_general}), the results of
Sec.~\ref{The_phase_diagram_of_massless_quarks} are more general
than those of Refs.~\cite{Alford_Kouvaris_Rajagopal1,
Alford_Kouvaris_Rajagopal2}. For example, I was able to
explicitly study the effects of the symmetric pairing channel,
described by the sextet gap parameters, that were neglected in
Refs.~\cite{Alford_Kouvaris_Rajagopal1,
Alford_Kouvaris_Rajagopal2}. As one might have expected, these
latter modify the quasiparticle dispersion relations only
slightly. This check was important, however, to see that the
zero-temperature phase transition from the CFL phase to the gCFL
phase, which is not associated with any symmetry, is robust
against such a deformation of the quark system.

In Sec.~\ref{The_phase_diagram_of_massless_quarks}, I also
studied the temperature dependence of the gap parameters and the
quasiparticle spectra. In particular, this study revealed that
there exist several different phases of neutral three-flavor
quark matter that have been predicted in the framework of the
Ginzburg-Landau-type effective theory in Ref.~\cite{Iida}. My
results extend the near-critical behavior discussed in
Ref.~\cite{Iida} to all temperatures. Also, I show how this
behavior evolves with changing the value of the strange quark
mass. The only real qualitative difference between my results
and the results of Refs.~\cite{Iida, Fukushima} is that, instead
of the dSC phase, I find the uSC phase in the phase diagram. In
Ref.~\cite{Shovkovy_Ruester_Rischke}, I confirmed the results of
Sec.~\ref{The_phase_diagram_of_massless_quarks} regarding the
existence of several different phases of neutral three-flavour
quark matter at nonzero temperature. I also confirmed the order
in which they appear. In particular, I observed the appearance
of the uSC phase as an intermediate state in the melting of the
(g)CFL phase into the 2SC phase. Formally, this is different
from the prediction of Ref.~\cite{Iida}. I find, however, that
the difference is connected with the choice of the model
parameters. In the NJL model with a cutoff parameter of 800~MeV
used in Ref.~\cite{Fukushima}, there is a
non-vanishing (although rather small) region of the dSC phase.
On the other hand, in the NJL model with a relatively small
cutoff parameter of 653.3~MeV, no sizeable window of the dSC
phase can be found. Therefore, I conclude that the appearence of
the uSC or dSC phase, repectively, is very sensitive to the
choice of the model parameters, or more precisely, it depends
strongly on the cutoff parameter.

The main result of
Sec.~\ref{The_phase_diagram_of_massless_quarks} is the complete
phase diagram of massless neutral three-flavor quark matter in
the $T$--$m_s^2 / \mu$ and $T$--$\mu$ plane, shown in
Fig.~\ref{phasediagram_massless}. In this figure, all
symmetry-related phase transitions are denoted by solid lines.
(The symmetries of all phases appearing in this figure were
discussed in Ref.~\cite{Iida}.) The dashed lines in
Fig.~\ref{phasediagram_massless} separate the mCFL and regular
2SC phases from the gCFL and g2SC phases. These cannot be real
phase transitions, but are at most smooth crossovers. At $T =
0$, there is an insulator-metal phase transition between the CFL
and the gCFL phase~\cite{Alford_Kouvaris_Rajagopal1,
Alford_Kouvaris_Rajagopal2}. At nonzero temperature, there
exists a similar insulator-metal-type transition between the CFL
and the mCFL phase, given by the dotted lines in
Fig.~\ref{phasediagram_massless}.

In Sec.~\ref{phase_diagram_self-consistent}, I studied the
$T$--$\mu$ phase diagram of neutral three-flavor quark matter
within the NJL model of Ref.~\cite{RKH} in which chiral symmetry
is broken explicitly by small but nonzero current quark masses.
As in the previous study in
Sec.~\ref{The_phase_diagram_of_massless_quarks}, I used the
mean-field approximation in the analysis. In contrast to
Sec.~\ref{The_phase_diagram_of_massless_quarks}, in
Sec.~\ref{phase_diagram_self-consistent}, the constituent quark
masses were treated self-consistently as dynamically generated
quantities. The main results were summarized in
Figs.~\ref{phasediagram} and~\ref{phasediagram_strong}.

By comparing the phase diagram of massless quarks shown in the
right panel of Fig.~\ref{phasediagram_massless} with the phase
diagram shown in Fig.~\ref{phasediagram}, I noticed several
important differences. First of all, I observed that a
self-consistent treatment of quark masses strongly influences
the competition between different quark phases. As was noticed
earlier in Ref.~\cite{Buballa_Oertel}, there exists a subtle
interplay between the two main effects. On the one hand, the
actual values of the quark masses directly influence the
competition between different normal and color-superconducting
phases. On the other hand, competing phases themselves determine
the magnitude of the masses. Very often, this leads to
first-order phase transitions in which certain regions in the
mass-parameter space become inaccessible.

Some differences to the results in
Sec.~\ref{The_phase_diagram_of_massless_quarks} and
Sec.~\ref{phase_diagram_self-consistent} are related to a
different choice of model parameters. Most importantly, the
value of the diquark coupling $G_D = \frac34 G_S$ is
considerably weaker in the NJL model of
Sec.~\ref{phase_diagram_self-consistent}. This can be easily
seen by comparing the magnitude of the zero-temperature gap
at a given value of the quark chemical potential, say at $\mu =
500$~MeV, in the two models. It is $\Delta_0^\mathrm{(500)}
\approx 140$~MeV in
Sec.~\ref{The_phase_diagram_of_massless_quarks} and
$\Delta_0^\mathrm{(500)} \approx 76$~MeV in
Sec.~\ref{phase_diagram_self-consistent}. (Note that the
strength of the diquark pairing in Ref.~\cite{Fukushima} is even
weaker, corresponding to $\Delta_0^\mathrm{(500)} \approx
20$~MeV.) It should be noted that even the strong coupling $G_D
= G_S$ which leads to $\Delta_0^\mathrm{(500)} \approx 120$~MeV,
is still slightly weaker than that in
Sec.~\ref{The_phase_diagram_of_massless_quarks}. In this case,
however, the corresponding results differ mostly because the
quark masses are treated very differently.

Because of the weaker diquark coupling strength, the Cooper
instabilities in Fig.~\ref{phasediagram} occur systematically
at higher values of the quark chemical potential than in the
right panel of Fig.~\ref{phasediagram_massless}. In particular,
this is most clearly seen from the critical lines of the
transition to the (g)CFL phase. Another consequence of the
weaker interaction is the possibility of a thermal enhancement
of the (g)2SC Cooper pairing at intermediate values of the quark
chemical potential. This kind of enhancement was studied in
detail in Refs.~\cite{g2SC_1, g2SC_2}. Making use of the same
arguments, one can tell immediately how the phase diagram in
Fig.~\ref{phasediagram} should change with increasing or
decreasing the diquark coupling strength.

In particular, with increasing (decreasing) the diquark coupling
strength, the region of the (g)2SC phase at intermediate values
of the quark chemical potential should expand (shrink) along the
temperature direction. The regions covered by the other (i.e.,
uSC and CFL) phases should have qualitatively the same shape,
but shift to lower (higher) values of the quark chemical
potential and to higher (lower) values of the temperature. In
the case of strong coupling, in particular, these general
arguments are confirmed by my numerical calculations. The
corresponding phase diagram is shown in
Fig.~\ref{phasediagram_strong}.

Several comments are in order regarding the choice of the NJL
model used in Sec.~\ref{phase_diagram_self-consistent}. The
model is defined by the set of parameters in
Eq.~(\ref{NJL_model_parameters}) which were fitted to reproduce
several important QCD properties in vacuum~\cite{RKH}. (Note
that the same model also was used in Ref.~\cite{SRP}.) It is
expected, therefore, that this is a reasonable effective model
of QCD that captures the main features of both, chiral and
color-superconducting pairing dynamics. Also, a relatively small
value of the cutoff parameter in the model, see
Eq.~(\ref{NJL_model_parameters}), should not necessarily be
viewed as a bad feature of the model. In fact, this might simply
mimic a natural property of the full theory in which the
coupling strength of relevant interactions is quenched at large
momenta.

In this relation, note that the approach of
Ref.~\cite{Fukushima} regarding the cutoff parameter in the NJL
model is very different. It is said there that a large value of
this parameter is beneficial in order to extract results which
are insensitive to a specific choice of the cutoff. However, I
do not find any physical argument that would support this
requirement. Instead, I insist on having an effective model that
describes reasonably well the QCD properties at zero quark
chemical potential. I do not pretend, of course, that a
na\"{\i}ve extrapolation of the model to large densities can be
rigorously justified. In absence of a better alternative,
however, this seems to be the only sensible choice.

The results of Sec.~\ref{phase_diagram_self-consistent} might be
relevant for understanding the physics of (hybrid) neutron stars
with quark cores, in which the deleptonization is completed. In
order to obtain a phase diagram that could be applied to
protoneutron stars, one has to generalize the analysis to take
into account neutrino trapping.

In Sec.~\ref{phase_diagram_neutrino_trapping}, I studied the
effect of neutrino trapping on the phase diagram of neutral
three-flavor quark matter within the NJL model of
Ref.~\cite{RKH}. The results were obtained in the mean-field
approximation, treating constituent quark masses as dynamically
generated quantities. The overall structure of the phase diagram
in the space of three parameters, namely temperature $T$, quark
chemical potential $\mu$ and lepton-number chemical potential
$\mu_{L_{e}}$, was summarized in Fig.~\ref{phase3d}. This was
further detailed in several two-dimensional slices of the phase
diagram, including phase diagrams in the plane of temperature
and quark chemical potential (see Figs.~\ref{phase200}
and~\ref{phase400}) and in the plane of temperature and
lepton-number chemical potential (see Figs.~\ref{phase_mu400}
and~\ref{phase_mu500}).

By making use of simple model-independent arguments, as well as
detailed numerical calculations in the framework of the NJL
model, I found that neutrino trapping helps Cooper pairing in
the 2SC phase and suppresses the CFL phase. In essence, this
is the consequence of satisfying the electric neutrality
constraint in the quark system. In two-flavor quark matter,
the (positive) lepton-number chemical potential $\mu_{L_e}$
helps to provide extra electrons without inducing a large
mismatch between the Fermi momenta of up and down quarks.
With reducing the mismatch, of course, Cooper pairing gets
stronger. This is in sharp contrast to the situation in the
CFL phase of quark matter, which is neutral in the absence
of electrons. Additional electrons due to large $\mu_{L_e}$
can only put extra stress on the system.

In application to protoneutron stars, my findings in
Sec.~\ref{phase_diagram_neutrino_trapping} strongly suggests
that the CFL phase is very unlikely to appear during the early
stage of the stellar evolution before the deleptonization is
completed. If color superconductivity occurs there, the 2SC
phase is the best candidate for the ground state. In view of
this finding, it might be quite natural to suggest that matter
inside protoneutron stars contains little or no
strangeness (just as the cores of the progenitor stars) during
the early times of their evolution. In this connection, it is
appropriate to recall that neutrino trapping also suppresses the
appearance of strangeness in the form of hyperonic matter and
kaon condensation~\cite{PNS1}. My finding, therefore, is a
special case of a generic property.

The authors of Ref.~\cite{Alford_Rajagopal} claimed that the 2SC
phase is \textit{absent} in compact stars and underpinned their
claim by using a very simple model-independent calculation. The
phase diagrams which were created by using more precise NJL-type
models and which are shown in my thesis are the best evidence
that there indeed could exist a 2SC phase in compact stars and
that the 2SC phase in general is not absent. This is always the
case in protoneutron stars and in cold neutron stars at large
diquark coupling constants. But in the limit of weak diquark
coupling and zero temperature in which the simple
model-independent calculation in Ref.~\cite{Alford_Rajagopal} is
valid, the statement seems to be true that the 2SC phase is
absent, cf.\ Fig.~\ref{phasediagram}.

After the deleptonization occurs, it is possible that the ground
state of matter at high density in the central region of neutron
stars is the CFL phase. This phase contains a large number of
strange quarks. Therefore, an abundant production of strangeness
should happen right after the deleptonization in protoneutron
stars. If realized in nature, in principle this scenario may
have observational signatures.
\section{Open questions and outlook}
\begin{figure}[H]
\begin{center}
\includegraphics[width=0.95\textwidth]{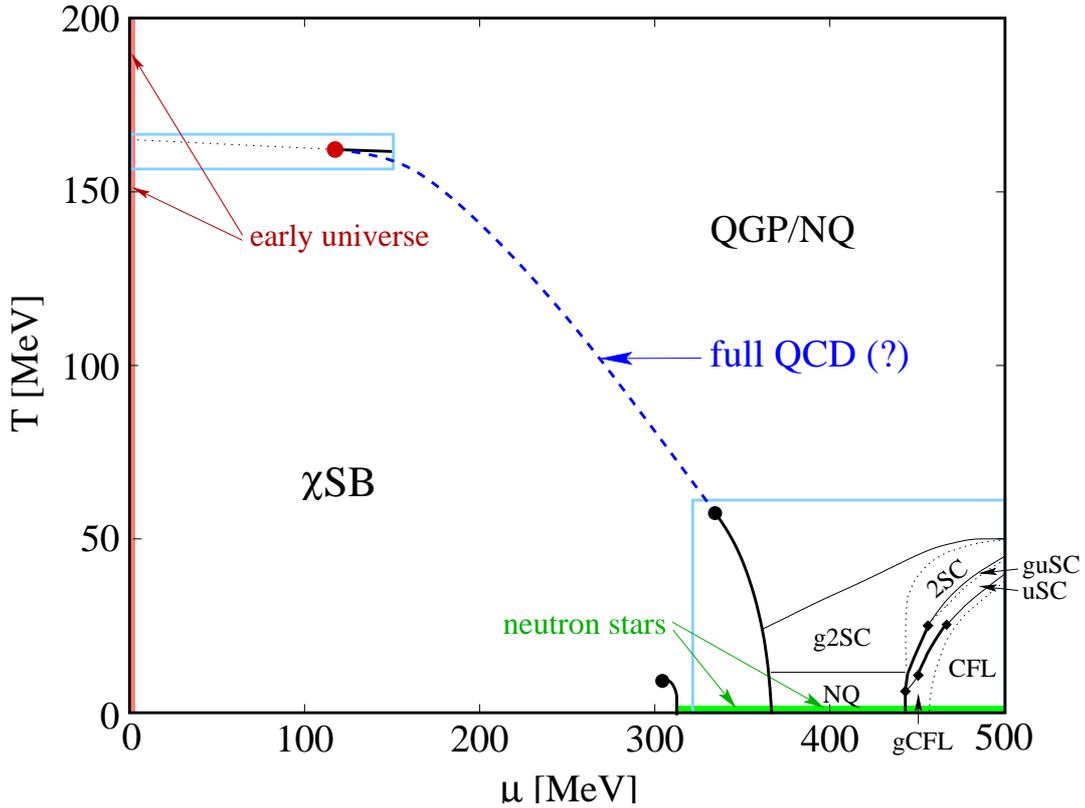}
\caption[The knowledge about the phase diagram of strongly
interacting matter in 2005.]{The knowledge about the phase
diagram of strongly interacting matter in 2005. The upper left
part shows the critical endpoint of the quark-hadron phase
transition which is obtained by lattice QCD
calculations~\cite{lattice}. The lower right part shows the
phase diagram of neutral three-flavor quark
matter~\cite{Ruester_Werth_Buballa_Shovkovy_Rischke1}, where
quark masses are treated self-consistently, see
Fig.~\ref{phasediagram}. The dashed line shows the discrepancy
between the critical endpoint of the quark-hadron phase
transition in the NJL model with that obtained in lattice QCD
calculations. This discrepancy could be resolved by using full
QCD. In addition, there is shown the state of the early
universe and of cold neutron stars after deleptonization.}
\label{phasediagram_2005}
\end{center}
\end{figure}

The quark sector of the phase diagram of strongly interacting
matter was poorly understood in 2003, see
Fig.~\ref{phase_diagram_schematic}. With my investigations of
this subject, I made the first attempt to illuminate the phase
structure in the quark sector of the phase diagram of strongly
interacting matter within the framework of an NJL model. In
Fig.~\ref{phasediagram_2005}, I finally show the status of
knowledge about the phase diagram of strongly interacting matter
in 2005.

With my investigations I took the first step in order to study
the phase diagram of neutral quark matter, see
Ref.~\cite{Ruester_Shovkovy_Rischke} and
Sec.~\ref{The_phase_diagram_of_massless_quarks}. I further
improved my result for the phase diagram of neutral quark matter
by properly incorporating the strange quark
mass~\cite{Shovkovy_Ruester_Rischke}. The next step in my
investigation was to treat the quark masses self-consistently as
dynamically generated quantities, see
Ref.~\cite{Ruester_Werth_Buballa_Shovkovy_Rischke1} and
Sec.~\ref{phase_diagram_self-consistent}, and finally to
incorporate a nonzero lepton-number chemical potential, see
Ref.~\cite{Ruester_Werth_Buballa_Shovkovy_Rischke2}
and Sec.~\ref{phase_diagram_neutrino_trapping}.

Here, I restricted my study to spin-zero color-superconducting
phases only which seem to be the most preferred
color-superconducting states because they have the highest
pressure. Therefore, I do not expect that other than these
color-superconducting states which I have investigated in this
thesis will occur in the phase diagram of neutral quark matter.
But if the pressures of one or more of these other
color-superconducting quark phases are larger than those of the
spin-zero color-superconducting quark phases so that these other
color-superconducting quark phases indeed will occur in the
phase diagram of neutral quark matter, then I expect that the
regions of such color-superconducting phases in the phase
diagram are small so that the phase diagram in
Fig.~\ref{phasediagram} will not be changed very much. In order
to avoid misunderstandings, I want to say that I do not expect
that such color-superconducting states occur in the phase
diagram of neutral quark matter. I do not say that such
color-superconducting phases are \textit{absent}.

An example for a possible occurrence of a color-superconducting
phase which I have not considered in my investigations is the
spin-one color-superconducting transverse CSL phase (the most
preferred spin-one color-superconducting quark
phase~\cite{Schmitt}). This phase has good chances to occur very
close to the axis of the quark chemical potential in the region
of the normal quark phase in the phase diagram of neutral quark
matter in Fig.~\ref{phasediagram}. If one will find that this is
really true then the phase diagram of neutral quark matter shown
in Fig.~\ref{phasediagram} will not be changed much because
spin-one color-superconducting quark phases like the transverse
CSL phase will be destroyed even at small temperatures.
Therefore, the line width of the axis of the quark chemical
potential is possibly thicker than the region of the spin-one
color-superconducting transverse CSL phase in the phase diagram
of neutral quark matter. Nevertheless, the CSL phase could occur
in cold neutron stars~\cite{Alford_Cowan}.
In Ref.~\cite{Hossein}, another possible spin-zero
color-superconducting quark phase, the A-phase, was checked if
it occurs in the phase diagram of neutral quark matter. The
result was that this phase always has a lower pressure than the
phases which I have investigated in my thesis. Therefore, the
A-phase does not appear in the phase diagram of neutral quark
matter.

However, it will be very important to check if other than these
color-superconducting phases which I have investigated in this
thesis indeed occur in the phase diagram of neutral quark
matter (see Ref.~\cite{Rajagopal_Schmitt}, where some of these
other color-superconducting quark phases are shown). Such an
investigation is outside the scope of my thesis. Much work
concerning this check has to be done in future. The pressures of
\textit{all} competing phases have to be compared. The phase
with the highest value of the pressure is the preferred one and
therefore appears in the phase diagram of neutral quark matter.

Despite the progress in understanding the phase diagram of
neutral quark matter, there still exists a fundamental problem.
Gapless color-superconducting quark phases are unstable in some
regions of the phase diagram of neutral quark matter because of
chromomagnetic instabilities~\cite{chromomagnetic_instability}
so that another phase will be the preferred state.
Chromomagnetic instabilities occur even in regular
color-superconducting quark phases. The author of
Ref.~\cite{Fukushima_unstable} shows that chromomagnetic
instabilities occur only at low temperatures in neutral
color-superconducting quark matter. The author of
Ref.~\cite{Huang} points out that the instabilities might be
caused by using BCS theory in mean-field approximation, where
phase fluctuations have been neglected. With the increase of the
mismatch between the Fermi surfaces of paired fermions, phase
fluctuations play more and more an important role, and soften
the superconductor. Strong phase fluctuations will eventually
quantum-disorder the superconducting state, and turn the system
into a phase-decoherent pseudogap state. By using an effective
theory of the CFL state, the author of Ref.~\cite{p_wave}
demonstrates that the chromomagnetic instability is resolved by
the formation of an inhomogeneous meson condensate. The authors
of Ref.~\cite{Gorbar} describe a new phase in neutral two-flavor
quark matter within the framework of the Ginzburg-Landau
approach, in which gluonic degrees of freedom play a crucial
role. They call it a gluonic phase. In this phase, gluon
condensates cure a chromomagnetic instability in the 2SC
solution and lead to spontaneous breakdown of the color gauge
symmetry, the $[ U(1)_\mathrm{em} ]$ and the rotational $SO(3)$
group. In other words, the gluonic phase describes an
anisotropic medium in which color and electric superconductivity
coexist. In Ref.~\cite{Giannakis}, it was suggested that the
chromomagnetic instability in gapless color-superconducting
phases indicates the formation of the LOFF phase~\cite{LOFF}
which is discussed in Ref.~\cite{LOFF_quark} in the context of
quark matter. Other possibilities could be the formation of a
spin-one color-superconducting quark phase, a mixed phase, or a
completely new state. The authors of
Ref.~\cite{Shovkovy_Hanauske_Huang} suggest that a mixed phase
composed of the 2SC phase and the normal quark phase may be more
favored if the surface tension is sufficiently
small~\cite{Reddy_Rupak}. The authors of
Ref.~\cite{Alford_Rajagopal_Reddy_Wilczek} suggest a single
first-order phase transition between CFL and nuclear matter.
Such a transition, in space, could take place either through a
mixed phase region or at a single sharp interface with
electron-free CFL and electron-rich nuclear matter in stable
contact. The authors of
Ref.~\cite{Alford_Rajagopal_Reddy_Wilczek} constructed a model
for such an interface.

Another possibility in order to avoid gapless
color-superconducting quark phases is to set the diquark
coupling constant to such high values that gapless phases will
not occur in the phase diagram~\cite{Abuki_Kunihiro}. But then
it is another question if such large diquark coupling constants
indeed occur in nature. My personal opinion is that one should
not increase the diquark coupling constant to such high values
in order to get rid of the problem but use the standard value of
the diquark coupling constant in Eq.~(\ref{G_D}) which is
predicted by the Fierz transformation in vacuum.

Whether gapless phases can indeed exist as color-superconducting
quark phases or whether they will be unstable because of
chromomagnetic instabilities is not known yet. Also, if they are
unstable, it is not yet known which states will be formed in the
respective regions of the phase diagram. Such an investigation
is outside the scope of my thesis, and much work has to be done
in future to solve this problem. The pressures of \textit{all}
competing quark phases have to be determined in these regions of
the phase diagram. The phase with the highest value of the
pressure is the preferred one.

Another problem which has to be solved is the discrepancy
between the critical endpoint of the quark-hadron phase
transition which is obtained by lattice QCD
calculations~\cite{lattice} and NJL
models~\cite{Buballa_Habilitationsschrift,
Ruester_Werth_Buballa_Shovkovy_Rischke1, Blaschke,
Ruester_Werth_Buballa_Shovkovy_Rischke2, Abuki_Kunihiro,
cr_point_NJL, Buballa_Oertel}, see also
Fig.~\ref{phasediagram_2005}. It is appropriate to mention here
that the location of the critical endpoint might be affected
very much by fluctuations of the composite chiral fields. These
are not included in the mean-field studies of the NJL model. In
fact, this is probably the main reason for their inability to
pin down the location of the critical endpoint consistent for
example, with lattice QCD calculations~\cite{lattice}. It is
fair to mention that the current lattice QCD calculations are
not very reliable at nonzero $\mu$ either. Therefore, the
predictions of my thesis, as well as of those in
Refs.~\cite{Buballa_Habilitationsschrift,
Ruester_Werth_Buballa_Shovkovy_Rischke1, Blaschke,
Ruester_Werth_Buballa_Shovkovy_Rischke2, Abuki_Kunihiro,
cr_point_NJL, Buballa_Oertel}, regarding the critical endpoint
cannot be considered as very reliable. One definitely will be
able to say where exactly the critical endpoint is located and
how the phase diagram of neutral quark matter looks like if one
uses full QCD. Such an investigation is outside the scope of my
thesis and has to be done in future. Since the NJL-type models
used in my thesis fulfill at least some important key features
of QCD, I expect that the phase diagram of neutral quark matter
which is shown in Fig.~\ref{phasediagram} will be changed only
quantitatively so that the overall phase structure does not
change if one computes a more realistic phase diagram of neutral
quark matter by using full QCD.

In the context of compact stars, it is not known yet if their
cores reach chemical potentials which make color
superconductivity possible. It is of great importance to resolve
this issue. Neutron stars are the most likely candidates for
color superconductivity in nature and therefore, one expects
that they contain color-superconducting quark cores. Therefore,
one has to find an experimental proof for color-superconducting
quark matter. This can be done by detecting neutrino and
$\gamma$-emissions of protoneutron stars and in colliders.

Since one does not know yet how large the lepton-number chemical
potential in protoneutron stars is during their evolution from
the supernova explosion until they become a cold neutron star,
it is difficult to say exactly, through which phases they go in
the three-dimensional phase diagram shown in Fig.~\ref{phase3d}.
Therefore, it would be very interesting to study the precise
evolution path of protoneutron stars in the three-dimensional
phase diagram.

I note that the analysis in this thesis is restricted to locally
neutral phases only. This automatically excludes, for example,
crystalline~\cite{LOFF_quark} and mixed~\cite{Reddy_Rupak,
Shovkovy_Hanauske_Huang, Neumann} phases. Also, in the mean
field approximation utilized here, I cannot get any phases with
meson condensates~\cite{Bedaque, Kaplan_Reddy,
Kryjevski_Kaplan_Schaefer}.

In nature, phase transitions can take place either through sharp
boundaries between pure phases which are located next to each
other or through mixed-phase regions. In this thesis, I
considered only the former possibility, where each phase has the
highest pressure in its region of the phase diagram and where
the pressures of the phases are equal only at the sharp phase
boundaries. Such phase transitions are obtained by so-called
Maxwell constructions. It would be interesting to study the
possibility of mixed phases by making so-called Gibbs
constructions~\cite{Weber_book}, in which the pressures and the
chemical potentials of the phases are equal and continuous in
contrast to Maxwell constructions.

Another interesting investigation in the context of compact
stars would be to create a phase-radius diagram that shows which
phase occurs at which radius of the compact star. Such a diagram
would directly show the composition and structure of a compact
star.
\appendix
\selectlanguage{american}
\chapter{Definitions of matrices}
In this chapter of the Appendix, I present some definitions such
as the Pauli and $\gamma$-matrices, and the generatores of the
$SU(3)$ group. I also show the chirality, energy,
energy-chirality, and the spin projectors as well as their
properties.
\section{The Pauli matrices}
The Pauli matrices are defined as:
\be
\begin{aligned}
\sigma_1 &= \left(
\begin{array}{r@{\hspace{4mm}}r}
0 & 1 \\
1 & 0
\end{array}
\right) \; , &
\sigma_2 &= \left( 
\begin{array}{rr}
0 & -i \\
i & 0
\end{array}
\right) \; , \\
\sigma_3 &= \left(
\begin{array}{rr}
1 & 0 \\
0 & -1
\end{array}
\right) \; , &
\fett{\sigma} &= \left( \sigma_1, \sigma_2, \sigma_3 \right) \;
.
\end{aligned}
\ee
The Pauli matrices are hermitian,
\be
\sigma_i^\dagger = \sigma_i \; .
\ee
The following relations are valid:
\be
\sigma_i^2 = 1 \; , \qquad
\sigma_i \sigma_j = i \sigma_k \; , \qquad
\left[ \sigma_i, \sigma_j \right] = 2 i \sigma_k \; , \qquad
\left\{ \sigma_i, \sigma_j \right\} = 2 \delta_{ij} \; .
\ee
\subsection{Spin projectors}
The spin projectors are defined as
\be
\label{spin_projectors}
\mathcal{P}_s \left( \fettu{k} \right) = \frac12 \, ( 1 + s
\fett{\sigma} \cdot \hat{\fettu{k}} ) \; ,
\ee
where $s = \pm$ corresponds to the projections onto spin up or
spin down states, respectively, and $\fett{\sigma}$ are the
Pauli matrices. The spin projectors fulfill the properties of
completeness and orthogonality,
\bsub
\bea
\mathcal{P}_+ \left( \fettu{k} \right) + \mathcal{P}_- \left(
\fettu{k} \right) &=& 1 \; , \\
\mathcal{P}_s \left( \fettu{k} \right) \mathcal{P}_{s'} \left(
\fettu{k} \right) &=& \delta_{ss'} \mathcal{P}_{s'} \left(
\fettu{k} \right) \; .
\eea
\esub
The spin projectors are hermitian,
\be
\mathcal{P}_s^\dagger \left( \fettu{k} \right) = \mathcal{P}_s
\left( \fettu{k} \right) \; .
\ee
The following relations are valid:
\bea
\mathcal{P}_+ \left( \fettu{k} \right) - \mathcal{P}_- \left(
\fettu{k} \right) &=& \fett{\sigma} \cdot \hat{\fettu{k}} \; ,
\\ 
\Tr \, \mathcal{P}_s \left( \fettu{k} \right) &=& 1 \; .
\eea
\section{Matrices in Dirac space}
\label{gamma_matrices}
The Dirac definition of the $\gamma$-matrices reads:
\be
\begin{aligned}
\gamma^0 &= \left(
\begin{array}{r@{\hspace{2mm}}r}
1 & 0 \\
0 & -1
\end{array}
\right) \; , &
\gamma^1 &= \left( 
\begin{array}{c@{\hspace{2mm}}l}
\ 0 & \sigma_1 \\
-\sigma_1 & 0
\end{array}
\right) \; , \\
\label{gamma2}
\gamma^2 &= \left( 
\begin{array}{c@{\hspace{2mm}}l}
\ 0 & \sigma_2 \\
-\sigma_2 & 0
\end{array}
\right) \; , &
\gamma^3 &= \left( 
\begin{array}{c@{\hspace{2mm}}l}
\ 0 & \sigma_3 \\
-\sigma_3 & 0
\end{array}
\right) \; , \\
\fett{\gamma} &= \left( 
\begin{array}{c@{\hspace{3mm}}l}
\ \ 0 & \fett{\sigma} \\
-\fett{\sigma} & 0
\end{array}
\right) \; , &
\gamma_5 &= \left( 
\begin{array}{r@{\hspace{3mm}}r}
0 & 1 \\
1 & 0
\end{array}
\right) = i \gamma^0 \gamma^1 \gamma^2 \gamma^3 \; .
\end{aligned}
\ee
The following relations are valid:
\be
\begin{gathered}
\gamma_0^2 = 1 \; , \qquad
\left( \gamma^i \right)^2 = - 1 \; , \qquad
\gamma_5^2 = 1 \; , \\
\gamma^\mu = \left( \gamma^0, \gamma^1, \gamma^2, \gamma^3
\right) = \left( \gamma^0, \fett{\gamma} \right) \; , \qquad
\gamma_0 \left( \gamma^\mu \right)^\dagger \gamma_0 = \gamma^\mu
\; , \qquad
\left\{ \gamma^\mu, \gamma^\nu \right\} = 2 g^{\mu\nu} \; ,
\end{gathered}
\ee
where
\be
g^{\mu\nu} = \diag \left( 1, -1, -1, -1 \right)
\ee
is the metric tensor. The charge-conjugation matrix in Dirac
representation is defined as
\be
C = i \gamma^2 \gamma_0 \; ,
\ee
where $\gamma^2$ is given by Eq.~(\ref{gamma2}).
The following relations are valid:
\bsub
\bea
C = - C^{-1} &=& - C^T = - C^\dagger \; , \\
C \gamma^\mu C^{-1} &=& - \left( \gamma^\mu \right)^T \; , \\
C \left( \gamma^\mu \right)^T C^{-1} &=& - \gamma^\mu \; .
\eea
\esub
With the charge-conjugation matrix, one can introduce the
charge-conjugate spinors in space-time,
\be
\label{charge-conjugate_spinors_space-time}
\begin{aligned}
\psi_C \left( X \right) &= C \bar\psi^T \left( X \right) \; ,
\\
\bar\psi_C \left( X \right) &= \psi^T \left( X \right) C \; ,
\\
\psi \left( X \right) &= C \bar\psi_C^T \left( X \right) \; ,
\\
\bar\psi \left( X \right) &= \psi_C^T \left( X \right) C \; ,
\end{aligned}
\ee
and in energy-momentum space,
\be
\begin{aligned}
\psi_C \left( K \right) &= C \bar\psi^T \left( -K \right) \; ,
\\
\bar\psi_C \left( K \right) &= \psi^T \left( -K \right) C \; ,
\\
\psi \left( -K \right) &= C \bar\psi_C^T \left( K \right) \; ,
\\
\bar\psi \left( -K \right) &= \psi_C^T \left( K \right) C \; .
\end{aligned}
\ee
\subsection{Projectors in Dirac space}
In the limit of vanishing mass, the chirality, energy, and
energy-chirality projectors are defined as:
\bsub
\bea
\mathcal{ P }_c &=& \textstyle \frac12 \displaystyle \, ( 1 + c
\gamma_5 ) \; , \\
\Lambda^e \left( \fettu{ k } \right) &=& \textstyle \frac12
\displaystyle \, ( 1 + e \gamma_0 \fett{ \gamma } \cdot
\hat{\fettu{k}} ) \; , \\
\label{energy-chirality_projectors}
\mathcal{P}_c^e \left( \fettu{k} \right) &\equiv& \mathcal{ P
}_c \, \Lambda^e \left( \fettu{ k } \right) = \textstyle \frac14
\displaystyle \, (1 + c \gamma_5 ) \, ( 1 + e \gamma_0 \fett{
\gamma } \cdot \hat{\fettu{k}} ) \; ,
\eea
\esub
where $c = \pm$ or $c = r, \ell$, respectively, stands for
right-handed and left-handed projections, and $e = \pm$ for
projections onto states of positive and negative energy. The
chirality and energy projectors fulfill the properties of
completeness and orthogonality,
\bsub
\bea
\mathcal{P}_r + \mathcal{P}_\ell &=& 1 \; , \\
\Lambda^+ \left( \fettu{k} \right) + \Lambda^- \left( \fettu{k}
\right)&=& 1 \; , \\
\mathcal{P}_c \mathcal{P}_{c'} &=& \delta_{c,c'}
\mathcal{P}_{c'} \; , \\
\Lambda^e \left( \fettu{k} \right) \Lambda^{e'} ( \fettu{k}) &=&
\delta^{e,e'} \Lambda^{e'} (\fettu{k}) \; .
\eea
\esub
These projectors are hermitian,
\bsub
\bea
\mathcal{P}_c^\dagger &=& \mathcal{P}_c \; , \\
\left[ \Lambda^e \right]^\dagger \left( \fettu{k} \right) &=&
\Lambda^e \left( \fettu{k} \right) \; ,
\eea
\esub
and they commute with each other,
\be
\left[ \mathcal{P}_c, \Lambda^e \left( \fettu{k} \right) \right]
= 0 \; .
\ee
The energy-chirality projectors fulfill the properties of
completeness and orthogonality,
\bsub
\bea
\sum_{c,e} \mathcal{P}_c^e \left( \fettu{k} \right) &=& 1 \; ,
\\
\mathcal{P}_c^e \left( \fettu{k} \right) \mathcal{P}_{c'}^{e'}
\left( \fettu{k} \right) &=& \delta_{c,c'} \delta^{e,e'}
\mathcal{P}_{c'}^{e'} \left( \fettu{k} \right) \; .
\eea
\esub
The energy-chirality projectors are hermitian,
\be
\left[ \mathcal{P}_c^e \right]^\dagger \left( \fettu{k} \right)
= \mathcal{P}_c^e \left( \fettu{k} \right) \; .
\ee
The following relations are valid:
\bea
\mathcal{P}_{-c}^{-e} \left( \fettu{k} \right) &=& \gamma_0
\mathcal{P}_c^e \left( \fettu{k} \right) \gamma_0 \; , \\
\Lambda^+ \left( \fettu{k} \right) - \Lambda^- \left( \fettu{k}
\right) &=& \gamma_0 \fett{\gamma} \cdot \hat{\fettu{k}} \; , \\
\Tr \, \mathcal{P}_c^e \left( \fettu{k} \right) &=& 1 \; , \\
\mathcal{P}_c \gamma^\mu &=& \gamma^\mu \mathcal{P}_{-c} \; .
\eea
\section{The generators of the \textit{SU}(3) group}
The generators of the $SU(3)$ group are defined as:
\be
\begin{aligned}
T_1 &= \frac12 \left( 
\begin{array}{r@{\hspace{4mm}}r@{\hspace{4mm}}r}
0 & 1 & 0 \\
1 & 0 & 0 \\
0 & 0 & 0
\end{array}
\right) \; , &
T_2 &= \frac12 \left( 
\begin{array}{rr@{\hspace{4mm}}r}
0 & -i & 0 \\
i &  0 & 0 \\
0 &  0 & 0
\end{array}
\right) \; , \\
T_3 &= \frac12 \left( 
\begin{array}{rr@{\hspace{4mm}}r}
1 &  0 & 0 \\
0 & -1 & 0 \\
0 &  0 & 0
\end{array}
\right) \; , &
T_4 &= \frac12 \left( 
\begin{array}{r@{\hspace{3mm}}r@{\hspace{3mm}}r}
0 & 0 & 1 \\
0 & 0 & 0 \\
1 & 0 & 0
\end{array}
\right) \; , \\
T_5 &= \frac12 \left( 
\begin{array}{r@{\hspace{3mm}}rr}
0 & 0 & -i \\
0 & 0 & 0 \\
i & 0 & 0
\end{array}
\right) \; , &
T_6 &= \frac12 \left( 
\begin{array}{r@{\hspace{3mm}}r@{\hspace{3mm}}r}
0 & 0 & 0 \\
0 & 0 & 1 \\
0 & 1 & 0
\end{array}
\right) \; , \\
T_7 &= \frac12 \left( 
\begin{array}{r@{\hspace{3mm}}rr}
0 & 0 & 0 \\
0 & 0 & -i \\
0 & i & 0
\end{array}
\right) \; , &
T_8 &= \frac{ 1 }{ 2 \sqrt{3} } \left( 
\begin{array}{r@{\hspace{3mm}}rr}
1 & 0 & 0 \\
0 & 1 & 0 \\
0 & 0 & -2
\end{array}
\right) \; .
\end{aligned}
\ee
The Gell-Mann matrices are defined by
\be
\label{Gell-Mann}
\lambda_a \equiv 2 T_a \; .
\ee
The following relations are valid:
\be
\Tr \, T_a = 0 \; , \qquad
T_a^\dagger = T_a \; , \qquad
\left[ T_a, T_b \right] = i \sum_{c=1}^8 f_{abc} T_c \; .
\ee
All $f_{abc} = 0$, except,
\bsub
\bea
f_{123} &=& 1 \; , \\
f_{147} &=& -f_{156} = f_{246} = f_{257} = f_{345} = -f_{367} =
\textstyle \frac12 \displaystyle \; , \\
f_{458}&=&f_{678}= \textstyle \frac12 \displaystyle \sqrt{3} \;
.
\eea
\esub
\chapter{Useful formulae}
Several relations which are used in this thesis are derived and
proven in this chapter of the Appendix.
\section{Non-interacting massless fermions and antifermions at
nonzero temperature}
\label{massless_fermions_at_nonzero_T}
The pressure of non-interacting massless fermions and
antifermions at nonzero temperature $T$ is given by
\be
p = \frac{g T}{2 \pi^2} \int_0^\infty \ud k \, k^2
\left\{ \ln \left[ 1 + \exp \left( - \frac{k - \mu}{ T } \right)
\right] + \ln \left[ 1 + \exp \left(- \frac{k + \mu}{ T }
\right) \right] \right\} \; ,
\ee
where $g$ is the degeneracy factor, $k \equiv \left| \fettu{k}
\right|$ is the momentum, and $\mu$ is the chemical potential of
the fermions. The first term corresponds to the pressure of
fermions while the second one corresponds to the pressure of
antifermions. By an integration by parts one obtains,
\be
p = \frac{g}{6 \pi^2} \int_0^\infty \ud k \, k^3 \left[ n_F
\left( \frac{k - \mu}{ T } \right) + n_F \left( \frac{k + \mu}{
T } \right) \right] \; ,
\ee
where
\be
n_F \left( x \right) \equiv \frac{1}{\e^x + 1} \; ,
\ee
is the Fermi-Dirac distribution function.

With the substitutions $x^\pm = (k \pm \mu) / T$ one computes,
\be
p = \frac{g T}{6 \pi^2} \left[ \int_{-\frac{\mu}{T}}^\infty \ud
x^- \left( T x^- + \mu \right)^3 n_F \left( x^- \right) +
\int_{\frac{\mu}{T}}^\infty \ud x^+ \left( T x^+ - \mu
\right)^3 n_F \left( x^+ \right) \right] \; .
\ee
The integrals can be split so that
\bea
p &=& \frac{g T}{6 \pi^2} \bigg[ \int_0^\infty \ud x^- \left( T
x^- + \mu \right)^3 n_F \left( x^- \right) + \int_0^\infty \ud
x^+ \left( T x^+ - \mu \right)^3 n_F \left( x^+ \right)
\nonumber \\
&& \hspace{5.5mm} + \int_{-\frac{\mu}{T}}^0 \ud x^- \left(
T x^- + \mu \right)^3 n_F \left( x^- \right) +
\int_{\frac{\mu}{T}}^0 \ud x^+ \left( T x^+ - \mu \right)^3 n_F
\left( x^+ \right) \bigg] \; .
\eea
The first two integrals can be joined. In the last integral,
one can use the substitution $x^+ = - \tilde x^+$. Then, also
the last two integrals can be joined,
\be
p = \frac{g T}{6 \pi^2} \left\{ \int_0^\infty \ud x \, \frac{2
T^3 x^3 + 6 T x \mu^2}{\e^x + 1} + \int_{-\frac{\mu}{T}}^0 \ud
x \, \left( T x + \mu \right)^3 \left[ n_F \left( x \right) +
n_F \left( -x \right) \right] \right\} \; ,
\ee
where I renamed $x^+$, $\tilde x^+$, and $x^-$ in $x$. With the
relation $n_F \, (x) + n_F \, (-x) = 1$, the
integrals~\cite{Maple, Mathematica},
\be
\int_0^\infty \ud x \, \frac{x^3}{\e^x + 1} = \frac{7}{120}
\pi^4 \; , \qquad \int_0^\infty \ud x \, \frac{x}{\e^x + 1} =
\frac{1}{12} \pi^2 \; ,
\ee
and the substitution, $z = T x + \mu$, one obtains for the
pressure of non-interacting massless fermions and antifermions
at nonzero temperature~\cite{Greiner},
\be
p = \frac{g}{24 \pi^2} \left( \mu^4 + 2 \pi^2 \mu^2 T^2 +
\frac{7}{15} \pi^4 T^4 \right) \; .
\ee
\section{The inverse Dirac propagator}
\label{The_inverse_Dirac_propagator}
The inverse tree-level propagator for quarks and
charge-conjugate quarks, respectively, is given by
\bsub
\bea
{[ \mathcal{G}_0^+ ]}^{-1} \left( X,Y \right) &\equiv& \, ( i
\covariant_X + \hat\mu \gamma_0 - \hat m ) \, \delta^{\left(
4 \right)} \left( X-Y \right) \; , \\
{[ \mathcal{G}_0^- ]}^{-1} \left( X,Y \right) &\equiv& \, ( i
\covariant_X^C - \hat\mu \gamma_0 - \hat m ) \, \delta^{\left( 4
\right)} \left( X-Y \right) \; ,
\eea
\esub
where $D_\mu = \partial_\mu - i g A_\mu^a T_a$ and $D_\mu^C =
\partial_\mu + i g A_\mu^a T_a^T$ are the covariant derivatives
for quarks and charge-conjugate quarks, respectively,
$\hat\mu$ is the color-flavor matrix of the chemical potentials,
and $\hat m$ is the quark-mass matrix. By using the
charge-conjugate
spinors~(\ref{charge-conjugate_spinors_space-time}), the
relation
\be
\bar\psi \left( X \right) \, [ \mathcal{G}_0^+ ]^{-1}  \left(
X,Y \right) \psi \left( Y \right) = \bar\psi_C \left( Y \right)
\, [ \mathcal{G}_0^- ]^{-1} \left( Y,X \right) \psi_C \left( X
\right)
\ee
can be proven:
\bea
\bar\psi \left( X \right) \, [ \mathcal{G}_0^+ ]^{-1} \left( X,Y
\right) \psi \left( Y \right)
&=& \psi_C^T \left( X \right)  C \, [ \mathcal{G}_0^+ ]^{-1}
\left( X,Y \right) C \bar\psi_C^T \left( Y \right) \nonumber \\
&=& - \psi_C^T \left( X \right) C \, [ i \covariant_X + \hat\mu
\gamma_0 - \hat m ] \, \delta^{\left( 4 \right)} \left( X-Y
\right) C^{-1} \bar\psi_C^T \left( Y \right) \nonumber \\
&=& - \psi_C^T \left( X \right) \, [ i C \gamma^\mu C^{-1} D_\mu
+ \hat\mu C \gamma_0 C^{-1} - \hat m ] \, \delta^{\left( 4
\right)} \left( X-Y \right)\bar \psi_C^T \left( Y \right)
\nonumber \\
&=& - \psi_C^T \left( X \right) \, [ - i \left( \gamma^\mu
\right)^T D_\mu - \hat\mu \gamma_0 - \hat m ] \, \delta^{\left(
4 \right)} \left( X-Y \right) \bar\psi_C^T \left( Y \right)
\nonumber \\
&=& \bar\psi_C \left( Y \right) \big\{ [ - i \left( \gamma^\mu
\right)^T D_\mu - \hat\mu \gamma_0 - \hat m ] \big\}^T
\delta^{\left( 4 \right)} \left( Y-X \right) \psi_C \left( X
\right) \nonumber \\
&=& \bar\psi_C \left( Y \right) \, [ i \covariant_X^C - \hat\mu
\gamma_0 - \hat m ] \, \delta^{\left( 4 \right)} \left( Y-X
\right) \psi_C \left( X \right) \nonumber \\ 
&=& \bar\psi_C \left( Y \right) \, [ \mathcal{G}_0^- ]^{-1}
\left( Y,X \right) \psi_C \left( X \right) \; .
\eea
\section{The tree-level quark propagator}
\label{The_tree-level_quark_propagator}
The inverse tree-level quark propagator in Nambu-Gorkov space is
given by
\be
\mathcal{S}^{-1} = \left( \begin{array}{cc}
[ \mathcal{G}_0^+ ]^{-1} + \Sigma^+ & \Phi^- \\
\Phi^+ & [ \mathcal{G}_0^- ]^{-1} + \Sigma^-
\end{array} \right) \; ,
\ee
where $[ \mathcal{G}_0^\pm ]^{-1}$ is the inverse tree-level
propagator for quarks or charge-conjugate quarks, respectively,
$\Sigma^\pm$ are the regular quark self-energies, and $\Phi^\pm$
are the gap matrices. From this expression, the tree-level quark
propagator in Nambu-Gorkov space,
\be
\mathcal{S} \equiv \left(
\begin{array}{cc}
\mathcal{G}^+ & \Xi^- \\
\Xi^+ & \mathcal{G}^-
\end{array}
\right) \; ,
\ee
can be obtained by using the relation $\mathcal{S}
\mathcal{S}^{-1} = \mathcal{S}^{-1} \mathcal{S} = 1$,
\be
\left( \begin{array}{cc}
\mathcal{G}^+ & \Xi^- \\
\Xi^+ & \mathcal{G}^-
\end{array} \right)
\left( \begin{array}{cc}
[ \mathcal{G}_0^+ ]^{-1} + \Sigma^+ & \Phi^- \\
\Phi^+ & [ \mathcal{G}_0^- ]^{-1} + \Sigma^-
\end{array} \right) =
\left( \begin{array}{@{\extracolsep{1mm}}cc}
1 & 0 \\
0 & 1
\end{array} \right) \; ,
\ee
where $\mathcal{G}^\pm$ is the propagator for quasiquarks or
charge-conjugate quasiquarks, respectively, and $\Xi^\pm$ are
the anomalous propagators. This can also be written in the form,
\bsub
\bea
\label{SA}
\mathcal{G}^+ \left( [ \mathcal{G}_0^+ ]^{-1} + \Sigma^+
\right) + \Xi^- \Phi^+ &=& 1 \; , \\
\label{SB}
\mathcal{G}^+ \Phi^- + \Xi^- \left( [
\mathcal{G}_0^- ]^{-1} + \Sigma^- \right) &=& 0 \; , \\
\label{SC}
\Xi^+ \left( [ \mathcal{G}_0^+ ]^{-1} + \Sigma^+
\right) + \mathcal{G}^- \Phi^+ &=& 0 \; , \\
\label{SD}
\Xi^+ \Phi^- + \mathcal{G}^- \left( [
\mathcal{G}_0^- ]^{-1} + \Sigma^- \right) &=& 1 \; .
\eea
\esub
From Eqs.~(\ref{SB}) and~(\ref{SC}) one gets
\be
\Xi^\pm = - \, \mathcal{G}^\mp \Phi^\pm \left( [
\mathcal{G}_0^\pm ]^{-1} + \Sigma^\pm \right)^{-1} \; .
\ee
By inserting these relations into Eqs.~(\ref{SA})
and~(\ref{SD}), one obtains
\be
\label{G+-}
\mathcal{G}^\pm = \left\{ [ \mathcal{G}_0^\pm ]^{-1} +
\Sigma^\pm - \Phi^\mp \left( [ \mathcal{G}_0^\mp ]^{-1} +
\Sigma^\mp \right)^{-1} \Phi^\pm \right\}^{-1} \; .
\ee
In the same way, one proceeds with
\be
\label{the_other_way_round}
\left( \begin{array}{cc}
[ \mathcal{G}_0^+ ]^{-1} + \Sigma^+ & \Phi^- \\
\Phi^+ & [ \mathcal{G}_0^- ]^{-1} + \Sigma^-
\end{array} \right)
\left( \begin{array}{cc}
\mathcal{G}^+ & \Xi^- \\
\Xi^+ & \mathcal{G}^-
\end{array} \right) =
\left( \begin{array}{@{\extracolsep{1mm}}cc}
1 & 0 \\
0 & 1
\end{array} \right) \; ,
\ee
and gets
\be
\Xi^\pm = - \left( [ \mathcal{G}_0^\mp ]^{-1} +
\Sigma^\mp \right)^{-1} \Phi^\pm \mathcal{G}^\pm \; .
\ee
By inserting these results into Eq.~(\ref{the_other_way_round}),
one again obtains for $\mathcal{G}^+$ and $\mathcal{G}^-$ 
the relations~(\ref{G+-}). For $\Xi^\pm$, one obtains
\be
\Xi^\pm = - \left( [ \mathcal{G}_0^\mp ]^{-1} + \Sigma^\mp
\right)^{-1} \Phi^\pm \mathcal{G}^\pm =  - \, \mathcal{G}^\mp
\Phi^\pm \left( [ \mathcal{G}_0^\pm ]^{-1} + \Sigma^\pm
\right)^{-1} \; .
\ee
\section{The Feynman gauged gluon propagator}
\label{The_Feynman_gauged_gluon_propagator}
The QCD Lagrangian density is given by
\be
\mathcal{L} = \bar \psi \left( i \covariant - \hat m \right)
\psi - \frac14 F_{\mu\nu}^a F_a^{\mu\nu} +
\mathcal{L}_\mathrm{gauge} \; .
\ee
In this section, I focus on the contribution of gluons to the
QCD Lagrangian density,
\be
\mathcal{L}_A \equiv - \frac14 F_{\mu\nu}^a F_a^{\mu\nu} \; .
\ee
The gluon field-strength tensor is defined as
\be
F_a^{\mu\nu} = \partial^\mu A_a^\nu - \partial^\nu A_a^\mu + g
f_{abc} A_b^\mu A_c^\nu \; .
\ee
By neglecting the gluon self-interaction, the gluon
field-strength tensor simplifies to
\be
F_a^{\mu\nu} = \partial^\mu A_a^\nu - \partial^\nu A_a^\mu \; ,
\ee
which breaks $[SU(3)_c]$ gauge invariance. The Abelian part of
the gluon Lagrangian density is
\bea
\mathcal{L}_A = - \frac14 F_{\mu\nu}^a F_a^{\mu\nu} &=& -
\frac14 \left( \partial_\mu A_\nu^a - \partial_\nu A_\mu^a
\right) \left( \partial^\mu A_a^\nu - \partial^\nu A_a^\mu
\right) \nonumber \\
&=& - \frac14 \left( \partial_\mu A_\nu^a \, \partial^\mu
A_a^\nu - \partial_\mu A_\nu^a \, \partial^\nu A_a^\mu -
\partial_\nu A_\mu^a \, \partial^\mu A_a^\nu + \partial_\nu
A_\mu^a \, \partial^\nu A_a^\mu \right) \nonumber \\
&=& - \frac12 \left( \partial_\mu A_\nu^a \, \partial^\mu
A_a^\nu - \partial_\mu A_\nu^a \, \partial^\nu A_a^\mu \right)
\; .
\eea
After integration by parts, and discarding surface terms, the
Abelian part of the gluon Lagrangian density reads~\cite{Ryder},
\be
\mathcal{L}_A = \frac12 \left( A_\nu^a \partial_\mu
\partial^\mu A_a^\nu - A_\nu^a \partial_\mu \partial^\nu A_a^\mu
\right) = \frac12 A_a^\mu \left( g_{\mu\nu} \quabla - \,
\partial_\mu \partial_\nu \right) A_a^\nu \; .
\ee
By choosing Lorentz gauge, $\partial_\mu A_a^\mu = 0$, one
obtains,
\be
\mathcal{L}_A = \frac12 A_a^\mu g_{\mu\nu} \quabla \, A_a^\nu \;
.
\ee
One gets the same result if one sets the gauge-fixing term of
the QCD Lagrangian density to
\be
\mathcal{L}_\mathrm{gauge} = - \frac12 \left( \partial_\mu
A_a^\mu \right)^2 \; ,
\ee
which leads to the QCD Lagrangian density in Feynman gauge.
The Feynman gauged gluon Lagrangian density can be rewritten as
\be
\mathcal{L}_A = \frac12 A_a^\mu \, [ D^{-1} ]_{\mu\nu}^{ab} \,
A_b^\nu \; .
\ee
The quantity
\be
[ D^{-1} ]_{\mu\nu}^{ab} = \delta^{ab} g_{\mu\nu} \quabla
\ee
is called the inverse Feynman gauged gluon propagator. Inverting
yields
\be
D_{\mu\nu}^{ab} = \delta^{ab} g_{\mu\nu} \quabla^{-1} \; .
\ee
In momentum space, the Feynman gauged gluon propagator reads
\be
\label{gluon_propagator}
D_{\mu\nu}^{ab} = - \delta^{ab} \frac{g_{\mu\nu}}{\Lambda^2} \;
.
\ee
\section{The determinant of the inverse quark propagator}
\label{The_determinant_of_the_inverse_quark_propagator}
In space-time, the contribution of the kinetic term to the
grand partition function reads up to irrelevant
constants~\cite{diploma_thesis_Ruester, Pisarski_Rischke3}
\be
\mathcal{Z}_\mathrm{kin} = \int \D \bar\psi \D
\psi \exp \left\{ I_\mathrm{kin} \left[ \bar\Psi, \Psi \right]
\right\} \; ,
\ee
where the action of the kinetic term is
\be
I_\mathrm{kin} \left[ \bar\Psi, \Psi \right] = \frac12
\int_{X,Y} \bar\Psi \left( X \right) S^{-1} \left( X,Y \right)
\Psi \left( Y \right) \; .
\ee
The inverse quark propagator (without regular self-energies)
\be
S^{-1} = \left( 
\begin{array}{cc}
[ G_0^+ ]^{-1} & \Phi^- \\
\Phi^+ & [ G_0^+ ]^{-1}
\end{array}
\right)
\ee
is a $2 \times 2$-matrix in Nambu-Gorkov space, where
\be
\bar\Psi \equiv
\left( \bar\psi, \bar\psi_C \right) \; , \qquad
\Psi \equiv \left(
\begin{array}{c}
\psi \\
\psi_C
\end{array}
\right) \; ,
\ee
are the Nambu-Gorkov quark spinors. The Fourier transforms of the
quark spinors read
\bsub
\bea
\psi \left( X \right) &=& \frac{1}{\sqrt V} \sum_K e^{ - i K X }
\psi \left( K \right) \; , \\
\bar \psi \left( X \right) &=& \frac{1}{\sqrt V} \sum_K e^{ i K
X } \bar\psi \left( K \right) \; , \\
\psi_C \left( X \right) &=& \frac{1}{\sqrt V} \sum_K e^{ - i K X
} \psi_C \left( K \right) \; , \\
\bar\psi_C \left( X \right) &=& \frac{1}{\sqrt V} \sum_K e^{ i K
X } \bar\psi_C \left( K \right) \; .
\eea
\esub
By assuming translational invariance, the Fourier transform of
the inverse quark propagator is given by
\be
S^{-1} \left( X,Y \right) = \frac{T}{V} \sum_K e^{ - i K
\left( X - Y \right) } S^{-1} \left( K \right) \; .
\ee
By Fourier transforming the kinetic term of the action, the
integration measure has to be rewritten,
\bea
\D \bar\psi \D \psi
&\equiv& \prod_K \ud \bar\psi \left( K \right) \ud \psi \left( K
\right) \nonumber \\
&=& \prod_{\left( K, -K \right)} \ud \bar\psi \left( K \right)
\ud \bar\psi \left( - K \right) \ud \psi \left( K \right) \ud
\psi \left( - K \right) \nonumber \\
&=& \tilde{\mathcal{N}} \prod_{\left( K, - K \right)} \ud
\bar\psi \left( K \right) \ud \psi_C \left( K \right) \ud \psi
\left( K \right) \ud \bar\psi_C \left( K \right) \; ,
\eea
where $\tilde{\mathcal{N}}$ is an irrelevant constant Jacobian
which arises from charge conjugation of quark spinors. The
Fourier-transformed kinetic term of the action is
\bea
I_\mathrm{kin} \left[ \bar\Psi , \Psi \right]
&=& \frac{T}{2 V^2} \sum_{K,P,Q} \int_{X,Y} \bar\Psi \left( K
\right) S^{-1} \left( Q \right) \Psi \left( P \right) \e^{ i K X
} \e^{ - i P Y } \e^{ - i Q \left( X - Y \right)} \nonumber \\
&=& \frac{T}{2 V^2} \sum_{K,P,Q} \int_X \e^{ i \left( K - Q
\right) X } \int_Y \e^{ i \left( Q - P \right) Y } \bar\Psi
\left( K \right) S^{-1} \left( Q \right) \Psi \left( P \right)
\nonumber \\
&=& \frac{T}{2 V^2} \sum_{K,P,Q} \frac{V}{T} \delta^{ \left( 4
\right) }_{K,Q} \frac{V}{T} \delta^{ \left( 4 \right) }_{Q,P}
\bar\Psi \left( K \right) S^{-1} \left( Q \right) \Psi \left( P
\right) \nonumber \\
&=& \frac12 \sum_K \bar\Psi \left( K \right) \frac{S^{-1} \left(
K \right)}{T} \Psi \left( K \right) \equiv \sum_{ \left( K, - K
\right) } \bar\Psi \left( K \right) \frac{S^{-1} \left( K
\right)}{T} \Psi \left( K \right) \; .
\eea
Therefore, the contribution of the kinetic term to the grand
partition function in energy-momentum space reads~\cite{Ryder}
\bea
\mathcal{Z}_\mathrm{kin} = \int \D \bar\psi \D \psi \exp \left\{
I_\mathrm{kin} \left[ \bar\Psi ,\Psi \right] \right\} \equiv
\tilde{\mathcal{N}} \det_{ \left( K, - K \right) } \left(
\frac{S^{-1}}{T} \right) = \tilde{\mathcal{N}} \left[ \det_K
\left( \frac{S^{-1}}{T} \right) \right]^{1/2} \; .
\eea
Up to irrelevant constants, the logarithm of the kinetic part of
the grand partition function is
\bea
\ln \mathcal{Z}_\mathrm{kin}
&\equiv& \ln \left[ \det_K \left( \frac{S^{-1}}{T} \right)
\right]^{1/2} = \frac12 \ln \det_K \left( \frac{S^{-1}}{T}
\right) = \frac12 \ln \left[ \prod_K \det \left(
\frac{S^{-1}}{T} \right) \right] \nonumber \\
&=& \frac12 \sum_K \ln \det \left( \frac{S^{-1}}{T} \right)
\equiv \frac12 \sum_K \Tr \ln \left( \frac{S^{-1}}{T} \right) \;
.
\eea
The relation in the last of line of this equation is proven in
Sec.~\ref{The_logarithm_of_the_determinant}.
\section{The logarithm of the determinant}
\label{The_logarithm_of_the_determinant}
The following relation is to be proven:
\be
\ln \det M = \Tr \ln M \; .
\ee
The quadratic matrix $M$ with dimension $d$ is diagonalizable if
a matrix $S$ exists, which fulfills the property~\cite{Fischer},
\be
S M S^{-1} = \diag \, \lambda_i \; ,
\ee
where $\lambda_i$ are the eigenvalues of the matrix $M$.
Therefore, one can write,
\be
\label{lndetTrlnlambda}
\ln \det M = \ln \det \, ( S M S^{-1} ) = \ln \prod_{i=1}^d
\lambda_i = \sum_{i=1}^d \ln \lambda_i = \Tr \ln \lambda_i \;
.
\ee
The logarithm can be expanded as
\be
\label{ln}
\ln \left( 1 + x \right) = \sum_{n=1}^\infty \frac{ \left( -1
\right)^{n-1} }{ n } \, x^n = x - \frac12 x^2 + \frac13 x^3 -
\frac14 x^4 \pm \ldots \; ,
\ee
so that Eq.~(\ref{lndetTrlnlambda}) becomes,
\bea
\ln \det M &=& \sum_{n=1}^\infty \frac{ \left( - 1 \right)^{n-1}
}{ n } \, \Tr \, \diag \left\{ \left( \lambda_i - 1 \right)^n
\right\} \nonumber \\
&=& \sum_{n=1}^\infty \frac{ \left( - 1 \right)^{n-1} }{ n } \,
\Tr \left\{ \left[ S \left( M - 1 \right) S^{-1} \right]^n
\right\} \nonumber \\
&=& \sum_{n=1}^\infty \frac{ \left( - 1 \right)^{n-1} }{ n } \,
\Tr \left[ S \left( M - 1 \right)^n S^{-1} \right] \; .
\eea
Since $\Tr \, ( A B ) = \Tr \, ( B A )$ and because of
Eq.~(\ref{ln}), one obtains that
\be
\ln \det M = \sum_{n=1}^\infty \frac{ \left( - 1 \right)^{n-1}
}{ n } \, \Tr \, \left( M - 1 \right)^n = \Tr \ln M \; .
\ee
\section{The Dirac trace}
\label{The_Dirac_trace}
By using the Minkowski space part of the gluon
propagator~(\ref{gluon_propagator}),
\be
D_{\mu\nu} = - \frac{g_{\mu\nu}}{\Lambda^2} \; ,
\ee
one obtains,
\be
\label{Dirac_trace_Appendix}
\Tr \, [ \mathcal{P}_c \Lambda^e \left( \fettu{k} \right)
\gamma^\mu \Lambda^{-e'} \left( \fettu{p} \right) \gamma^\nu ]
\, {D}_{\mu\nu} = \frac{ 1 }{ \Lambda^2 } \left\{ - \Tr \, [
\mathcal{P}_c \Lambda^e \left(\fettu{k}\right) \Lambda^{e'}
\left(\fettu{p}\right) ] + \Tr \, [ \mathcal{P}_c \Lambda^e
\left(\fettu{k}\right) \gamma^i \Lambda^{-e^\prime}
\left(\fettu{p}\right) \gamma^i ] \right\} \; .
\ee
Expanding the projectors yields,
\bea
\mathcal{P}_c \Lambda^e \left(\fettu{k}\right) \Lambda^{\pm e'}
\left(\fettu{p}\right)
&=& \textstyle \frac18 \displaystyle \, ( 1 \pm e' \gamma_0
\fett{\gamma} \cdot \hat{\fettu{p}} + e \gamma_0 \fett{\gamma}
\cdot \hat{\fettu{k}} \mp e e' \fett{\gamma} \cdot
\hat{\fettu{k}} \cdot \fett{\gamma} \cdot \hat{\fettu{p}}
\nonumber \\
&& \hspace{1mm} + \, c \gamma_5 \pm c e' \gamma_5 \gamma_0
\fett{\gamma} \cdot \hat{\fettu{p}} + c e \gamma_5 \gamma_0
\fett{\gamma} \cdot \hat{\fettu{k}} \mp c e e' \gamma_5
\fett{\gamma} \cdot \hat{\fettu{k}} \cdot \fett{\gamma} \cdot
\hat{\fettu{p}} ) \; .
\eea
With the relation,
\be
\label{relation_gamma}
- \fett{\gamma} \cdot \hat{\fettu{k}} \cdot \fett{\gamma}
\cdot \hat{\fettu{p}} = - \gamma^i \gamma^j \hat k^i \hat p^j
= - \textstyle \frac12 \displaystyle \left( \{ \gamma^i,
\gamma^j \} + [ \gamma^i, \gamma^j ] \right) \hat k^i \hat p^j =
\hat{\fettu{k}} \cdot \hat{\fettu{p}} - \textstyle \frac12
\displaystyle \, [ \gamma^i, \gamma^j ] \, \hat k^i \hat p^j \;
,
\ee
one computes the trace of the product of these projectors,
\be
\label{Tr1}
\Tr \, [ \mathcal{P}_c \Lambda^e \left(\fettu{k}\right)
\Lambda^{\pm e'} \left( \fettu{p} \right) ]
= \textstyle \frac12 \displaystyle \, ( 1 \pm ee'
\hat{\fettu{k}} \cdot \hat{\fettu{p}} ) \; .
\ee
Herewith, the first trace in Eq.~(\ref{Dirac_trace_Appendix}) is
solved. From the second trace in
Eq.~(\ref{Dirac_trace_Appendix}), one gets,
\bea
\Tr \, [ \mathcal{P}_c \Lambda^e \left(\fettu{k}\right) \gamma^i
\Lambda^{-e'} \left(\fettu{p}\right) \gamma^i ]
&=& \textstyle \frac12 \displaystyle \Tr \, [ \mathcal{P}_c
\Lambda^e \left( \fettu{k} \right) \, ( \gamma^i - e' \gamma^i
\gamma_0 \fett{\gamma} \cdot \hat{\fettu{p}} ) \gamma^i ]
\nonumber \\
&=& \textstyle \frac12 \displaystyle \Tr \, [ \mathcal{P}_c
\Lambda^e \left(\fettu{k}\right) \, ( \gamma^i + e' \gamma_0
\gamma^i \gamma^j \hat p^j ) \gamma^i ] \; ,
\eea
which can be rewritten as
\bea
&& \textstyle \frac12 \displaystyle \Tr \, [ \mathcal{P}_c
\Lambda^e \left(\fettu{k}\right) ( \gamma^i + e' \gamma_0 \{
- \gamma^j \gamma^i - 2 \delta_{ij} \} \hat p^j ) \gamma^i ]
\nonumber \\
&=& \textstyle \frac12 \displaystyle \Tr \, [ \mathcal{P}_c
\Lambda^e \left(\fettu{k}\right) ( \gamma^i \gamma^i - e'
\gamma_0 \fett{\gamma} \cdot \hat{\fettu{p}} \, \gamma^i
\gamma^i - 2 e' \gamma_0 \delta_{ij} \hat p^j \gamma^i ) ] \; .
\eea
Summing over all $i$ yields,
\be
\Tr \, [ \mathcal{P}_c \Lambda^e \left(\fettu{k}\right) \, ( - 3
\Lambda^{-e'} \left( \fettu{p} \right) - e' \gamma_0
\fett{\gamma} \cdot \hat{\fettu{p}} ) ]
= - 3 \, \Tr \, [ \mathcal{P}_c \Lambda^e \left( \fettu{k}
\right) \Lambda^{ - e'} \left(\fettu{p}\right) ]
- \Tr \, [ \mathcal{P}_c \Lambda^e \left( \fettu{k} \right) e'
\gamma_0 \fett{\gamma} \cdot \hat{\fettu{p}} ] \; .
\ee
The first of these two traces is already known by
Eq.~(\ref{Tr1}). The second one is determined by using the
relation~(\ref{relation_gamma}),
\bea
&& \Tr \, [ \mathcal{P}_c \Lambda^e \left(\fettu{k}\right) e'
\gamma_0 \fett\gamma \cdot \hat{\fettu{p}} ] \nonumber \\
&=& \textstyle \frac14 \displaystyle \Tr \, [ ( 1 + c
\gamma_5 ) \, ( 1 + e \gamma_0 \fett\gamma \cdot \hat{ \fettu{k}
} ) e' \gamma_0 \fett\gamma \cdot \hat{\fettu{p}} ] \nonumber \\
&=& \textstyle \frac14 \displaystyle \Tr \, [ ( 1 + e
\gamma_0 \fett \gamma \cdot \hat{\fettu{k}} + c \gamma_5 + c e
\gamma_5 \gamma_0 \fett\gamma \cdot\hat{\fettu{k}} ) e' \gamma_0
\fett\gamma \cdot \hat{\fettu{p}} ] \nonumber \\
&=& \textstyle \frac14 \displaystyle \Tr \, [ ( e' \gamma_0
\fett\gamma \cdot \hat{\fettu{p}} - e e' \fett{\gamma} \cdot
\hat{\fettu{k}} \cdot \fett{\gamma} \cdot \hat{\fettu{p}} + c e'
\gamma_5 \gamma_0 \fett\gamma \cdot \hat{\fettu{p}} - c e e'
\gamma_5 \fett{\gamma} \cdot \hat{\fettu{k}} \cdot \fett{\gamma}
\cdot \hat{\fettu{p}} ) ] \nonumber \\
&=& e e' \hat{\fettu{k}} \cdot \hat{\fettu{p}} \; .
\eea
Herewith, one obtains the second trace in
Eq.~(\ref{Dirac_trace_Appendix}),
\be
\Tr \, [ \mathcal{P}_c \Lambda^e \left(\fettu{k}\right) \gamma^i
\Lambda^{ - e' } \left(\fettu{p}\right) \gamma^i ] = -
\textstyle \frac12 \displaystyle \, ( 3 - e e' \hat{\fettu{k}}
\cdot \hat{\fettu{p}} ) \; .
\ee
Therefore, one calculates for the Dirac
trace~(\ref{Dirac_trace_Appendix}),
\be
\Tr \, [ \mathcal{P}_c \Lambda^e \left(\fettu{k}\right)
\gamma^\mu \Lambda^{ - e'} \left(\fettu{p}\right) \gamma^\nu ]
\, {D}_{\mu\nu} \left( K - P \right) = - \frac{ 2 }{ \Lambda^2 }
\; .
\ee
\section{The trace of the logarithm}
\label{The_trace_of_the_logarithm}
The relation,
\be
\Tr \ln \sum_{i=1}^m a_i \mathcal{P}_i = \sum_{i=1}^m
\ln a_i \Tr \, \mathcal{P}_i \; ,
\ee
is to be proven. Each trace in this section is taken over a $d
\times d$ matrix. The $\mathcal{P}_i$ are projectors, which
fulfill the properties of completeness, $\sum_i \mathcal{P}_i =
1$, and orthogonality, $\mathcal{P}_i \mathcal{P}_j =
\delta_{ij} \mathcal{P}_j$, and the $a_i$ are factors, where $1
\le i \le m$ and $1 \le j \le m$. Because of the completeness of
the projectors, one gets,
\be
\Tr \ln \sum_{i=1}^m a_i \mathcal{P}_i
= \Tr \ln \left( a_1 \mathcal{P}_1 + \sum_{i=2}^m a_i
\mathcal{P}_i \right)
= \Tr \ln \left[ a_1 \left( 1 - \sum_{i=2}^m
\mathcal{P}_i \right) + \sum_{i=2}^m a_i \mathcal{P}_i
\right] \; ,
\ee
which can be rewritten as
\bea
\label{Trlnappendix}
\Tr \ln \sum_{i=1}^m a_i \mathcal{P}_i
&=&\Tr \ln \left[ a_1 + \sum_{i=2}^m \left( a_i - a_1
\right) \mathcal{P}_i \right] \nonumber \\
&=&\Tr \ln \left\{ a_1 \left[ 1 + \sum_{i=2}^m \left(
\frac{a_i}{a_1} - 1 \right) \mathcal{P}_i \right] \right\}
\nonumber \\
&=&\Tr \ln a_1 + \Tr \ln \left[ 1 + \sum_{i=2}^m \left(
\frac{a_i}{a_1} - 1 \right) \mathcal{P}_i \right] \; .
\eea
With the expansion of the logarithm,
\be
\ln \left( 1 + x \right) = \sum_{n=1}^\infty \frac{ \left( -1
\right)^{n-1} }{ n } \, x^n \; ,
\ee
Eq.~(\ref{Trlnappendix}) reads,
\bea
\Tr \ln \sum_{i=1}^m a_i \mathcal{P}_i
&=&d \ln a_1 + \Tr \sum_{n=1}^\infty \frac{ \left( -1
\right)^{n-1} }{ n } \left[ \sum_{i=2}^m \left( \frac{ a_i
}{ a_1 } - 1 \right) \mathcal{P}_i \right]^n \nonumber \\ 
&=&\ln a_1^d + \Tr \sum_{n=1}^\infty \frac{ \left( -1
\right)^{n-1} }{ n } \sum_{i=2}^m \left( \frac{ a_i }{ a_1 }
- 1 \right)^n \mathcal{P}_i \nonumber \\ 
&=&\ln a_1^d + \sum_{n=1}^\infty \frac{ \left( -1 \right)^{n-1}
}{ n } \sum_{i=2}^m \left( \frac{ a_i }{ a_1 } - 1 \right)^n
\Tr \, \mathcal{P}_i \; .
\eea
This equation can be rewritten as
\bea
\Tr \ln \sum_{i=1}^m a_i \mathcal{P}_i
&=& \ln a_1^d + \sum_{i=2}^m \Tr \, \mathcal{P}_i \ln \frac{
a_i }{ a_1 } \nonumber \\
&=& \ln a_1^d + \sum_{i=2}^m \ln \left[ \left( \frac{ a_i }{
a_1 } \right)^{\Tr \, \mathcal{P}_i} \right]
= \ln \left[ a_1^d \prod_{i=2}^m \left( \frac{ a_i }{ a_1 }
\right)^{\Tr \, \mathcal{P}_i} \right] \; .
\eea
Further calculations yield
\bea
\Tr \ln \sum_{i=1}^m a_i \mathcal{P}_i
&=&\ln \left( a_1^d a_1^{- \sum_{i=2}^m \Tr \, \mathcal{P}_i}
\prod_{i=2}^m a_i^{\Tr \, \mathcal{P}_i} \right) \nonumber \\
&=&\ln \left( a_1^d a_1^{- \Tr \sum_{i=2}^m \mathcal{P}_i}
\prod_{i=2}^m a_i^{\Tr \, \mathcal{P}_i} \right) \nonumber \\
&=&\ln \left( a_1^d a_1^{- \Tr \, \left( 1 - \mathcal{P}_1
\right)} \prod_{i=2}^m a_i^{\Tr \, \mathcal{P}_i} \right) \;
.
\eea
Finally, one obtains,
\be
\Tr \ln \sum_{i=1}^m a_i \mathcal{P}_i
= \ln \left( a_1^d a_1^{-d} a_1^{\Tr \, \mathcal{P}_1}
\prod_{i=2}^m a_i^{\Tr \, \mathcal{P}_i} \right)
= \ln \prod_{i=1}^m a_i^{\Tr \, \mathcal{P}_i}
= \sum_{i=1}^m \ln a_i \Tr \, \mathcal{P}_i \; .
\ee
\section{Cubic equations}
\label{Cubic_equations}
In this section, I show Cardano's formulae which are the roots
of the general cubic equation,
\be
\label{general_cubic_equation}
A_3 x^3 + A_2 x^2 + A_1 x + A_0 = 0 \; ,
\ee
where $A_3 \neq 0$~\cite{numrec, cubic_equations,
Taschenbuch_Mathe}.

Recall that a polynomial of degree $n$ has $n$ roots. The roots
can be real or complex, and they might not be distinct. If the
coefficients of the polynomial are real, then complex roots will
occur in pairs that are complex conjugates, i.e., if $x_1 = a +
ib$ is a root then $x_2 = a - ib$ will also be a root. When the
coefficients are complex, the complex roots need not be related.

The solution to the cubic (as well as the quartic) equation was
published by Gerolamo Cardano (1501--1576) in his treatise
\textit{Ars Magna}. However, Cardano was not the original
discoverer of either of these results. The hint for the cubic
equation had been provided by Niccol\`o Tartaglia, while the
quartic equation had been solved by Ludovico Ferrari. However,
Tartaglia himself had probably caught wind of the solution from
another source. The solution was apparently first arrived at by
a little-remembered professor of mathematics at the University
of Bologna by the name of Scipione del Ferro (ca.\ 1465--1526).
While del Ferro did not publish his solution, he disclosed it to
his student Antonio Maria Fior~\cite{Boyer_Merzbach}. This is
apparently where Tartaglia learned of the solution around 1541.

The general cubic equation always can be brought into the normal
form,
\be
x^3 + a_2 x^2 + a_1 x + a_0 = 0 \; ,
\ee
where
\be
a_2 \equiv \frac{A_2}{A_3} \; , \qquad a_1 \equiv
\frac{A_1}{A_3} \; , \qquad a_0 \equiv \frac{A_0}{A_3} \; .
\ee
With the substitution $x = y + s$, the cubic equation in normal
form assumes the form,
\be
y^3 + \left( 3 s + a_2 \right) y^2 + \left( 3 s^2 + 2 a_2 s +
a_1 \right) y + s^3 + a_2 s^2 + a_1 s + a_0 = 0 \; .
\ee
By setting $s = -a_2 / 3$ the quadratic term disappears.
Thereby, one obtains with the substitution,
\be
\label{substitution_s}
x = y - \frac{ a_2 }{ 3 } \; ,
\ee
the cubic equation in standard form,
\be
y^3 + q y + r = 0 \; ,
\ee
where
\bsub
\bea
q &=& a_1 - \frac13 a_2^2 \; , \\
r &=& \frac{ 2 }{ 27 } a_2^3 - \frac13 a_2 a_1 + a_0 \; .
\eea
\esub
The cubic equation in standard form can be rewritten as
\be
\label{cubic_equation}
y^3 + 3 Q y - 2 R = 0 \; ,
\ee
where
\be
Q \equiv \frac{q}{3} \; , \qquad R \equiv - \frac{r}{2} \; .
\ee
Let $B$ and $C$ be, for the moment, arbitrary constants. The
cubic equation in standard form~(\ref{cubic_equation}) can be
decomposed into a linear and a quadratic term,
\be
\label{linear_quadratic}
( y - B ) \, ( y^2 + B y + C ) = y^3 + y \, ( C - B^2 ) - B C =
0 \; ,
\ee
so that one easiliy recognizes by comparing the coefficients
that
\bea
\label{3Q}
3 Q &=& C - B^2 \; , \\
\label{2R}
2 R &=& B C \; .
\eea
Solving Eq.~(\ref{3Q}) for $C$ yields,
\be
\label{expression_for_C}
C = B^2 + 3 Q \; .
\ee
Inserting this result into Eq.~(\ref{2R}) gives,
\be
B^3 + 3 Q B = 2 R \; .
\ee
Therefore, if one can find an expression for $B$ which satisfies
this identity, one has factored a linear term from the cubic
equation, thus reducing it to a quadratic equation. The trial
solution accomplishing this miracle turns out to be the
symmetrical expression,
\be
B \equiv S^+ + S^- \; ,
\ee
where
\be
S^\pm \equiv \sqrt[3]{R \pm \sqrt{D}} \; .
\ee
The discriminant is given by
\be
D \equiv R^2 + Q^3 \; .
\ee
Now, the remaining quadratic equation has to be solved.
Inserting Eq.~(\ref{expression_for_C}) into the quadratic part
of Eq.~(\ref{linear_quadratic}) leads to
\be
y^2 + B y + B^2 + 3 Q = 0 \; .
\ee
Solving this quadratic equation gives the solutions,
\be
y = - \frac{B}{2} \pm \frac12 \sqrt{3} \, i \sqrt{ B^2 + 4 Q} \;
.
\ee
With the definition,
\be
A \equiv S^+ - S^- \; ,
\ee
one can find out that
\be
A^2 = B^2 + 4 Q \; .
\ee
Therefore, with the substitution~(\ref{substitution_s}), the
roots of the general cubic
equation~(\ref{general_cubic_equation}), which are called
Cardano's formulae, are given by
\bsub
\bea
x_1 &=& - \frac{a_2}{3} + B \; , \\
x_2 &=& - \frac{a_2}{3} - \frac{B}{2} + \frac12 i \sqrt{3} A \;
, \\
x_3 &=& - \frac{a_2}{3} - \frac{B}{2} - \frac12 i \sqrt{3} A \;
.
\eea
\esub
In the case of real-valued coefficients in the general cubic
equation~(\ref{general_cubic_equation}), one can distinguish the
following three cases:
\begin{description}
\item[$\fett{D > 0}$:] one real and two complex solutions,
\be
x_1 \in \mathbb{R} \; , \qquad x_2, x_3 \in \mathbb{C} \; ,
\qquad  x_3 = x_2^* \; ,
\ee
\item[$\fett{D = 0}$:] three real solutions,
\begin{description}
\item[$\fett{R \neq 0}$:] one single and one double solution,
\be
x_1 = - \frac{a_2}{3} + 2 \sqrt[3]{R} \; , \qquad x_2 = x_3 = -
\frac{a_2}{3} - \sqrt[3]{R} \; ,
\ee
\item[$\fett{R = 0}$:] one triple solution,
\be
x_1 = x_2 = x_3 = - \frac{a_2}{3} \; ,
\ee
\end{description}
\item[$\fett{D < 0}$:] \textit{casus irreducibilis}, three
different real solutions,
\be
x_i = 2 \sqrt{-Q} \cos \left( \frac{\varphi + 2 \left( i - 1
\right) \pi}{3} \right) - \frac{a_2}{3} \; ,
\ee
where
\be
\cos \varphi \equiv \frac{R}{\sqrt{-Q^3}} \; .
\ee
\end{description}
The equations for $D < 0$ in which $i = 1,2,3$ first appear in
Chapter~VI of Fran\c{c}ois Vi\`ete's treatise \textit{De
emendatione}, published in 1615.
\section{Summation over the fermionic Matsubara frequencies}
\label{Summation_over_the_fermionic_Matsubara_frequencies}
\begin{figure}[H]
\begin{center}
\makebox[0.99\textwidth][l]{
\includegraphics[width=0.47\textwidth]{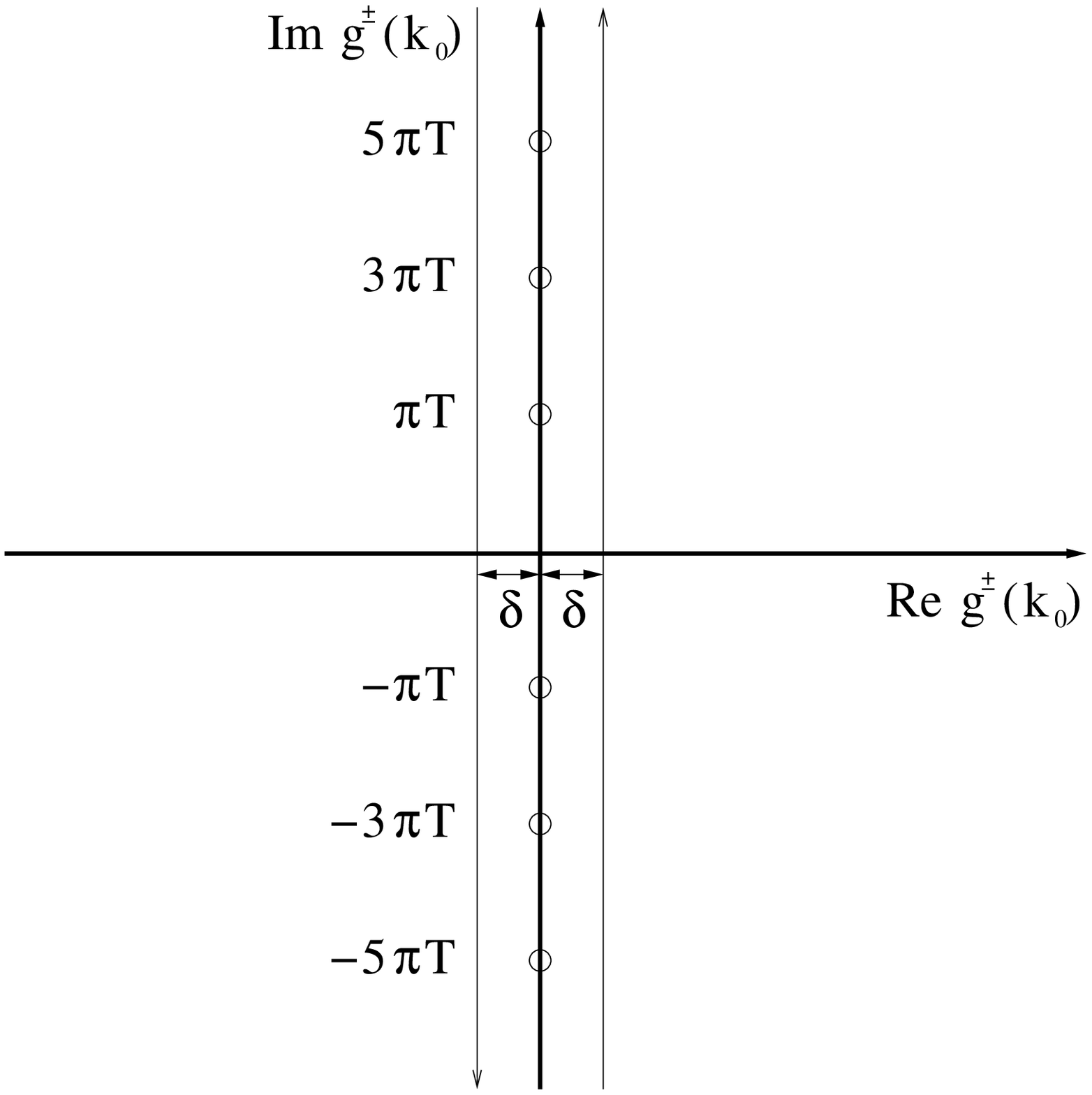}
\hspace{0.005\textwidth}
\includegraphics[width=0.47\textwidth]{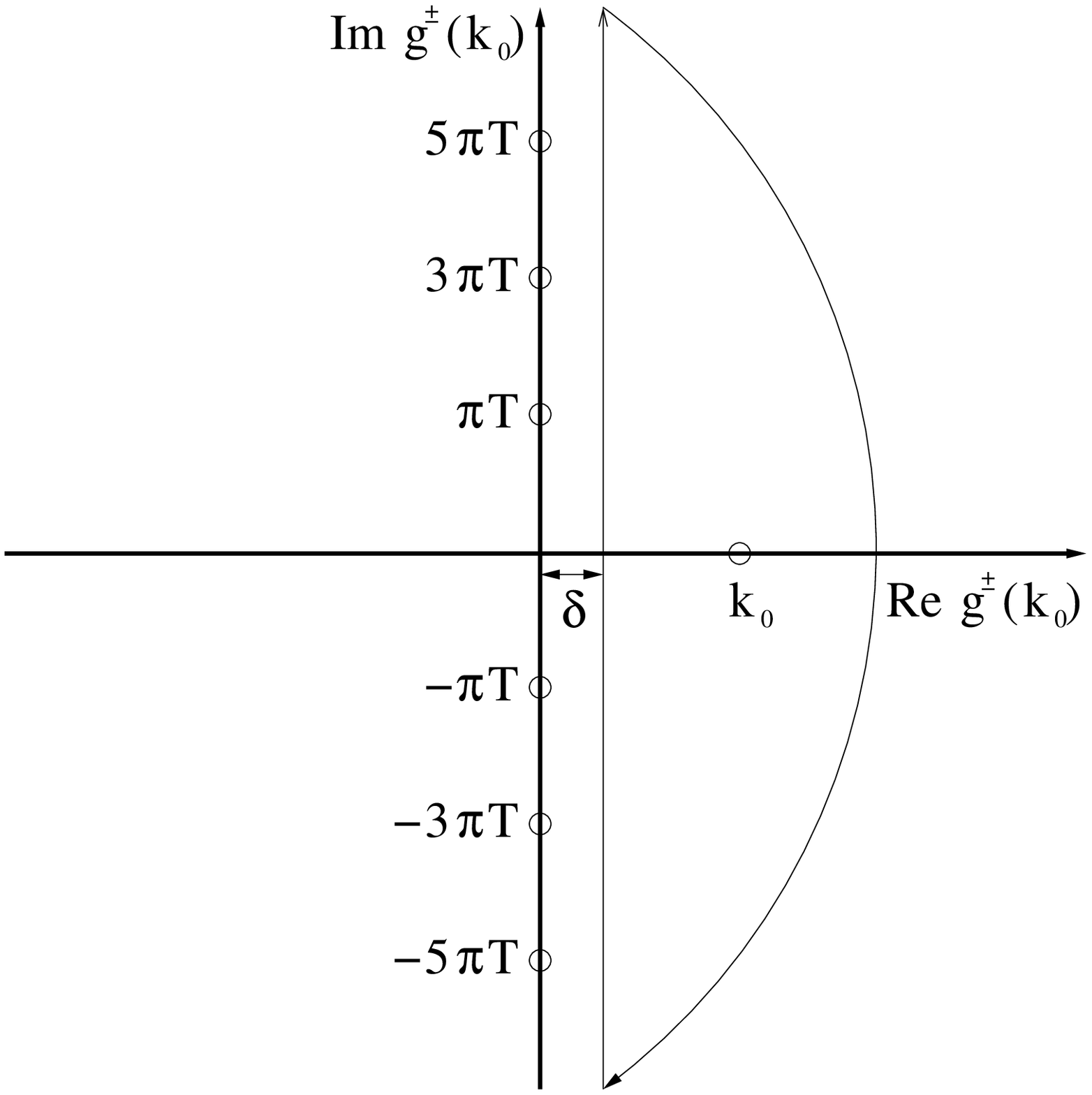}}
\caption[Integration over the fermionic poles.]{Integration over
the fermionic poles. Left panel: integration over the fermionic
poles on the imaginary axis. Right panel: integration over the
fermionic pole on the positive real axis.}
\label{residua}
\end{center}
\end{figure}

The following kinetic term of a fermionic logarithmic grand
partition function is given:
\be
\ln \mathcal{Z}_\mathrm{kin} = \sum_K \ln \left[ \frac{ ( k_0 -
\delta\mu )^2 - \epsilon^2 }{ T^2 } \right] \; ,
\ee
which can be split into two parts,
\be
\ln \mathcal{Z}_\mathrm{kin} = \sum_K \left[ \ln \left( \frac{
k_0 - \delta\mu + \epsilon }{ T } \right) + \ln \left( \frac{
k_0 - \delta\mu - \epsilon }{ T } \right) \right] \; .
\ee
This is up to irrelevant constants,
\be
\label{gk0}
\ln \mathcal{Z}_\mathrm{kin} = \sum_K \left( \int_1^{b^-} \frac{
\ud x }{ \frac{ k_0 }{ T } + x } - \int_1^{b^+} \frac{ \ud x }{
\frac{ k_0 }{ T } - x } \right) \; ,
\ee
where
\be
b^\pm = \frac{ \epsilon \pm \delta\mu }{ T } \; .
\ee
One uses the relation,
\be
T \sum_n g^\pm \left (k_0 \right) = \frac{ 1 }{ 4 \pi i } \oint
\ud k_0 \tanh \left( \frac{ k_0 }{ 2 T } \right) g^\pm \left(
k_0 \right) \; ,
\ee
where
\be
\label{gpm}
g^\pm \left( k_0 \right) = \frac{ \e^{-\eta x} }{ k_0 \pm T x }
\; ,
\ee
and integrates over the fermionic poles,
\be
k_0 = - i \left( 2 n + 1 \right) \pi T = - i \omega_n \; ,
\ee
on the imaginary axis. The $\omega_n$ are called fermionic
Matsubara frequencies. The factor $\e^{-\eta x}$ in
Eq.~(\ref{gpm}) is needed as a damping factor so that the sum
over all $n$ does not diverge, $\eta > 0$. In the end of the
calculation, one sets $\eta = 0$. Because of the symmetry around
the imaginary axis, see left panel in Fig.~\ref{residua}, one
obtains,
\be
T \sum_n g^\pm \left( k_0 \right) = \frac{ 1 }{ 4 \pi i }
\oint_{-i \infty + \delta}^{+i \infty + \delta} \ud k_0 \tanh
\left( \frac{ k_0 }{ 2 T } \right) \left[ g^\pm \left( k_0
\right) + g^\pm \left( -k_0 \right) \right] \; .
\ee
With the relation,
\be
\tanh \left( \frac{x}{2} \right) = 1 - 2 n_F \left( x \right) \;
,
\ee
where
\be
n_F \left( x \right) = \frac{1}{\e^x + 1}
\ee
is the Fermi-Dirac distribution function, one gets
\be
\label{contour_integration}
T \sum_n g^\pm \left( k_0 \right) = \frac{ 1 }{ 2 \pi i }
\oint_{-i \infty + \delta}^{+i \infty + \delta} \ud k_0 \left[
\frac12 - n_F \left( \frac{ k_0 }{ T } \right) \right] \left(
\frac{ \e^{ - \eta x} }{ k_0 \pm T x } + \frac{ \e^{ - \eta x}
}{ - k_0 \pm T x } \right) \; .
\ee
In this equation, one uses the theorem of residues of first
order,
\be
\frac{ 1 }{ 2 \pi i } \oint f \left( z \right) = \sum_i \lim_{z
\rightarrow z_i} \left[ f \left( z \right) \left( z - z_i
\right) \right] = \sum_i \Res \, f \left( z \right)
\Big|_{z = z_i} \; .
\ee
In this connection, one has to note that contour integrations
are done anticlockwise. But the contour integration in
Eq.~(\ref{contour_integration}) has to be done clockwise, see
right panel in Fig.~\ref{residua}. Therefore, the result has to
be multiplied by minus one. One also has to note that the
respective pole is on the positive real axis. Therefore, an
extra Heaviside function has to be included in the result,
\bsub
\bea
\Res \, g^\pm \left( k_0 \right)
\big|_{k_0 = T x} &=& - \Heaviside \left( T x \right) \left[
\frac12 - n_F \left( x \right) \right] \nonumber \\
&& \times \left[ \frac{\e^{-\eta x}}{k_0 \pm T x} \left( k_0 - T
x \right) + \frac{\e^{-\eta x}}{-k_0 \pm T x} \left( k_0 - T x
\right) \right] \Bigg|_{k_0 = T x} \nonumber \\
&\stackrel{\eta \rightarrow 0}{=}& \pm \Heaviside \left( T x
\right) \left[ \frac12 - n_F \left( x \right) \right]
= \pm \frac12 \tanh \left( \frac{x}{2} \right) \Heaviside \left(
T x \right) \; , \\
\Res \, g^\pm \left( k_0 \right)
\big|_{k_0 = - T x} &=& - \Heaviside \left( - T x \right) \left[
\frac12 - n_F \left( -x \right) \right] \nonumber \\
&& \times \left[ \frac{\e^{-\eta x}}{k_0 \pm T x} \left( k_0 + T
x \right) + \frac{\e^{-\eta x}}{-k_0 \pm T x} \left( k_0 + T x
\right) \right] \Bigg|_{k_0 = - T x} \nonumber \\
&\stackrel{\eta \rightarrow 0}{=}& \mp \Heaviside \left( - T x
\right) \left[ \frac12 - n_F \left( -x \right) \right]
= \mp \frac12 \tanh \left( -\frac{x}{2} \right) \Heaviside
\left( -T x \right) \nonumber \\
&=& \pm \frac12 \tanh \left( \frac{x}{2} \right) \Heaviside
\left( - T x \right) \; .
\eea
\esub
Summing up the two residua for each sign in
Eq.~(\ref{contour_integration}) leads to the result,
\be
T \sum_n g^\pm \left( k_0 \right) = \pm \left[ \frac12 - n_F
\left( x \right) \right] \; .
\ee
Therefore, one obtains for Eq.~(\ref{gk0})
\be
\ln \mathcal{Z}_\mathrm{kin} = \sum_\fettu{k} \left\{
\int_1^{b^-} \ud x \left[ \frac12 - n_F \left( x \right) \right]
+ \int_1^{b^+} \ud x \left[ \frac12 - n_F \left( x \right)
\right] \right\} \; .
\ee
Up to irrelevant constants, one computes,
\be
\ln \mathcal{Z}_\mathrm{kin} = \sum_\fettu{k} \left\{ \frac{
\epsilon }{ T } + \ln \left[ 1 + \exp \left( - \frac{ \epsilon -
\delta\mu }{ T } \right) \right] + \ln \left[ 1 + \exp \left( -
\frac{ \epsilon + \delta\mu }{ T } \right) \right] \right\} \; .
\ee
\selectlanguage{german}
\chapter{Zusammenfassung}
Das Phasendiagramm von neutraler Quarkmaterie war, bis ich mit
meinen Forschungen auf diesem Gebiet im Jahr 2003 begonnen
hatte, nur unzureichend bekannt,
siehe Abb.~\ref{phase_diagram_schematic}. F\"ur meine
Dissertation war dieses Thema daher besonders gut geeignet.
Meine Aufgabe war es, das Phasendiagramm von neutraler
Quarkmaterie zu erstellen.

Im Kapitel~\ref{Introduction} wird man in das Themengebiet
Quarkmaterie, Farbsupraleitung und Sterne eingef\"uhrt.
Neutronensterne sind in der Natur die wohl wahrscheinlichsten
Pl\"atze, in denen farbsupraleitende Quarkmaterie vorkommen
k\"onnte. Daher gehe ich in Kapitel~\ref{Introduction} nicht nur
auf die wichtigsten Phasen der Farbsupraleitung, sondern auch
auf das Thema Sternentwicklung ein, um dann schlie{\ss}lich die
Entstehung von Neutronensternen zu erkl\"aren, deren Kerne
m\"oglicherweise aus sehr dichter, neutraler, farbsupraleitender
Quarkmaterie bestehen. Bei extrem hohen Dichten, wie sie
wahrscheinlich in Neutronensternkernen vorkommen, herrscht eine
attraktive Wechselwirkung in Quarkmaterie vor, die die Bildung
von Quark-Cooper-Paaren bewirkt~\cite{Barrois, Frautschi,
Bailin_Love}. Dies verursacht Farbsupraleitung. Da Quarks als
Spin-$\frac12$ Fermionen in verschiedener Farbe und Flavor
vorkommen, gibt es viele M\"oglichkeiten, um Quark-Cooper-Paare
zu bilden. Daher gibt es viele unterschiedliche
farbsupraleitende Quarkphasen.

In der \textbf{2}-flavor
color-\textbf{s}uper\textbf{c}onducting (2SC)-Phase paaren sich
rote up mit gr\"unen down Quarks und rote down mit gr\"unen up
Quarks und bilden antiblaue Quark-Cooper-Paare. Die blauen up
und down Quarks bleiben ungepaart und sind daher gaplose
Quasiteilchen, die einen gro{\ss}en Beitrag zur spezifischen
W\"arme, zur elektrischen Leitung und zur W\"armeleitung
leisten. Die gaplosen blauen Quarks sind au{\ss}erdem f\"ur eine
starke Neutrinoemission verantwortlich, die durch
$\beta$-Prozesse verursacht wird. 

Die Beitr\"age der vier gegappten Quasiteilchen zu allen
Transport- und vielen thermodynamischen Gr\"o{\ss}en ist bei
niedrigen Temperaturen durch den exponentiell kleinen Faktor
$\exp \left( - \Delta / T \right)$
unterdr\"uckt~\cite{lecturesShovkovy}. Die Gluonen sind Bosonen,
und daher ist deren Teilchenzahldichte bei niedrigen
Temperaturen klein. In der 2SC-Phase ist die $[ SU(3)_c ]$
Eichsymmetrie zu $[ SU(2)_c ]$ gebrochen. (Eichsymmetrien sind
durch eckige Klammern gekennzeichnet.) Daher gibt es in der
2SC-Phase $8 - 3 = 5$ gebrochene Generatoren. Diese f\"uhren zu
f\"unf massiven Gluonen in der 2SC-Phase. Daher haben diese
Gluonen nur geringe Einfl\"usse auf die Eigenschaften der
Quarkmaterie in der 2SC-Phase. Die ungepaarten blauen Quarks
sind f\"ur die Abwesenheit von Baryonsuperfluidit\"at
verantwortlich. Nur die antiblauen Quasiteilchen tragen eine von
Null verschiedene Baryonenzahl. In der 2SC-Phase existiert kein
elektromagnetischer Mei{\ss}nereffekt, und das ist der Grund
daf\"ur, warum ein Magnetfeld aus der farbsupraleitenden Region
nicht herausgedr\"angt werden w\"urde. Die 2SC-Phase ist ein
sogenannter $\tilde Q$-Leiter, weil ihre elektrische
Leitf\"ahigkeit aufgrund der ungepaarten blauen up Quarks
gro{\ss} ist.

Wenn das chemische Potential der strange Quarks die strange
Quarkmasse \"ubersteigt, dann treten auch die strange Quarks in
der Quarkmaterie bei der Temperatur $T = 0$ auf. Dadurch ist es
m\"oglich, da{\ss} auch die strange Quarks bei der Bildung von
Quark-Cooper-Paaren beteiligt sind. Das Kondensat bricht $[
SU(3)_c ] \times SU(3)_{r + \ell}$ zur vektoriellen Untergruppe
$SU(3)_{c + r + \ell}$ und ist noch immer invariant unter
Vektortransformationen im Farb- und Flavorraum. Das bedeutet,
da{\ss} eine Transformation in Farbe eine gleichzeitige
Transformation in Flavor voraussetzt, um die Invarianz des
Kondensates zu bewahren. Deswegen nannten die
Entdecker~\cite{CFL_discoverers} diese Phase auch
\textbf{c}olor--\textbf{f}lavor-\textbf{l}ocked (CFL)-Phase. 

Bei hohen Dichten und niedrigen Temperaturen ist die CFL-Phase
der wahre Grundzustand von Quarkmaterie, da sie den h\"ochsten
Druck hat und weil alle Quarks gegappt sind. Im Gegensatz zur
2SC-Phase mit ihren Antitriplettgaps $\Delta$ existieren in der
CFL-Phase neben ihren Antitriplettgaps $\Delta_{\left( \bar 3,
\bar 3 \right)}$ auch symmetrische Sextettgaps $\Delta_{\left(
6,6 \right)}$. In den Quasiteilchenspektra treten sowohl der
Singlettgap $\Delta_1$, als auch der Oktettgap $\Delta_2$ auf.
Wenn man den kleinen repulsiven Sextettgap vernachl\"assigt,
dann findet man heraus, da{\ss} $\Delta_1 = 2 \Delta_2 \equiv 2
\Delta$ ist. In der CFL-Phase existieren keine gaplosen Quarks.

Die Beitr\"age der Quasiteilchen zu allen Transport- und vielen
thermodynamischen Gr\"o{\ss}en ist bei niedrigen Temperaturen
durch den exponentiell kleinen Faktor $\exp \left( - \Delta / T
\right)$ unterdr\"uckt~\cite{lecturesShovkovy}. Der Einflu{\ss}
der Gluonen ist vernachl\"assigbar, da in der CFL-Phase alle
Gluonen gegappt sind. Im Gegensatz zur 2SC-Phase ist die
CFL-Phase superfluid, weil die $U(1)_B$ Baryonzahlsymmetrie
gebrochen ist. Aber die CFL-Phase besitzt eine ungebrochene $[
U(1)_\mathrm{em} ]$ Eichsymmetrie und ist daher, wie die
2SC-Phase, kein elektromagnetischer Supraleiter. Daher dr\"angt
die CFL-Phase auch kein Magnetfeld aus ihrem Inneren heraus. Die
CFL-Phase ist ein $\tilde Q$-Isolator, da alle Quarks gegappt
sind und keine elektrische Ladung verbleibt, wie das bei den
ungegappten blauen up Quarks in der 2SC-Phase der Fall ist.
Daher ist die CFL-Phase elektrisch ladungsneutral. Bei $T = 0$
sind in der CFL-Phase keine Elektronen
vorhanden~\cite{enforced_neutrality}. Bei niedrigen Temperaturen
wird die elektrische Leitf\"ahigkeit in der CFL-Phase durch
thermisch angeregte Elektronen und Positronen gew\"ahrleistet.
Ein Phasen\"ubergang zu einer CFL-Phase mit einem
Mesonkondensat ist m\"oglich, falls $M_s \gtrsim M_u^{1/3}
\Delta^{2/3}$~\cite{Schaefer, Hong, Bedaque, Kaplan_Reddy,
Kryjevski_Kaplan_Schaefer}, wobei $M_u$ und $M_s$ die up
bzw.\ strange Quarkmasse sind.

Die 2SC- und die CFL-Phase sind farbsupraleitende Phasen, deren
Kondensate einen Spin $J = 0$ haben. Die Bildung von
Quark-Cooper-Paaren bei $J = 0$ f\"ur nur einen Flavor ist wegen
des Pauliprinzips nicht erlaubt. Aber dies ist m\"oglich bei $J
= 1$, d.h.\ bei Spin-Eins-Farbsupraleitung.
Spin-Eins-Farbsupraleitung ist jedoch viel schw\"acher als
Spin-Null-Farbsupraleitung. W\"ahrend der Gap in den Phasen der
Spin-Null-Farbsupraleitung ca.\ 100~MeV gro{\ss} ist, betr\"agt
er in den Phasen der Spin-Eins-Farbsupraleitung nur ca.\
100~keV. Solch ein kleiner Gap hat keine gro{\ss}en Einfl\"usse
auf Transport- und viele thermodynamische Eigenschaften von
Quarkmaterie~\cite{lecturesShovkovy}. Spin-Eins-Farbsupraleitung
ist weniger energetisch bevorzugt als
Spin-Null-Farbsupraleitung, da die letztere einen h\"oheren
Druck besitzt. Daher ist es auch nicht zu erwarten, da{\ss}
Spin-Eins-Farbsupraleitung im Phasendiagramm von neutraler
Quarkmaterie auftritt. Jedoch k\"onnte
Spin-Eins-Farbsupraleitung bevorzugt werden, wenn der
Unterschied zwischen den Fermioberfl\"achen bei
Spin-Null-Farbsupraleitung zu gro{\ss} ist.

Die wichtigsten Phasen der Spin-Eins-Farbsupraleitung sind die
A-Phase, die
\textbf{c}olor--\textbf{s}pin-\textbf{l}ocked (CSL)-Phase, die
polare Phase und die planare Phase. In
Spin-Eins-Farbsupraleitern kann, im Gegensatz zu
Spin-Null-Farbsupraleitern, ein elektromagnetischer
Mei{\ss}nereffekt auftreten. Dies ist z.B.\ der Fall in der
CSL-Phase. Die am meisten energetisch bevorzugte Phase der
Spin-Eins-Farbsupraleitung ist die transverse CSL-Phase, da sie
den h\"ochsten Druck hat~\cite{Schmitt}.

Sterne entstehen in interstellaren Gaswolken. Sie sind nukleare
Fusionsfabriken, da in ihnen leichtere Kerne zu schwereren
Kernen verbrannt werden. Nachdem alle Fusionsreaktionen
abgeschlossen sind, werden Sterne entweder zu Wei{\ss}en
Zwergen, Neutronensternen oder Schwarzen L\"ochern. Ein
Neutronenstern entsteht, wenn ein Roter \"Uberriese seine
Fusionsreaktionen beendet hat. Dann kollabiert dieser, weil der
Innendruck fehlt und daher die Gravitation \"uberwiegt, bis
schlie{\ss}lich der Druck von entarteten Neutronen den Kollaps
zum Erliegen bringt. Die \"au{\ss}eren Schichten fallen auf den
entstandenen Protoneutronenstern, prallen zur\"uck und erzeugen
dadurch eine Supernova vom Typ~II. Durch den Kollaps wird eine
riesige Anzahl von Neutrinos produziert, die im
Protoneutronenstern zun\"achst gefangen sind. Dies wird als
Neutrinotrapping bezeichnet. Der zun\"achst hei{\ss}e
Protoneutronenstern k\"uhlt durch Neutrinoemission und sp\"ater
durch Photoemission ab. Neutronensterne werden als Pulsare
detektiert. Diese senden wegen des sogenannten Leuchtturmeffekts
sehr pr\"azise Radiopulse aus und besitzen ein starkes
Magnetfeld. Ihre Rotationsperiode liegt im Millisekunden- bis
Sekundenbereich.

Neutronensterne bestehen nicht etwa, wie der Name vermuten
l\"a{\ss}t, nur aus Neutronen. Sie bestehen jedoch
haupts\"achlich aus Neutronen. Es gibt in Neutronensternen bei
verschiedenen Massendichten unterschiedliche Grundzust\"ande.
Die Materie in Neutronensternen ist im Gleichgewicht bez\"uglich
der schwachen Wechselwirkung, im sogenannten
$\beta$-Gleichgewicht. Zudem sind Neutronensterne
ladungsneutral, d.h.\ sie tragen keine elektrische Ladung und
keine Farbladung, denn ansonsten w\"urden sie instabil sein.

Neutronensterne besitzen eine Atmosph\"are, die aus Elektronen,
Kernen und Atomen besteht. Falls der Neutronenstern eine
Temperatur $T \gtrsim 100$~eV besitzt, dann ist seine
Oberfl\"achenschicht fl\"ussig. Sie besteht aus Kernen und
Elektronen. Bei einer Massendichte $\rho \sim 10^4$~g/cm$^3$
beginnt die \"au{\ss}ere Kruste von Neutronensternen, die aus
Elektronen und Kernen besteht, wobei letztere in einem
\textbf{b}ody-\textbf{c}entered \textbf{c}ubic (bcc) lattice
angeordnet sind, wodurch ein g\"unstigerer Energiezustand
eingenommen wird. Mit zunehmender Massendichte werden die Kerne
immer neutronenreicher, bis bei $\rho \simeq 4.3 \cdot
10^{11}$~g/cm$^3$ die sogenannte Neutrondripline erreicht wird,
bei der die Neutronen beginnen, sich von den Kernen zu l\"osen.
Das Ende der \"au{\ss}eren Kruste ist damit erreicht. Die innere
Kruste von Neutronensternen besteht aus Kernen, Neutronen und
Elektronen. Bei h\"oheren Massendichten zerfallen die Kerne in
ihre Bestandteile, Protonen und Neutronen. Diese werden
superfluid. Auch Myonen treten in diesen Neutronensternschichten
auf. In welcher Phase sich die Kerne von Neutronensternen
befinden, h\"angt von deren Zentraldichte ab. Neutronensterne
mit geringerer Zentraldichte bestehen in ihrem Inneren nur aus
Protonen, Neutronen, Elektronen und Myonen. Neutronensterne mit
gr\"o{\ss}erer Zentraldichte k\"onnten in ihrem Inneren aus
Pionen und Hyperonen bestehen, w\"ahrend sich bei
Neutronensternen mit riesigen Zentraldichten die Protonen,
Neutronen und Hyperonen in ihre Bestandteile aufl\"osen: Quarks,
die in den Kernen von Neutronensternen wahrscheinlich
farbsupraleitend sind.

Bei gro{\ss}en strange Quarkmassen kann neutrale Quarkmaterie
bestehend aus up und down Quarks im $\beta$-Gleichgewicht einen
eher ungew\"ohnlichen Grundzustand einnehmen, den man gaplose
2SC (g2SC)-Phase nennt. Die Symmetrie im g2SC-Grundzustand ist
dieselbe wie in der regul\"aren 2SC-Phase. Jedoch weist das
Quasiteilchenspektrum in der g2SC-Phase eine gapfreie Mode auf.
Gapfreie Moden entstehen, wenn der Unterschied zwischen den
Fermiimpulsen von sich paarenden Quarks gro{\ss} wird. Deswegen
existiert auch eine gaplose CFL (gCFL)-Phase. Jedoch weisen
sowohl die gaplosen, als auch die regul\"aren farbsupraleitenden
Quarkphasen bei niedrigen Temperaturen chromomagnetische
Instabilit\"aten auf~\cite{Fukushima_unstable}.

Um nun vorhersagen zu k\"onnen, welche Phasen in der Natur
energetisch bevorzugt werden, ist es notwendig, das
Phasendiagramm von neutraler Quarkmaterie zu erstellen, in das
diejenigen Phasen eingetragen werden, die bei vorgegebenen
Temperaturen und quarkchemischen Potentialen vorherrschen. In
diesem Zusammenhang ist es wichtig, \textit{neutrale}
Quarkmaterie zu betrachten, da ansonsten repulsive
Coulombkr\"afte auftreten w\"urden, die die Neutronensterne zur
Explosion zwingen w\"urden. Zudem mu{\ss} Quarkmaterie auch
farbladungsneutral sein, weil farbgeladene Objekte noch niemals
in der Natur gesehen worden sind und auch farbgeladene
Neutronensterne instabil w\"aren. Zudem befindet sich die
Materie in Neutronensternen im $\beta$-Gleichgewicht. Das
Phasendiagramm von neutraler Quarkmaterie, bestehend aus up,
down und strange Quarks im $\beta$-Gleichgewicht, habe ich im
Kapitel~\ref{phase_diagram} erstellt.

Im Abschnitt~\ref{The_phase_diagram_of_massless_quarks} habe ich
masselose neutrale Quarkmaterie bestehend aus up, down und
strange Quarks bei hohen Dichten untersucht. Die Effekte der
strange Quarkmasse $m_s$ sind durch eine Verschiebung im
chemischen Potential der strange Quarks ber\"ucksichtigt worden,
$\mu_s \rightarrow \mu_s - m_s^2 / \left( 2 \mu \right)$.
Diese Verschiebung spiegelt die Reduktion der Fermiimpulse der
strange Quarks wegen der strange Quarkmasse wider. Solch eine
N\"aherung ist sicherlich zuverl\"assig bei kleinen Werten der
strange Quarkmasse. In Ref.~\cite{Shovkovy_Ruester_Rischke}, wo
die strange Quarkmasse richtig miteinbezogen wurde, best\"atige
ich, da{\ss} sie auch qualitativ korrekte Ergebnisse bei
gro{\ss}en strange Quarkmassen liefert. Ich habe eine sehr
mannigfaltige Phasenstruktur durch Variieren der strange
Quarkmasse, des quarkchemischen Potentials und der Temperatur
erhalten.

Bei $T = 0$ gibt es zwei M\"oglichkeiten f\"ur den Grundzustand
von Quarkmaterie: die CFL- und die gCFL-
Phase~\cite{Alford_Kouvaris_Rajagopal1,
Alford_Kouvaris_Rajagopal2}. Ich best\"atige, da{\ss} das
farbchemische Potential $\mu_8$ daf\"ur verantwortlich ist,
da{\ss} die CFL-Phase bei einem kritischen Wert der strange
Quarkmasse in die gCFL-Phase \"ubergeht. Da ich einen
neunparametrischen Ansatz f\"ur die Gapmatrix benutze, bin ich
in der Lage, auch die Effekte der repulsiven Sextettgaps zu
untersuchen. Wie zu erwarten war, modifizieren diese die
Quasiteilchendispersionsrelationen nur geringf\"ugig. Jedoch ist
diese Untersuchung wichtig gewesen, um zu kontrollieren, ob die
Sextettgaps vielleicht ungeahnte Auswirkungen auf den \"Ubergang
von der CFL- zur gCFL-Phase haben.

Im Abschnitt~\ref{The_phase_diagram_of_massless_quarks} habe ich
auch die Temperaturabh\"angigkeit der Gapparameter und der
Quasiteilchenspektra untersucht. Ich habe gezeigt, da{\ss} viele
unterschiedliche Phasen existieren, die in Ref.~\cite{Iida}
vorhergesagt wurden. Jedoch erhalte ich, im Gegensatz
zu den Autoren von Ref.~\cite{Iida, Fukushima}, die die
dSC-Phase erhalten, in der alle down Quarks gepaart sind, die
uSC-Phase, in der alle up Quarks gepaart sind. In
Ref.~\cite{Shovkovy_Ruester_Rischke} habe ich mein Resultat
best\"atigt, da{\ss} die uSC-Phase und nicht die dSC-Phase im
Phasendiagramm von neutraler Quarkmaterie auftritt. Das
Auftreten der dSC-Phase h\"angt stark von den Modellparametern,
genauer gesagt vom Cutoff, ab. W\"ahlt man n\"amlich einen
gr\"o{\ss}eren Cutoff als in Ref.~\cite{Iida, Fukushima}, dann
erh\"alt man in der Tat die dSC-Phase, und die uSC-Phase kommt
im Phasendiagramm von neutraler Quarkmaterie im
$\beta$-Gleichgewicht nicht mehr vor.

Das wichtigste Resultat in
Abschnitt~\ref{The_phase_diagram_of_massless_quarks} ist das
Phasendiagramm von masseloser neutraler Quarkmaterie bestehend
aus up, down und strange Quarks im $\beta$-Gleichgewicht, siehe
Abb.~\ref{phasediagram_massless}. In dieser Abbildung sind alle
symmetriebezogenen Phasen\"uberg\"ange mit durchgezogenen Linien
gekennzeichnet. Die gestrichelten Linien kennzeichnen die
\"Uberg\"ange von regul\"aren zu gaplosen Phasen. Dies sind
jedoch keine echten Phasen\"uberg\"ange, sondern weiche
\"Uberg\"ange. Bei $T = 0$ existiert ein
Isolator-Metall-Phasen\"ubergang zwischen der CFL- und der
gCFL-Phase~\cite{Alford_Kouvaris_Rajagopal1,
Alford_Kouvaris_Rajagopal2}. Bei h\"oheren Temperaturen ist ein
\"ahnlicher Isolator-Metall-Phasen\"ubergang zwischen der CFL-
und der metallischen CFL (mCFL)-Phase vorhanden. Dieser ist in
Abb.~\ref{phasediagram_massless} durch die gepunkteten Linien
gekennzeichnet.

Im Abschnitt~\ref{phase_diagram_self-consistent} habe ich das
Phasendiagramm von neutraler Quarkmaterie mit Hilfe des
Nambu--Jona-Lasinio (NJL)-Modells aus Ref.~\cite{RKH}
untersucht. Wie in der vorangegangenen Untersuchung benutze ich
eine Mean-Field-N\"aherung in der Analyse. Im Gegensatz zu
Abschnitt~\ref{The_phase_diagram_of_massless_quarks} behandele
ich jedoch in Abschnitt~\ref{phase_diagram_self-consistent} die
Quarkmassen selbstkonsistent als dynamisch generierte
Gr\"o{\ss}en. Die wichtigsten Resultate sind in
Abb.~\ref{phasediagram} und Abb.~\ref{phasediagram_strong}
zusammengefa{\ss}t.

Beim Vergleichen des Phasendiagramms f\"ur masselose Quarks,
siehe Abb.~\ref{phasediagram_massless} rechts, mit dem
Phasendiagramm in Abb.~\ref{phasediagram} bemerkt man einige
wichtige Unterschiede. Es f\"allt auf, da{\ss} die
selbstkonsistente Behandlung von Quarkmaterie das Auftreten der
unterschiedlichen Quarkphasen stark
beeinflu{\ss}t~\cite{Buballa_Oertel}. Zudem bestimmen aber auch
die Phasen ihrerseits wiederum die Gr\"o{\ss}e der Quarkmassen.
Dies f\"uhrt sehr oft zu Phasen\"uberg\"angen erster Ordnung.

Viele Unterschiede zu den Resultaten in den
Abschnitten~\ref{The_phase_diagram_of_massless_quarks}
und~\ref{phase_diagram_self-consistent} sind auf das Verwenden
von unterschiedlichen Modellparametern zur\"uckzuf\"uhren. Der
Wert der Diquarkkopplungskonstante $G_D = \frac34 G_S$ in
Abschnitt~\ref{phase_diagram_self-consistent} ist schw\"acher
als jener in
Abschnitt~\ref{The_phase_diagram_of_massless_quarks}, was man
sehr leicht sehen kann, indem man die Werte der Gaps bei $T = 0$
und $\mu = 500$~MeV vergleicht. Das sind $\Delta_0^{(500)}
\approx 140$~MeV im
Abschnitt~\ref{The_phase_diagram_of_massless_quarks} und 
$\Delta_0^{(500)} \approx 76$~MeV in
Abschnitt~\ref{phase_diagram_self-consistent}. Im Vergleich
hierzu ist der Wert der Diquarkkopplungskonstante
in Ref.~\cite{Fukushima} noch viel schw\"acher, wo
$\Delta_0^{(500)} \approx 20$~MeV ist. Im Fall von starker
Kopplung im Abschnitt~\ref{phase_diagram_self-consistent} mit
$G_D = G_S$, wodurch $\Delta_0^{(500)} \approx 120$~MeV ist, ist
$\Delta_0^{(500)}$ immer noch schw\"acher als im
Abschnitt~\ref{The_phase_diagram_of_massless_quarks}. In diesem
Fall unterscheiden sich die Resultate haupts\"achlich dadurch,
da{\ss} die Quarkmassen unterschiedlich behandelt werden.

Wegen der niedrigen Diquarkkopplung geschehen die
Cooperinstabilit\"aten in Abb.~\ref{phasediagram} bei
gr\"o{\ss}eren Werten des quarkchemischen Potentials als in
Abb.~\ref{phasediagram_massless} rechts. Das kann man am besten
am \"Ubergang zur (g)CFL-Phase erkennen. Eine andere Konsequenz
der schw\"acheren Wechselwirkung ist die M\"oglichkeit eines
thermal verursachten Anstiegs des (g)2SC-Cooperpaarens bei
mittleren Werten des quarkchemischen Potentials. Diese Art des
Anstiegs wurde in Ref.~\cite{g2SC_1, g2SC_2} im Detail
untersucht. Mit denselben Argumenten kann man sofort sagen, wie
sich das Phasendiagramm in Abb.~\ref{phasediagram} durch die Zu-
oder Abnahme der Diquarkkopplungskonstante \"andern sollte.

Insbesondere sollte durch Zunahme (Abnahme) der
Diquarkkopplungskonstante die Region der (g)2SC-Phase bei
mittleren Werten des quarkchemischen Potentials expandieren
(schrumpfen). Die Regionen, die durch die anderen (d.h.\ uSC-
und CFL-) Phasen abgedeckt werden, sollten qualitativ dieselbe
Form haben, sich jedoch zu niedrigeren (h\"oheren) Werten des
quarkchemischen Potentials und zu h\"oheren (niedrigeren)
Temperaturwerten verschieben.

Das NJL-Modell, das ich in
Abschnitt~\ref{phase_diagram_self-consistent} verwende, ist
definiert durch einen Parametersatz, dessen Werte so
angepa{\ss}t wurden, da{\ss} sie wichtige QCD-Eigenschaften im
Vakuum reproduzieren. (Das gleiche Modell wurde auch
in Ref.~\cite{SRP} verwendet.) Es ist daher zu erwarten, da{\ss}
es ein gutes effektives Modell der QCD ist, das die beiden
wichtigen Merkmale, chirale und farbsupraleitende Dynamik,
abdeckt. Auch der relativ niedrige Wert des Cutoffs sollte nicht
notwendigerweise als ein schlechtes Merkmal des Modells
angesehen werden. Vielmehr k\"onnte das eine Nachahmung der
nat\"urlichen Eigenschaft der vollst\"andigen Theorie sein, in
der die Kopplungsst\"arke von Wechselwirkungen bei gro{\ss}en
Impulsen gering ist.

In diesem Zusammenhang ist es wichtig darauf hinzuweisen,
da{\ss} die N\"aherung in Ref.~\cite{Fukushima} bez\"uglich des
Cutoffs im NJL-Modell im Vergleich zu meiner sehr
unterschiedlich ist. Dort wird n\"amlich behauptet, da{\ss} ein
gro{\ss}er Wert f\"ur den Cutoff vorteilhaft sei, um Resultate
zu erhalten, die unempfindlich auf eine spezielle Wahl des
Cutoffs sind. Ich finde jedoch kein physikalisches Argument, das
diese Voraussetzung unterst\"utzt. Vielmehr bestehe ich darauf,
ein effektives Modell zu verwenden, das die QCD-Eigenschaften
bei verschwindendem quarkchemischen Potential gut beschreibt.
Nat\"urlich will ich damit nicht behaupten, da{\ss} eine
Extrapolation des Modells zu gro{\ss}en Dichten damit zu
rechtfertigen ist. Jedoch scheint mir das momentan die beste
L\"osung zu sein.

Die Resultate aus Abschnitt~\ref{phase_diagram_self-consistent}
k\"onnten wichtig f\"ur das physikalische Verst\"andnis von
(hybriden) Neutronensternen mit Quarkkern sein, in denen die
Deleptonisierung stattgefunden hat. Um ein Phasendiagramm zu
erhalten, das auf Protoneutronensterne angewandt werden
kann, mu{\ss} man den Effekt von Neutrinotrapping
ber\"ucksichtigen.

Im Abschnitt~\ref{phase_diagram_neutrino_trapping} habe ich den
Effekt von Neutrinotrapping auf das Phasendiagramm von neutraler
Quarkmaterie bestehend aus up, down und strange Quarks im
$\beta$-Gleichgewicht mit Hilfe des NJL-Modells aus
Ref.~\cite{RKH} untersucht. Die Resultate habe ich in der
Mean-Field-N\"aherung erhalten. Die Quarkmassen sind als
dynamisch generierte Gr\"o{\ss}en behandelt worden. Die
Gesamtstruktur der Phasen wird als dreidimensionales Diagramm im
Raum von Temperatur $T$, quarkchemischem Potential $\mu$ und dem
chemischen Potential der Leptonen $\mu_{L_e}$ in
Abb.~\ref{phase3d} dargestellt. Dieses Phasendiagramm wird zudem
anhand von s\"amtlichen zweidimensionalen Teildiagrammen
desselben detailliert diskutiert, siehe
Abb.~\ref{phase200},~\ref{phase400},~\ref{phase_mu400}
und~\ref{phase_mu500}.

Sowohl anhand von einfachen modellunabh\"angigen Argumenten, als
auch anhand von detaillierten numerischen Berechnungen mit Hilfe
des NJL-Modells habe ich herausgefunden, da{\ss}
Neutrinotrapping das Cooper-Paaren in der 2SC-Phase beg\"unstigt
und die CFL-Phase unterdr\"uckt. Haupts\"achlich kommt das
durch das Erf\"ullen der elektrischen Neutralit\"atsbedingung
zustande. In Quarkmaterie, die nur aus up und down Quarks im
$\beta$-Gleichgewicht besteht, hilft das chemische Potential der
Leptonen $\mu_{L_e}$, zus\"atzliche Elektronen zur Verf\"ugung
zu stellen, ohne einen gro{\ss}en Unterschied zwischen den
Fermiimpulsen zu verursachen. Durch Verringern der
Fermiimpulsunterschiede wird das Cooperpaaren nat\"urlich
st\"arker. Dies ist eine v\"ollig andere Situation als in der
CFL-Phase, in der die Quarkmaterie in Abwesenheit von Elektronen
neutral ist. Zus\"atzliche Elektronen w\"urden, wegen des
gro{\ss}en chemischen Potentials der Leptonen, die CFL-Phase
belasten.

Wendet man dieses Resultat auf Protoneutronensterne an, die
farbsupraleitende Quarkkerne besitzen, dann bedeutet das,
da{\ss} die CFL-Phase wohl kaum in Protoneutronensternen
auftritt bevor die Deleptonisierung komplett ist. Daher ist die
2SC-Phase der wahrscheinlichste Grundzustand in
Protoneutronensternen mit farbsupraleitenden Quarkkernen. Zudem
habe ich festgestellt, da{\ss} Protoneutronensterne kaum strange
Quarks enthalten.

Die Autoren von Ref.~\cite{Alford_Rajagopal} behaupten, da{\ss}
die 2SC-Phase in kompakten Sternen nicht vorkommt und
untermauern ihre Behauptung mit Hilfe einer einfachen
modellunabh\"angigen Berechnung. Die Phasendiagramme, die ich
mit Hilfe des genaueren NJL-Modells erstellt habe und in meiner
Dissertation zeige, sind der beste Beweis daf\"ur, da{\ss} die
2SC-Phase wirklich in kompakten Sternen vorkommen kann. Falls
kompakte Sterne farbsupraleitende Kerne besitzen, dann k\"ame
die 2SC-Phase auf jeden Fall in Protoneutronensternen und in
kalten Neutronensternen mit gro{\ss}er Diquarkkopplungskonstante
vor. F\"ur kleine oder mittlere Werte der
Diquarkkopplungskonstante und bei verschwindender Temperatur,
f\"ur die die einfache modellunabh\"angige Berechnung
in Ref.~\cite{Alford_Rajagopal} g\"ultig ist, scheint es jedoch
richtig zu sein, da{\ss} die 2SC-Phase nicht auftritt. Wie dem
auch sei, man sollte mit der Aussage, da{\ss} eine Phase nicht
auftritt, sehr vorsichtig sein.

Nachdem die Deleptonisierung stattgefunden hat, ist es
m\"oglich, da{\ss} der Grundzustand im Inneren von
farbsupraleitenden Neutronensternen die CFL-Phase ist. Diese
Phase besitzt eine gro{\ss}e Anzahl von strange Quarks. Deswegen
sollte eine gro{\ss}e Produktion von strange Quarks sofort nach
der Deleptonisierung einsetzen. Dieser Vorgang k\"onnte
beobachtbar sein.

Im Kapitel~\ref{Conclusions} fasse ich meine Resultate zusammen
und diskutiere sie. Zudem zeige ich offengebliebene Fragen auf
und gebe einen Ausblick auf m\"ogliche zuk\"unftige
Untersuchungen.

Im Anhang befinden sich n\"utzliche Formeln und Definitionen,
die in meiner Dissertation verwendet werden.
\selectlanguage{american}

\selectlanguage{german}
\chapter*{Lebenslauf}
\addcontentsline{toc}{chapter}{Lebenslauf}
\makebox[\textwidth][r]
{\raisebox{-1.3cm}[0cm][0cm]{\includegraphics{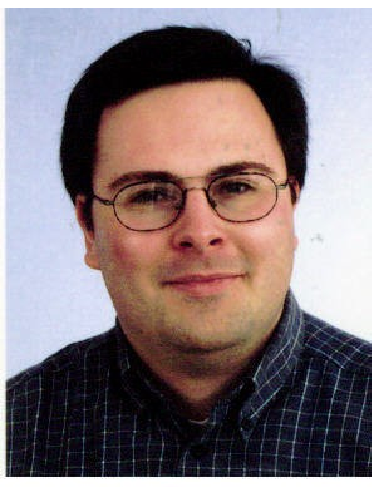}}}
\begin{tabular}{ll}
Name: & Stefan Bernhard R\"uster \\
Geburtsdatum: & 28.\ Juni 1978 \\
Geburtsort: & Alzenau-Wasserlos \\
Mutter: & Maria-Rita (Sekret\"arin) \\
Vater: & Karl Georg J\"urgen (technischer Angestellter) \\
Wohnort: & Friedhofstra{\ss}e 12, 63579 Freigericht-Horbach \\
Grundschule: & Schule der bunten Raben, Freigericht-Horbach
(1984--1988) \\
Gymnasium: & Kopernikusschule Freigericht (1988--1994) \\
Oberstufe: & Kopernikusschule Freigericht (1994--1997) \\
Leistungskurse: & Mathematik und Physik \\
Jugend Forscht: & Zweiter Platz in Hessen und Sonderpreis
(1996) \\
Abitur: & 27.\ Juni 1997 \\
Immatrikulation: & 4.\ August 1997 \\
Universit\"at: & J.\ W.\ Goethe~-~Universit\"at, Frankfurt am
Main \\
Studienfach: & Physik (Diplom) \\
Vordiplom: & 22.\ September 1999 \\
Institut: & Institut f\"ur Theoretische Physik \\
Arbeitsgruppe: & Farbsupraleitung \\ 
Diplomarbeitsthema: & Farbsupraleitung in
Quarksternen \\
Betreuer: & Prof.\ Dr.\ Rischke \\
Diplom: & 12.\ Februar 2003 \\
Diplomgesamtnote: & Sehr gut \\
Wissenschaftlicher Mitarbeiter: & 1.\ April 2003 \\
Arbeitsgruppen: & Farbsupraleitung und
Astronomie/Astrophysik \\
Doktorand: & 29.\ April 2003 \\
Promotionsthema: & Das Phasendiagramm von neutraler Quarkmaterie \\
Betreuer: & Prof.\ Dr.\ Rischke \\
Promotion: & 14.\ Dezember 2006 \\
Promotionsgesamtnote: & Sehr gut \\
& \\
& \\
\textbf{Akademische Lehrer:} & \\
& \\
Prof.\ Dr.\ Rischke & HD PD Dr.\ Schaffner-Bielich \\
Prof.\ Dr.\ A{\ss}mus & Prof.\ Dr.\ Becker \\
Prof.\ Dr.\ Constantinescu & Prof.\ Dr.\ Elze \\
Prof.\ Dr.\ Fried & Prof.\ Dr.\ Jelitto \\
Prof.\ Dr.\ Kegel & Prof.\ Dr.\ Klein \\
Prof.\ Dr.\ Lynen & Prof.\ Dr.\ Maruhn \\
Prof.\ Dr.\ Mohler & Prof.\ Dr.\ Roskos \\
Prof.\ Dr.\ Schaarschmidt & Prof.\ Dr.\ Schmidt-B\"ocking \\
Prof.\ Dr.\ Stock & Prof.\ Dr.\ Tetzlaff
\end{tabular}
\end{document}